\documentclass[12pt]{elsarticle}

\usepackage{amssymb}
\usepackage{amsmath}
\usepackage{subfigure}

\usepackage{dcolumn}
\usepackage{makecell}
\usepackage{lipsum}
\usepackage{bm}
\usepackage{adjustbox}
\usepackage{booktabs,makecell}
\usepackage{nomencl}
\usepackage[nameinlink,capitalize]{cleveref}
\usepackage{booktabs}
\usepackage{array}
\biboptions{sort&compress}

\usepackage{graphicx}
\usepackage{epstopdf}
\usepackage{dcolumn}
\usepackage{bm}

\usepackage[utf8]{inputenc}
\usepackage[T1]{fontenc}
\usepackage{mathptmx}
\usepackage{etoolbox}

\makeatletter
\def\ps@pprintTitle{%
 \let\@oddhead\@empty
 \let\@evenhead\@empty
 \def\@oddfoot{}%
 \let\@evenfoot\@oddfoot}
\makeatother

\usepackage[symbol]{footmisc}

\usepackage{gensymb} 
\usepackage[numbers]{natbib}

\begin{document}

\begin{frontmatter}



\title{Impact of an Arc-shaped Control Plate on Flow and Heat Transfer around a Isothermally Heated Rotating Circular Cylinder}

\def\correspondingauthor{\footnote[1]{Corresponding author: rajendra@iitmandi.ac.in}}
\author[inst1]{Amarjit Haty}


\author[inst1]{Rajendra K. Ray\correspondingauthor{}}

\affiliation[inst1]{organization={School of Mathematical and Statistical Sciences, Indian Institute of Technology Mandi},
            city={Mandi},
            postcode={175005}, 
            state={Himachal Pradesh},
            country={India}}

\begin{abstract}
The main objective of this paper is to study the flow characteristics of a rotating, isothermally heated circular cylinder with a vertical arc-shaped control plate placed downstream. Stream function-Vorticity ($\psi-\omega$) formulation of two dimensional (2-D) Navier-Stokes (N-S) equations is considered as the governing equation and the simulations are performed for different distances of the control plate ($0.5$, $1$, $2$, $3$), rotational rates ($0.5$, $1$, $2.07$, $3.25$) at Prandtl number $0.7$ and Reynolds number $150$. The governing equations are discretized using the Higher Order Compact (HOC) scheme and the system of algebraic equations, arising from HOC discretization, is solved using the Bi-Conjugate Gradient Stabilized approach. Present computed results show that the vortex shedding plane is shifted upward from the centerline of the flow domain by the cylinder's rotational motion. The structure of the wake varies based on the plate's position. The size of vortices is greatly reduced when the control plate is set at $d/R_0=3$ and the rotational rate is very high. At greater rotational rates, the impact of varied positions of the arc-shaped control plate is very significant. The rotation of the cylinder and the location of the plate can be used to lower or enhance the values of drag and lift coefficients as well as the heat transfer from the surface of the cylinder. The maximum value of the drag coefficient, which is about $3$, is achieved for $d/R_0=2$ and $\alpha=3.25$.
\end{abstract}




\end{frontmatter}


\section{Introduction\protect} \label{Introduction}
Active control of flow past a rotating circular cylinder has always been an interesting topic in fluid dynamics. The wake behaviour for flow past a rotating cylinder is more complicated than for flow past a stationary cylinder because the rotation of the cylinder separates the shear layer and modifies the boundary layer. In 1928, Bickley \cite{bickley1928influence} was among the first to attempt analytical study of the viscous flow over a rotating cylinder. He considered the potential flow created by a vortex in the vicinity of a cylinder. The wake structure in flow past a cylinder is complicated due to interactions between a boundary layer, a separating free shear layer, and a wake. It has huge significance in engineering as the alternating shedding pattern of the vortices in the wake causes considerable fluctuating pressure forces in a direction transverse to the fluid flow, which can produce structural vibrations, acoustic noise, or resonance, and in certain situations, structural collapse. In 1966, Gerrard \cite{gerrard1966mechanics} experimentally studied the flow past bluff bodies along with flow past circular cylinder with splitter plates for high Reynolds numbers. He found that the shear layer was drawn by the vortex formation from the opposite side of the wake across the center line of the wake, cutting off the vorticity supply to the expanding vortex. He found that the width of the gap between the cylinder and a splitter plate parallel to the flow, is the only relevant parameter than the position of the trailing edge of the plate. He studied the effect of a plate normal to the flow and found that the length of the effective vortex formation area equalled the distance of the plate from the domain boundary. He observed a substantial cross-flow velocity created near the plate when a vortex grew close behind it, facilitating the shedding process and increasing the frequency. Pralits et al. \cite{pralits2010instability} numerically studied the flow past rotary cylinder and found that the increased rotational speed caused two distinct instability in the flow. Kang et al. \cite{kang1999laminar} found that the vortex shedding was stopped completely when the cylinder rotation rate was set at twice the velocity of free stream fluid. Diaz et al. \cite{diaz1983vortex} have experimentally studied the flow past rotary cylinder for Reynolds number $9000$. They saw a decrease in periodic vortex activity and a rise in random modulation of the shedding process, which he attributed to the relocation of the stagnation point and the thickening of the spinning fluid layer near the cylinder surface. They discovered that when the rotating speed equals the free-stream speed, a regular periodic vortex shedding occurs, and that the periodic vortex shedding is suppressed at large velocity ratios. For velocity ratios equal to or greater than $1.5$, they concluded that rotation considerably alters the traditional Karman vortex shedding. Similar findings were produced by Massons et al. \cite{massons1989image} for flow past rotating cylinder. Stojkovic et al. \cite{stojkovic2002effect} studied the flow at greater rotation rates and discovered a second shedding mode in a limited interval $[4.85,\ 5.15]$ of rotation rate where the shedding frequency was substantially lower than that of the traditional Von-Karman vortex shedding. At a high Reynolds number ($Re = 10^5$), Roshko \cite{roshko1955wake} investigated the impact of a splitter plate positioned downstream of a bluff body and parallel to the free stream. By bringing the plate closer to the cylinder, he observed that the shedding frequency and base suction were reduced. Bearman \cite{bearman1965investigation} found that the separating shear flow on the top of the surface is pushed to rejoin if the circular cylinder with an end plate downstream is spun at a constant pace. As a result, the effects and vibrations caused by boundary-layer development are diminished, and the vortex formation is suppressed. Apelt et al. \cite{Apelt1973} used a horizontal splitter plate with varied lengths to diameter ratios less than $2$ to investigate the flow past a circular cylinder for $10^4< Re<5\times10^4$. The splitter plate considerably reduces drag by stabilising separation points, lowers the Strouhal number, and increases base pressure by roughly $50\%$, according to their research. They also discovered that when using a splitter plate instead of a cylinder without one, the wake pattern narrows. Kwon and Choi \cite{kwon1996} indicated that there is a critical length of splitter plate that causes vortex shedding to totally disappear, and that this critical length is proportional to the Reynolds number. They also discovered that the Strouhal number rises as the plate's length increases until it equals the cylinder's diameter. Bao and Tao \cite{bao2013passive} analyzed the flow past a circular cylinder with twin parallel plates attached and discovered that optimal positioning can outperform the standard splitter plate. More studies with control plate can be found in \cite{akilli2005suppression,lu2014numerical,liu2016experimental,bouzari2017unsteady}.\\

Along with studying the wake structure and pressure forces, force convective heat transfer from rotating cylinders has been widely investigated by many researchers for its many real-life applications and scientific interests. Drying cylindrical items \cite{kaya2007numerical}; cylindrical cooling devices in the plastics and glass industries; drying and coating of papers using a hot spinning cylinder; chemical and food processing industries; textile and paper manufacturing, and so on are some examples of real-world uses. In an experiment, Anderson and Saunders \cite{anderson1953convection} explored heat convection in a confined room filled with air using an isothermally heated rotating circular cylinder. Temperatures were elevated to $140$ degrees Fahrenheit above the ambient temperature while air pressure was maintained at $4$ $atm$. The experiment used three distinct cylinders, each with varying diameters ($1$, $1.8$, and $3.9$ inches) but the same length ($2$ feet). They determined that heat exchange is nearly steady when rotational speed is between $0$ to a crucial value of $0.9$, and past that point, heat exchange increases in proportion to the rotational speed's $2/3$ power. Badr and Dennis \cite{badr1985laminar} conducted a numerical research on force convective heat transfer from an unconfined rotary cylinder, concluding that increasing rotational speed reduces overall rate of heat transfer because the cylinder is isolated from the stream by the spinning fluid layer. Mohanty et al. \cite{mohanty1995heat} performed experimental study on heat transfer from rotating cylinder for high Reynolds numbers. They discovered that rotational motion increases average heat transmission by roughly $30\%$ when compared to a fixed cylinder with a fixed Reynolds number. They also discovered that as compared to stationary cylinders, rotational motion caused a lower heat transfer rate at the front stagnation point. An analytical study was attempted by Kendoush \cite{kendoush1996approximate} and a formula, $Nu=0.6366(RePr)^{1/2}$, was proposed to compute the local Nusselt number ($Nu$) for low Prandtl numbers ($Pr$), where $Re$ denotes the Reynolds number. With the help of the finite volume technique, Paramane and Sharma \cite{paramane2009numerical} studied the heat transfer and fluid flow across a rotating cylinder for Prandtl number of $0.7$, low Reynolds numbers ranging from $20$ to $160$, and rotary speeds of $0\leq\alpha\leq 6$. They discovered that when rotary speeds rise, the average Nusselt number falls while the Reynolds number rises. It was concluded that the rotation could be employed to reduce drag and suppress heat transmission from the cylinder. Sufyan et al. \cite{sufyan2015free} discovered that low and medium rotary speeds immediately reduce heat transmission, but that at higher rotational rates, the increased size of the enclosing vortex causes even more heat transfer reduction. A few more studies on this subject can be found on \cite{jalil2007heat,paramane2010heat,sharma2012heat,sufyan2014heat}.\\

After an extensive literature survey, it is found that many researchers worked on heat transfer and flow across a rotating circular cylinder. There are numerous works on the flow across a circular cylinder with splitter plates and attached fins. Effect of curved fins and plates are studied by few researchers for missiles \cite{eastman1985aerodynamics} and formula$-1$ cars \cite{martins2021influence} and these are being used in real life. There are some researchers who tried to study the wake structure and base pressure after applying the rotation to the cylinder with attached splitter plates or fins, but the effect of both rotation and the presence of control plates on the process of heat transfer is not tested. Considering the importance, the current investigation is centred on the impact of a control plate on forced convective heat transfer and flow across a rotating circular cylinder. It can be useful in electronic equipment cooling and processing industries. We have taken into account an arc-shaped plate with a vertical orientation since we are considering a polar coordinate system with non-uniform grids. For this investigation, the Reynolds number is fixed at $150$ and the Prandtl number is fixed at $0.7$. The plate distance to cylinder radius ratio varies between $0.5$ and $3$, while the rotational rates range from $0.5$ to $3.25$. The two-dimensional unsteady Navier-Stokes equations and energy equation are first non-dimensionalized and then discretized by using a Higher Order Compact (HOC) scheme \cite{kalita2009transformation,ray2011transformation} based on non-uniform polar grids. Temporal accuracy of $2^{nd}$ order and spatial accuracy of atleast $3^{rd}$ order are obtained through the application of the finite difference scheme. To obtain a solution from a discretized system, the Bi-conjugate Gradient Stabilized method approach is employed.\\

The paper is arranged as follows: in Section 2, we discuss the governing equations and initial and boundary conditions related to the current problem; in Section 3, the numerical scheme is described as well as the independence tests and validity of the numerical scheme are produced; results are discussed in Section 4; and finally, we conclude our remarks in Section 5.
 
\begin{table*}
\setlength{\tabcolsep}{6pt} 
\begin{tabular}{|cc|}
\hline
\textbf{Nomenclature}&\\
&\\
$Re$ & \thead{Reynolds number ($=2R_0U_{\infty}/\nu$)}\\

$Pr$ & \thead{Prandtl number ($=\nu/\beta)$}\\

$R_0$& \thead{Radius of the circular cylinder}\\

$R_{\infty}$& \thead{Radius of the far field boundary}\\

$d$& \thead{Dimensional distance of the control plate \\from the cylinder surface}\\

$U_{\infty}$ & \thead{The free-stream fluid's velocity}\\

$T_{\infty}$& \thead{The free-stream fluid's temperature}\\

$\hat{t}$, $t$ & \thead{Time in dimensional and nondimensional form}\\

$T_s$ & \thead{Surface temperature of the cylinder in dimensional form}\\

$\hat{\alpha}$, $\alpha$ & \thead{Rotational velocity in dimensional and \\nondimensional form ($\alpha=\hat{\alpha}R_0/U_\infty)$}\\

$d$ & \thead{Distance of control plate from the surface of the cylinder}\\

$Nu$, $\overline{Nu}$, $\overline{Nu}_t$ & \thead{Nusselt number (local, average, and time-averaged total)}\\


$h$, $h_{avg}$ & \thead{Coefficients of heat transfer (local and average)}\\

$\nu$ & \thead{The fluid's kinematic viscosity}\\

$K$ & \thead{The fluid's thermal conductivity}\\

$\beta$ & \thead{The fluid's thermal diffusivity}\\

$Q''$ & \thead{Radial heat flux on the surface (Local)}\\

$\hat{\psi}$, $\psi$ & \thead{Stream function in dimensional and nondimensional form}\\

$\hat{\omega}$, $\omega$ & \thead{Vorticity in dimensional and nondimensional form}\\

$T$, $\phi$ & \thead{Temperature in dimensional and nondimensional form}\\

$\hat{u}$, $u$ & \thead{Radial velocity in dimensional and nondimensional form}\\

$\hat{v}$, $v$ & \thead{Tangential velocity in dimensional and nondimensional form}\\

$\hat{r}$, $r$ & \thead{Radius in dimensional and nondimensional form}\\

\hline

\end{tabular}
\end{table*}

\section{The governing equations and the problem\protect}\label{The governing equations and the problem}

\begin{figure}[!t]
       \centering
 \includegraphics[width=0.9\textwidth,trim={0.5cm 0.5cm 0.5cm 0.5cm},clip]{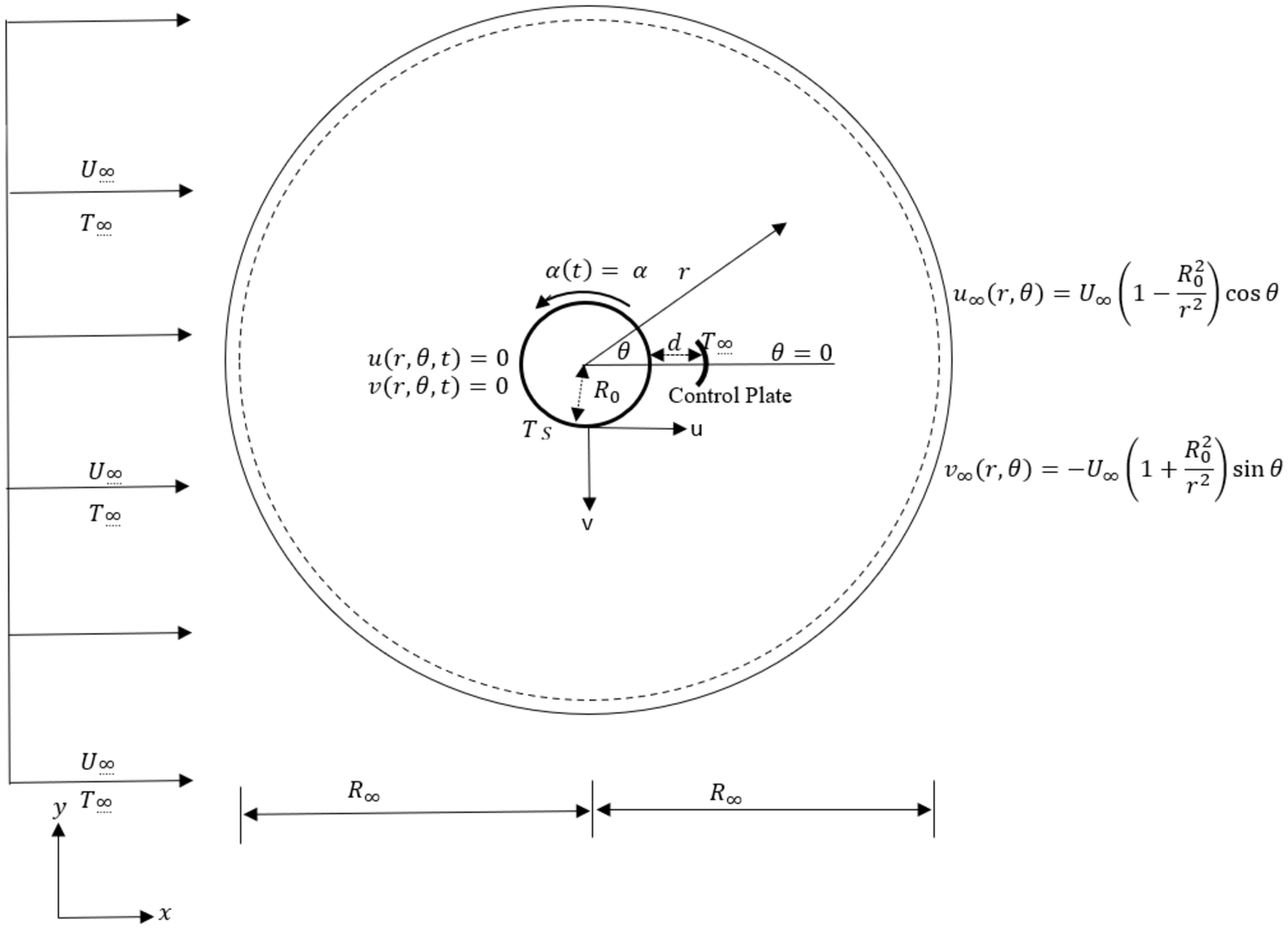}
       \caption{The schematic illustration of the current problem.}
       \label{fig:diagram}
    \end{figure}

The considered system is represented in \cref{fig:diagram} as a two-dimensional unsteady, incompressible, laminar, and viscous flow of a Newtonian fluid over an isothermally heated circular cylinder of radius $R_0$. At $\hat{t}=0$, the cylinder acquires the surface temperature $T_s$ impulsively. The following formulas are used to transform dimensional parameters to dimensionless form: $t=\frac{\hat{t}U_\infty}{R_0}$, $r=\frac{\hat{r}}{R_0}$, $u=\frac{\hat{u}}{U_\infty}$, $v=\frac{\hat{v}}{U_\infty}$, $\psi=\frac{\hat{\psi} U_\infty}{R_0}$, $\omega=\frac{\hat{\omega}R_0}{U_\infty}$, $\phi=\frac{(T-T_\infty)}{(T_s-T_\infty)}$. The control plate has unit arc length and a constant thickness roughly equal to $0.18$ times the cylinder radius and is situated at a distance $d$ from the cylinder surface. On the surface of the control plate, impermeability and no-slip boundary conditions are considered. The control plate is kept constant at the same temperature as the free stream fluid.\\

The nondimensional stream-function-vorticity formulation of the 2-D Navier-Stokes equations and energy equation in polar coordinates $(r,\theta)$ are given as,
\begin{equation}\label{eq:1}
\dfrac{\partial^2 \omega}{\partial r^2}+\frac{1}{r}\dfrac{\partial \omega}{\partial r}+\frac{1}{r^2}\dfrac{\partial^2 \omega}{\partial \theta^2}=\frac{Re}{2}\left(u\dfrac{\partial \omega}{\partial r}+\frac{v}{r}\dfrac{\partial \omega}{\partial \theta}+\dfrac{\partial \omega}{\partial t} \right)
\end{equation}
\begin{equation}\label{eq:2}
\dfrac{\partial^2 \psi}{\partial r^2}+\frac{1}{r}\dfrac{\partial \psi}{\partial r}+\frac{1}{r^2}\dfrac{\partial^2 \psi}{\partial \theta^2}=-\omega
\end{equation}
\begin{equation}\label{eq:3}
\dfrac{\partial^2 \phi}{\partial r^2}+\frac{1}{r}\dfrac{\partial \phi}{\partial r}+\frac{1}{r^2}\dfrac{\partial^2 \phi}{\partial \theta^2}=\frac{RePr}{2}\left(u\dfrac{\partial \phi}{\partial r}+\frac{v}{r}\dfrac{\partial \phi}{\partial \theta}+\dfrac{\partial \phi}{\partial t} \right)
\end{equation}
The velocities, $v$ and $u$ can be expressed as 
\begin{equation}\label{eq:4}
v=-\dfrac{\partial \psi}{\partial r} \hspace{0.5cm} \text{and} \hspace{0.5cm} u=\frac{1}{r}\dfrac{\partial \psi}{\partial \theta}
\end{equation}
$\omega$ can be written as
\begin{equation}\label{eq:5}
\omega=\frac{1}{r}\left[\dfrac{\partial}{\partial r}(vr)-\dfrac{\partial u}{\partial \theta}\right]
\end{equation}

The boundary conditions on the cylinder's surface include impermeability, no-slip, and constant temperature, i.e.
\begin{equation}\label{eq:6}
\psi=0,\hspace{0.5cm}\frac{\partial \psi}{\partial r}=-\alpha\hspace{0.5cm}\text{and}\hspace{0.5cm}\phi=1.0\hspace{0.5cm}\text{when}\hspace{0.5cm}r=1
\end{equation}

The condition of surface vorticity is provided by
\begin{equation}\label{eq:7}
\omega=-\dfrac{\partial^2 \psi}{\partial r^2} \hspace{0.5cm} \text{when} \hspace{0.5cm}r=1
\end{equation}

In the distant field, $R_\infty$, the vorticity's resulting decay and the free-stream condition are taken to constitute the boundary conditions, i.e.
\begin{equation}\label{eq:8}
\begin{split}
\psi\,\to\, \left(r-\frac{1}{r}\right)sin\ \theta,\ \ \ \dfrac{\partial \psi}{\partial r}\,\to\,\left(1+\frac{1}{r^2}\right)sin\ \theta\\ \text{and}\ \ \phi\,\to\, 0 \ \ \ \text{as}\ \ \ r\,\to\,\frac{R_{\infty}}{R_0}
\end{split}
\end{equation}
\begin{equation}\label{eq:9}
\omega\,\to\,0\ \ \ \text{as}\ \ \ r\,\to\,\frac{R_{\infty}}{R_0}
\end{equation}

The criteria \cref{eq:6,eq:7,eq:8,eq:9} must be followed by all the parameters with $0\leq\theta\leq2\pi$ for all $\theta$. In addition, all of the parameters are functions of $\theta$ with a period of $2\pi$. The initial conditions for the stream function are given by \cref{eq:8,eq:9}. The vorticity in the distant field is initially assumed to be zero. \cref{eq:4,eq:8} provide the initial requirements for the velocities as follows:
\begin{equation}\label{eq:10}
u=\left(1-\frac{1}{r^2}\right)cos\ \theta\ \ \ \text{and}\ \ \ v=-\left(1+\frac{1}{r^2}\right)sin\ \theta
\end{equation}

\section{Numerical Scheme\protect}\label{Numerical Scheme}

Using a temporally second order accurate and spatially atleast third order accurate higher order compact (HOC) finite difference technique \cite{kumar2015numerical,ray2017numerical,kumar2019structural,kumar2020structural}, the governing equations of motion and the energy equation are discretized on non-uniform polar grids in the circular region $([R_0,R_\infty]\times [0,2\pi])$ with grid points $(r_i,\theta_j)$. The non-uniform grid concentrated around the cylinder is generated using the stretching function $r_i=exp\left(\dfrac{\lambda \pi i}{i_{max}}\right),\ \ 0\leq i\leq i_{max}$. The function $\theta_j$ is given by, $\theta_j=\dfrac{2\pi j}{j_{max}}$, $0\leq j\leq j_{max}$. The discretized equations can be written as \cite{kalita2009transformation,mittal2017numerical,mittal2018numerical}:
\begin{equation}
\begin{split}
[X1_{ij}\delta^2_r+X2_{ij}\delta^2_\theta+X3_{ij}\delta_r+X4_{ij}\delta_r\delta_\theta+X5_{ij}\delta_r\delta^2_\theta\\+X6_{ij}\delta^2_r\delta_\theta+X7_{ij}\delta^2_r\delta^2_\theta]\psi_{ij}=G_{ij}
\end{split}
\end{equation}
\begin{equation}
\begin{split}
[Y11_{ij}\delta^2_r+Y12_{ij}\delta^2_\theta+Y13_{ij}\delta_r+Y14_{ij}\delta_\theta+Y15_{ij}\delta_r\delta_\theta\\+Y16_{ij}\delta_r\delta^2_\theta+Y17_{ij}\delta^2_r\delta_\theta+Y18_{ij}\delta^2_r\delta^2_\theta]\omega^{n+1}_{ij}\\=[Y21_{ij}\delta^2_r+Y22_{ij}\delta^2_\theta+Y23_{ij}\delta_r+Y24_{ij}\delta_\theta+Y25_{ij}\delta_r\delta_\theta\\+Y26_{ij}\delta_r\delta^2_\theta+Y27_{ij}\delta^2_r\delta_\theta+Y28_{ij}\delta^2_r\delta^2_\theta]\omega^{n}_{ij}
\end{split}
\end{equation}
and
\begin{equation}
\begin{split}
[Z11_{ij}\delta^2_r+Z12_{ij}\delta^2_\theta+Z13_{ij}\delta_r+Z14_{ij}\delta_\theta+Z15_{ij}\delta_r\delta_\theta\\+Z16_{ij}\delta_r\delta^2_\theta+Z17_{ij}\delta^2_r\delta_\theta+Z18_{ij}\delta^2_r\delta^2_\theta]\phi^{n+1}_{ij}\\=[Z21_{ij}\delta^2_r+Z22_{ij}\delta^2_\theta+Z23_{ij}\delta_r+Z24_{ij}\delta_\theta+Z25_{ij}\delta_r\delta_\theta\\+Z26_{ij}\delta_r\delta^2_\theta+Z27_{ij}\delta^2_r\delta_\theta+Z28_{ij}\delta^2_r\delta^2_\theta]\phi^{n}_{ij}
\end{split}
\end{equation}
The coefficients $X1_{ij}$, $X2_{ij}$,$...$, $X7_{ij}$; $G_{ij}$;  $Y11_{ij}$, $Y12_{ij}$,$...$, $Y18_{ij}$; $Y21_{ij}$, $Y22_{ij}$,$...$, $Y28_{ij}$; $Z11_{ij}$, $Z12_{ij}$,$...$, $Z18_{ij}$ and $Z21_{ij}$, $Z22_{ij}$,$...$, $Z28_{ij}$  are the functions of the parameters $r$ and $\theta$. \cite{kalita2009transformation,mittal2017numerical,mittal2018numerical} provide the expressions for the non-uniform central difference operators $\delta_{\theta}$, $\delta^2_{\theta}$, $\delta_r$ and $\delta^2_r$, as well as the notations $\theta_f$, $\theta_b$, $r_f$, $r_b$ and the coefficients. The Bi-conjugate Gradient Stabilized approach is employed in order to solve the discretized problem.

\subsection{\textbf{Drag and lift coefficients}}

The forces acting on a circular cylinder submerged in fluids for uniform flow are generally caused by surface friction and surface pressure distribution. The expressions for drag ($C_D$) and lift ($C_L$) coefficients are adopted from \cite{kalita2009transformation, mittal2017numerical}. The expressions are as follows,
\begin{equation}
C_D=\frac{1}{Re}\int^{2\pi}_0\left[ \left(\frac{\partial \omega}{\partial r}\right)_{R_0}-\omega_{R_0}\right]\cos{\theta}d\theta
\end{equation}

\begin{equation}
C_L=\frac{1}{Re}\int^{2\pi}_0\left[ \left(\frac{\partial \omega}{\partial r}\right)_{R_0}-\omega_{R_0}\right]\sin{\theta}d\theta
\end{equation}

The integral values are calculated using Simpson's $1/3$ method. The time-averaged drag, $\overline{C}_D$ is expressed as,
\begin{equation}
\overline{C}_D=\frac{1}{t_1-t_2}\int_{t_1}^{t_2}{C}_Ddt
\end{equation}
When the flow achieves a periodic mode and executes numerous cycles, the time span between $t_1$ and $t_2$ is selected.

\subsection{\textbf{The heat transfer parameters}}

Initially, heat conduction happens from the cylinder surface to the adjacent fluid, and subsequently it convects away with the flow. The heat conduction path follows the radius of the cylinder surface. The dimensionless local heat flux in the radial direction is the local Nusselt number, $Nu$, defined by,
\begin{equation}
Nu=\frac{2hR_0}{k}=\frac{Q''(2R_0)}{k(T_s-T_\infty)}
\end{equation}
where $h$ represents the local heat transfer coefficient, $k$ represents the thermal conductivity of the fluid, and $Q''$ represents the surface local radial heat flux. $Q''$ is expressed as, $Q''=-k\frac{\partial T}{\partial r}|_{r=R_0}$.
The average Nusselt number, denoted by $\overline{Nu}$, used to represent the dimensionless heat transfer from the cylinder's surface, is expressed as
\begin{equation}
\overline{Nu}=\frac{2h_{avg}R_0}{k}=\frac{1}{2\pi}\int_0^{2\pi}Nud\theta
\end{equation}
The average heat transfer coefficient ($h_{avg}$) is expressed as $h_{avg}=$ $\frac{1}{2\pi}\int_0^{2\pi}hd\theta$. $\overline{Nu}_t$, the time-averaged total Nusselt number is given as,
\begin{equation}
\overline{Nu}_t=\frac{1}{t_1-t_2}\int_{t_1}^{t_2}\overline{Nu}dt
\end{equation}
When the flow achieves a periodic mode and executes numerous cycles, the time span between $t_1$ and $t_2$ is selected.

\subsection{\textbf{Validation}}

\begin{figure*}[!t]
\centering
\subfigure[]{
\includegraphics[width=0.45\textwidth,trim={0.1cm 0.1cm 0.1cm 0.1cm},clip]{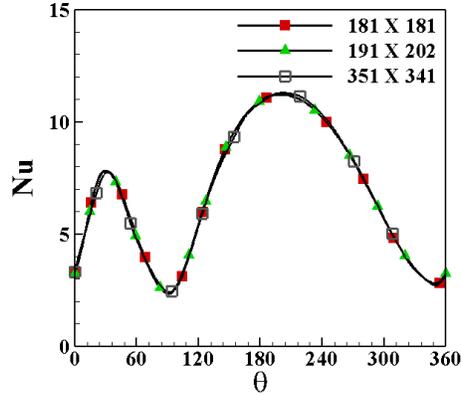}\label{fig:Grid_Independence}
}
\\
\subfigure[]{\includegraphics[width=0.45\textwidth,trim={0.1cm 0.1cm 0.1cm 0.1cm},clip]{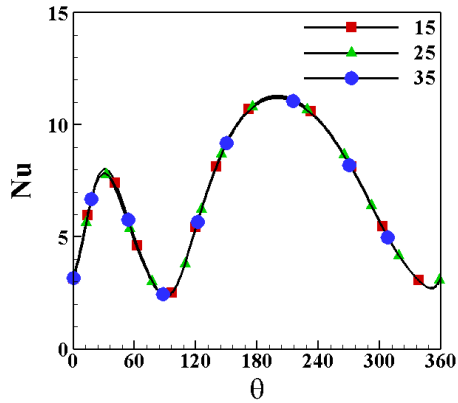}\label{fig:Space_Independence}}
\quad
\subfigure[]{\includegraphics[width=0.45\textwidth,trim={0.1cm 0.1cm 0.1cm 0.1cm},clip]{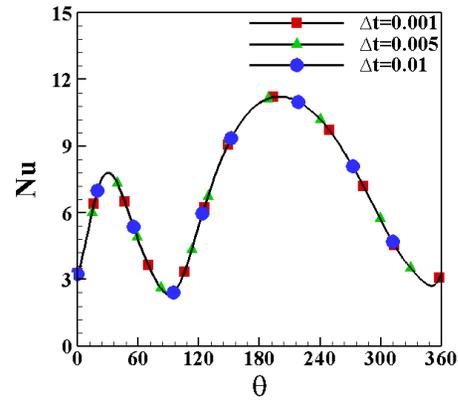}\label{fig:Time_Independence}}
\caption{Variation of local Nusselt number distribution, $Nu$ (a) grid independence test with grid sizes $181\times181$, $191\times202$, $351\times341$, (b) space independence test with outer boundary radius $15$, $25$, $35$ and (c) time independence test with time steps $0.001$, $0.005$, $0.01$ at instant $t = 10$ for $Re = 150$, $Pr=0.7$, $\alpha=1$ and $d=1$.}
\label{fig:Independence}
\end{figure*}

The computational domain is discretized using non-uniform grids. The grid independence test is performed in \cref{fig:Independence}\subref{fig:Grid_Independence} with three different grid sizes $(181\times181)$, $(191\times202)$  and $(351\times341)$, with a set time step $\Delta t=0.01$, a fixed $25:1$ domain-to-cylinder-radius ratio, $Pr=0.7$, $Re=150$, $\alpha=1$ and $d/R_0=1$. All the grid sizes seem to produce almost same results. The grid size $(191\times202)$ is chosen for future computations. For grid size $(181\times181)$ and time step $\Delta t=0.01$, the domain independence test is performed in \cref{fig:Independence}\subref{fig:Space_Independence} for three distinct radii, $15$, $25$ and $35$ of the outer boundary, other parameter values, on the other hand, are treated the same as the grid independence test. This test demonstrates that a domain radius of $25$ is adequate to provide the best possible results. Finally, with a set grid size $(181\times181)$ and the far field border defined at $25:1$ domain-to-cylinder-radius ratio, the time independence test is conducted in \cref{fig:Independence}\subref{fig:Time_Independence} for time increments $\Delta t=0.001,\ 0.005,\ 0.01,\ 0.02$. For later computations, we used $\frac{R_\infty}{R_0}=25$ and $\Delta t=0.01$, as suggested by these test findings.\\

\begin{table}[!t]
        \centering
        \caption{Comparison of the current computation with the equivalent time-averaged total Nusselt number computed by Paramane \& Sharma \cite{paramane2009numerical} for $Re =40,\ 100$, $Pr=0.7$, $\alpha=1$, and isothermally heated cylinder.}
    \label{tab:valid_Paramane}
    \begin{tabular}{|c|c|c|}
    \hline
     $Re$ & $40$ & $100$\\
     \hline
    $\overline{Nu}_t$ (Current) & $3.276112$ & $4.936597$ \\
    \hline
    $\overline{Nu}_t$ (Paramane \& Sharma) & $3.213$  & $4.991$ \\
    \hline
    $Difference (\%)$ & $1.964\%$ & $1.09\%$ \\
    \hline
    \end{tabular}    
  \end{table}
  
 
\begin{table*}[!htbp]
\centering
\caption{Comparison between time-averaged drag results from the current study to Kwon and Choi's work \cite{kwon1996} for $Re=160$.}\label{tab:Validation}
\begin{tabular}{ |c|c|c| } 
\hline
Length of splitter plate & $1$ & $2$\\ 
\hline
Time-averaged Drag (Present Study) & $1.133021$ & $1.056131$\\ 
\hline
Time-averaged Drag (Kwon and Choi) & $1.10162$ & $1.08812$ \\ 
\hline
$Difference (\%)$ & $2.85\%$ & $2.94\%$ \\ 
\hline
\end{tabular}
\end{table*}

There hasn't been any research towards controlling heat and flow transfer from a rotating cylinder using an arc-shaped vertical control plate placed across a free stream of uniform flow. To prove the correctness of our code and model, we begin by comparing our findings to those of previous studies of heat transfer from rotating cylinders \cite{paramane2009numerical}, then it is compared with the results of flow past circular cylinder with a splitter plate attached \cite{kwon1996}. When the flow becomes periodic, the mean drag coefficients are used to determine the time-averaged drag coefficient on the cylinder surface in \cref{tab:Validation}. According to \cref{tab:valid_Paramane}, the maximum difference of time-averaged total Nusselt number is $1.964\%$, which is within a reasonable range. Also, \cref{tab:Validation} shows the maximum difference of time-averaged Drag coefficients from the results of the current study and the previously published works is $2.94\%$ which is also within a considerable range. As a result, the current findings are consistent with earlier studies.

\section{Results and Discussions\protect}\label{Results and Discussions}

\begin{figure*}[!t]
\centering

\includegraphics[width=0.3\textwidth,trim={0.5cm 0.25cm 0.5cm 0.5cm},clip]{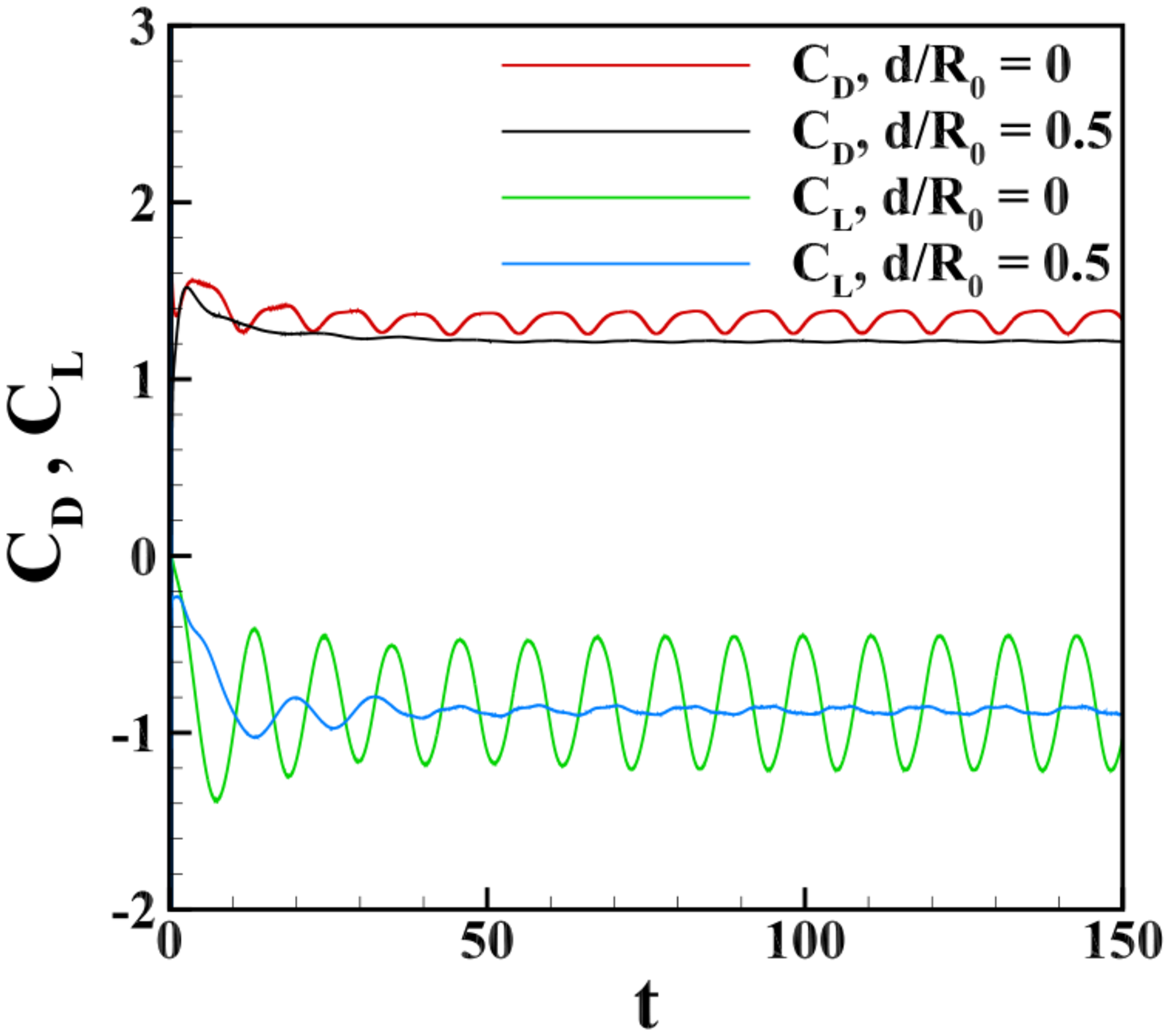}
\\
\hspace{0.5em}\scriptsize{(a)}
\\
\includegraphics[width=0.3\textwidth,trim={0.5cm 0.25cm 0.5cm 0.5cm},clip]{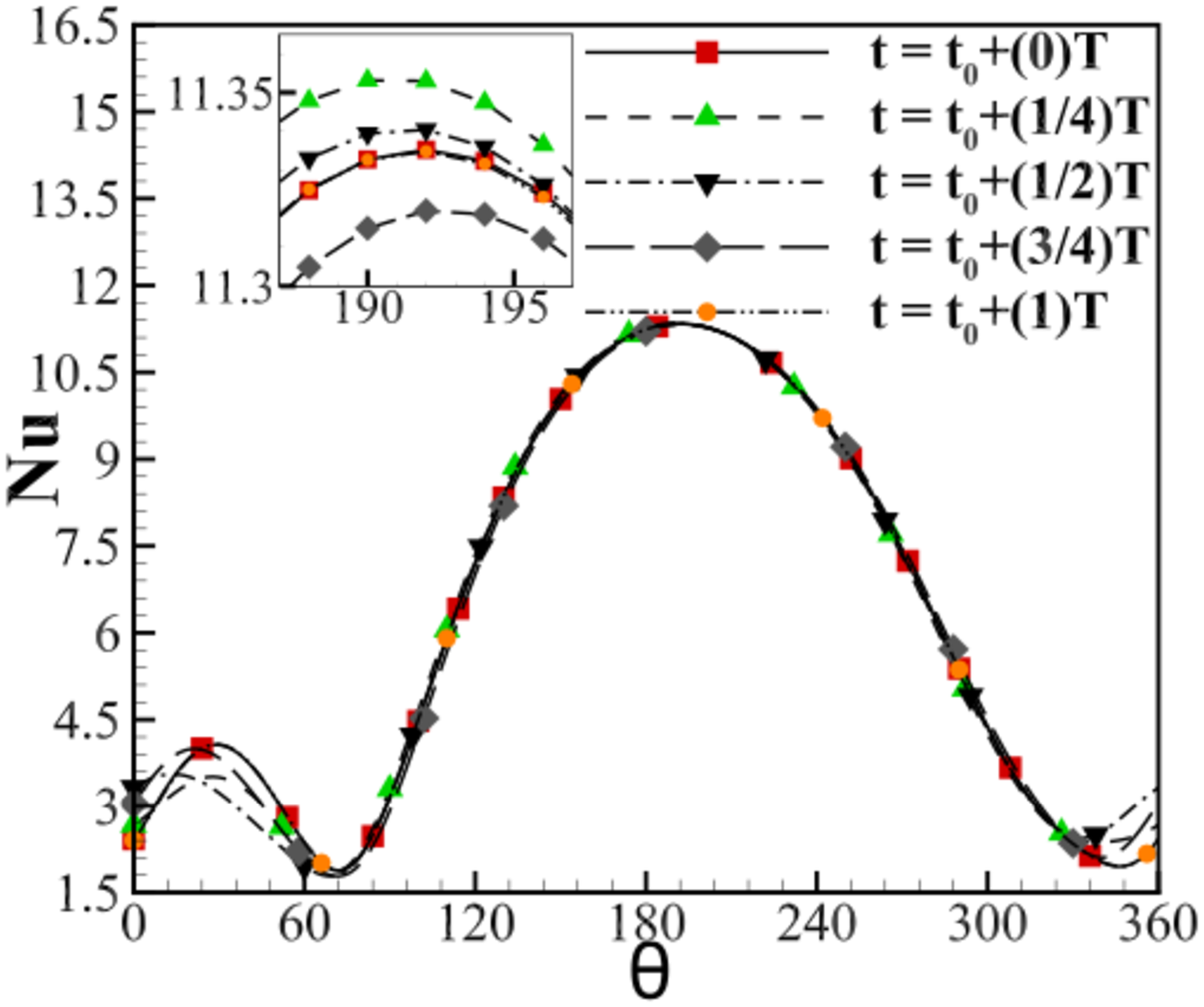}\hspace{1cm}%
\includegraphics[width=0.3\textwidth,trim={0.5cm 0.25cm 0.5cm 0.5cm},clip]{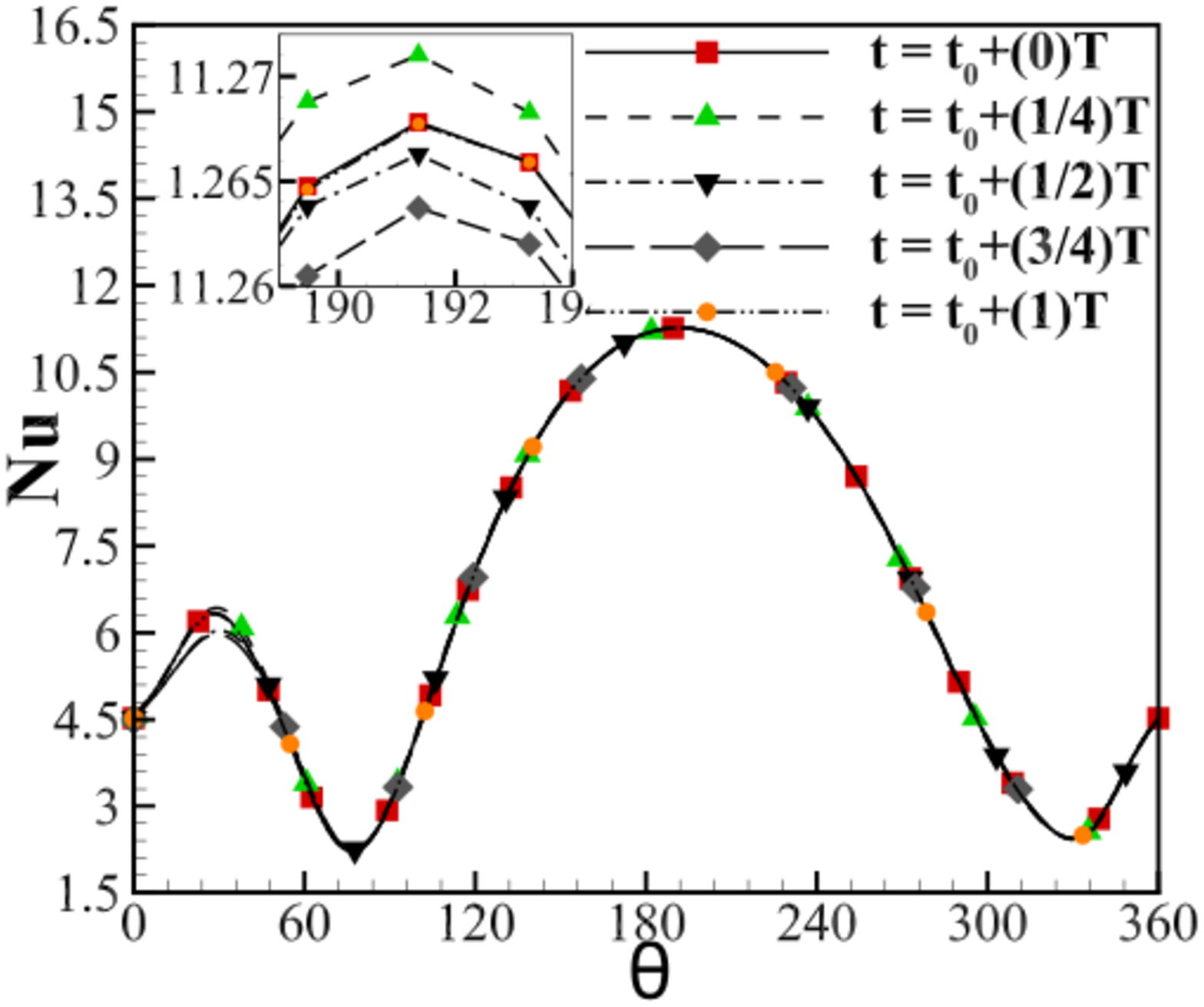}
\\
\hspace{2cm}(b) \hspace{4cm}(c)\hspace{2cm}
 \caption{(a) Drag ($C_D$) and lift ($C_L$) coefficients, (b) variation of local Nusselt number ($Nu$) for $d/R_0=0$ and (c) variation of local Nusselt number ($Nu$) for $d/R_0=0.5$  with fixed $\alpha=0.5$.}
 \label{fig:lift-drag-Nu}
\end{figure*}

Reynolds number ($Re$), Prandtl number ($Pr$), angular velocity of the cylinder ($\alpha$), and control plate distance ($d/R_0$) are all well-known factors that influence flow and heat fields. \cref{fig:lift-drag-Nu} exhibits the Drag ($C_D$) and lift ($C_L$) coefficients, as well as the variation of local Nusselt number ($Nu$) for $d/R_0=0$ and $0.5$ with fixed $\alpha=0.5$, $Re=150$, $Pr=0.7$. The parameter value $d/R_0=0$ corresponds to the case without the arc-shaped control plate. \cref{fig:lift-drag-Nu}(a) clearly demonstrates that, the peak value of $C_D$ is significantly reduced as well as the amplitude of $C_L$ with the introduction of control plate at a distance $d/R_0=0.5$ downstream. $d/R_0=0$, indicating flow across the cylinder in the absence of the control plate. By comparing \cref{fig:lift-drag-Nu}(b) and \cref{fig:lift-drag-Nu}(c), it is found that the introduction of the control plate slightly reduced the peak value of $Nu$ approximately from $11.35$ to $11.27$ at $\theta\approx 192\degree$, but the local maximum peak is significantly increased at $\theta\approx 30\degree$. It means, although the heat transfer near the front stagnation is slightly decreased by the control plate, the heat transfer is significantly increased near the rear stagnation point, which eventually increases the overall heat transfer from the upper half of the cylinder surface. The control plate alters the vortex shedding process, which in turn affects the thermal boundary layer and causes this effect. Realizing the importance of the arc-shaped control plate, the current studies are performed for $Re=150$, $0.5\leq \alpha \leq 3.25$ and $0.5\leq d/R_0 \leq 3$, while $Pr$ is maintained at $0.7$. The values of $\alpha$ are typically chosen in accordance with \cite{ray2011transformation}.\\

\begin{figure*}[!t]
\centering
\scriptsize{$t=t_0+(0)T$}
\\
\includegraphics[width=0.3\textwidth,trim={0.5cm 0.3cm 0.5cm 0.3cm},clip]{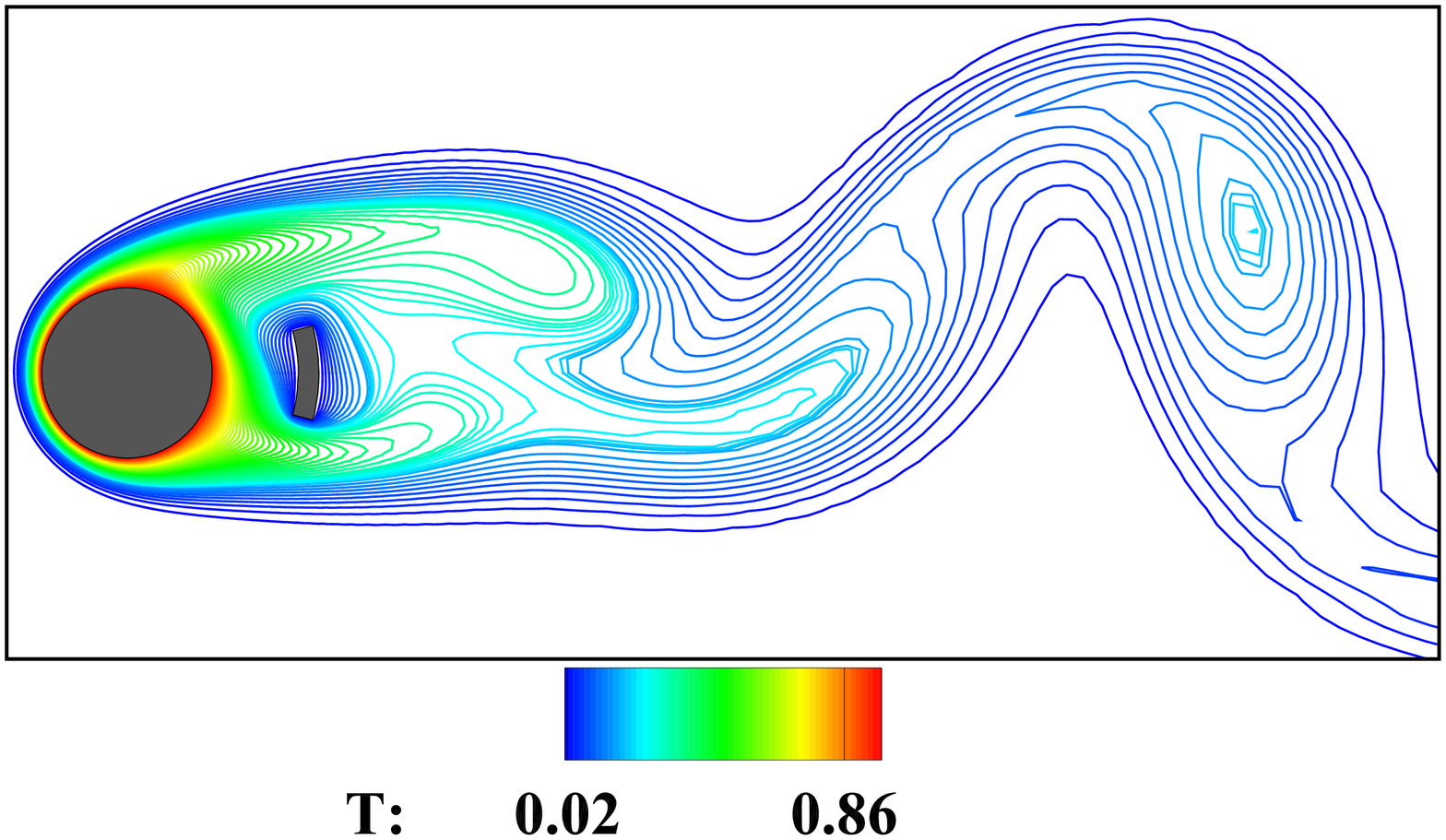}
\includegraphics[width=0.3\textwidth,trim={0.5cm 0.3cm 0.3cm 0.3cm},clip]{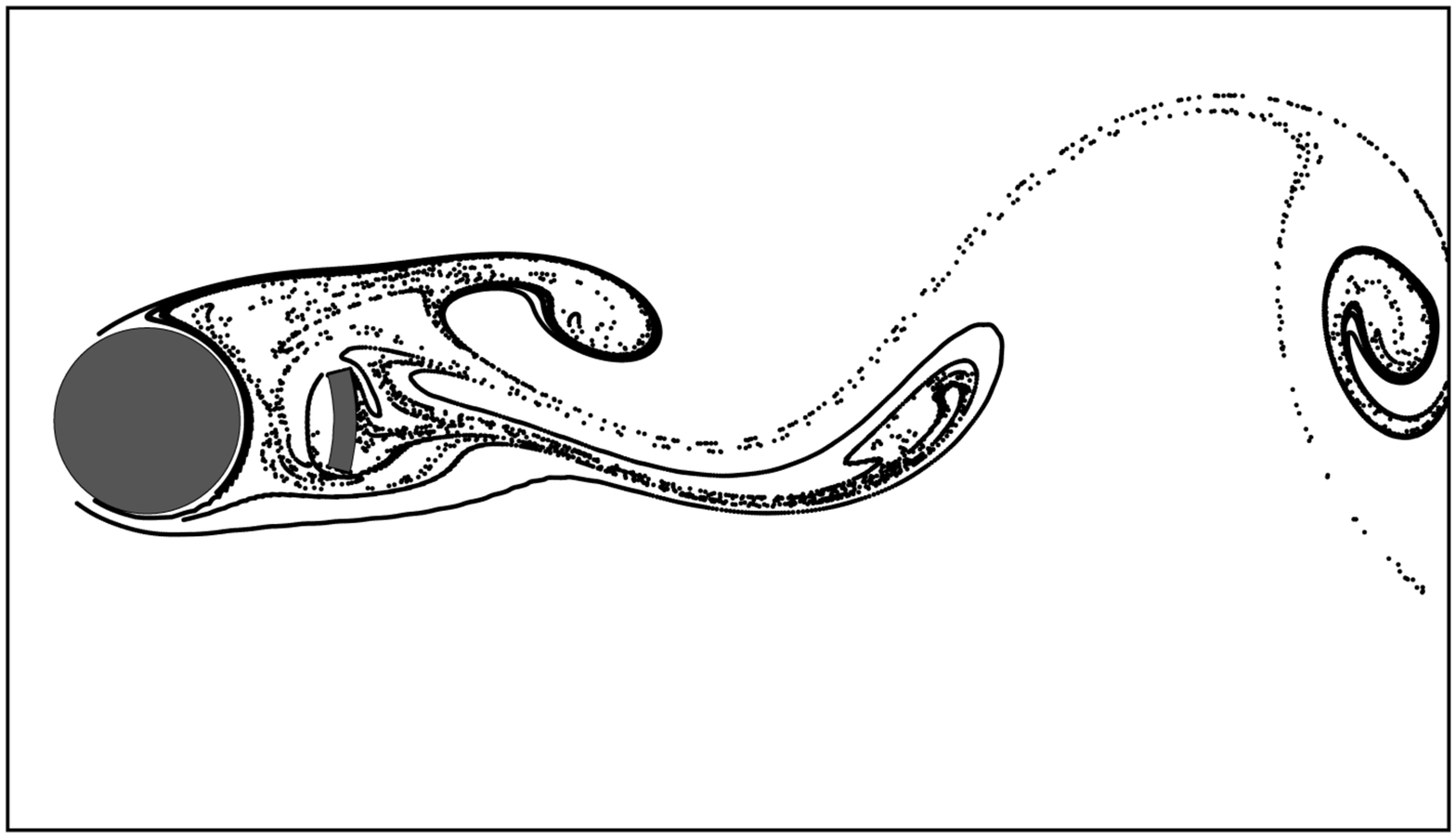}
\includegraphics[width=0.3\textwidth,trim={0.5cm 0.3cm 0.3cm 0.3cm},clip]{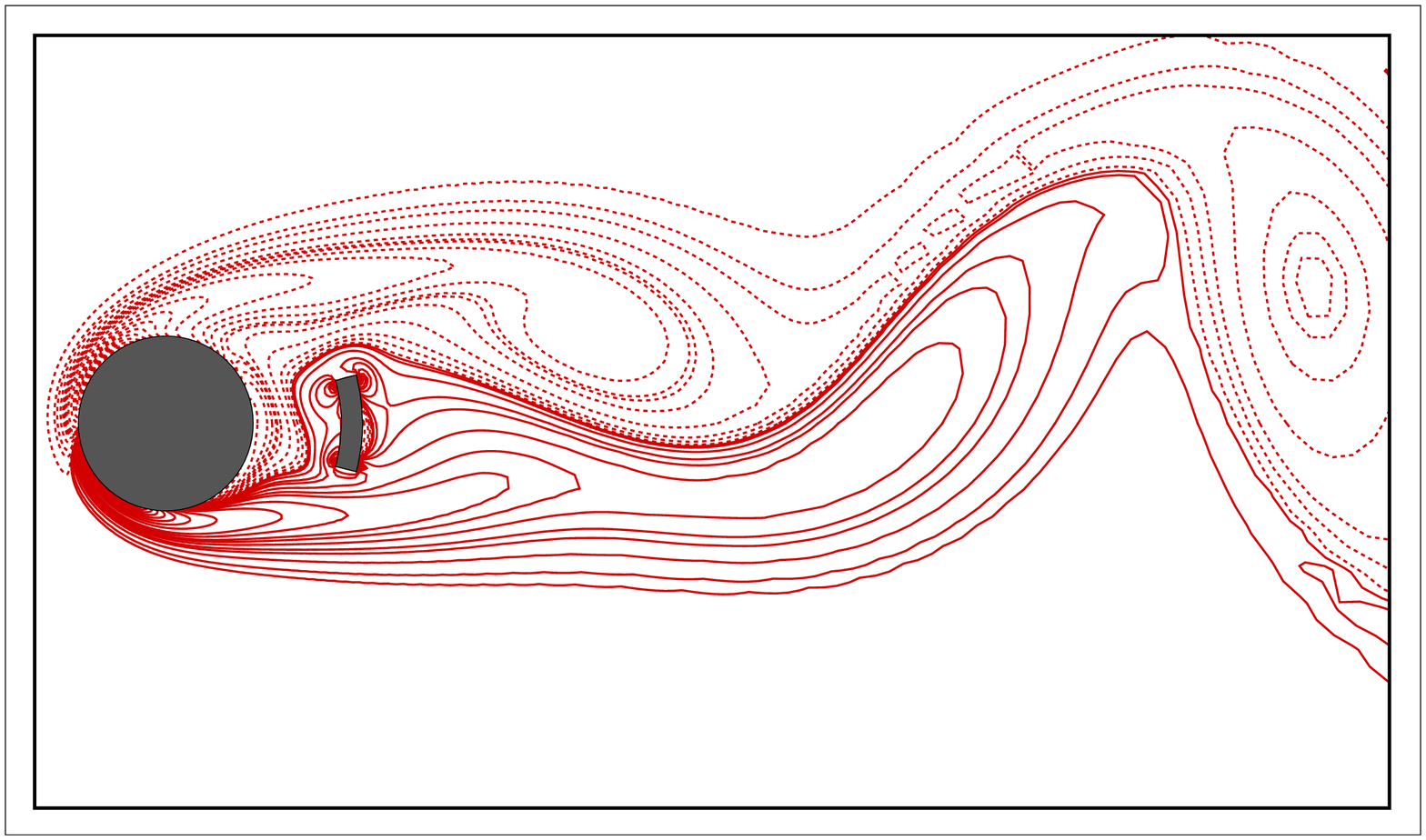}
\\
\hspace{0.5em}\scriptsize{$t=t_0+(1/4)T$}
\\
\includegraphics[width=0.29\textwidth,trim={0.5cm 0.3cm 0.5cm 0.3cm},clip]{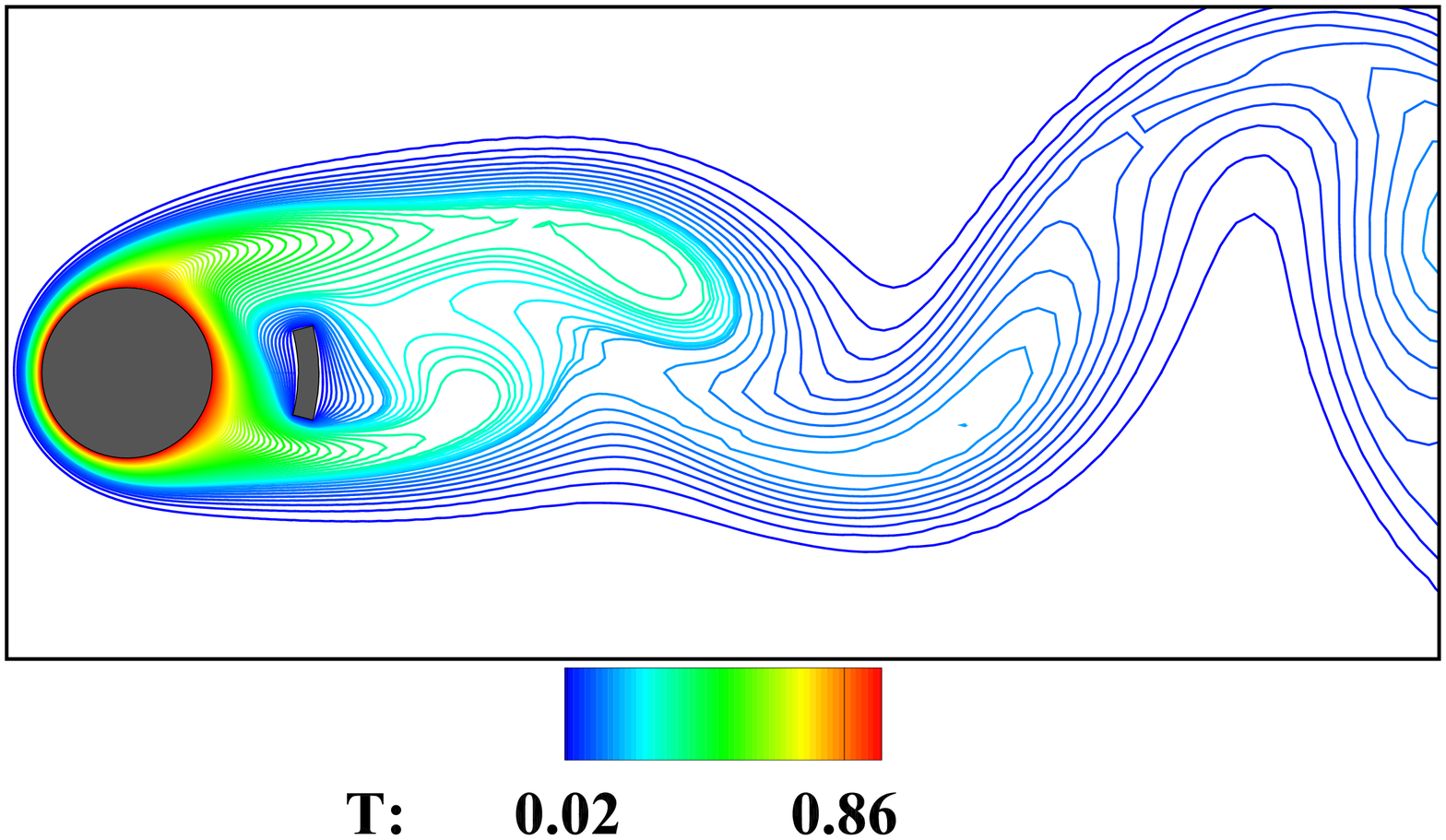}
\includegraphics[width=0.3\textwidth,trim={0.5cm 0.3cm 0.3cm 0.3cm},clip]{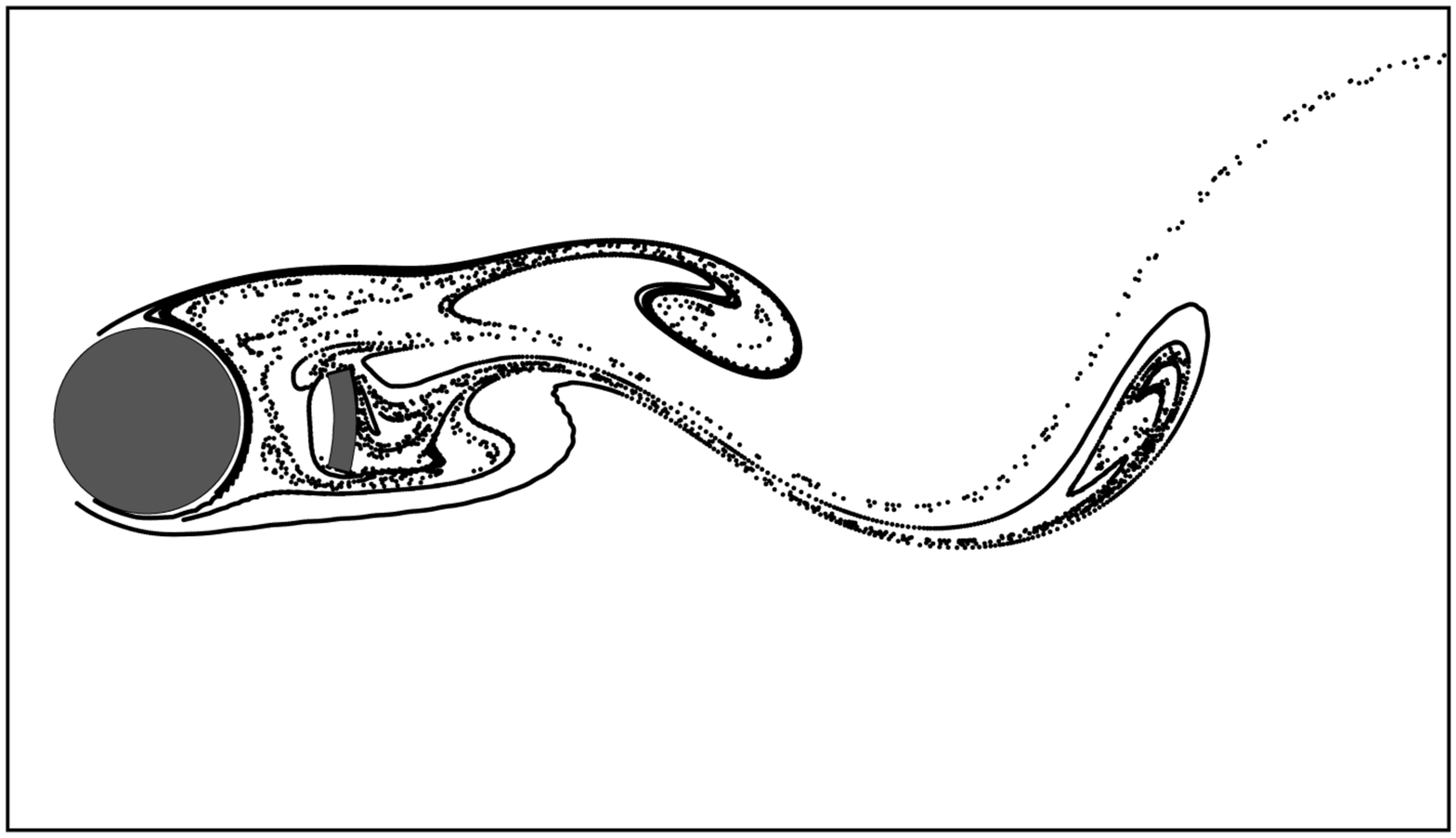}
\includegraphics[width=0.3\textwidth,trim={0.5cm 0.3cm 0.3cm 0.3cm},clip]{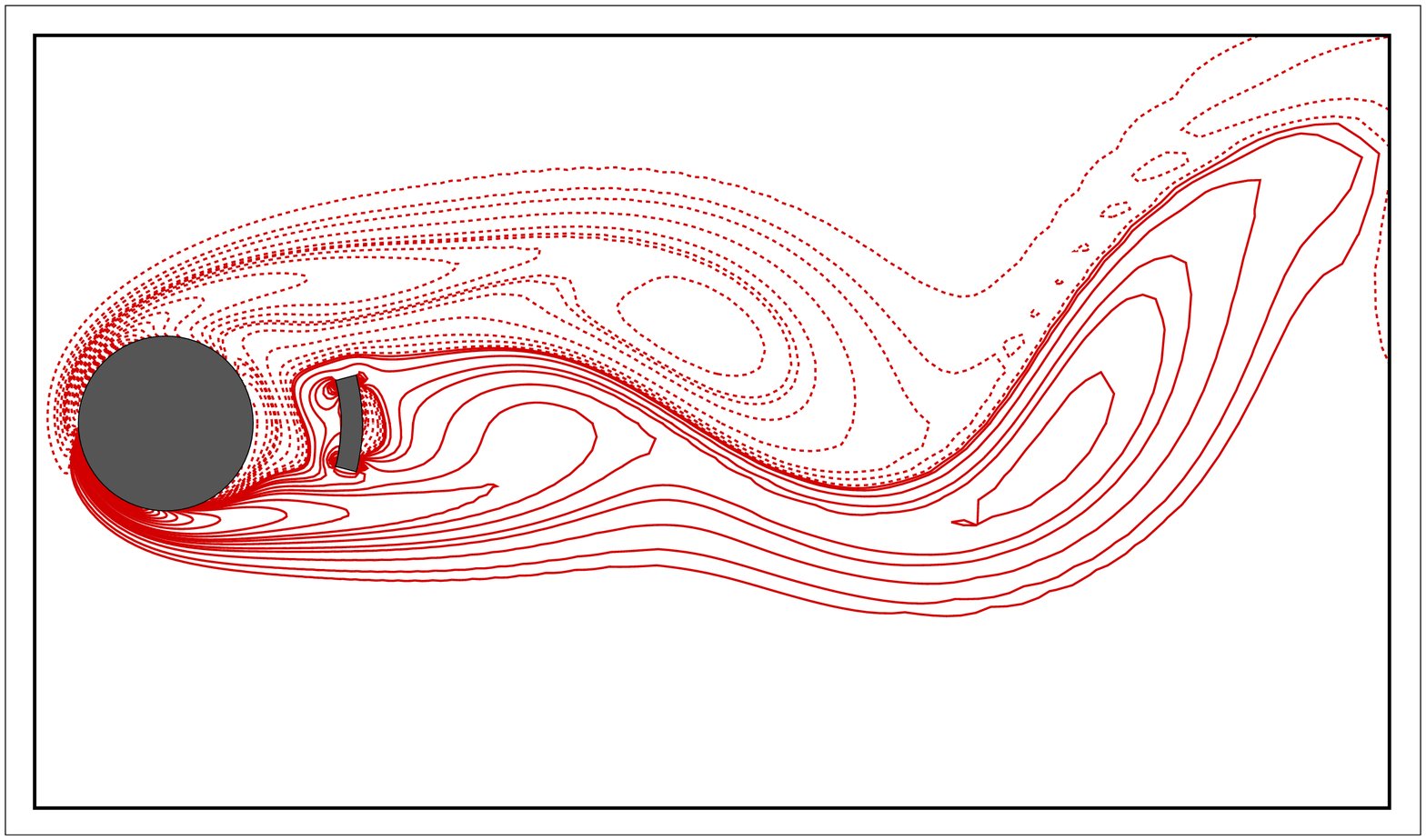}
\\
\hspace{0.5em}\scriptsize{$t=t_0+(1/2)T$}
\\
\includegraphics[width=0.29\textwidth,trim={0.5cm 0.3cm 0.5cm 0.3cm},clip]{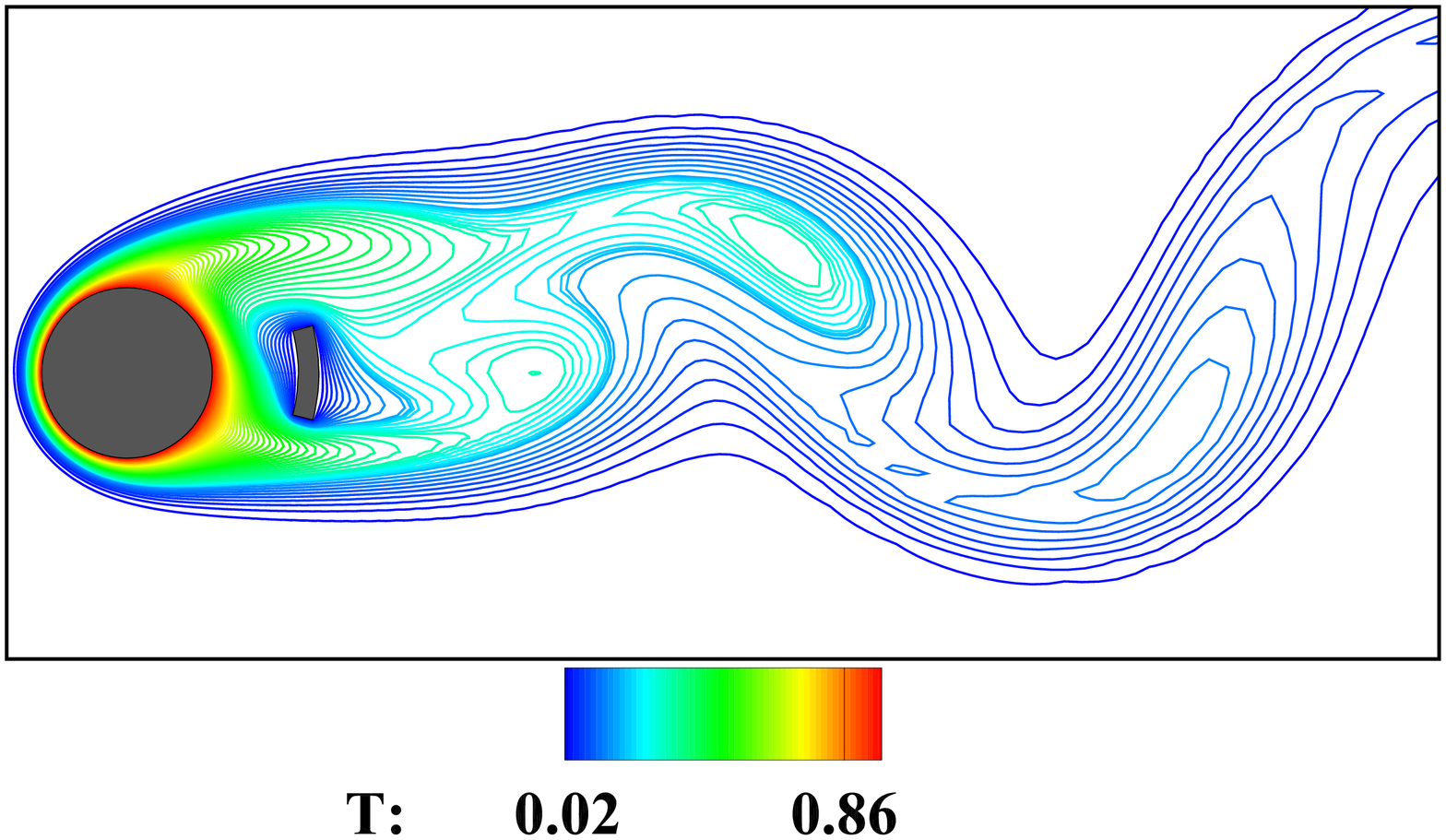}
\includegraphics[width=0.3\textwidth,trim={0.5cm 0.3cm 0.3cm 0.3cm},clip]{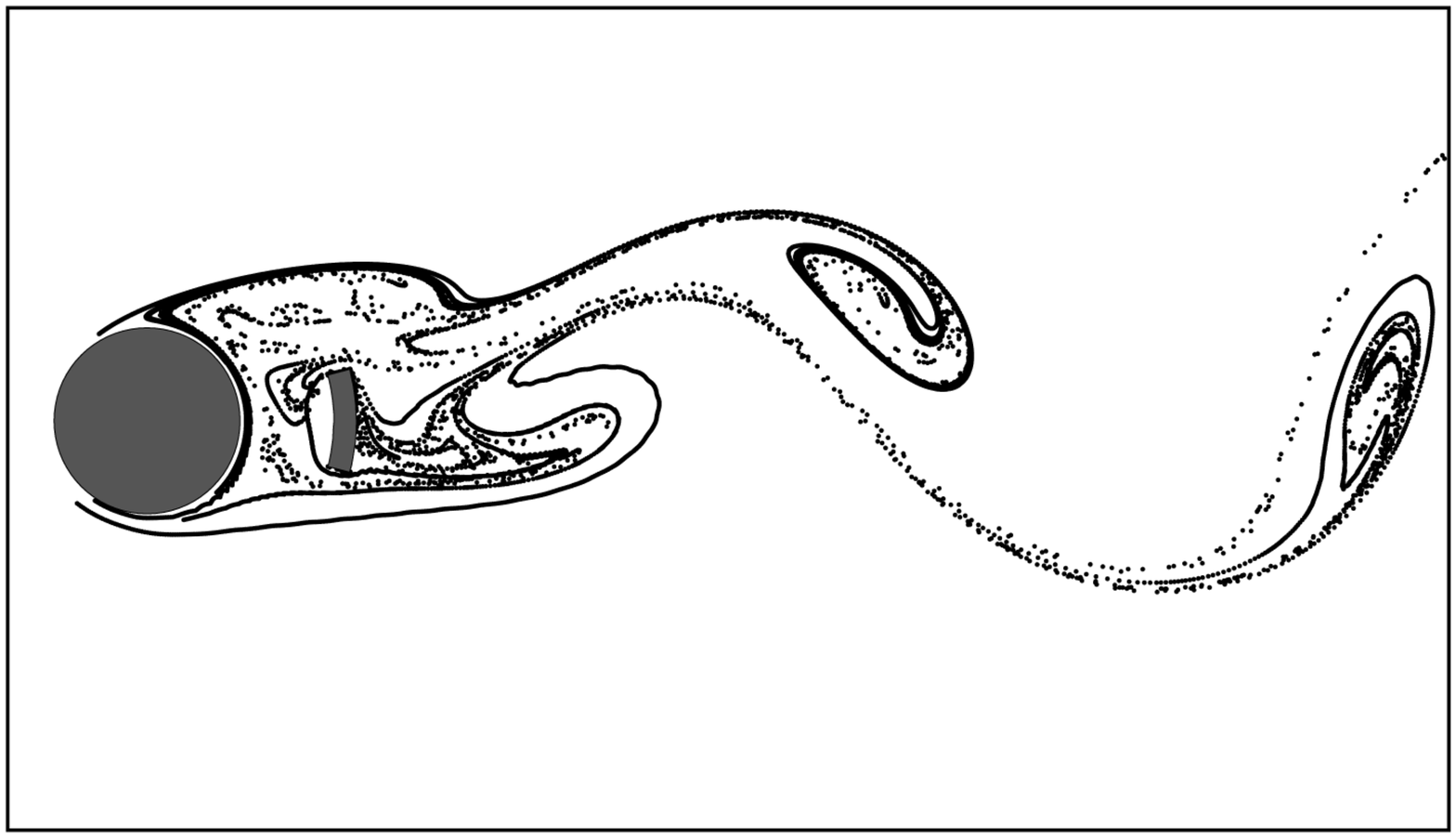}
\includegraphics[width=0.3\textwidth,trim={0.5cm 0.3cm 0.3cm 0.3cm},clip]{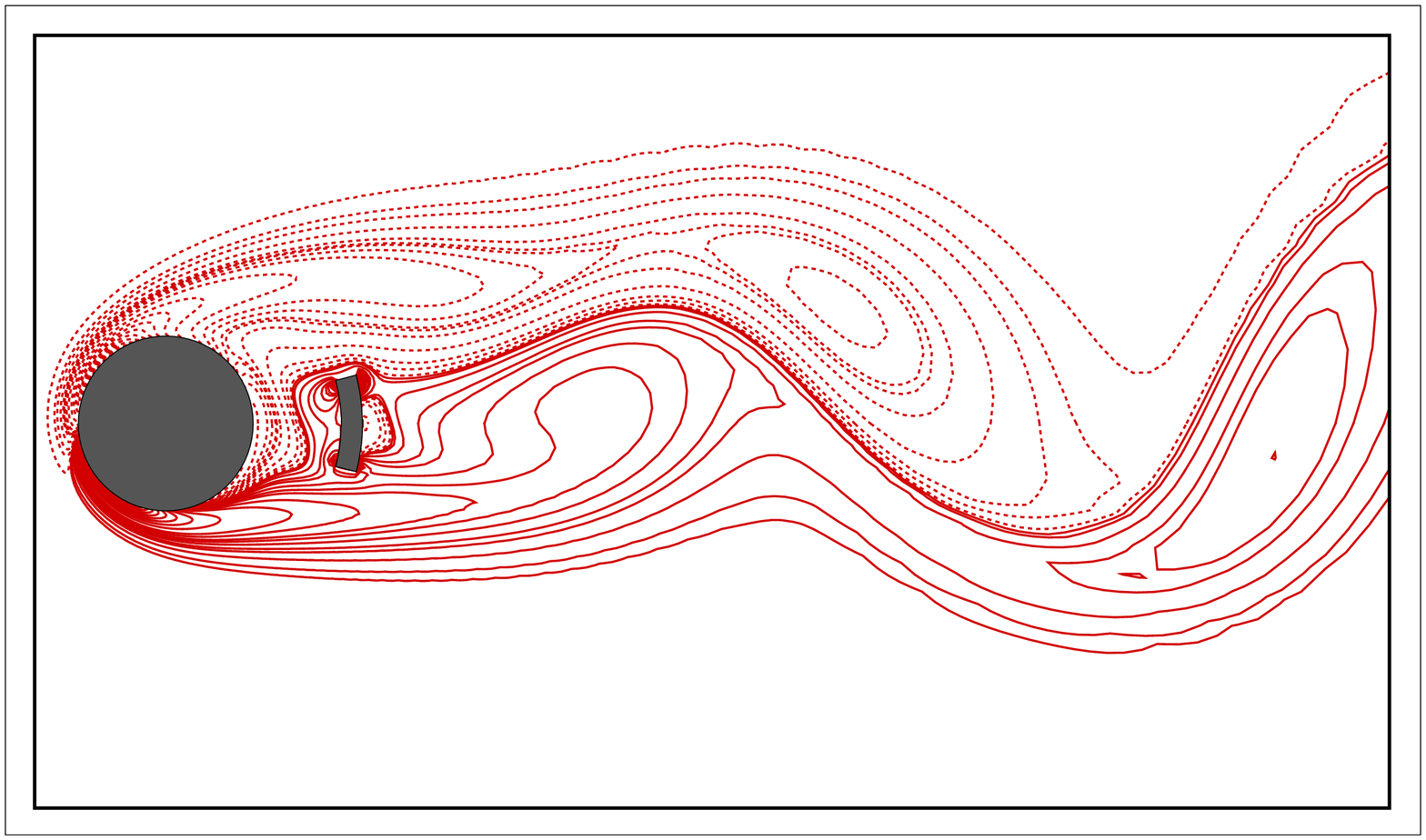}
\\
\hspace{0.5em}\scriptsize{$t=t_0+(3/4)T$}
\\
\includegraphics[width=0.29\textwidth,trim={0.5cm 0.3cm 0.5cm 0.3cm},clip]{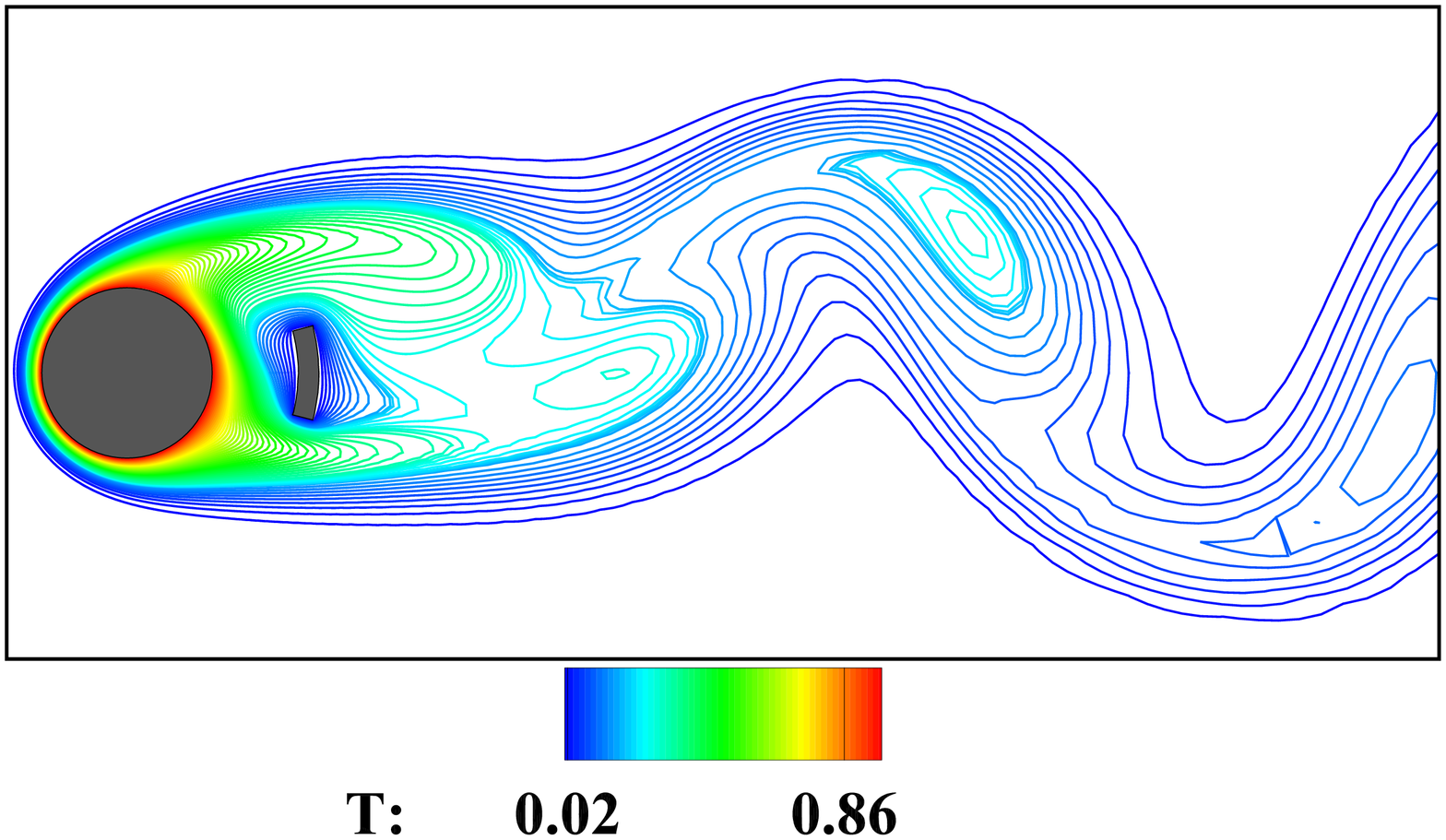}
\includegraphics[width=0.3\textwidth,trim={0.5cm 0.3cm 0.3cm 0.3cm},clip]{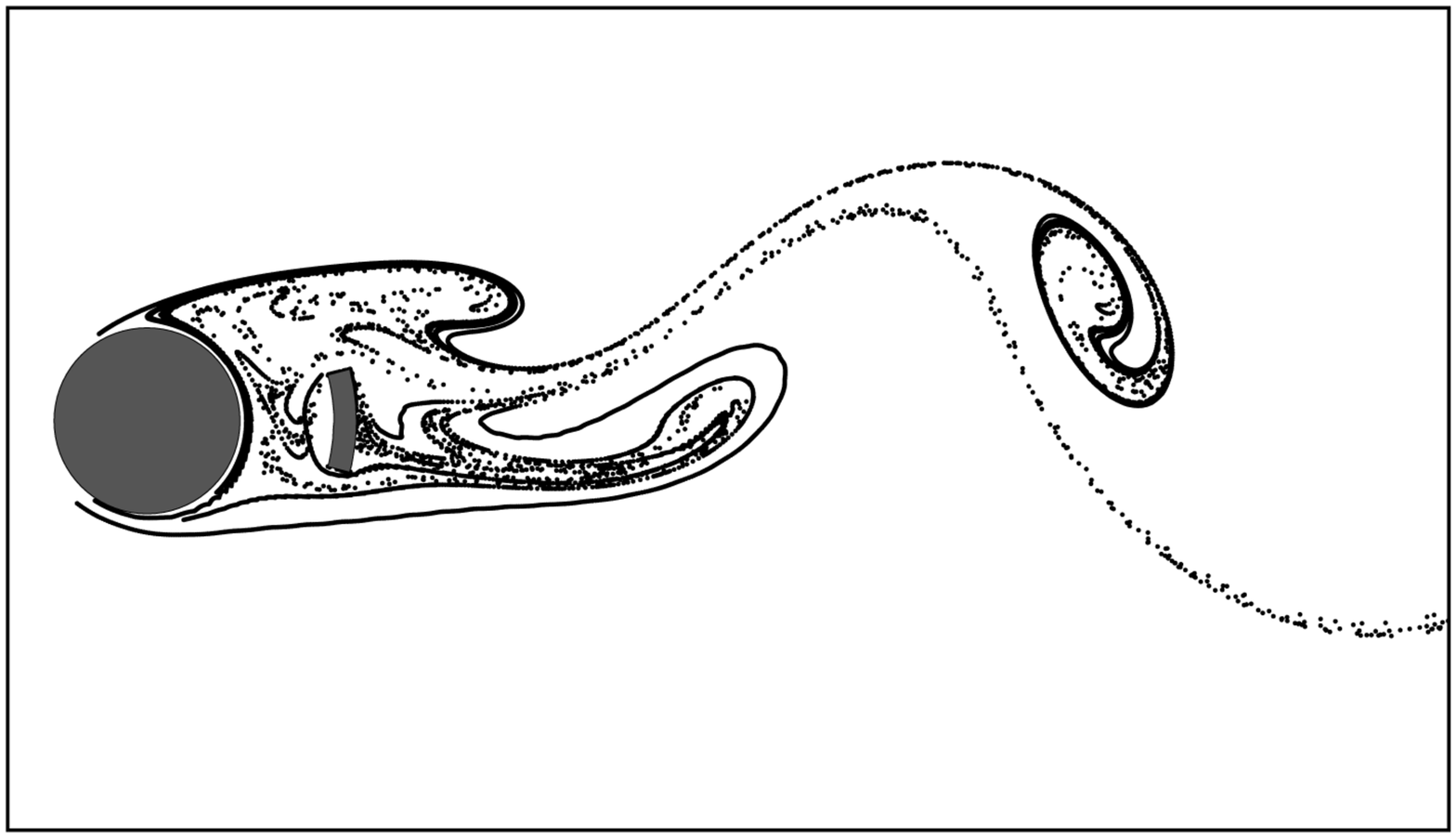}
\includegraphics[width=0.3\textwidth,trim={0.5cm 0.3cm 0.3cm 0.3cm},clip]{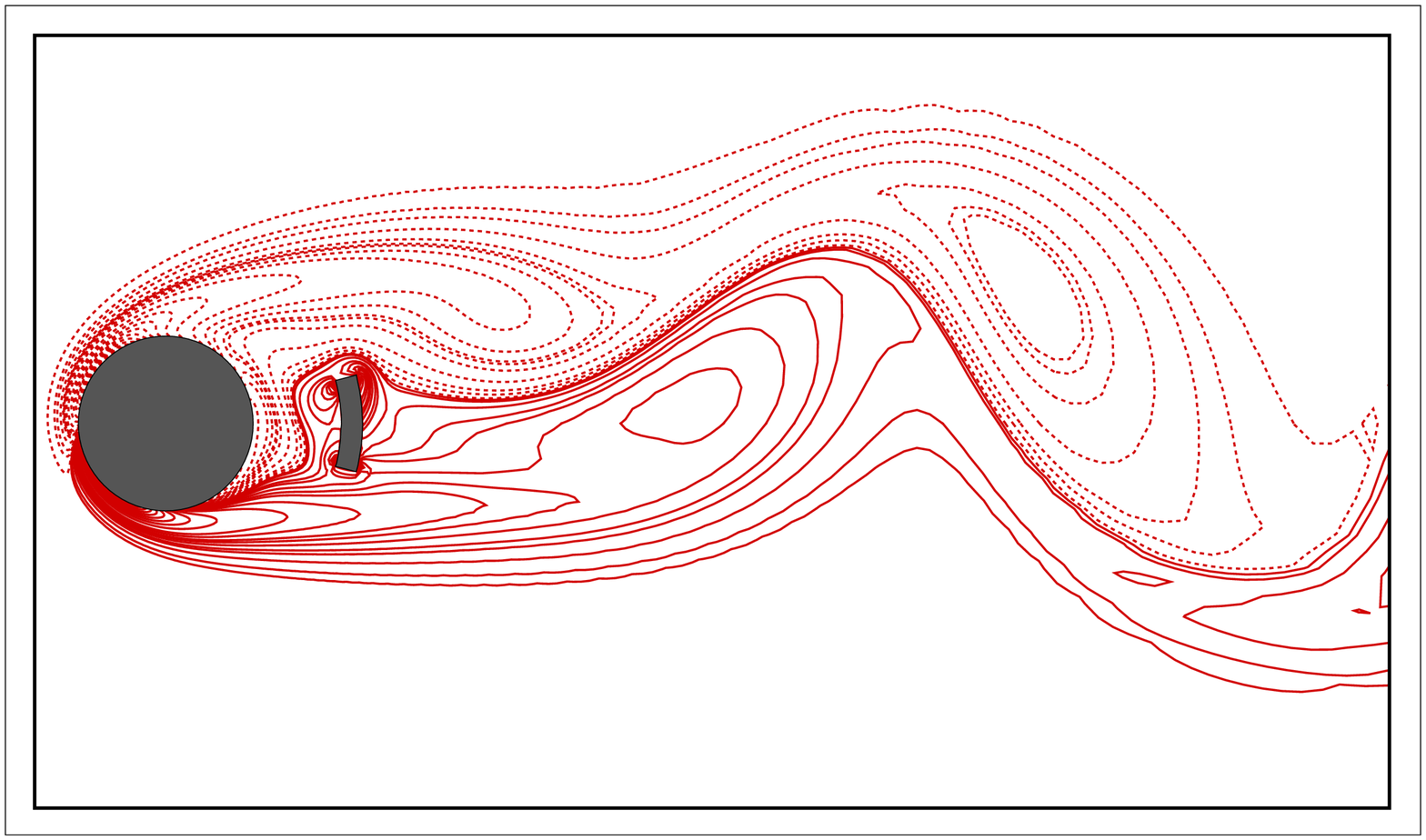}
\\
\hspace{0.5em}\scriptsize{$t=t_0+(1)T$}
\\
\includegraphics[width=0.29\textwidth,trim={0.5cm 0.3cm 0.5cm 0.3cm},clip]{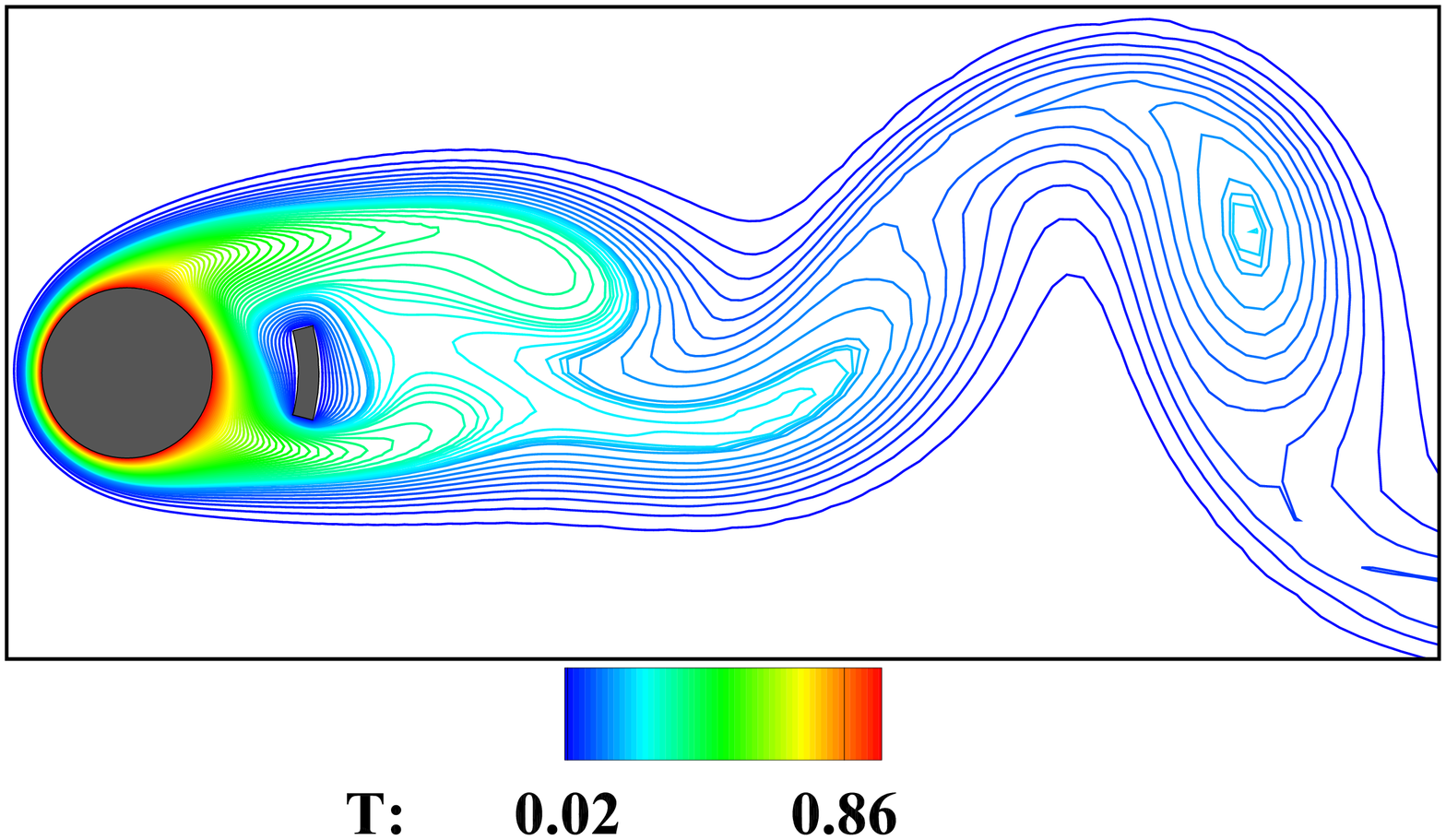}
\includegraphics[width=0.3\textwidth,trim={0.5cm 0.3cm 0.3cm 0.3cm},clip]{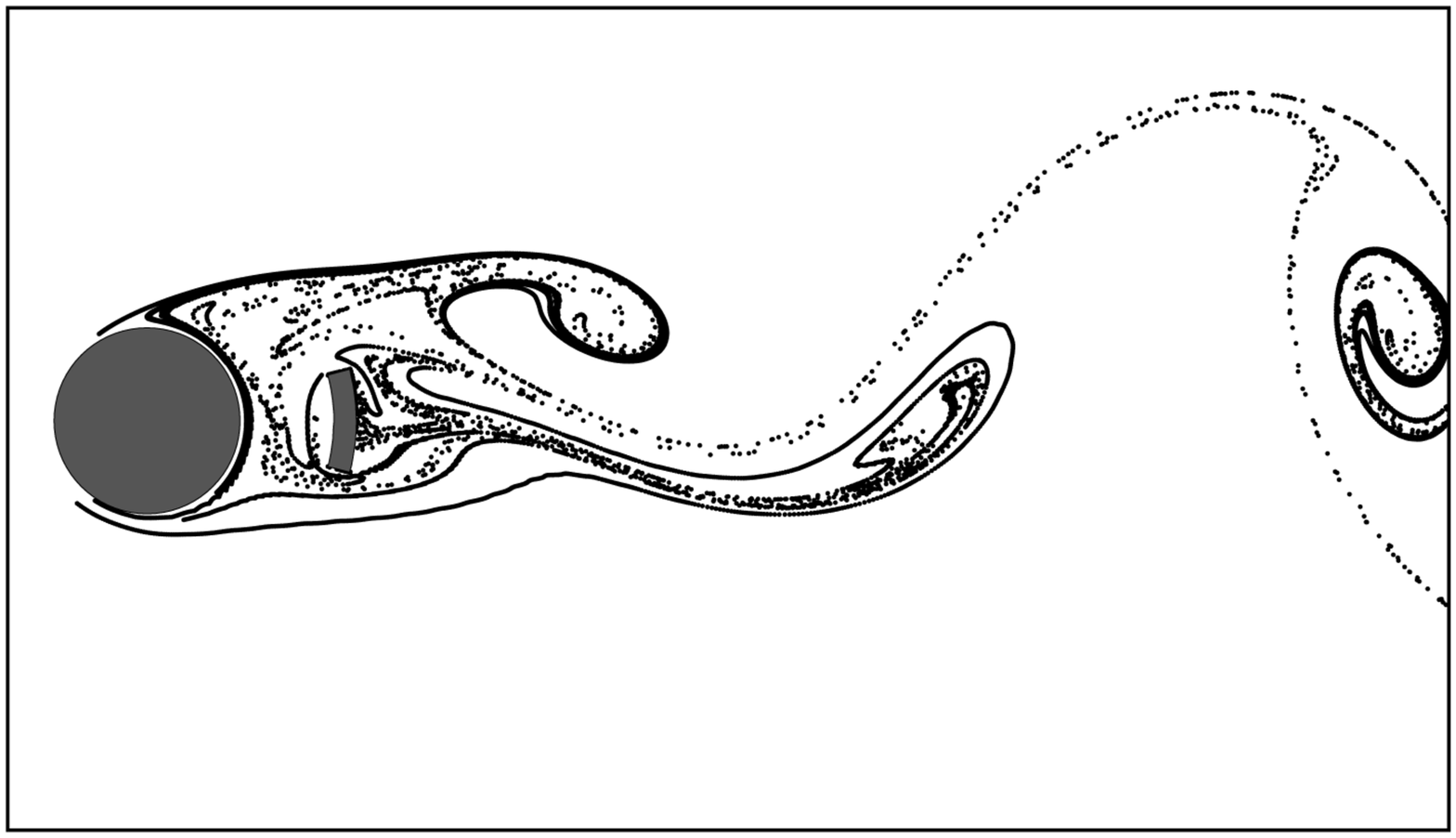}
\includegraphics[width=0.3\textwidth,trim={0.5cm 0.3cm 0.3cm 0.3cm},clip]{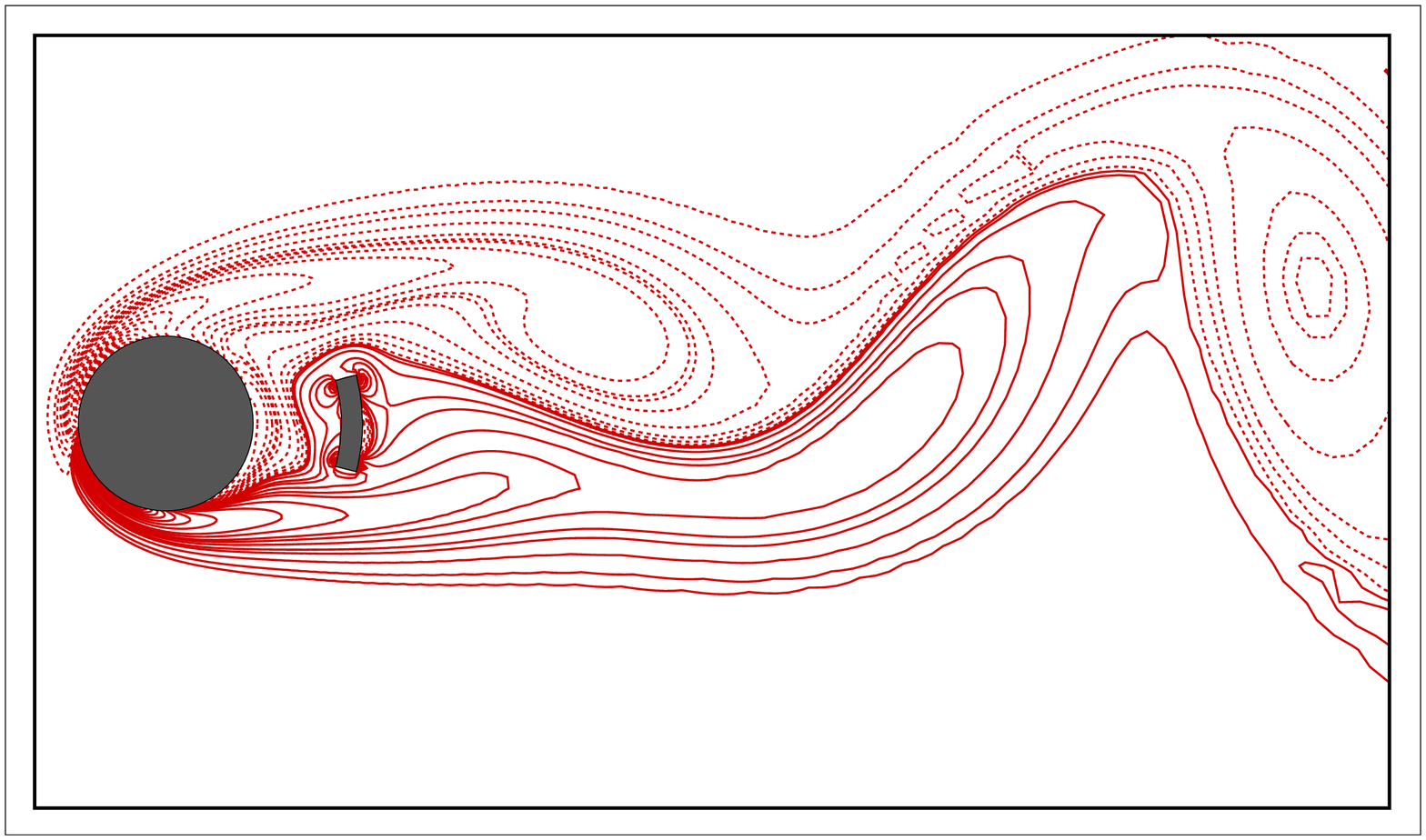}
\\
\hspace{2cm}(a) \hspace{4cm}(b) \hspace{4cm}(c)\hspace{2cm}
 \caption{(a) Isotherm, (b) streakline and (c) vorticity contour for $Pr=0.7$, $Re=150$, $\alpha=0.5$ and $d/R_0=1$ at different phases.}
 \label{fig:d_1_a_0-5}
\end{figure*}

\begin{figure*}[!t]
\centering
\scriptsize{$t=t_0+(0)T$}
\\
\includegraphics[width=0.3\textwidth,trim={0.5cm 0.3cm 0.5cm 0.3cm},clip]{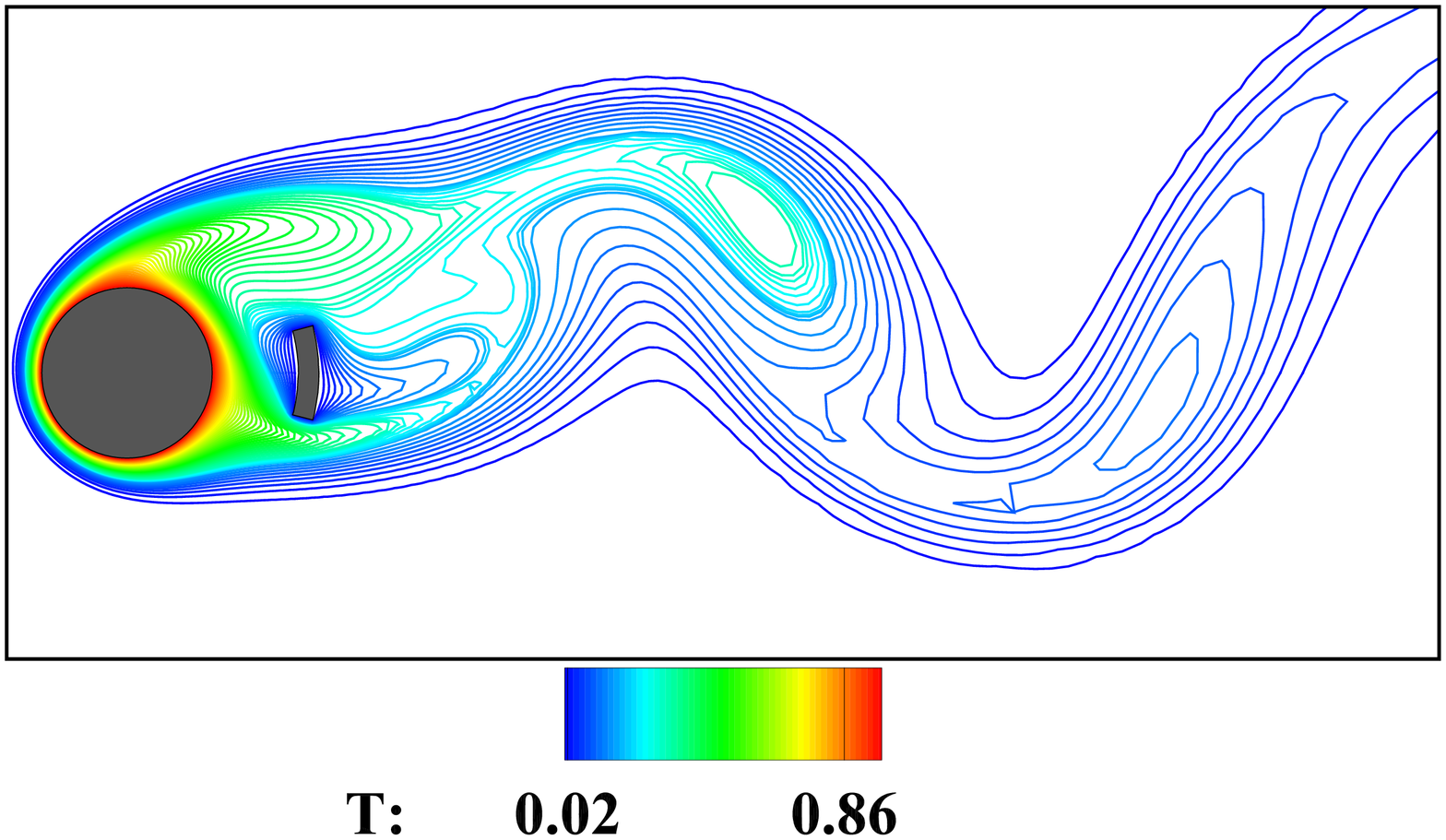}
\includegraphics[width=0.3\textwidth,trim={0.5cm 0.3cm 0.3cm 0.3cm},clip]{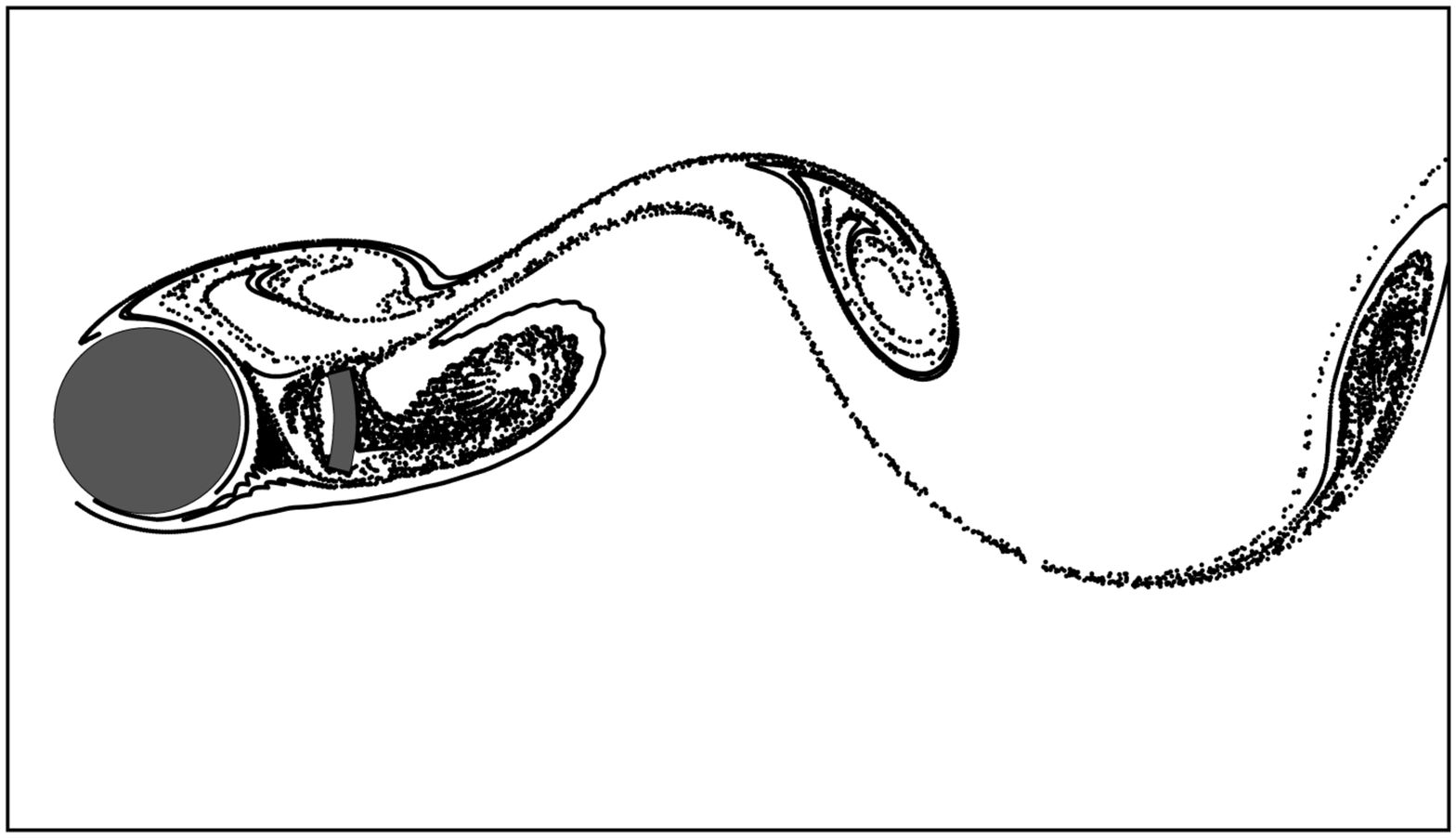}
\includegraphics[width=0.3\textwidth,trim={0.5cm 0.3cm 0.3cm 0.3cm},clip]{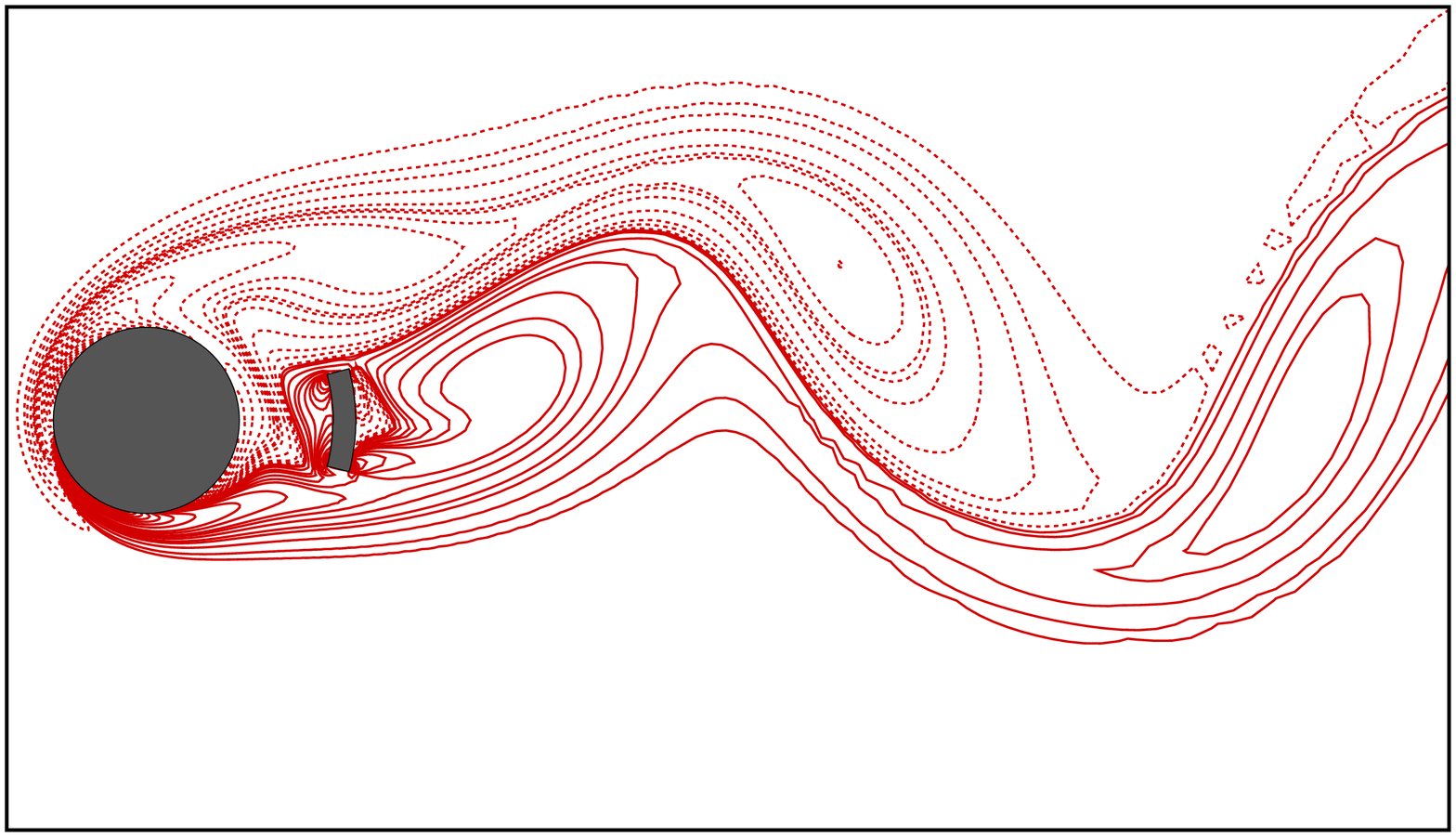}
\\
\hspace{0.5em}\scriptsize{$t=t_0+(1/4)T$}
\\
\includegraphics[width=0.29\textwidth,trim={0.5cm 0.3cm 0.5cm 0.3cm},clip]{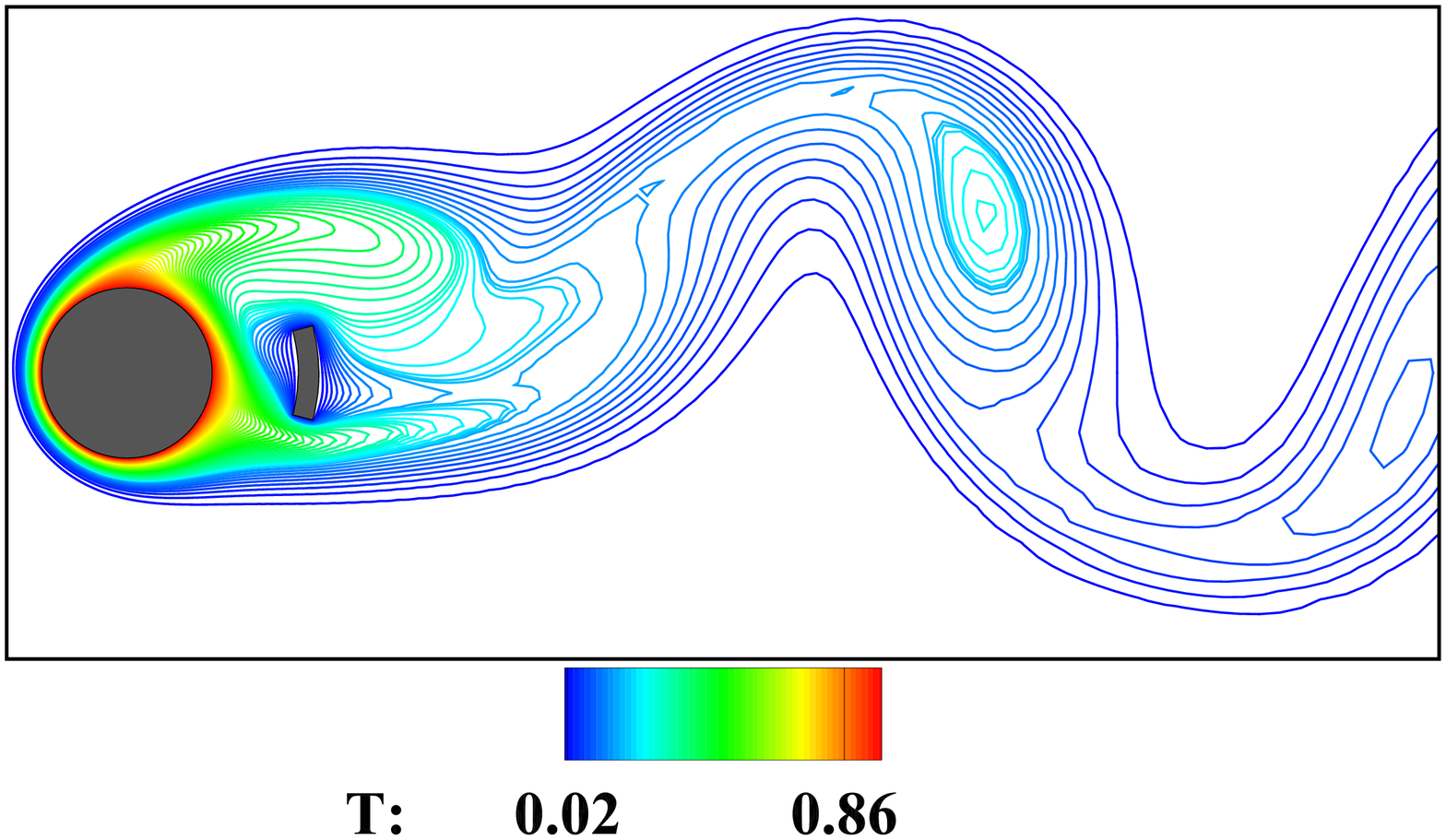}
\includegraphics[width=0.3\textwidth,trim={0.5cm 0.3cm 0.3cm 0.3cm},clip]{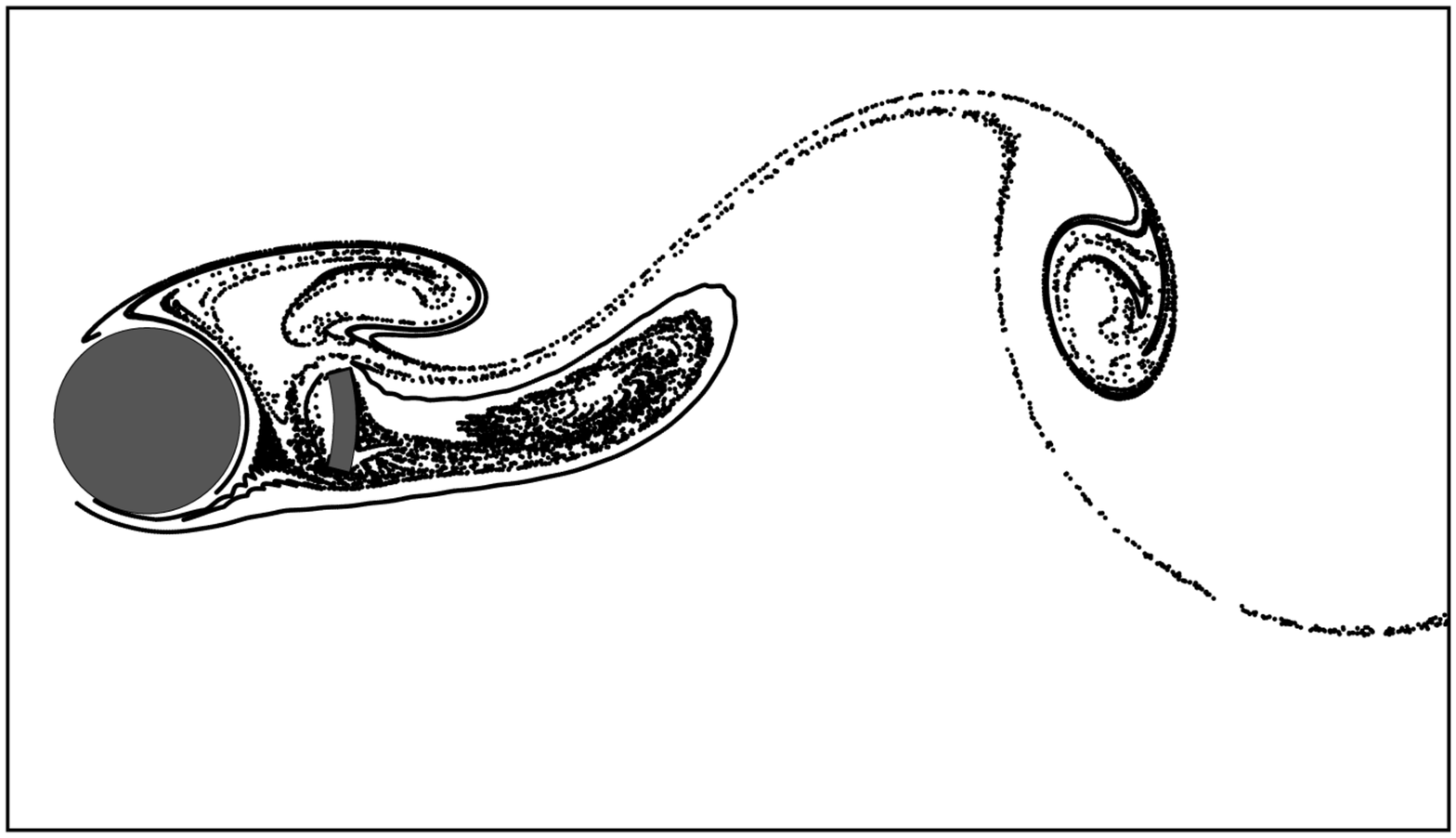}
\includegraphics[width=0.3\textwidth,trim={0.5cm 0.3cm 0.3cm 0.3cm},clip]{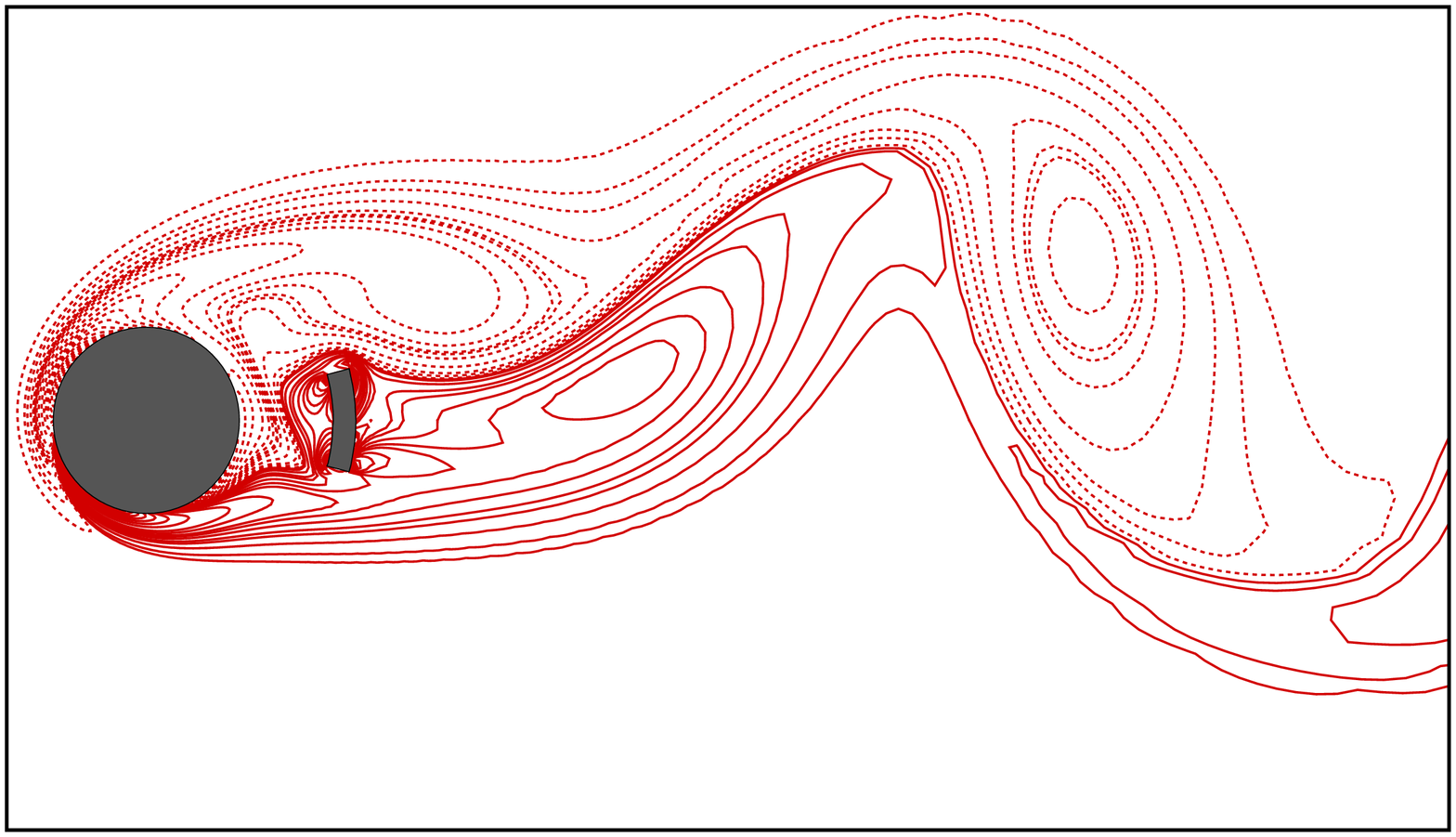}
\\
\hspace{0.5em}\scriptsize{$t=t_0+(1/2)T$}
\\
\includegraphics[width=0.29\textwidth,trim={0.5cm 0.3cm 0.5cm 0.3cm},clip]{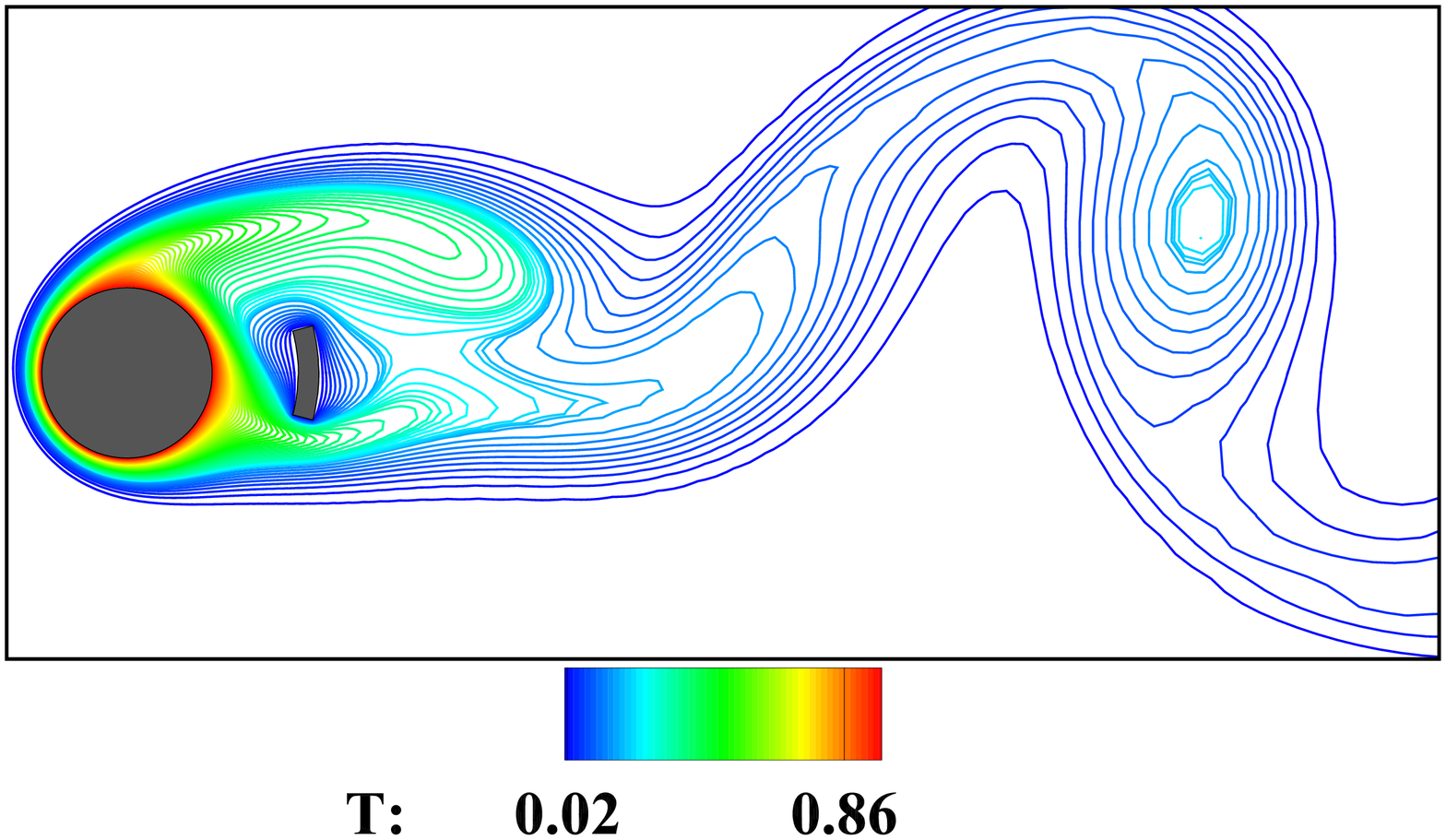}
\includegraphics[width=0.3\textwidth,trim={0.5cm 0.3cm 0.3cm 0.3cm},clip]{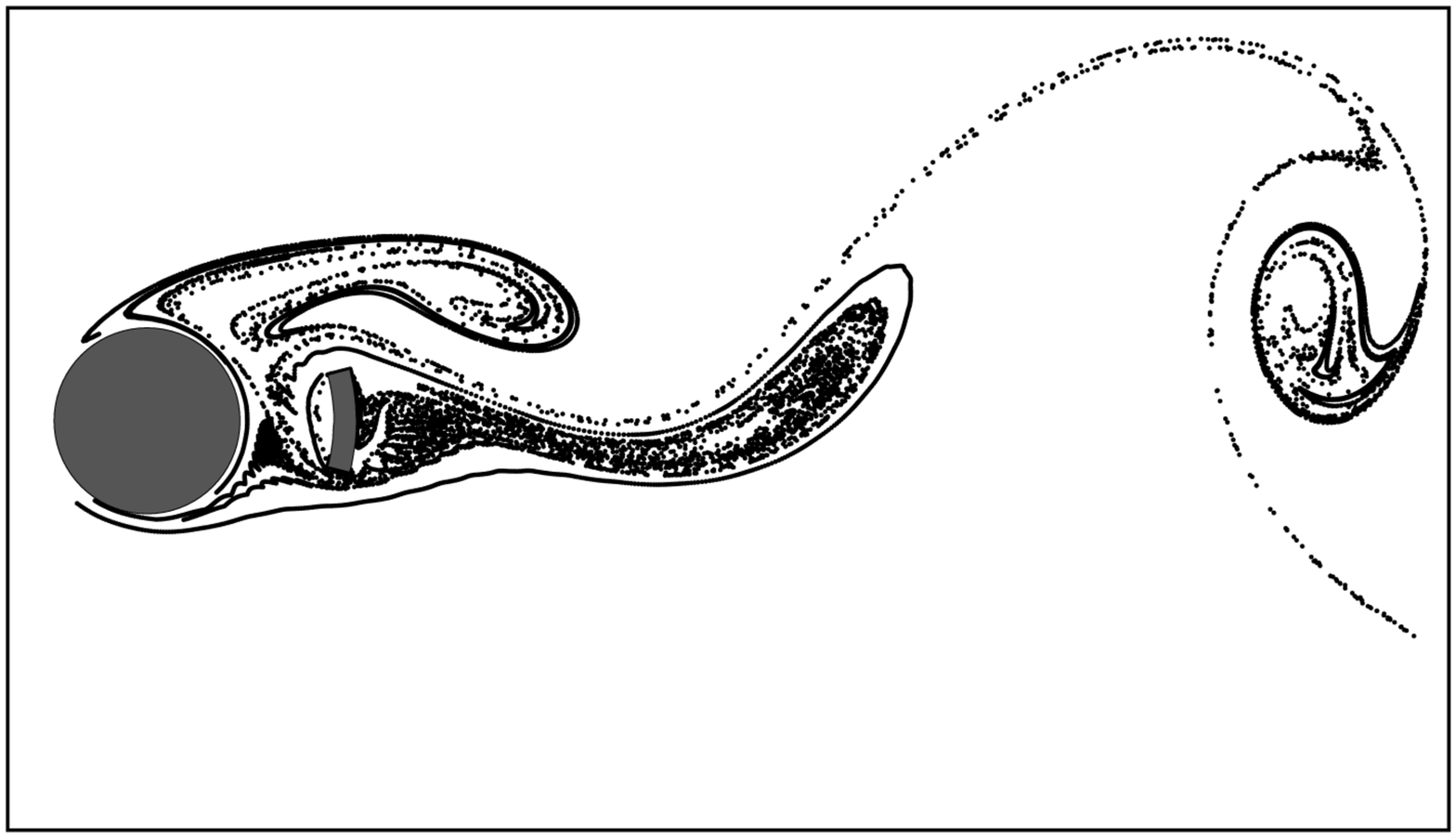}
\includegraphics[width=0.3\textwidth,trim={0.5cm 0.3cm 0.3cm 0.3cm},clip]{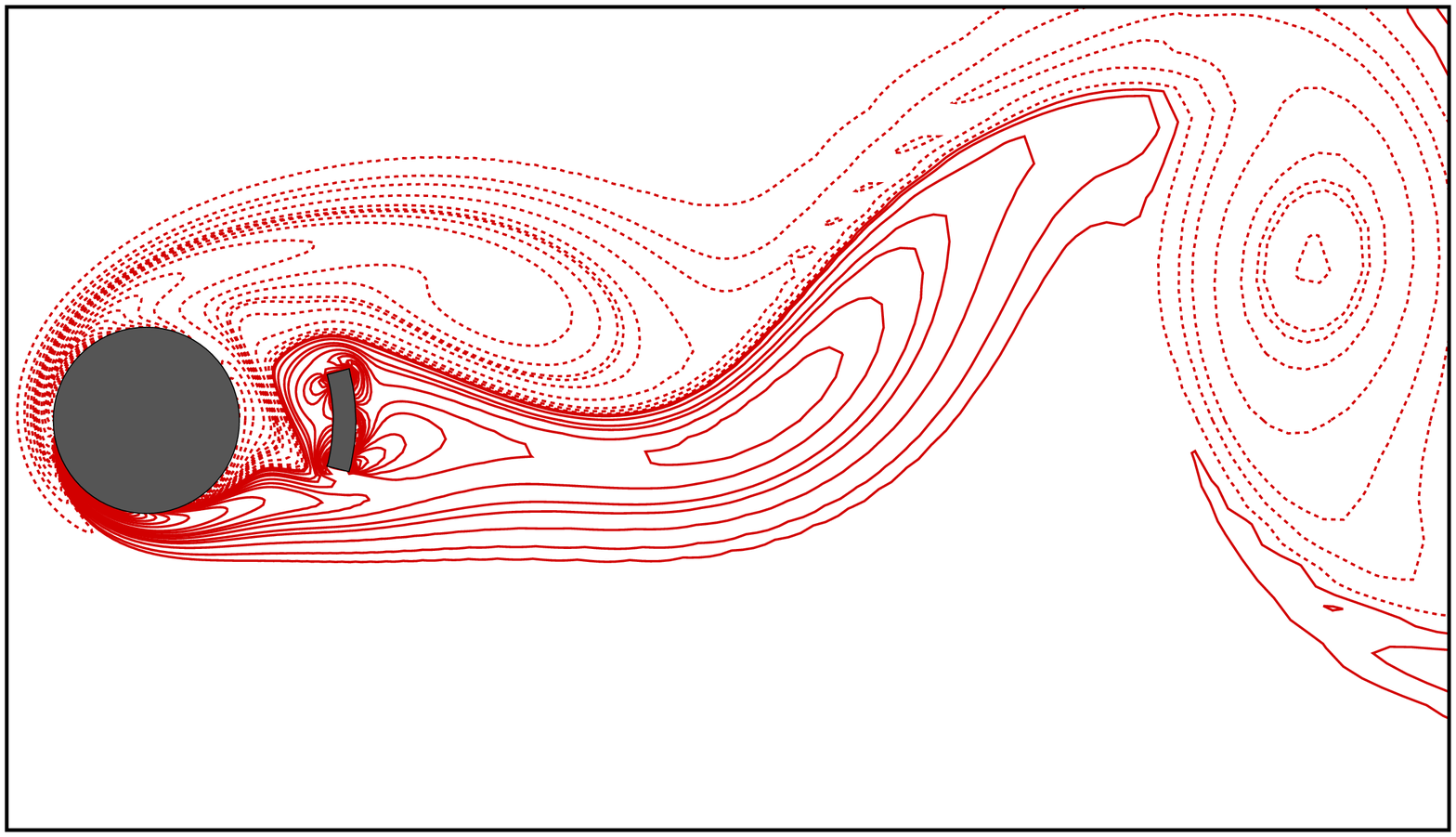}
\\
\hspace{0.5em}\scriptsize{$t=t_0+(3/4)T$}
\\
\includegraphics[width=0.29\textwidth,trim={0.5cm 0.3cm 0.5cm 0.3cm},clip]{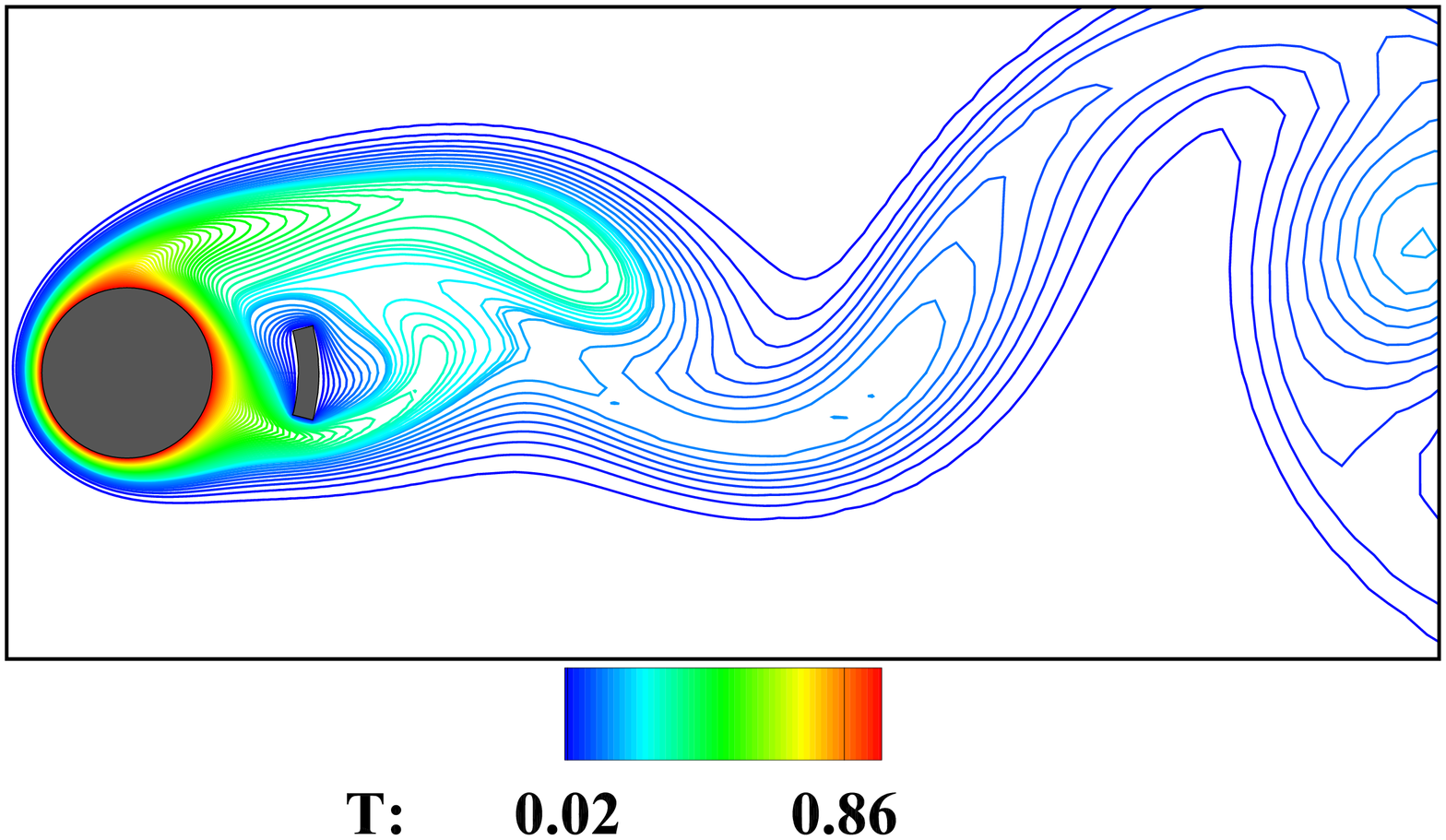}
\includegraphics[width=0.3\textwidth,trim={0.5cm 0.3cm 0.3cm 0.3cm},clip]{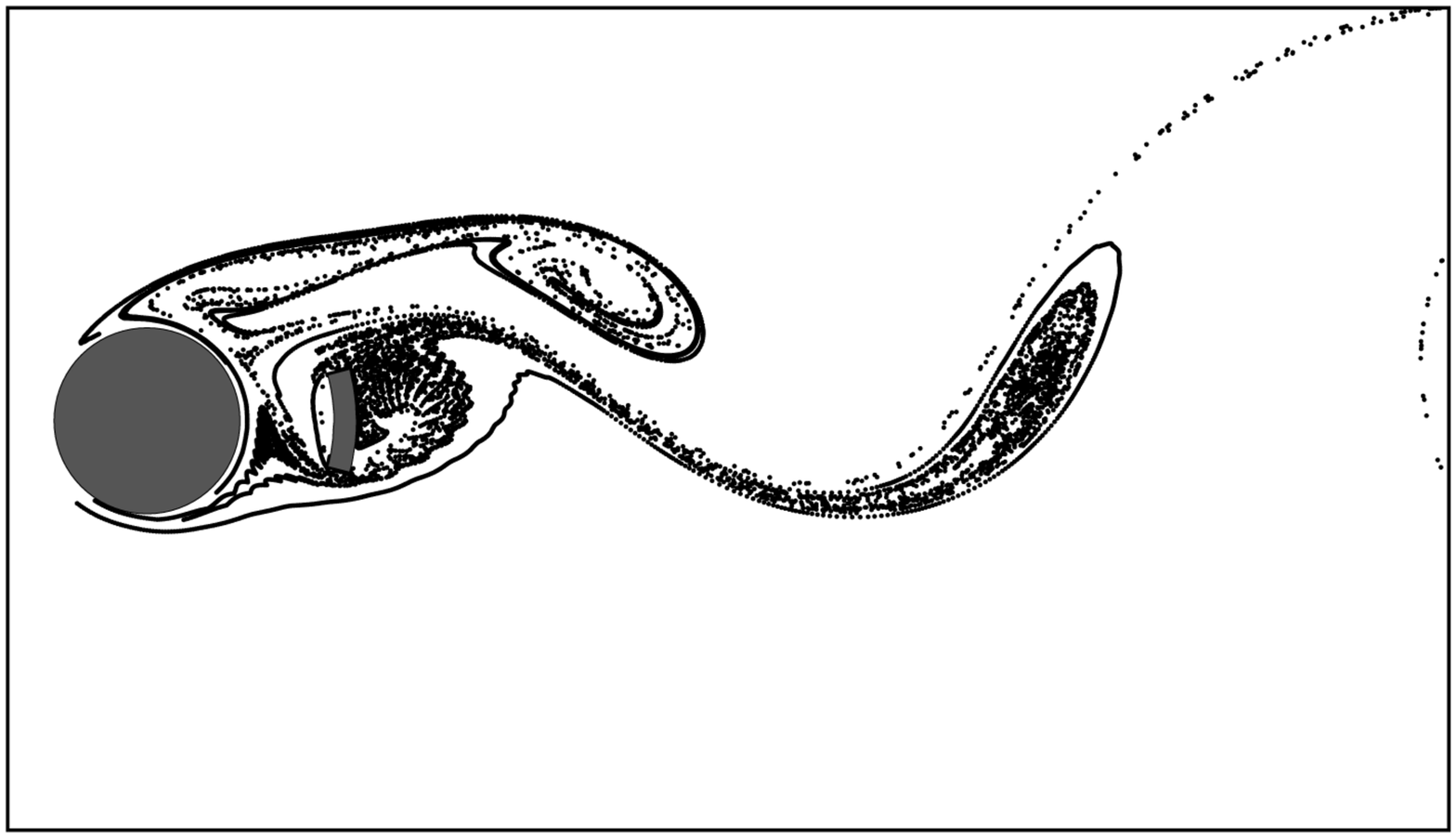}
\includegraphics[width=0.3\textwidth,trim={0.5cm 0.3cm 0.3cm 0.3cm},clip]{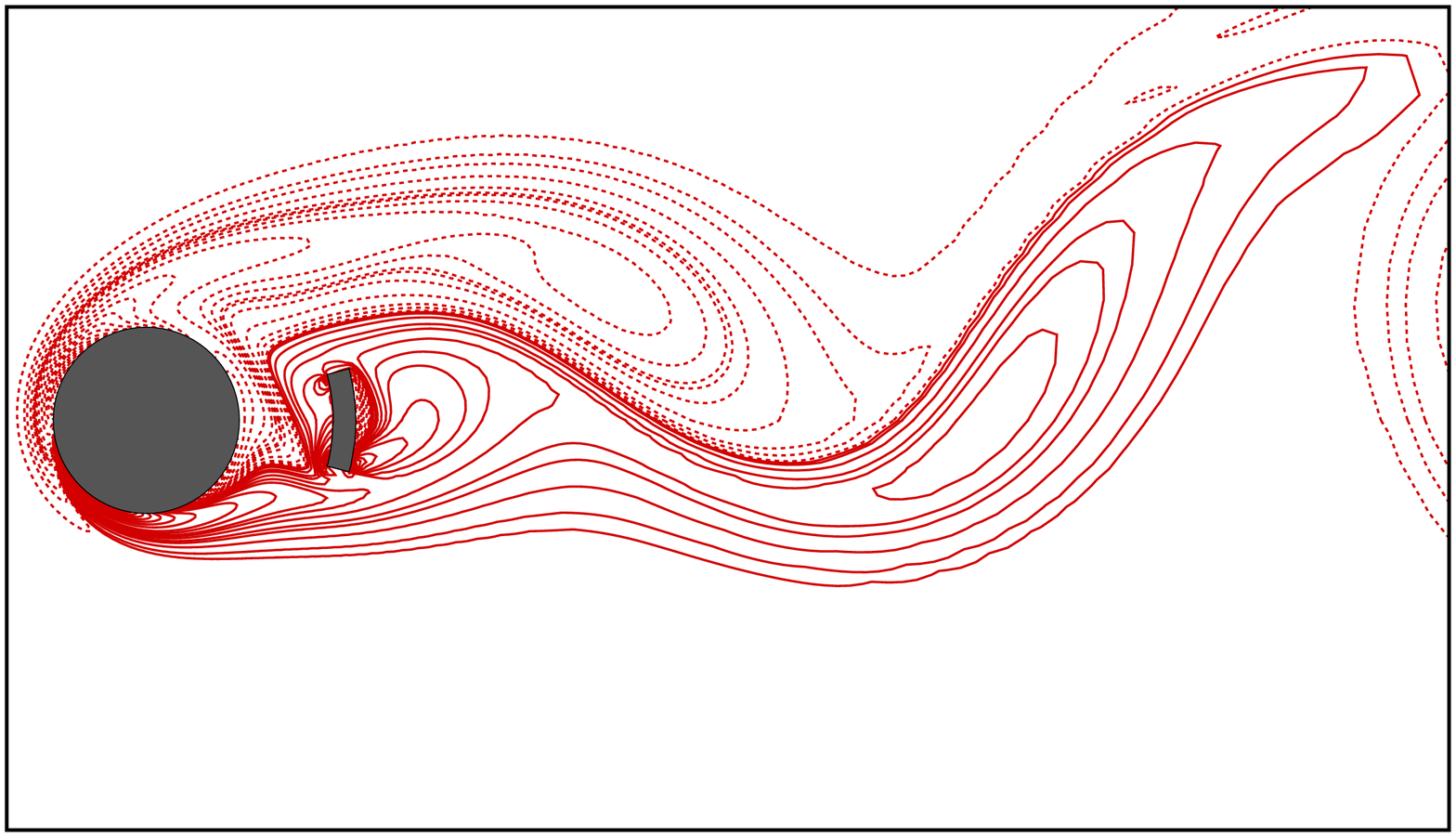}
\\
\hspace{0.5em}\scriptsize{$t=t_0+(1)T$}
\\
\includegraphics[width=0.29\textwidth,trim={0.5cm 0.3cm 0.5cm 0.3cm},clip]{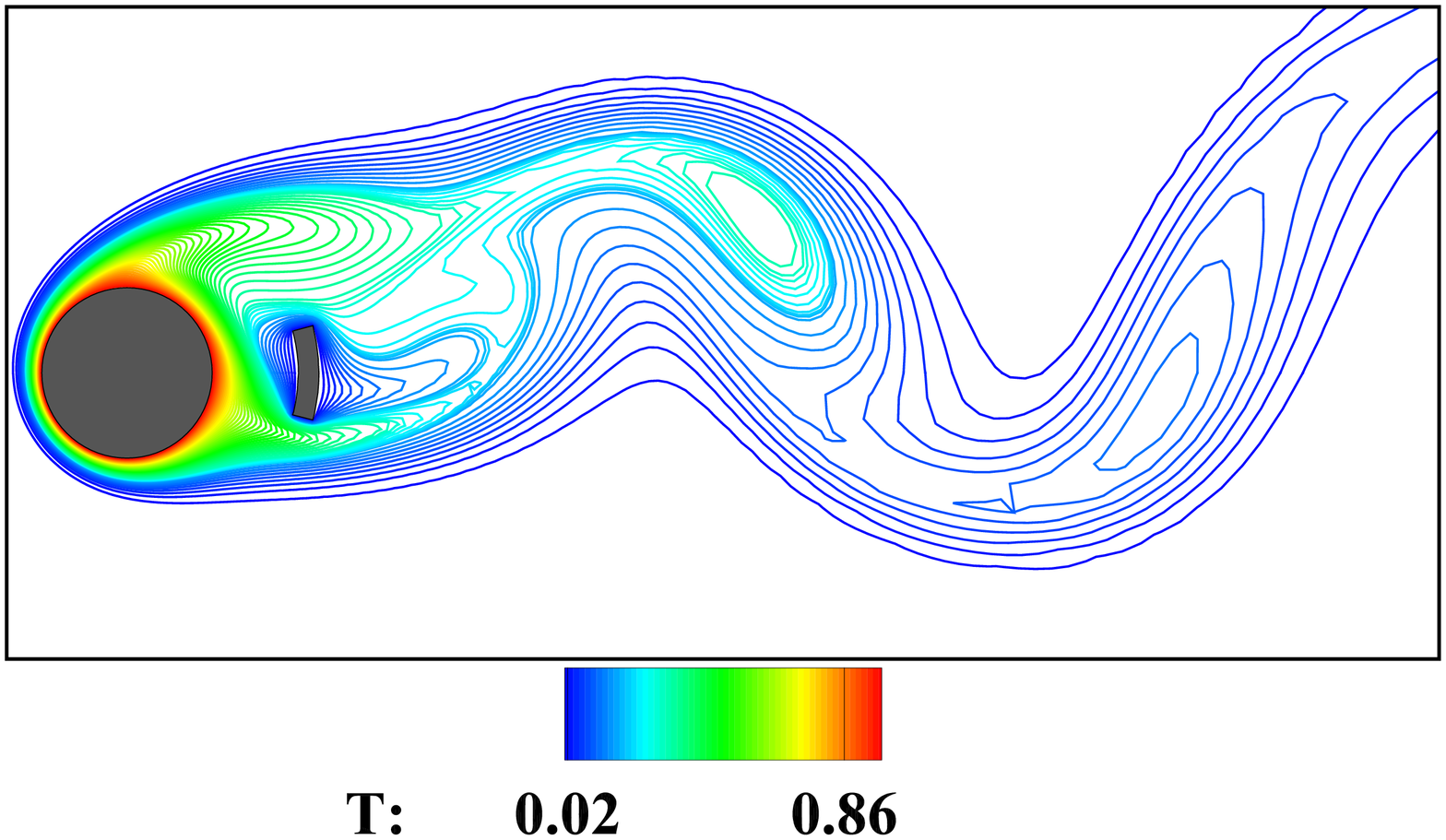}
\includegraphics[width=0.3\textwidth,trim={0.5cm 0.3cm 0.3cm 0.3cm},clip]{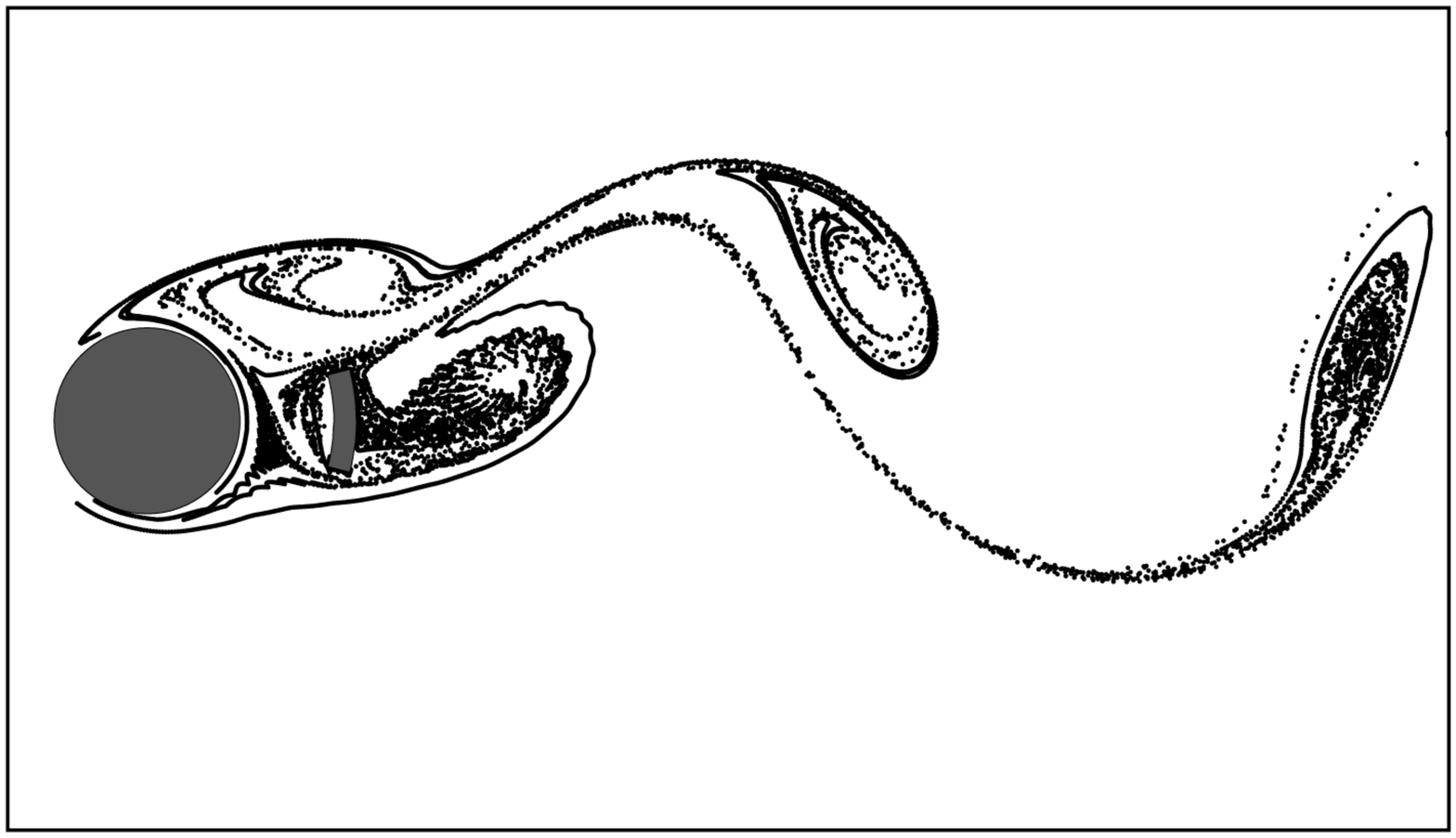}
\includegraphics[width=0.3\textwidth,trim={0.5cm 0.3cm 0.3cm 0.3cm},clip]{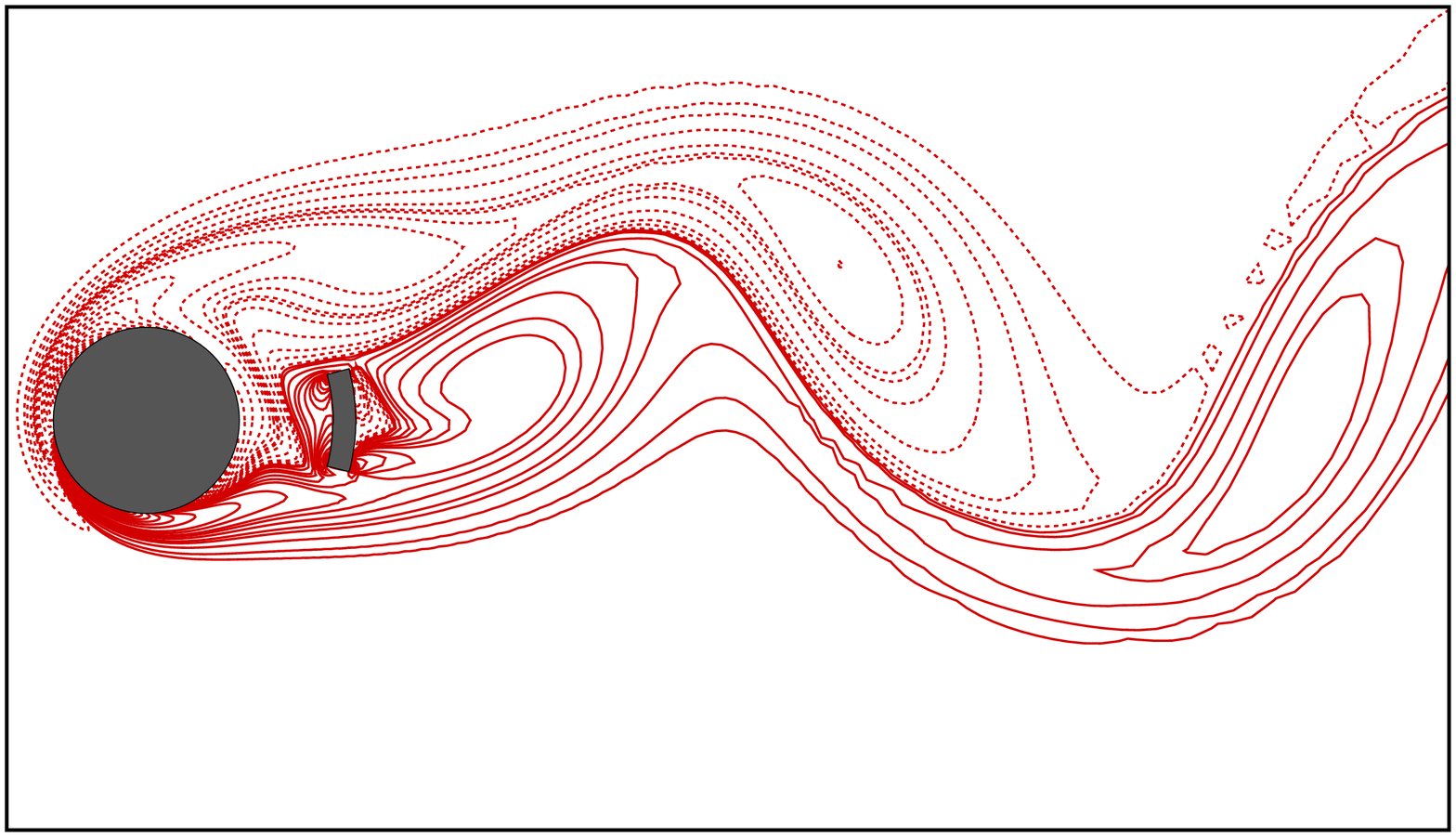}
\\
\hspace{2cm}(a) \hspace{4cm}(b) \hspace{4cm}(c)\hspace{2cm}
 \caption{(a) Isotherm, (b) streakline and (c) vorticity contour for $Pr=0.7$, $Re=150$, $\alpha=1$ and $d/R_0=1$ at different phases.}
 \label{fig:d_1_a_1-0}
\end{figure*}

\begin{figure*}[!t]
\centering
\scriptsize{$t=t_0+(0)T$}
\\
\includegraphics[width=0.3\textwidth,trim={0.5cm 0.3cm 0.5cm 0.3cm},clip]{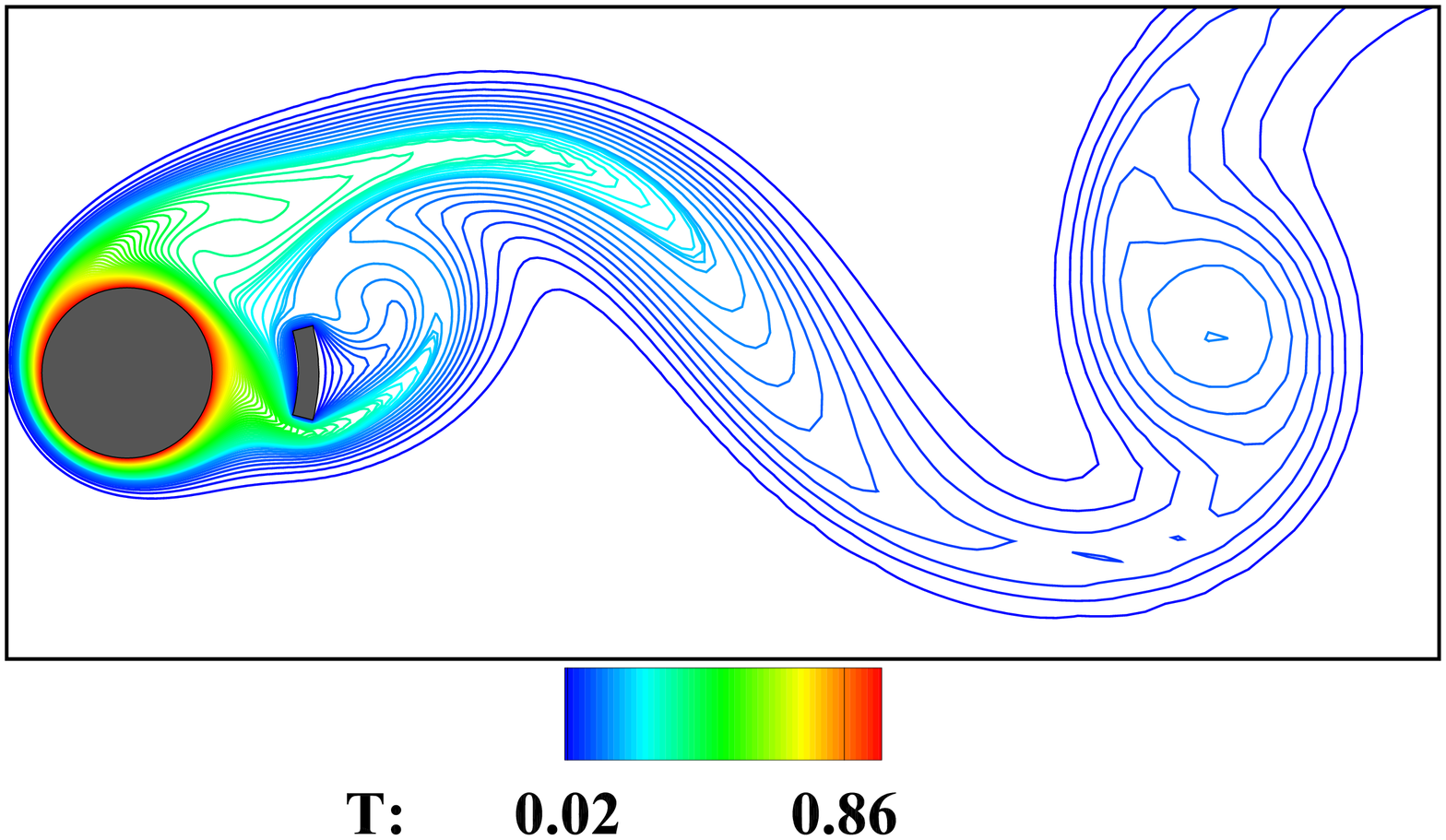}
\includegraphics[width=0.3\textwidth,trim={0.5cm 0.3cm 0.3cm 0.3cm},clip]{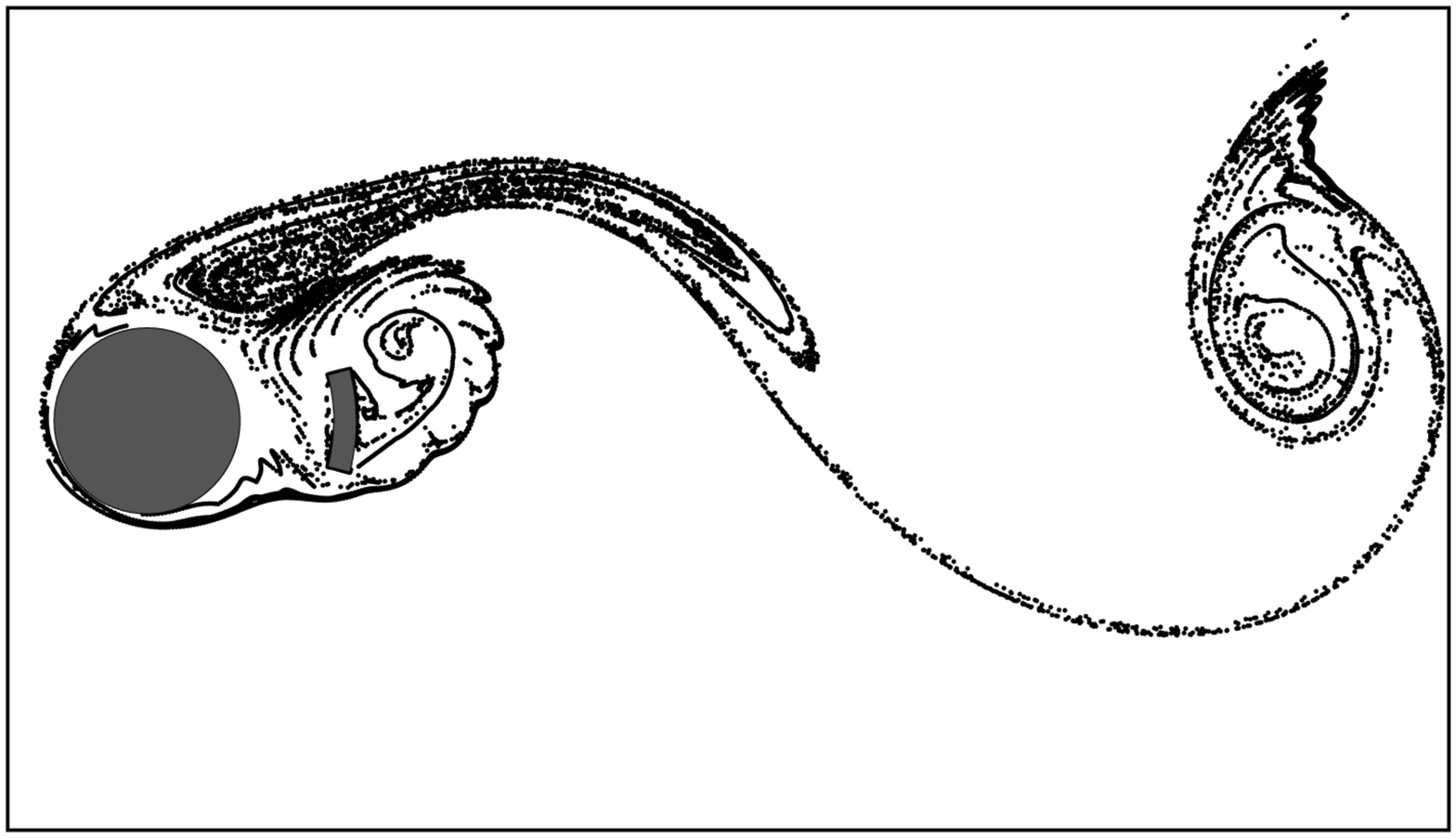}
\includegraphics[width=0.3\textwidth,trim={0.5cm 0.3cm 0.3cm 0.3cm},clip]{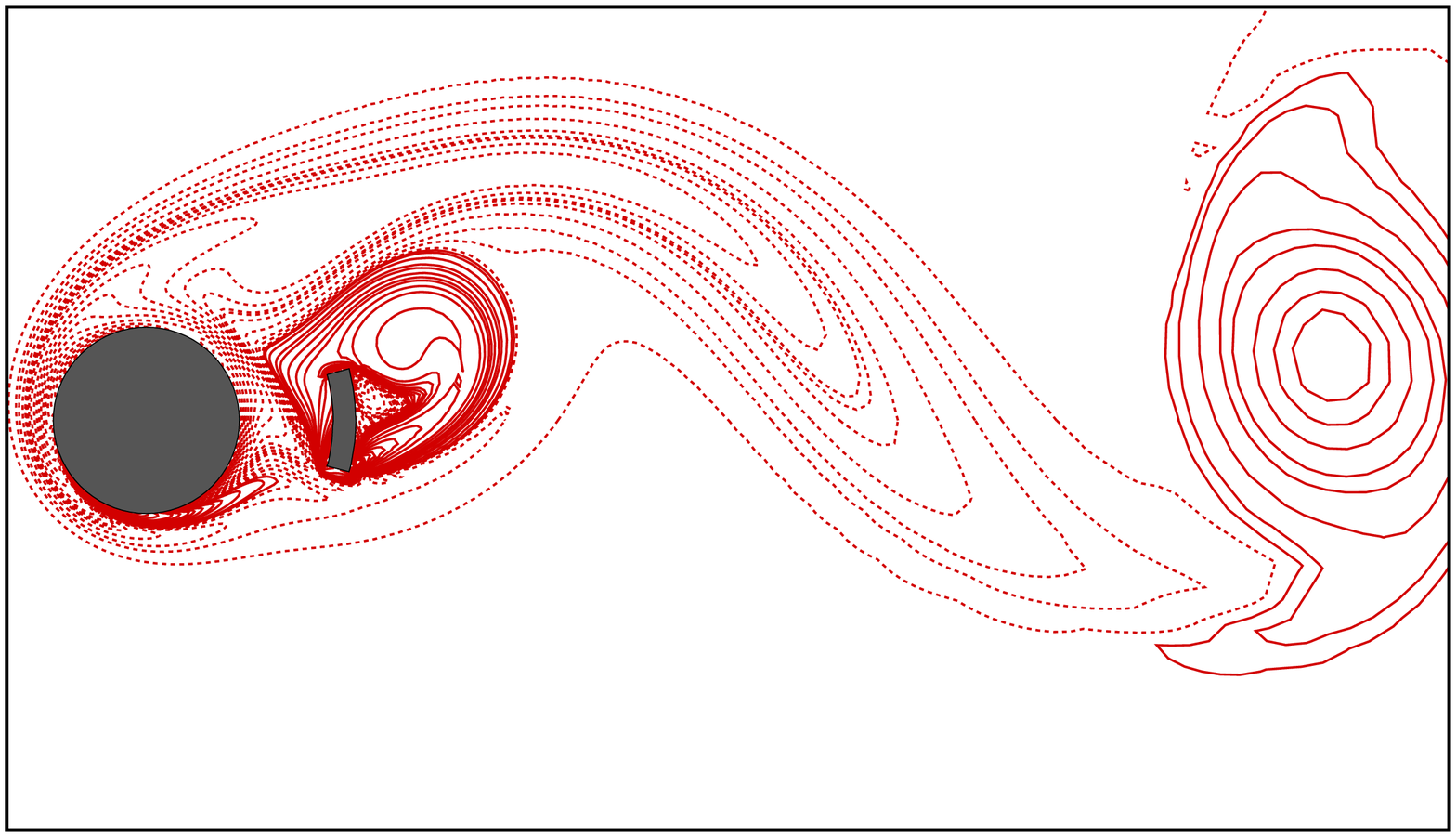}
\\
\hspace{0.5em}\scriptsize{$t=t_0+(1/4)T$}
\\
\includegraphics[width=0.29\textwidth,trim={0.5cm 0.3cm 0.5cm 0.3cm},clip]{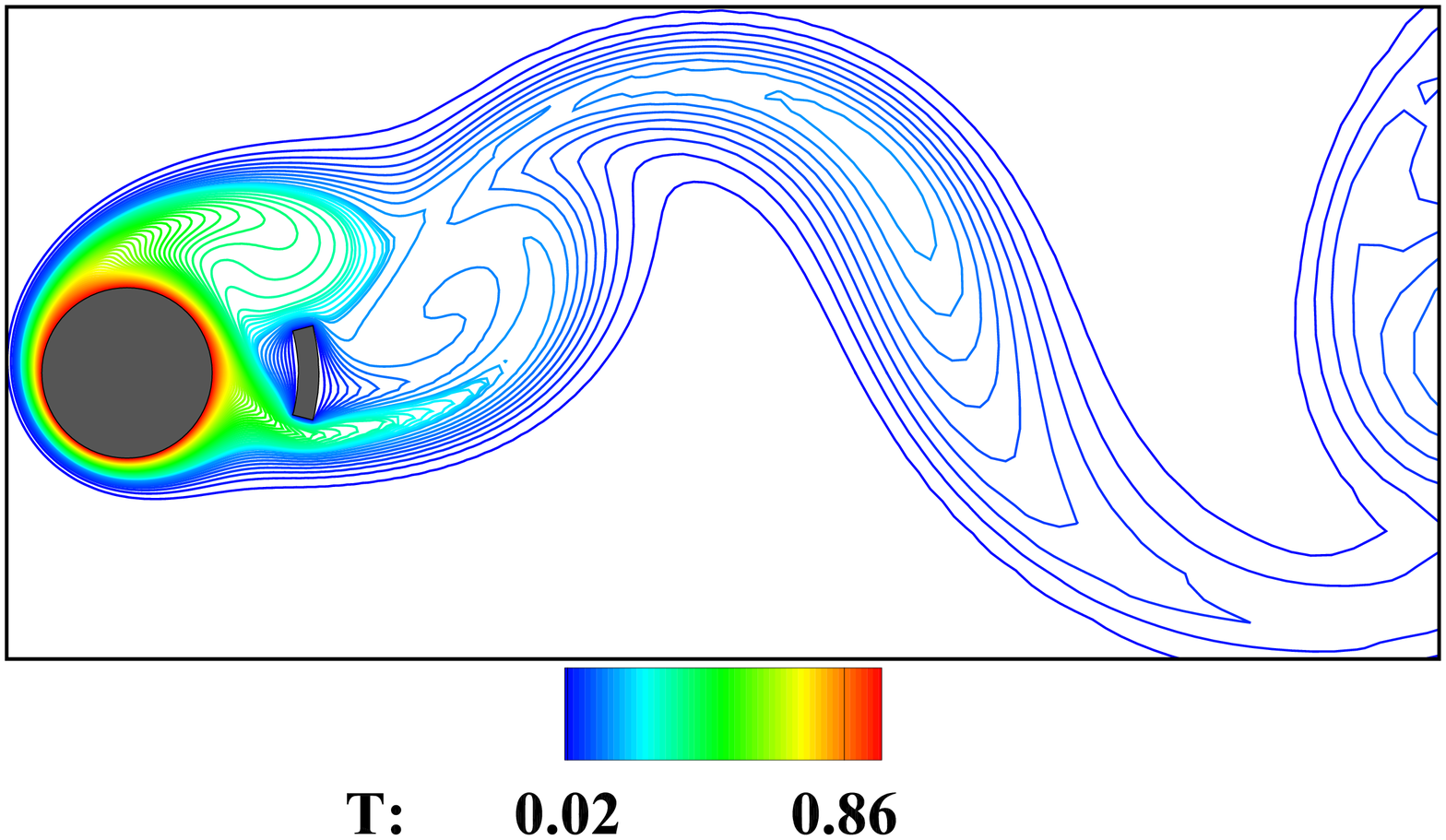}
\includegraphics[width=0.3\textwidth,trim={0.5cm 0.3cm 0.3cm 0.3cm},clip]{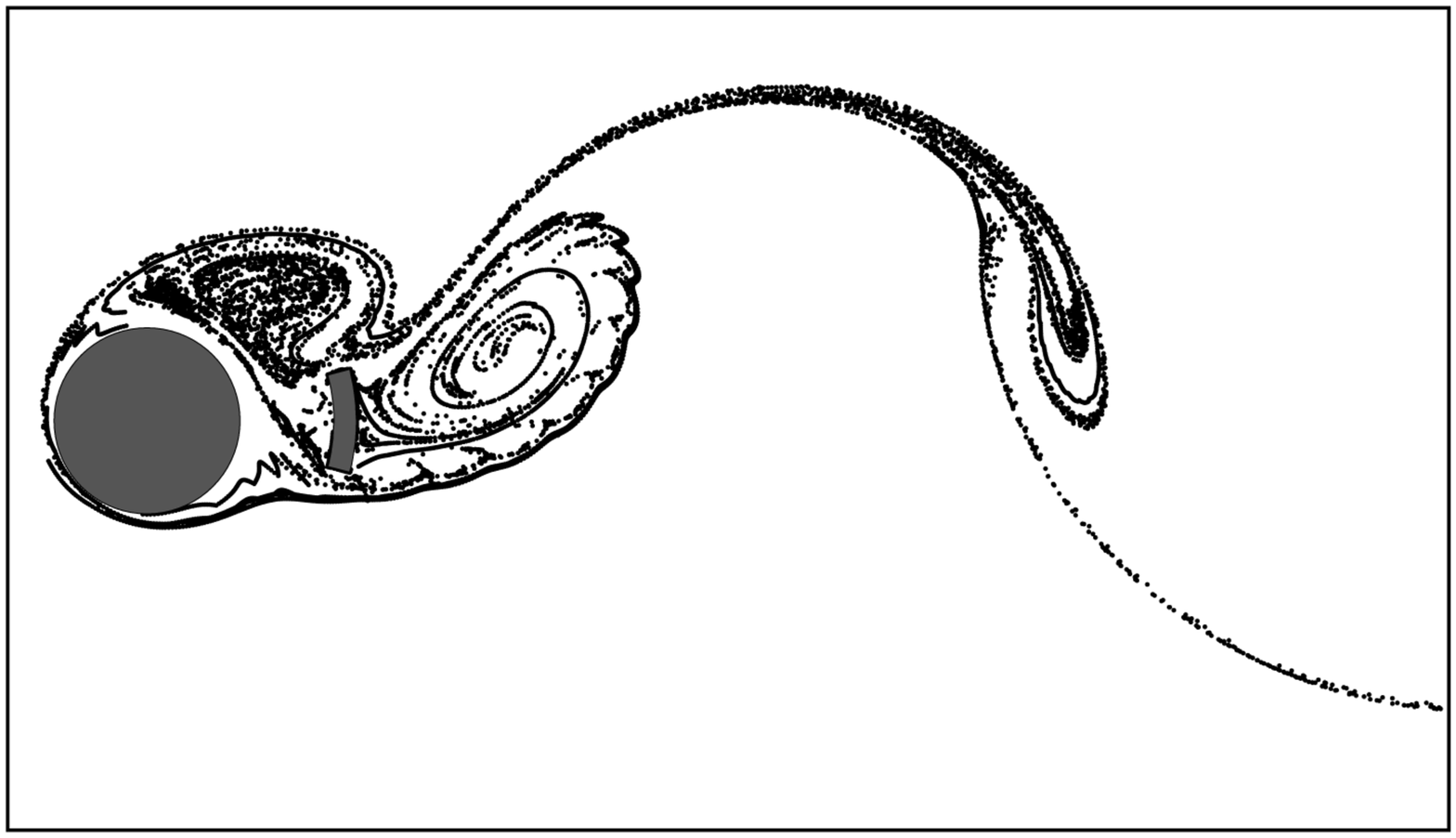}
\includegraphics[width=0.3\textwidth,trim={0.5cm 0.3cm 0.3cm 0.3cm},clip]{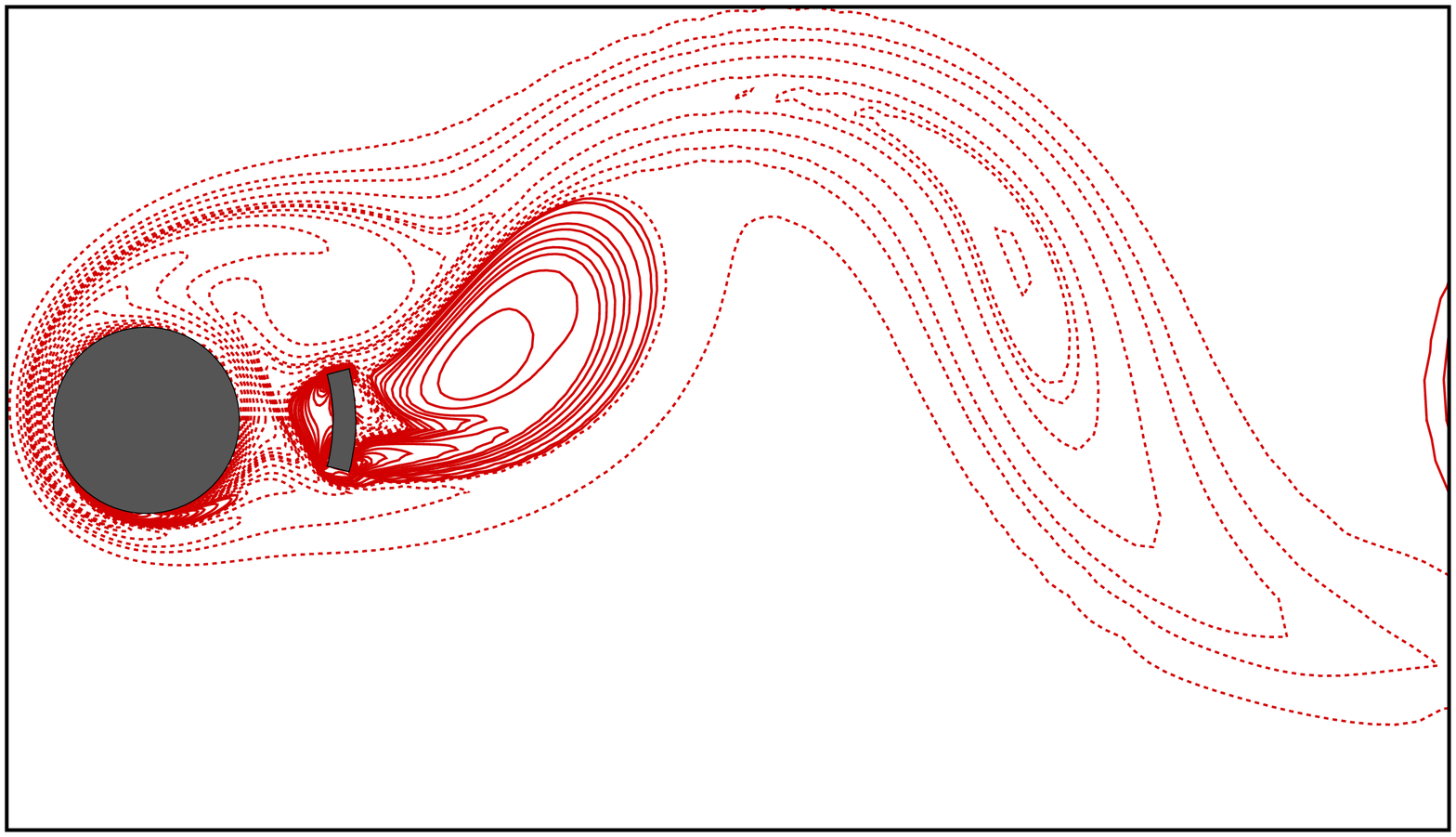}
\\
\hspace{0.5em}\scriptsize{$t=t_0+(1/2)T$}
\\
\includegraphics[width=0.29\textwidth,trim={0.5cm 0.3cm 0.5cm 0.3cm},clip]{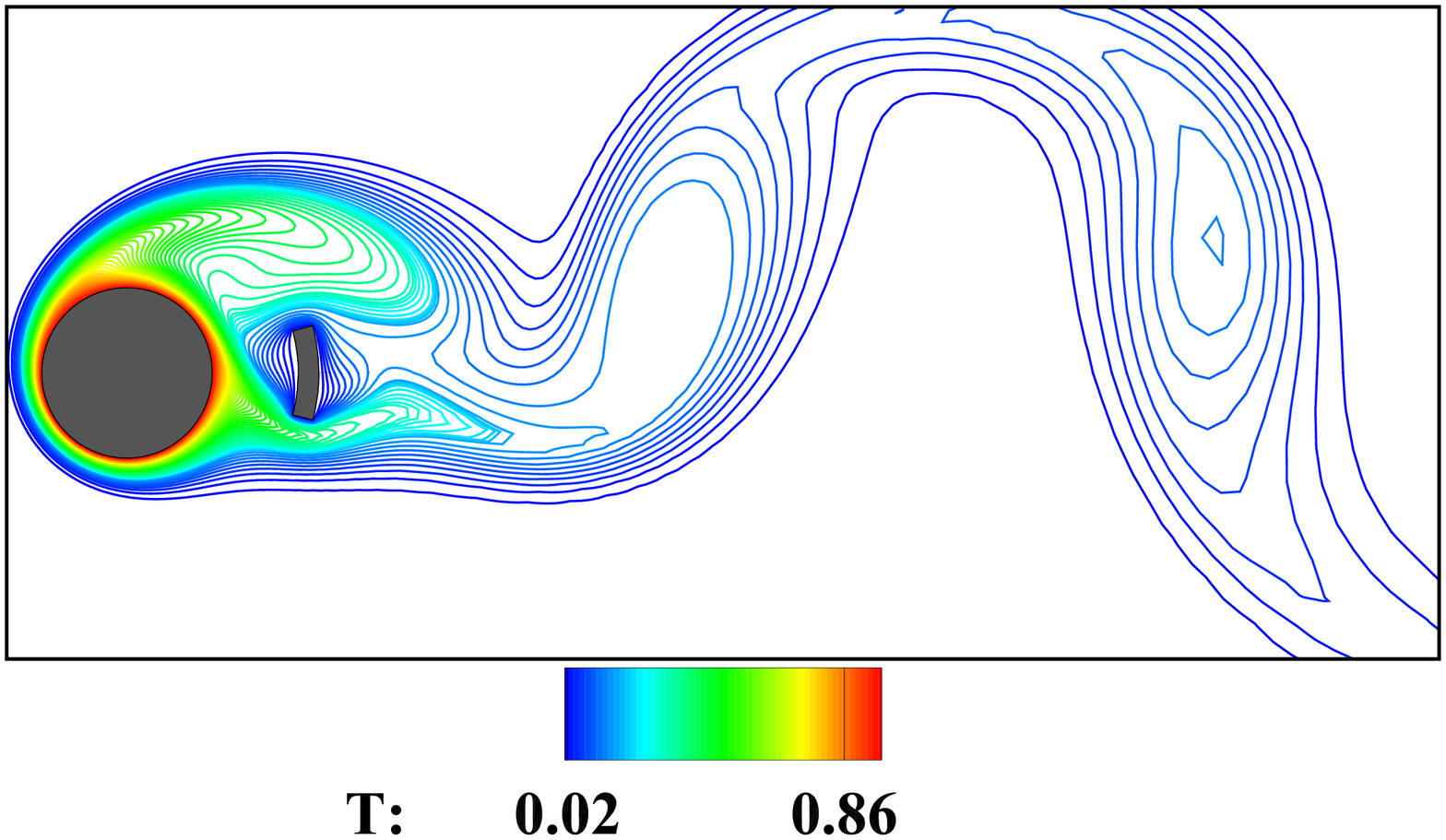}
\includegraphics[width=0.3\textwidth,trim={0.5cm 0.3cm 0.3cm 0.3cm},clip]{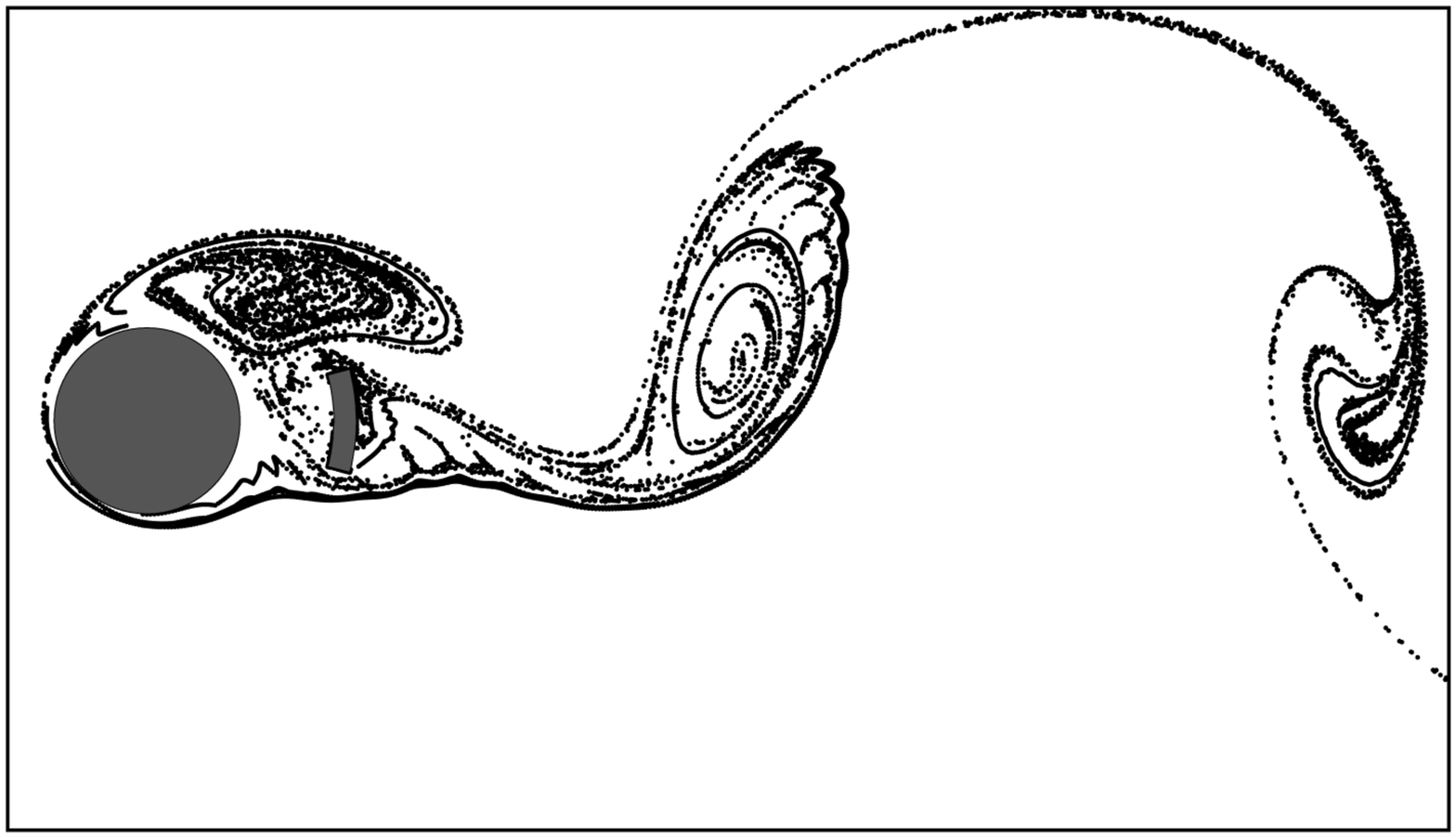}
\includegraphics[width=0.3\textwidth,trim={0.5cm 0.3cm 0.3cm 0.3cm},clip]{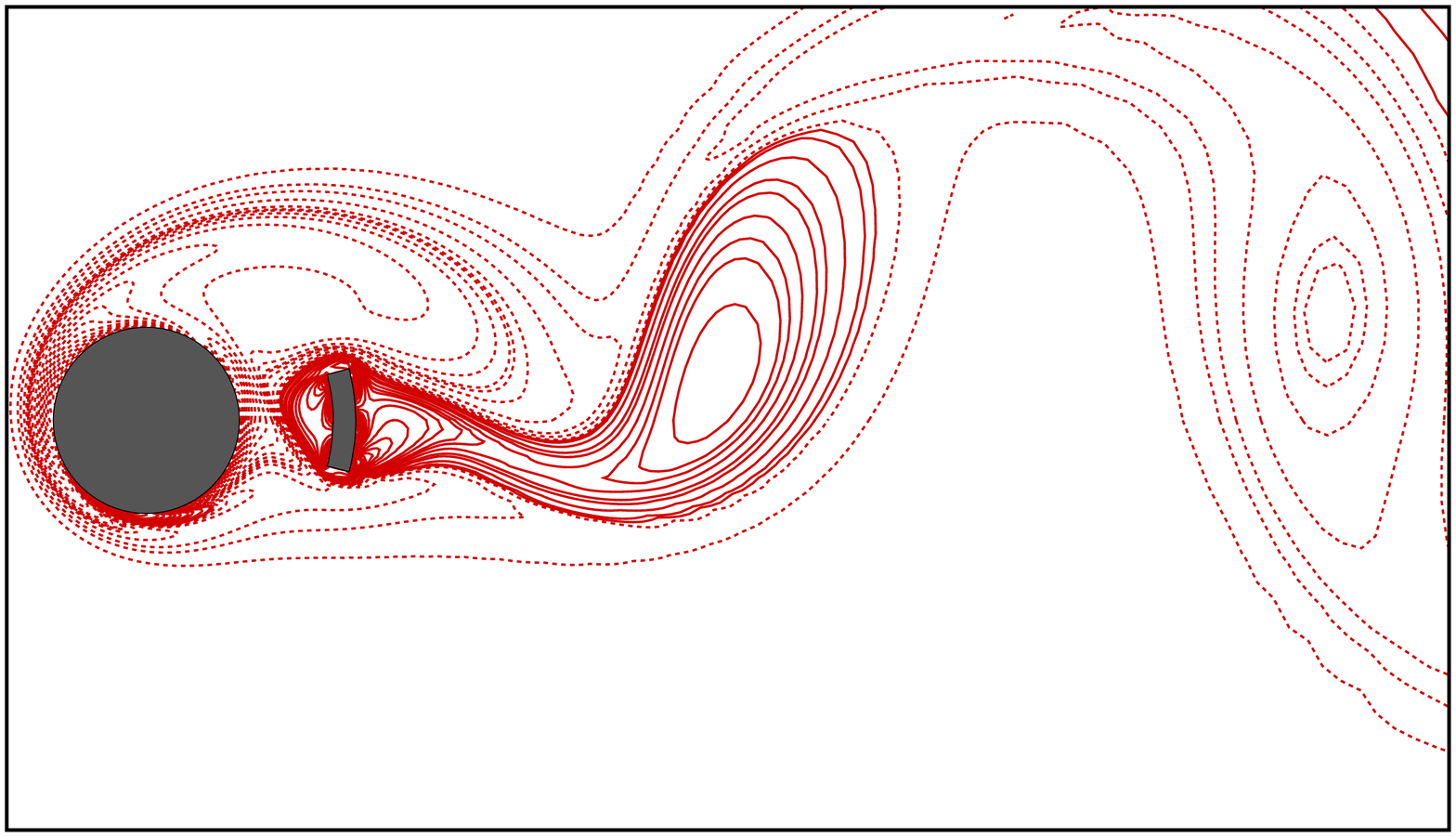}
\\
\hspace{0.5em}\scriptsize{$t=t_0+(3/4)T$}
\\
\includegraphics[width=0.29\textwidth,trim={0.5cm 0.3cm 0.5cm 0.3cm},clip]{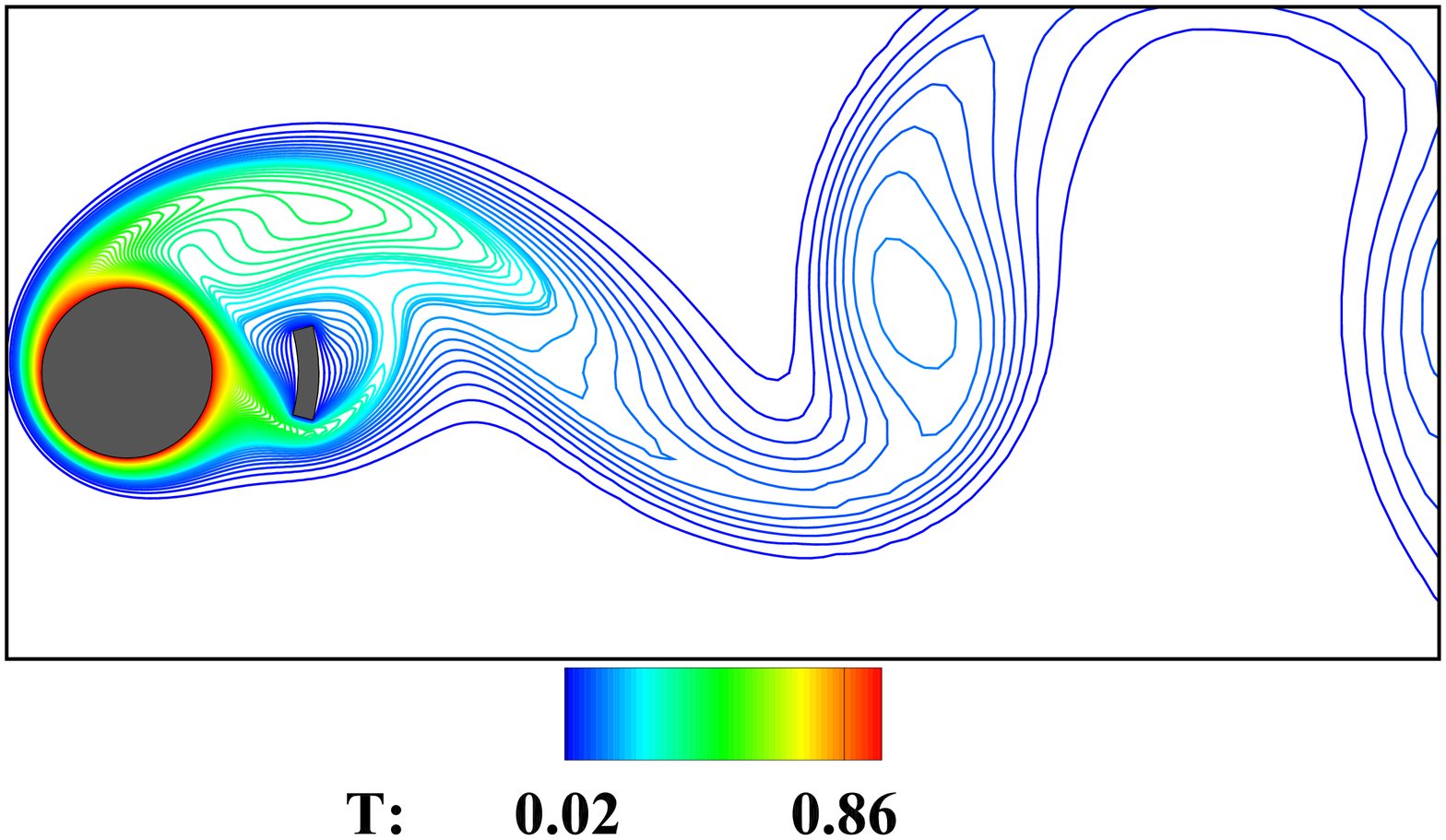}
\includegraphics[width=0.3\textwidth,trim={0.5cm 0.3cm 0.3cm 0.3cm},clip]{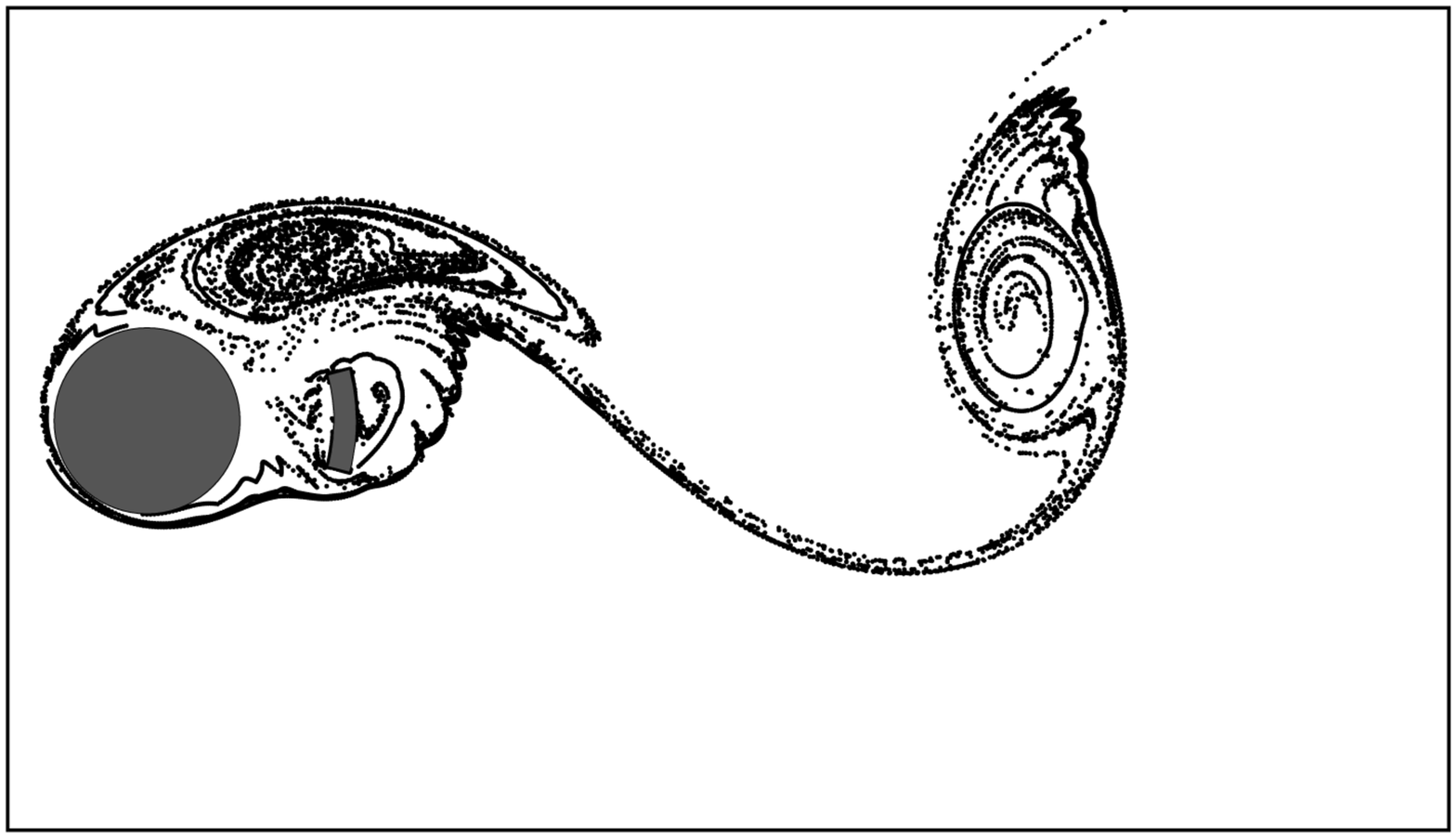}
\includegraphics[width=0.3\textwidth,trim={0.5cm 0.3cm 0.3cm 0.3cm},clip]{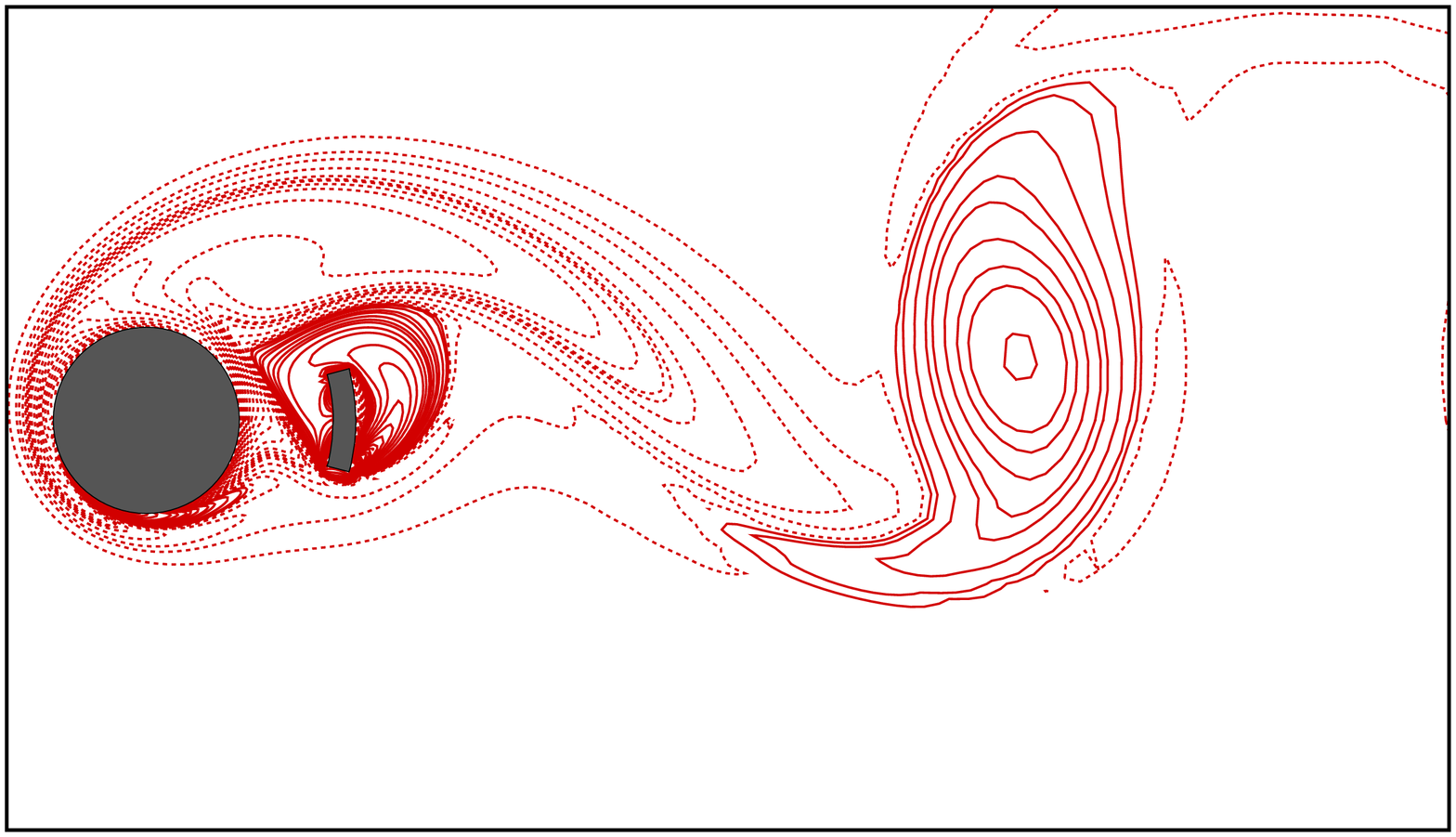}
\\
\hspace{0.5em}\scriptsize{$t=t_0+(1)T$}
\\
\includegraphics[width=0.29\textwidth,trim={0.5cm 0.3cm 0.5cm 0.3cm},clip]{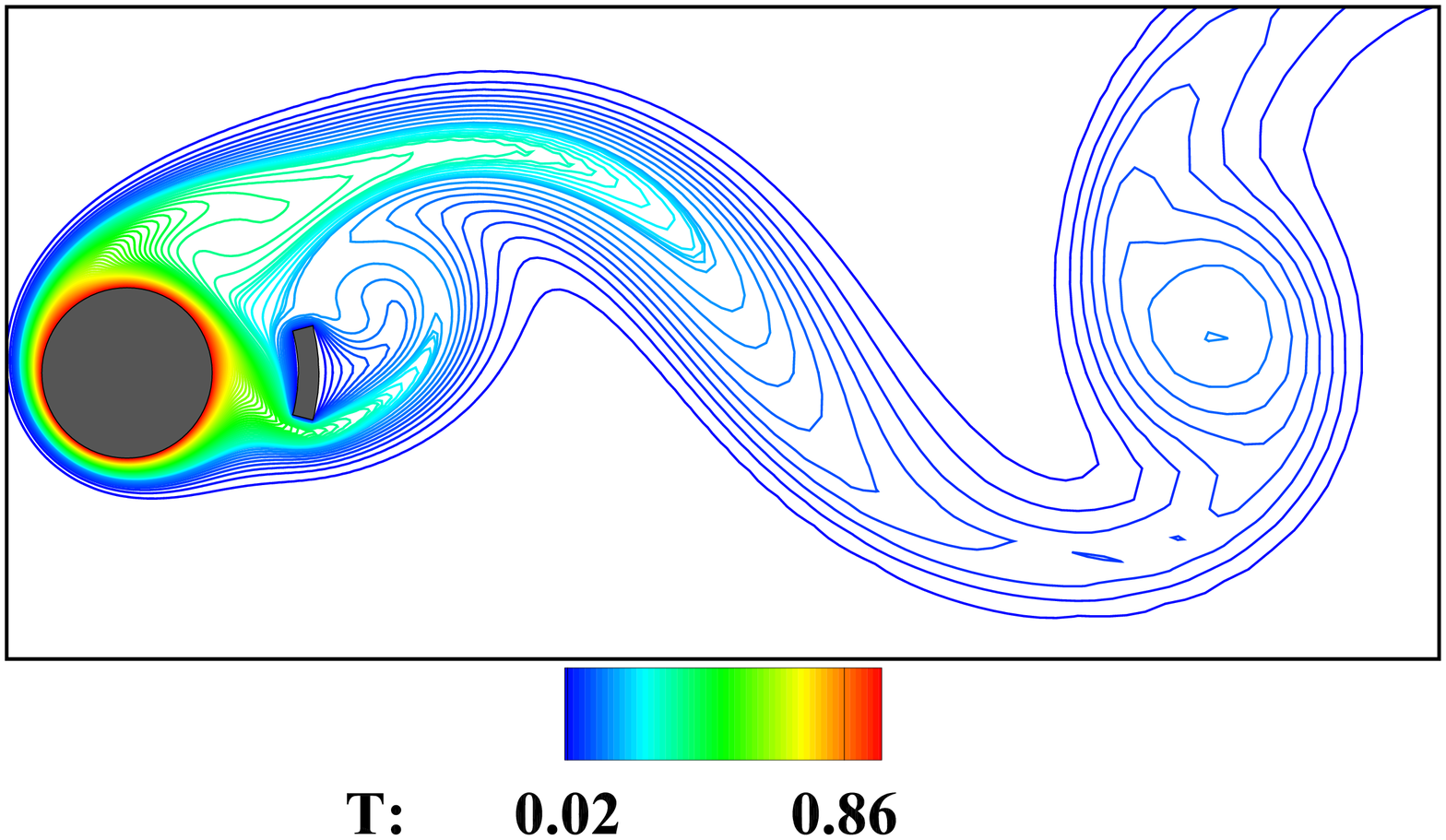}
\includegraphics[width=0.3\textwidth,trim={0.5cm 0.3cm 0.3cm 0.3cm},clip]{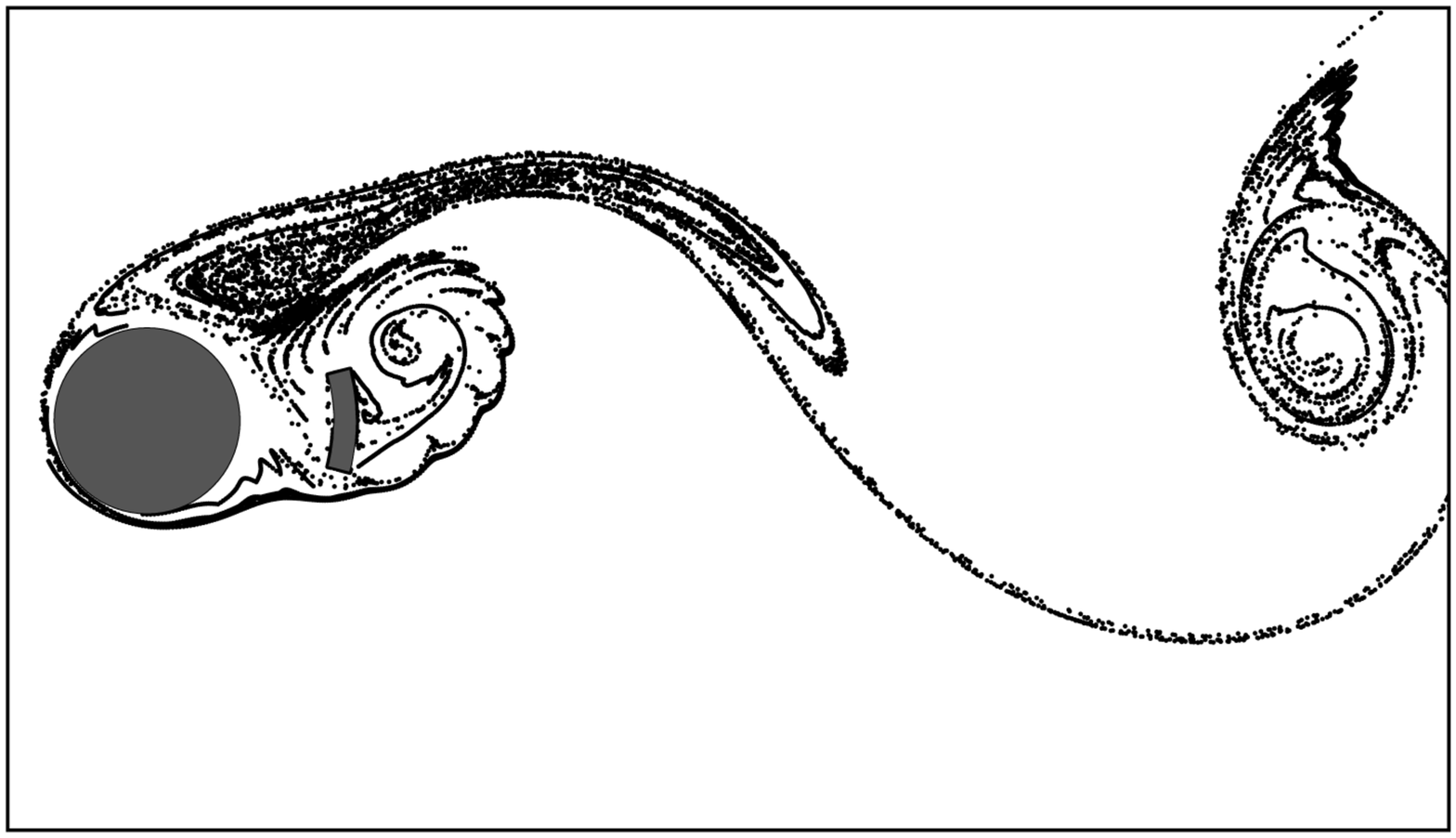}
\includegraphics[width=0.3\textwidth,trim={0.5cm 0.3cm 0.3cm 0.3cm},clip]{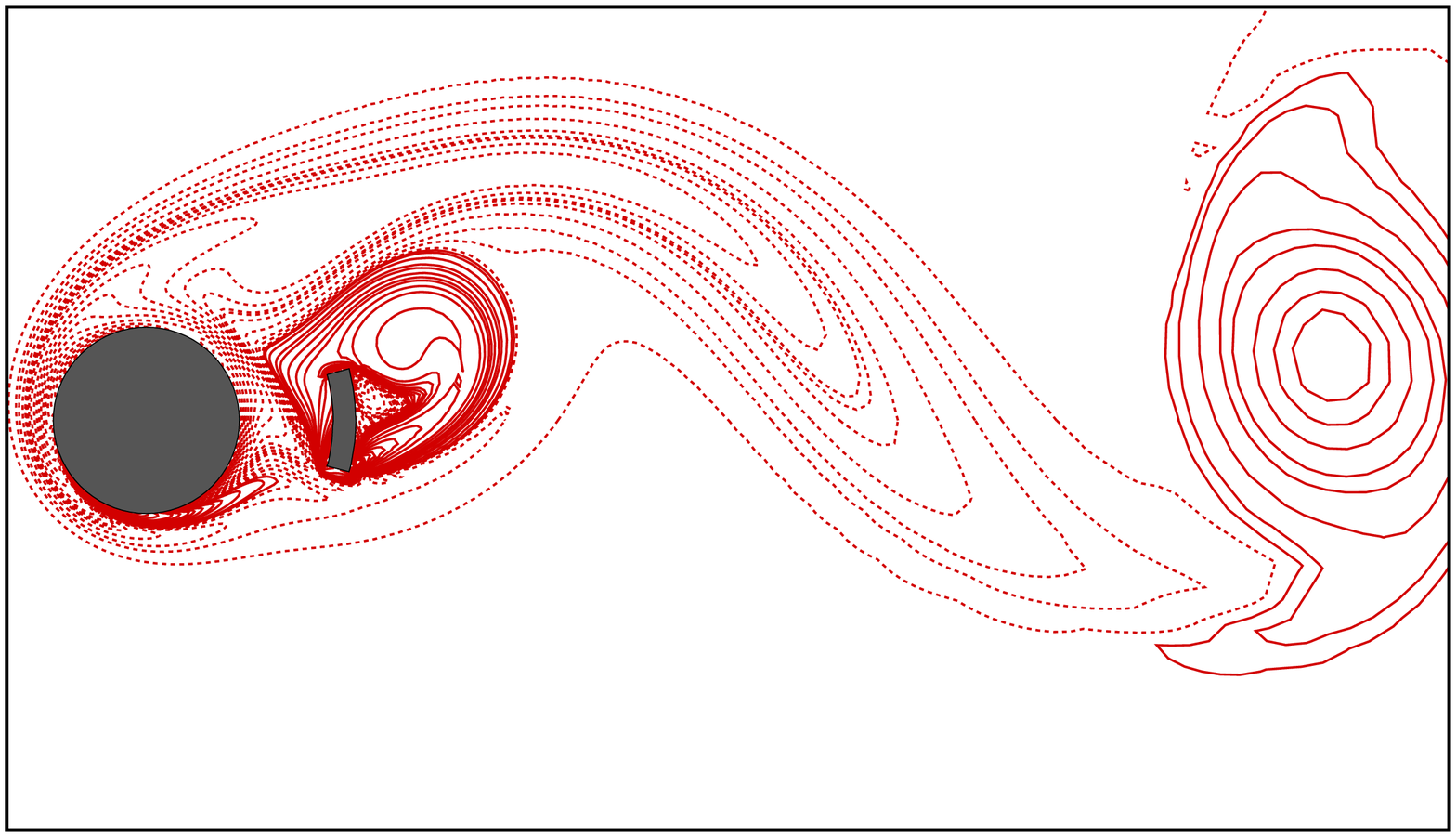}
\\
\hspace{2cm}(a) \hspace{4cm}(b) \hspace{4cm}(c)\hspace{2cm}
 \caption{(a) Isotherm, (b) streakline and (c) vorticity contour for $Pr=0.7$, $Re=150$, $\alpha=2.07$ and $d/R_0=1$ at different phases.}
 \label{fig:d_1_a_2-07}
\end{figure*}

\begin{figure*}[!t]
\centering
\scriptsize{$t=t_0+(0)T$}
\\
\includegraphics[width=0.29\textwidth,trim={0.5cm 0.3cm 0.5cm 0.3cm},clip]{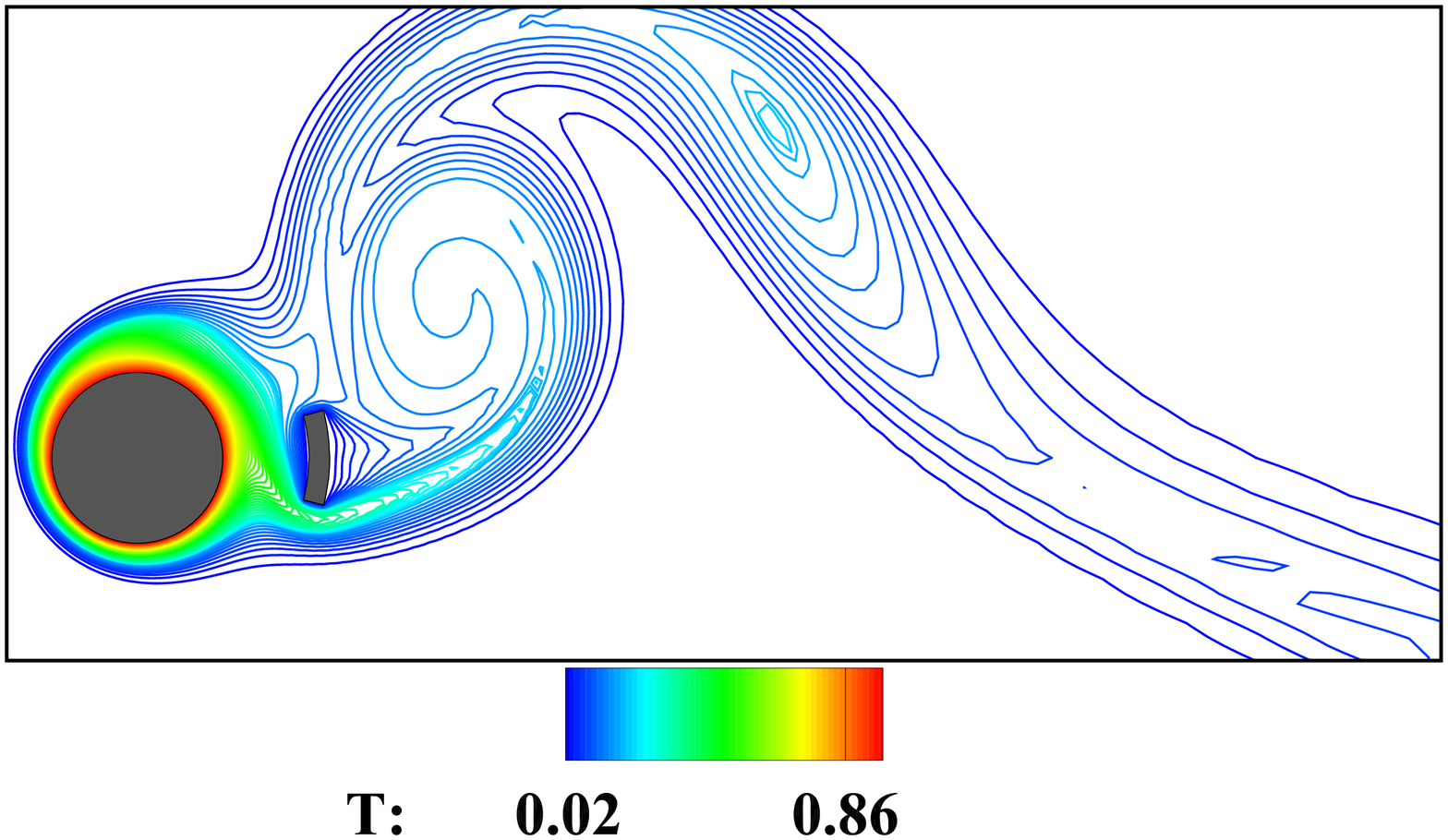}
\includegraphics[width=0.3\textwidth,trim={0.5cm 0.3cm 0.3cm 0.3cm},clip]{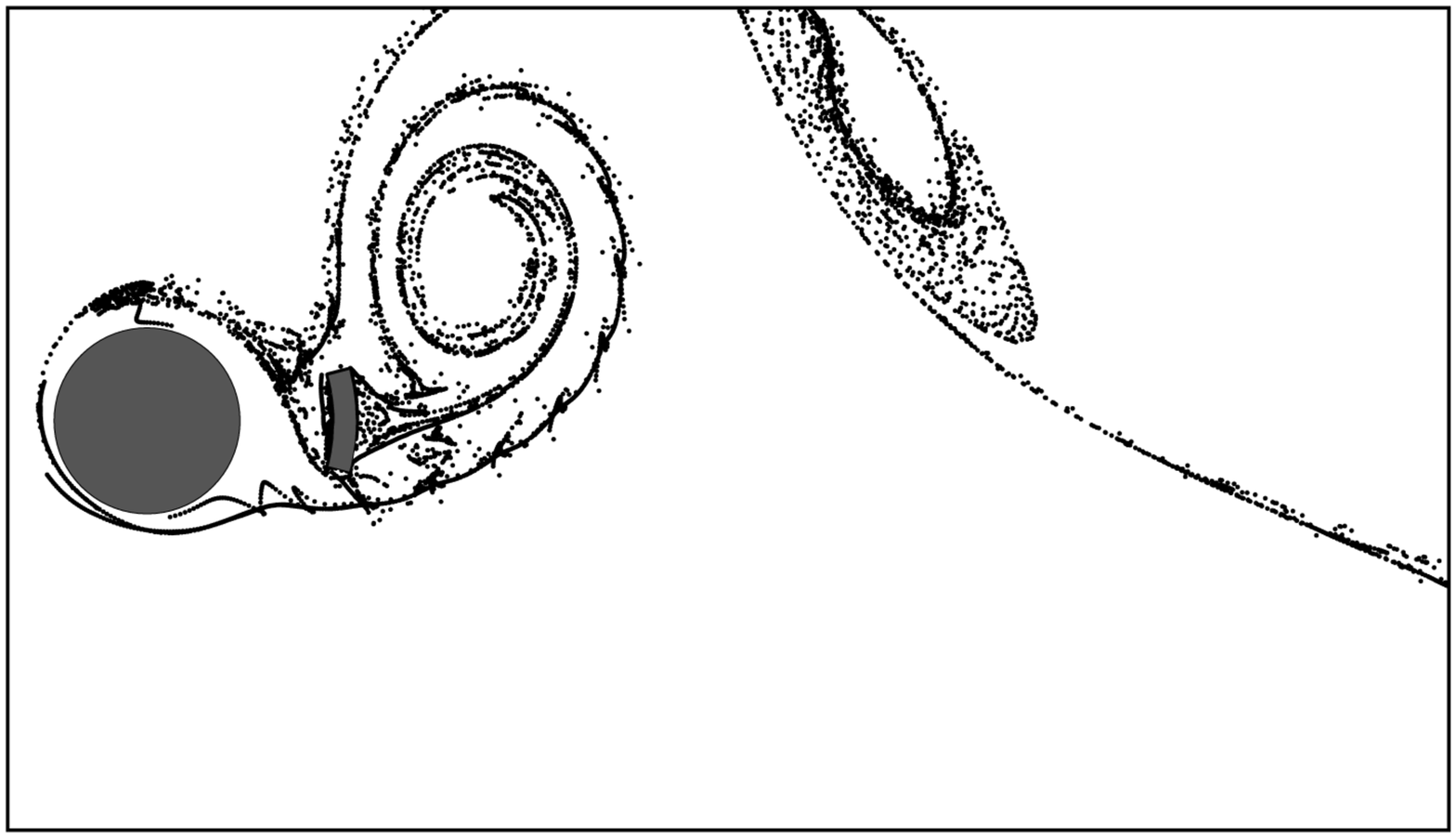}
\includegraphics[width=0.3\textwidth,trim={0.5cm 0.3cm 0.3cm 0.3cm},clip]{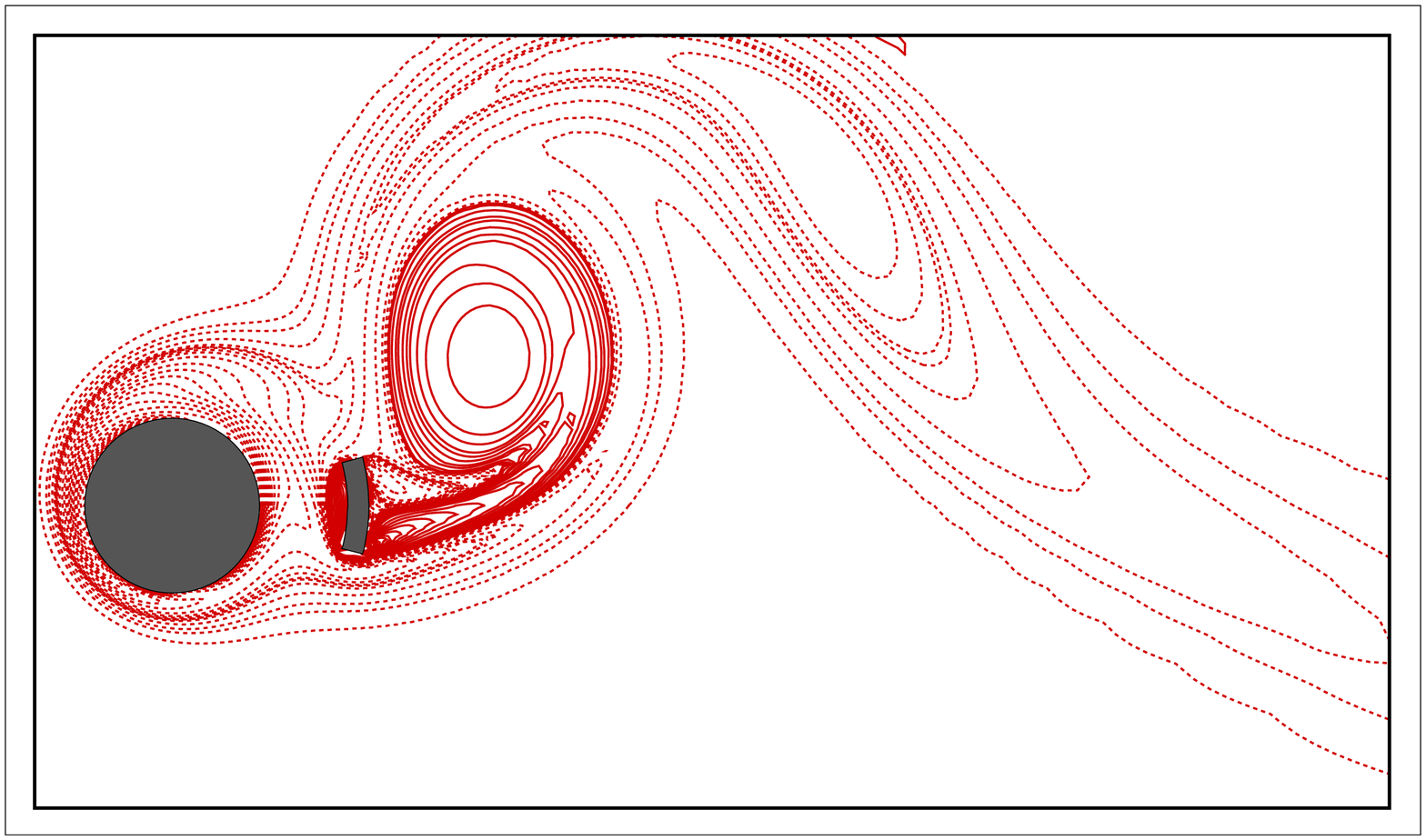}
\\
\hspace{0.5em}\scriptsize{$t=t_0+(1/4)T$}
\\
\includegraphics[width=0.29\textwidth,trim={0.5cm 0.3cm 0.5cm 0.3cm},clip]{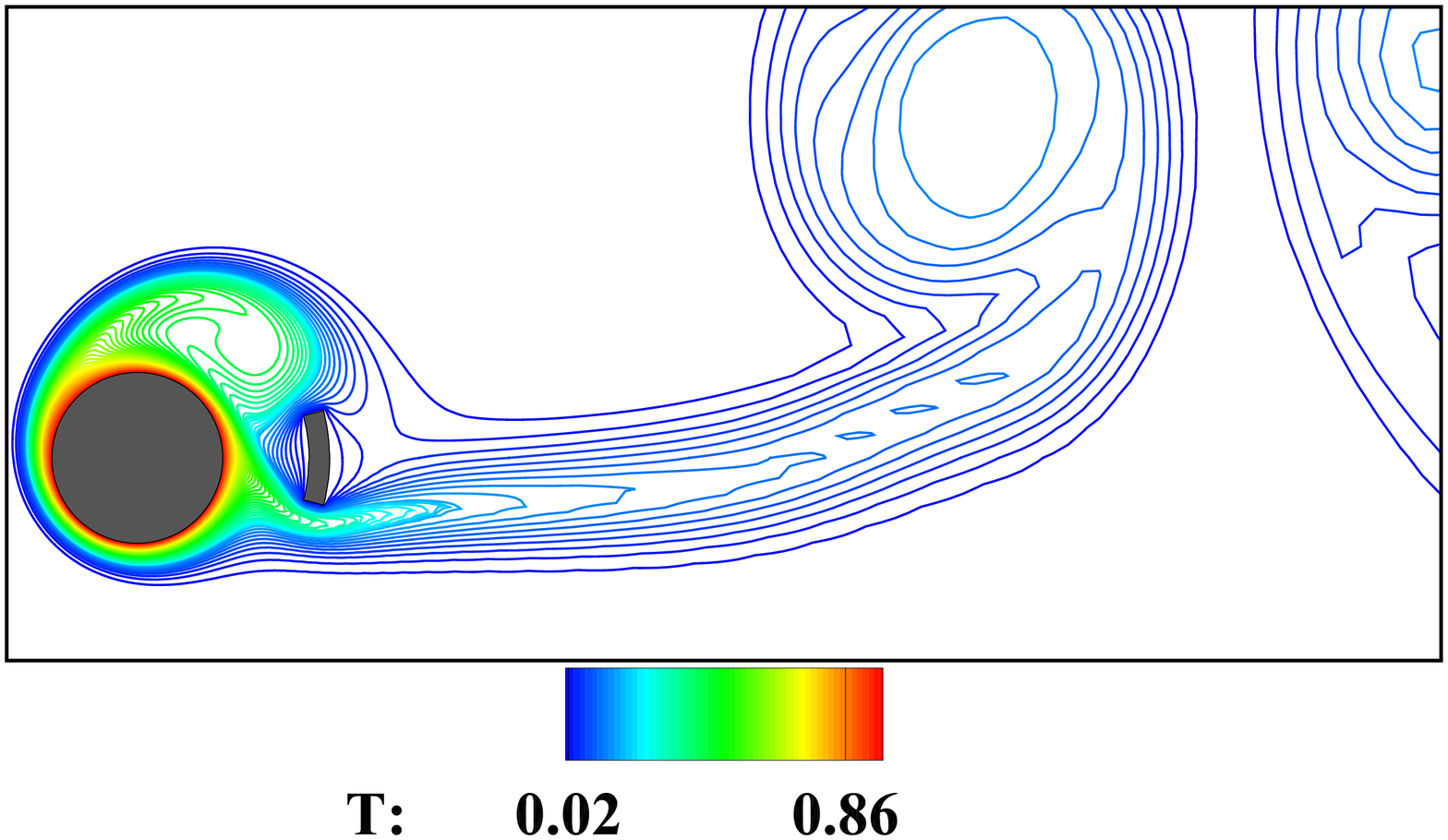}
\includegraphics[width=0.3\textwidth,trim={0.5cm 0.3cm 0.3cm 0.3cm},clip]{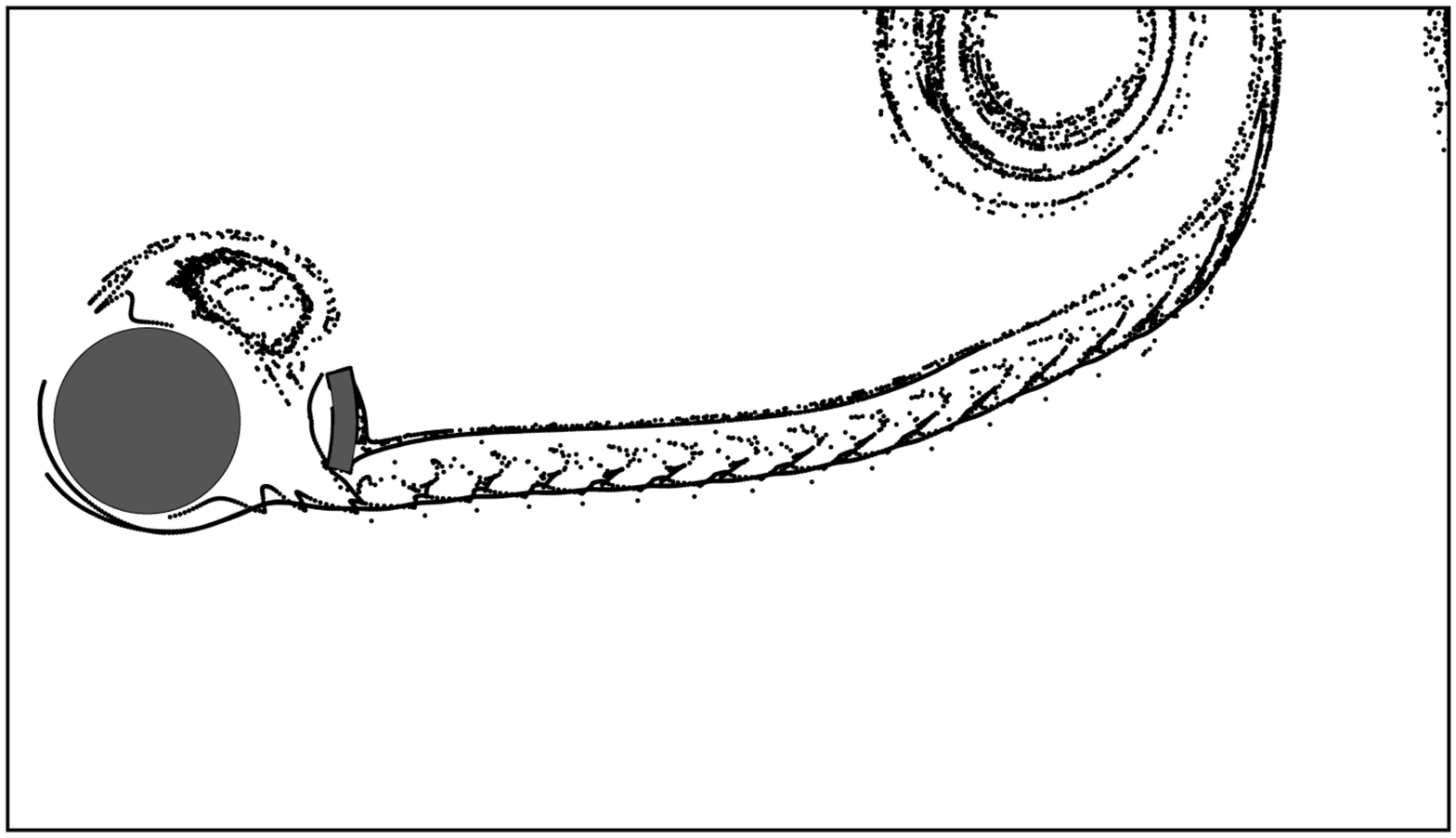}
\includegraphics[width=0.3\textwidth,trim={0.5cm 0.3cm 0.3cm 0.3cm},clip]{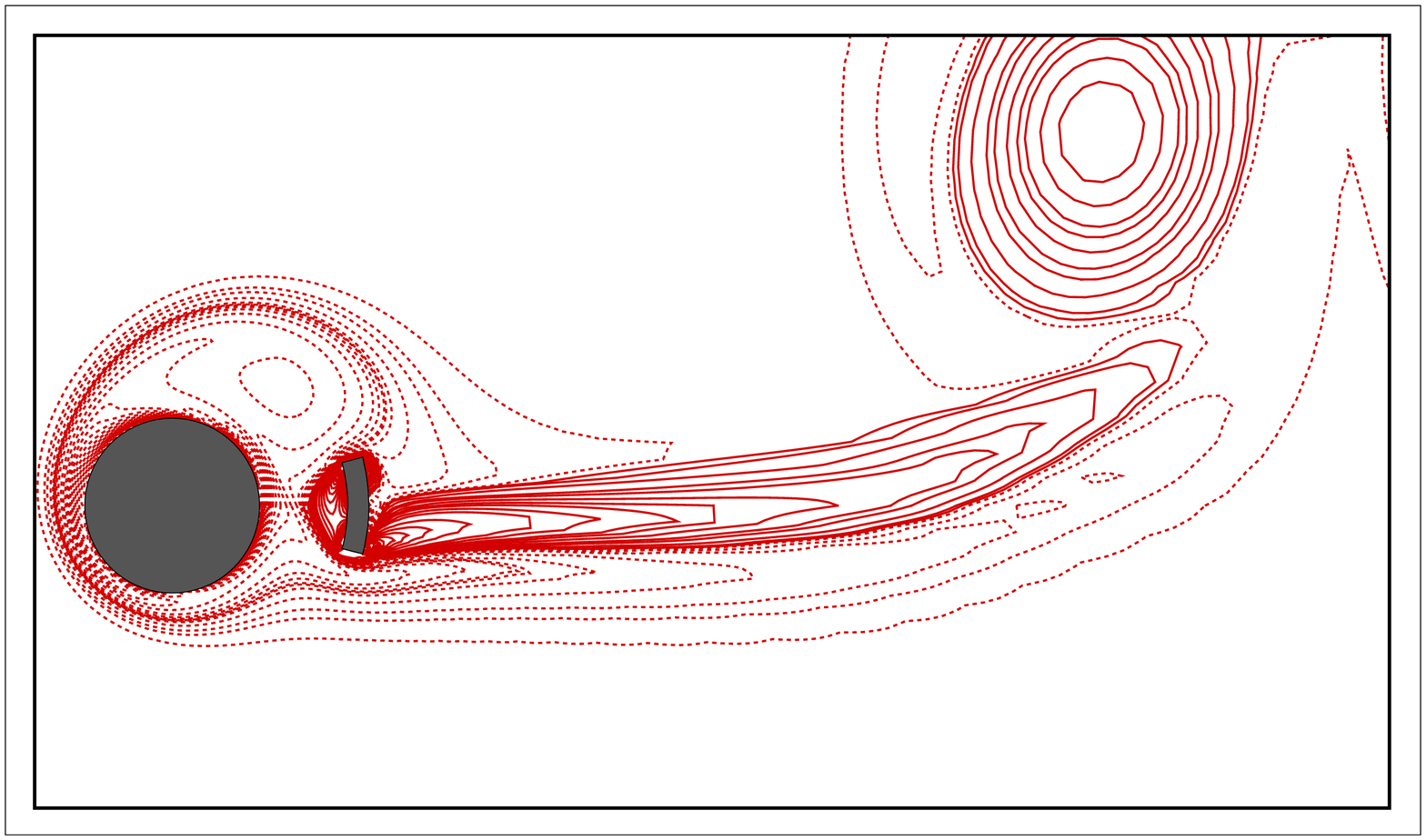}
\\
\hspace{0.5em}\scriptsize{$t=t_0+(1/2)T$}
\\
\includegraphics[width=0.29\textwidth,trim={0.5cm 0.3cm 0.5cm 0.3cm},clip]{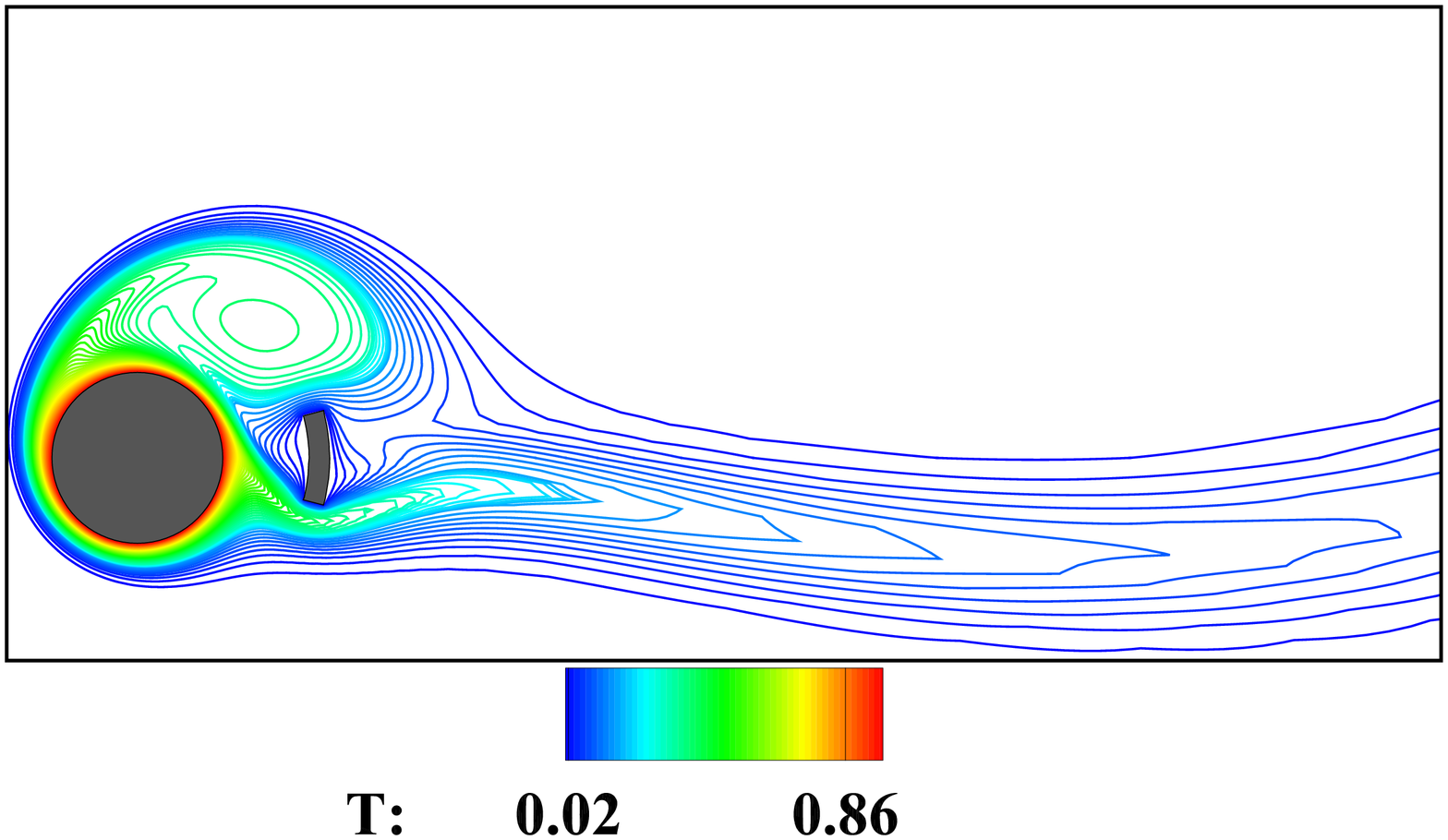}
\includegraphics[width=0.3\textwidth,trim={0.5cm 0.3cm 0.3cm 0.3cm},clip]{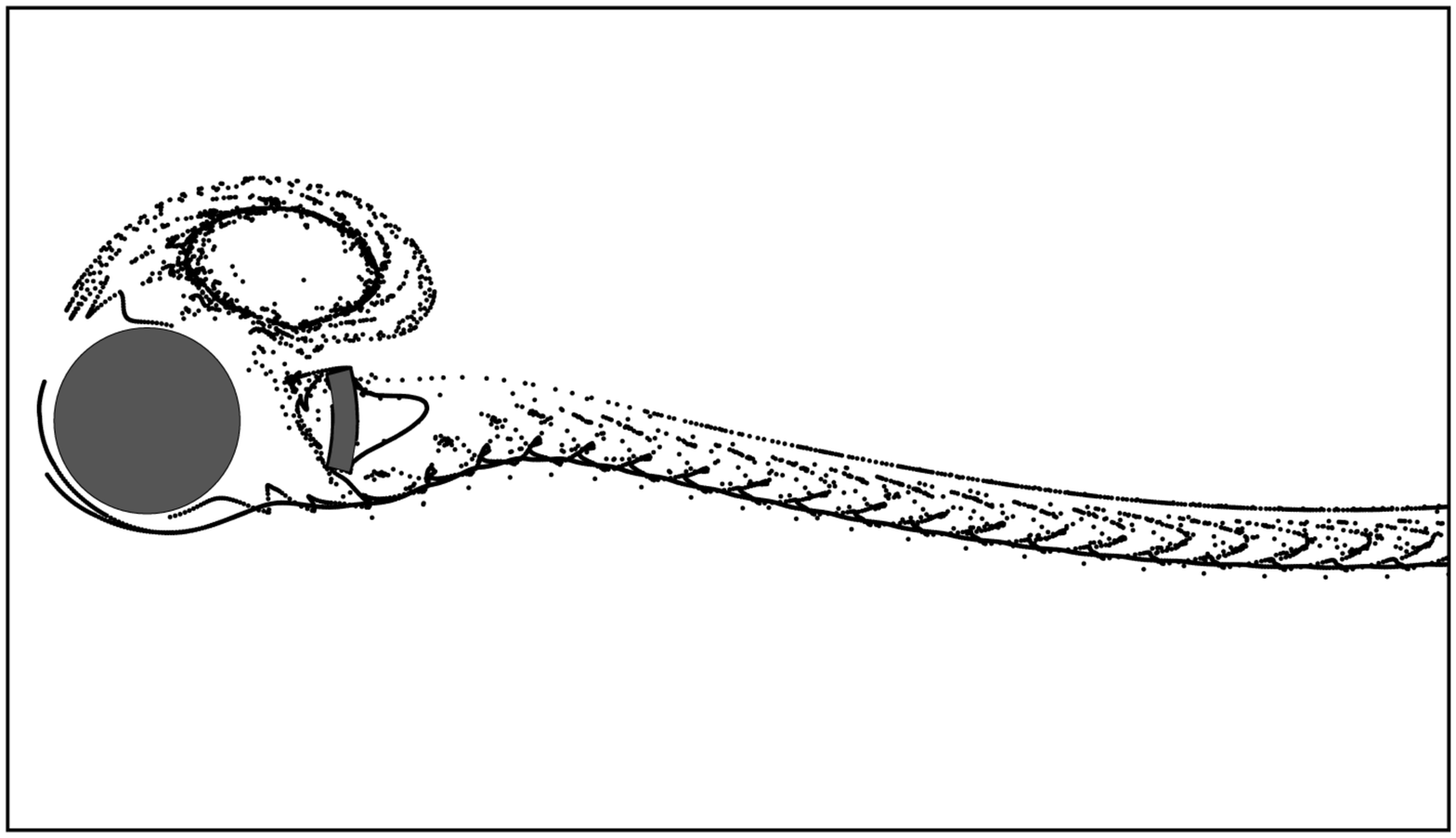}
\includegraphics[width=0.3\textwidth,trim={0.5cm 0.3cm 0.3cm 0.3cm},clip]{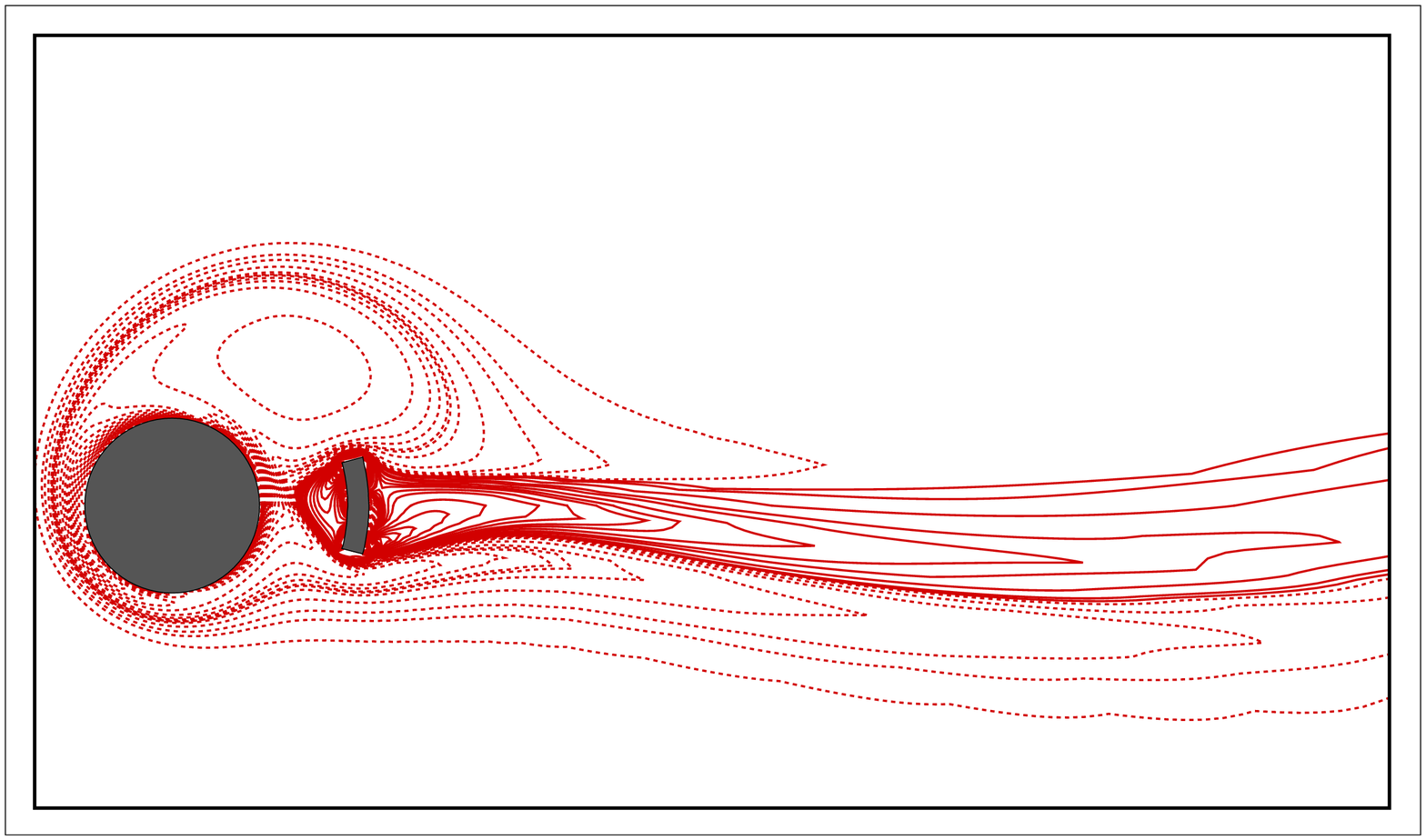}
\\
\hspace{0.5em}\scriptsize{$t=t_0+(3/4)T$}
\\
\includegraphics[width=0.29\textwidth,trim={0.5cm 0.3cm 0.5cm 0.3cm},clip]{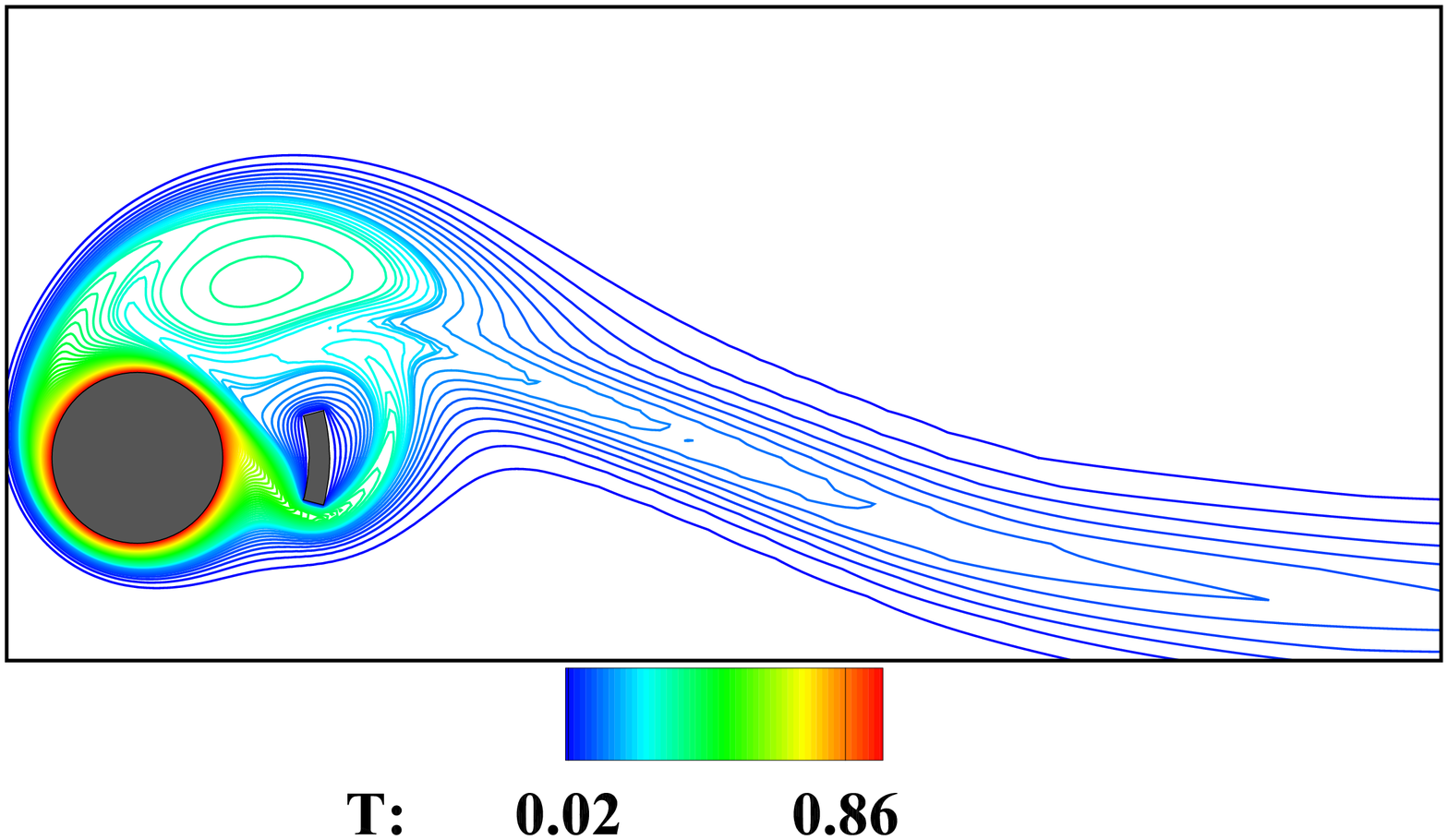}
\includegraphics[width=0.3\textwidth,trim={0.5cm 0.3cm 0.3cm 0.3cm},clip]{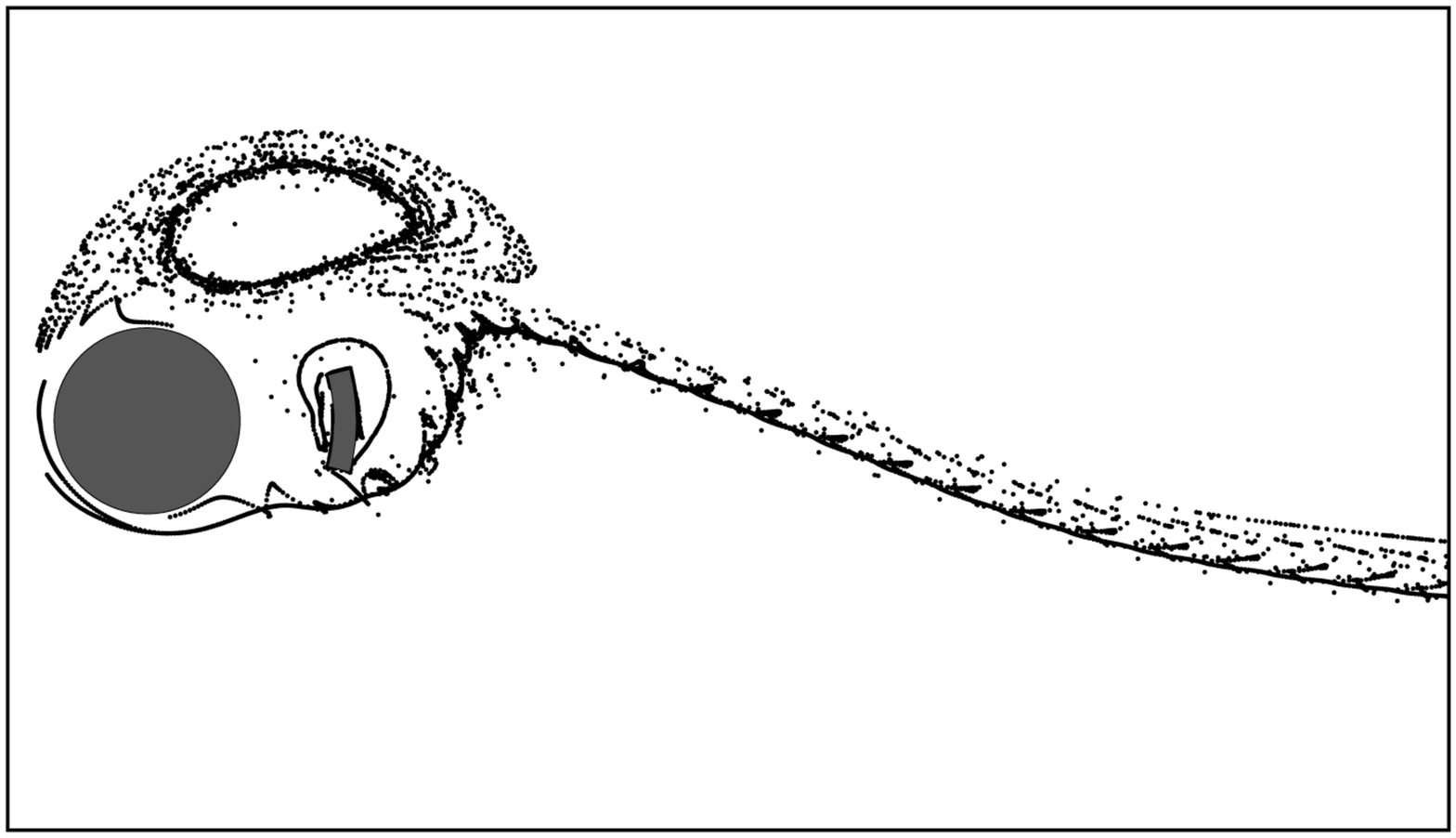}
\includegraphics[width=0.3\textwidth,trim={0.5cm 0.3cm 0.3cm 0.3cm},clip]{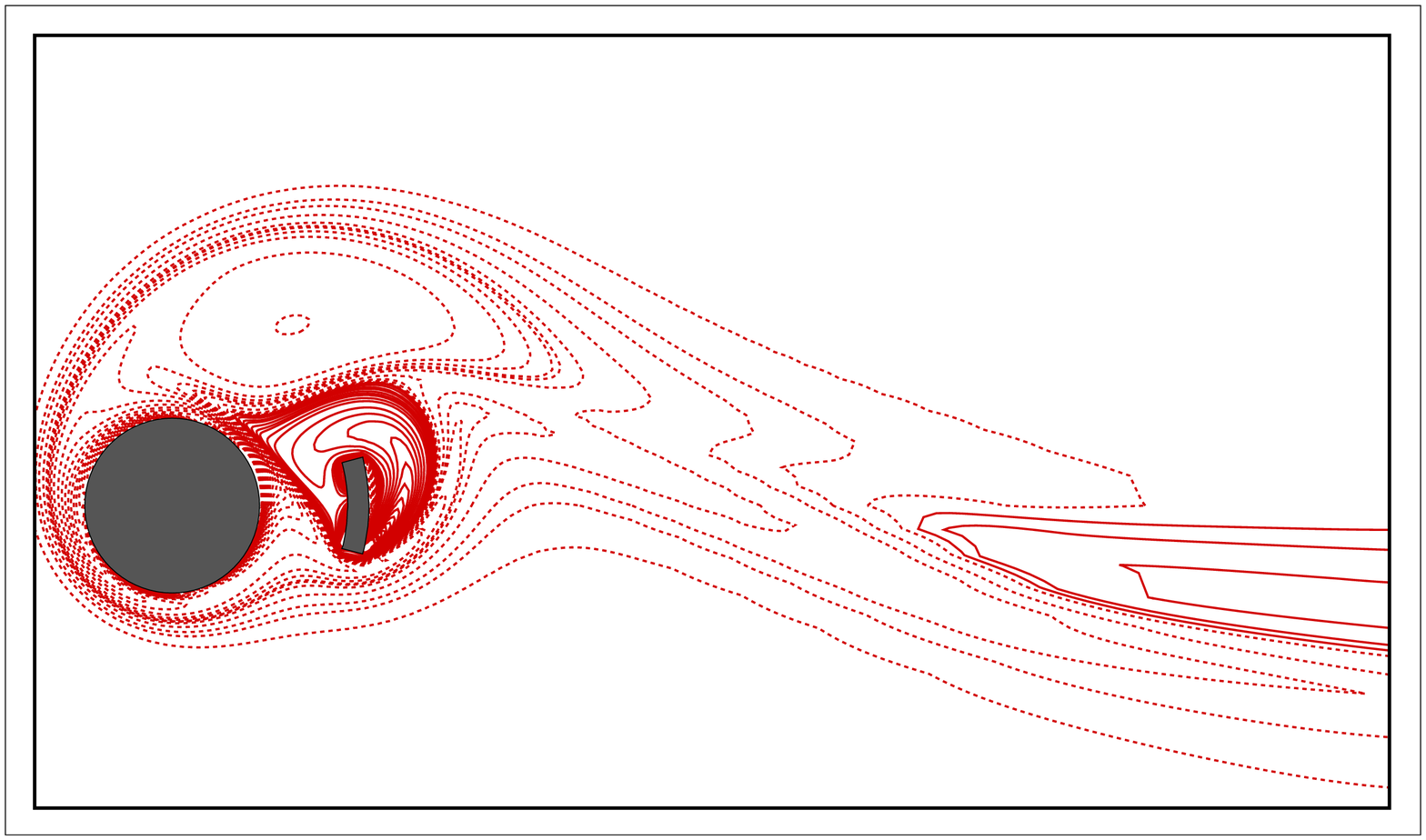}
\\
\hspace{0.5em}\scriptsize{$t=t_0+(1)T$}
\\
\includegraphics[width=0.29\textwidth,trim={0.5cm 0.3cm 0.5cm 0.3cm},clip]{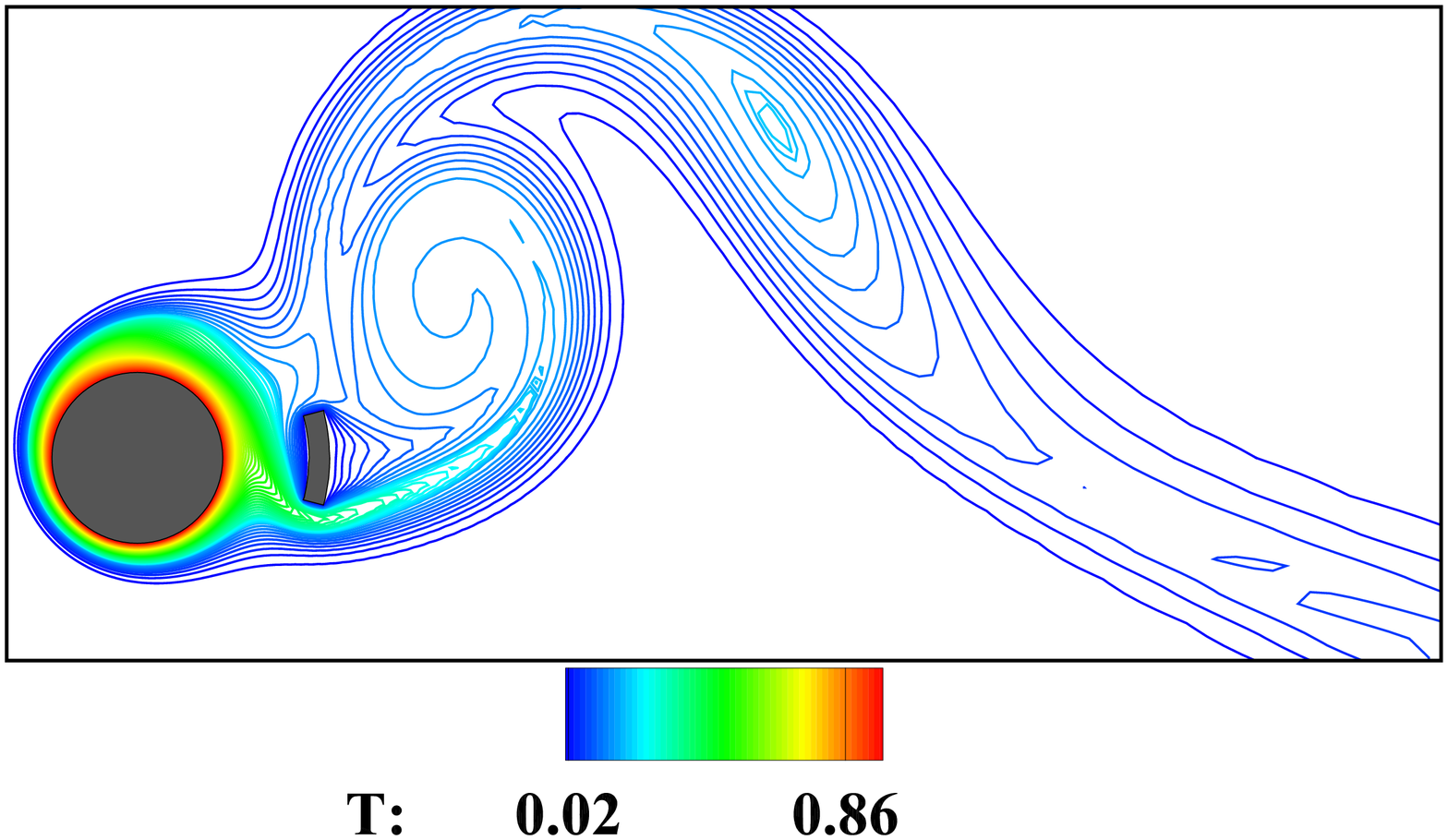}
\includegraphics[width=0.3\textwidth,trim={0.5cm 0.3cm 0.3cm 0.3cm},clip]{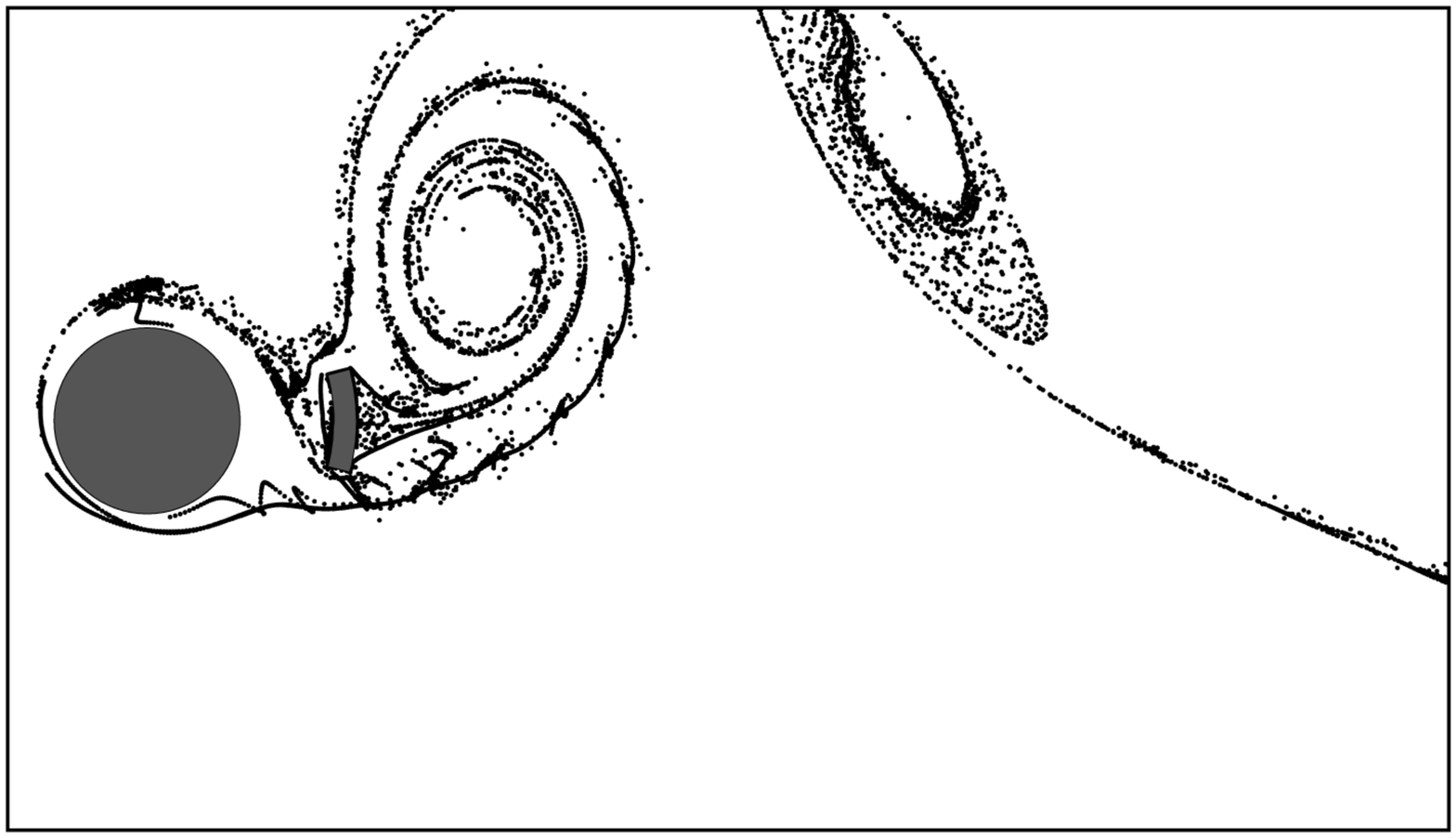}
\includegraphics[width=0.3\textwidth,trim={0.5cm 0.3cm 0.3cm 0.3cm},clip]{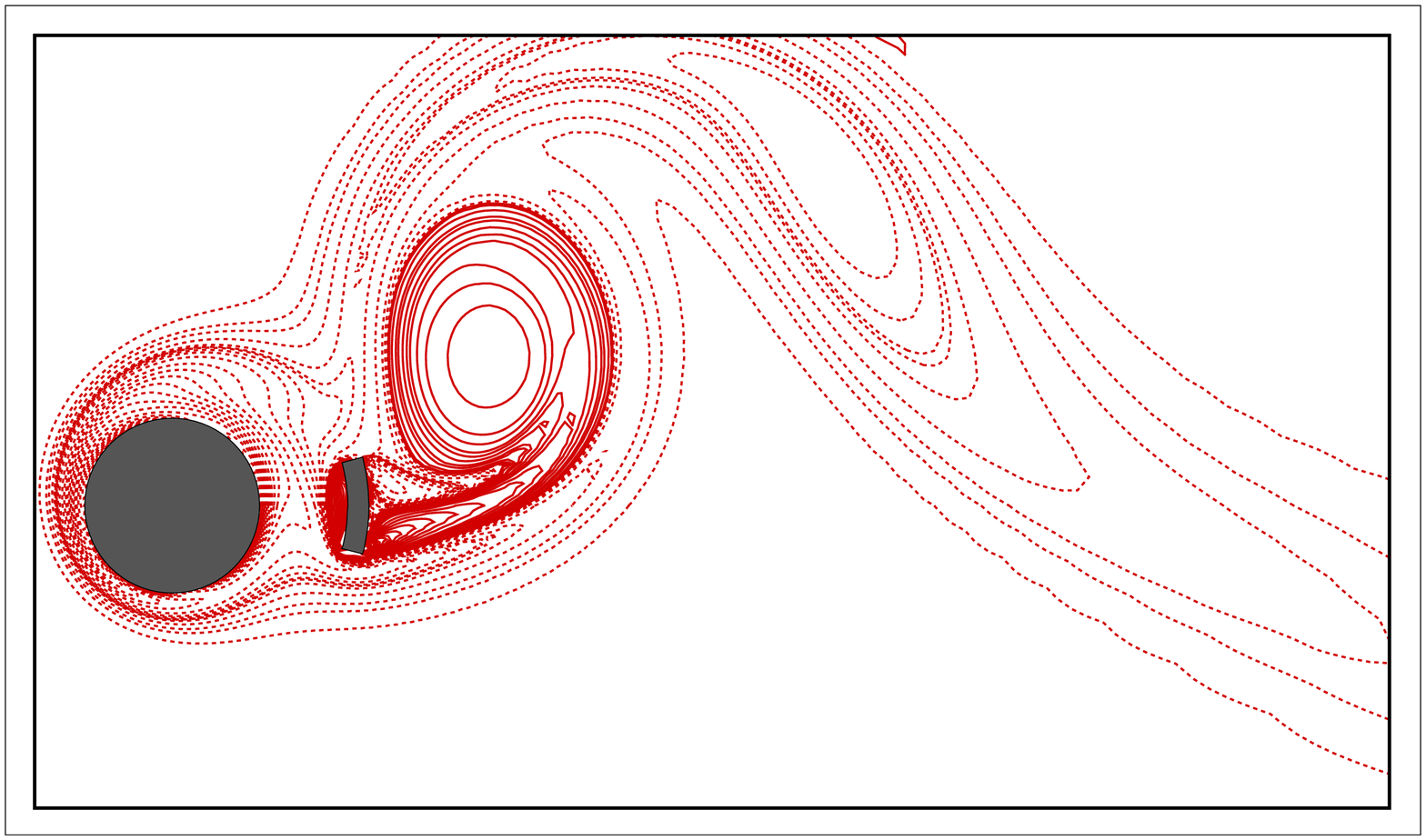}
\\
\hspace{2cm}(a) \hspace{4cm}(b) \hspace{4cm}(c)\hspace{2cm}
 \caption{(a) Isotherm, (b) streakline and (c) vorticity contour for $Pr=0.7$, $Re=150$, $\alpha=3.25$ and $d/R_0=1$ at different phases.}
 \label{fig:d_1_a_3-25}
\end{figure*}

For $\alpha=0.5$ and $d/R_0=1$, \cref{fig:d_1_a_0-5} exhibits the isotherm, streamline and vorticity at periodic phases. Two vortices are shed periodically from the upper and lower sides of the cylinder, according to the vorticity and streamline. The upper vortex is slightly larger than the lower vortex. The continuous and dashed lines indicate the positive and negative contours, respectively. The vortex shedding plane is shifted by approximately $\theta=20\degree$ from the centerline or the x-axis due to the rotation of the cylinder. The shear layer around the plate changes the negative equi-vorticity lines that come from the surface of the cylinder, but the positive equi-vorticity lines from the cylinder merge with the shear layer due to the rotation of the cylinder. No recirculation zone or vortex is observed between the cylinder and the plate. Two large vortices as lumps of hot fluid shed periodically from the upper and bottom sides of the cylinder according to the isotherm contours. The isotherm density is high near the front stagnation point, which indicates the higher heat transfer rate in this region. Isotherm, streakline and vorticity are displayed in \cref{fig:d_1_a_1-0} for $\alpha=1$ and $d/R_0=1$. Streakline and vorticity indicate that two vortices are periodically shed from the upper side and lower side of the cylinder. The increase in rotational rate, increases the movement of the fluid around the control plate which leads to thickening of shear layer around the control plate. It affects the vorticity contour coming from the cylinder. Positive equi-vorticity lines from the cylinder and the plate get merged together to shed a sleek, elongated vortex. The positive equi-vorticity lines from the cylinder completely cover the control plate, also dragging the negative equi-vorticity lines towards the bottom of the cylinder. This increases the density in thermal boundary layein the upper half of the cylinder, increasing the heat transfer. The upper vortex is much wider as compared to the sleek bottom vortex. Because of the increased $\alpha$, the vortex shedding plane is shifted by approximately $\theta=23\degree$ from the centerline. The isotherm contours suggest that two warm blobs convect away periodically from the upper and lower sides of the cylinder. There is no vortex or recirculation zone found between the cylinder and the plate. Isotherm, streakline and vorticity for $\alpha=2.07$ and $d/R_0=1$ are shown in \cref{fig:d_1_a_2-07}. The streakline and vorticity suggest that two vortices are periodically shed in the flow domain. One vortex is shed from the top of the cylinder and the other one is shed from the back of the plate. The lower vortex pushes the upper vortex due to the high rotation rate of the cylinder. As a result, the upper vortex is shed much earlier than at lower rotational rates. Also, the upper vortex becomes sleek and the bottom vortex becomes wide. Negative equi-vorticity lines coming from the cylinder completely cover the control plate as well as the positive vortex. Due to the high movement of fluid around the control plate, the shear layers get thickened and drag the negative equi-vorticity lines from the cylinder to the bottom of the plate. This affects the thermal boundary layer of the cylinder by thinning around the rear stagnation point. As a result, the heat transfer is increased in this region. However, the high rotation of the cylinder thickens the thermal boundary layer around the front stagnation point, leading to a decrease in heat transfer rate. Here, The vortex shedding plane is shifted by approximately $\theta=37\degree$ from the centerline. The isotherm contours suggest that two warm blobs convect away periodically by the vortices generated in the flow domain. \cref{fig:d_1_a_3-25} shows the isotherm, streakline and vorticity for $\alpha=3.25$ and $d/R_0=1$. Due to very high rotational rate, the negative equi-vorticity lines completely cover the cylinder as well as the positive equi-vorticity lines originated from the control plate. Two vortices are shed periodically, one from the top of the cylinder and another from the back of the control plate. Due to the high rotational speed, the bottom vortex pushes the upper vortex. As a result, the upper vortex is shed much earlier. Also, the bottom vortex is much larger than the upper vortex. One small negative vortex is formed behind the control plate, but it gets dissolved into the positive vortex. After the negative vortex is shed, the shear layer from the cylinder splits on top and bottom of the shear layer from the control plate. It gradually merges and creates an elongated negative vortex. Between the cylinder and the plate, no vortex or recirculation zone forms. The moving fluid around the cylinder drags the shear layer from the control plate towards the top of the cylinder, which leads to the increased density of the isotherm contour. As a result, heat transfer is boosted in this region. The vortex shedding plane is displaced from the centerline by approximately $\theta=50\degree$ at this rotational rate. The isotherm contours suggest that the density of the isotherm around the cylinder becomes less than at the lower rotational rates, which means that the high rotation rate is suppressing the heat transfer rate from the cylinder surface. Additionally, two warm blobs periodically convect away from the cylinder's upper side and the plate's rear. The top blob is sleek and the bottom one is wide, similar to the vortices. \cref{fig:d_1_a_0-5,fig:d_1_a_1-0,fig:d_1_a_2-07,fig:d_1_a_3-25} show that increasing rotational rates increased the size of vortices as well as the angle of vortex shedding plane from the centerline for a fixed $d/R_0=1$.\\

\begin{figure*}[!t]
\centering
\scriptsize{$t=t_0+(0)T$}
\\
\includegraphics[width=0.3\textwidth,trim={0.5cm 0.3cm 0.5cm 0.3cm},clip]{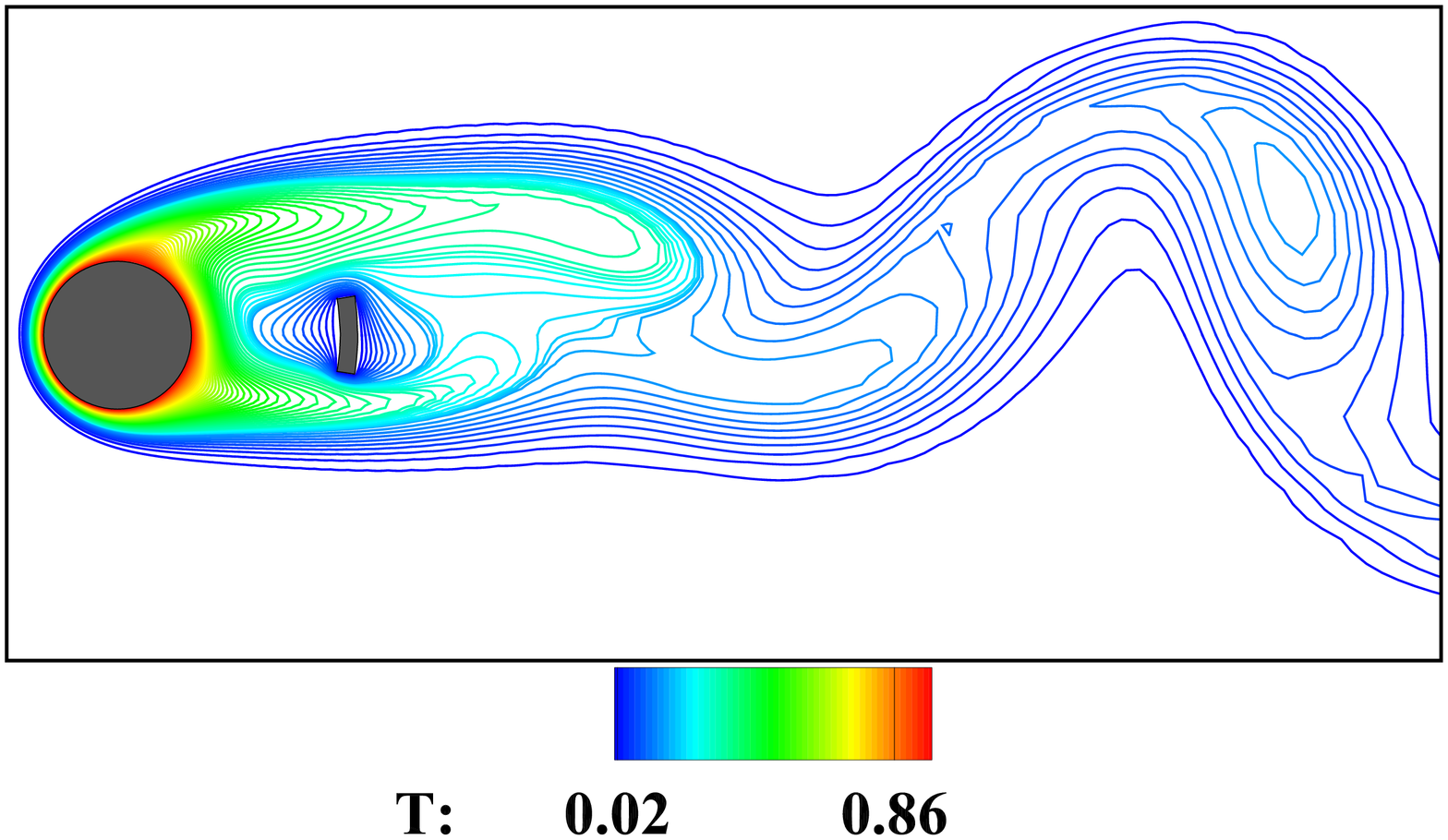}
\includegraphics[width=0.3\textwidth,trim={0.5cm 0.3cm 0.3cm 0.3cm},clip]{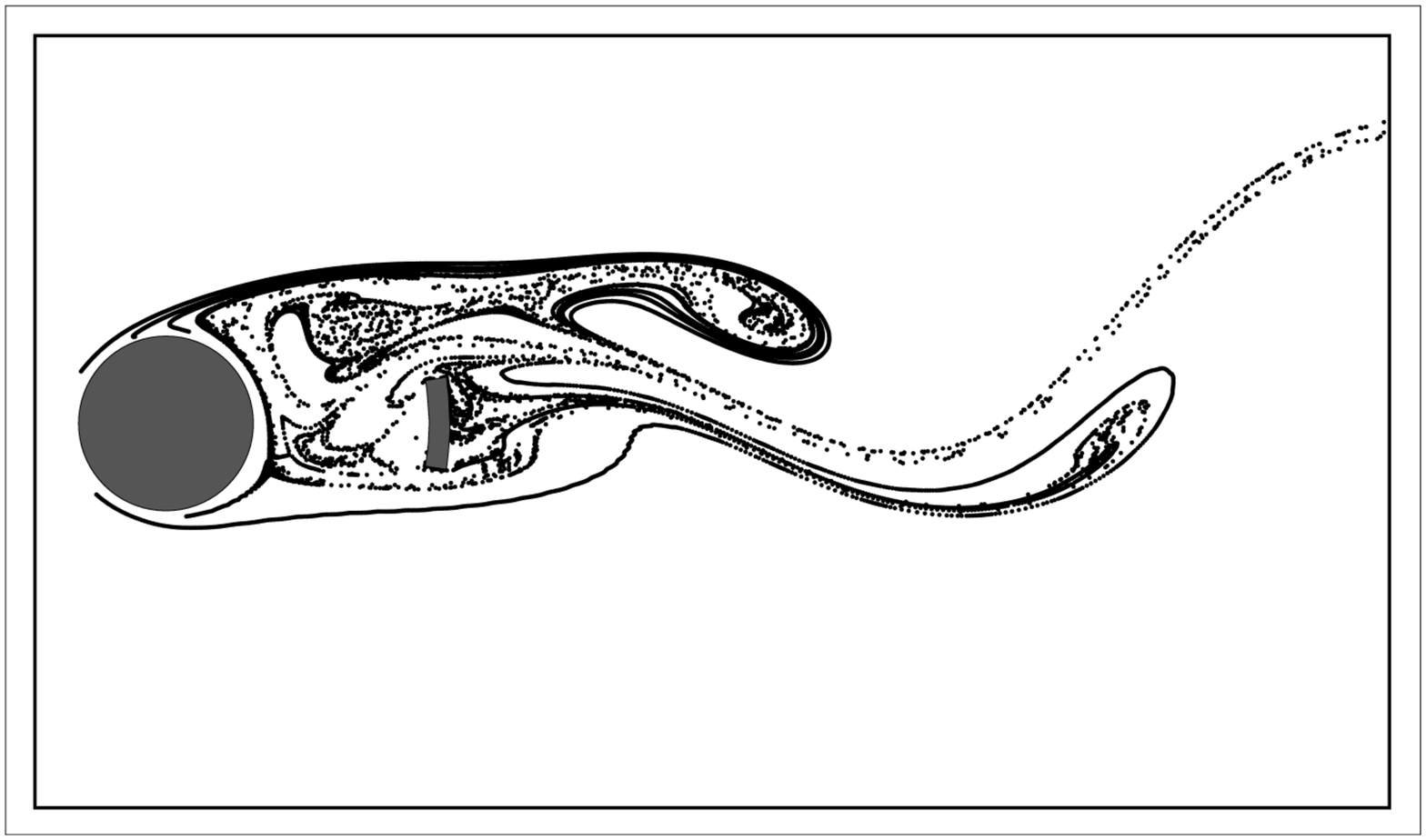}
\includegraphics[width=0.3\textwidth,trim={0.5cm 0.3cm 0.3cm 0.3cm},clip]{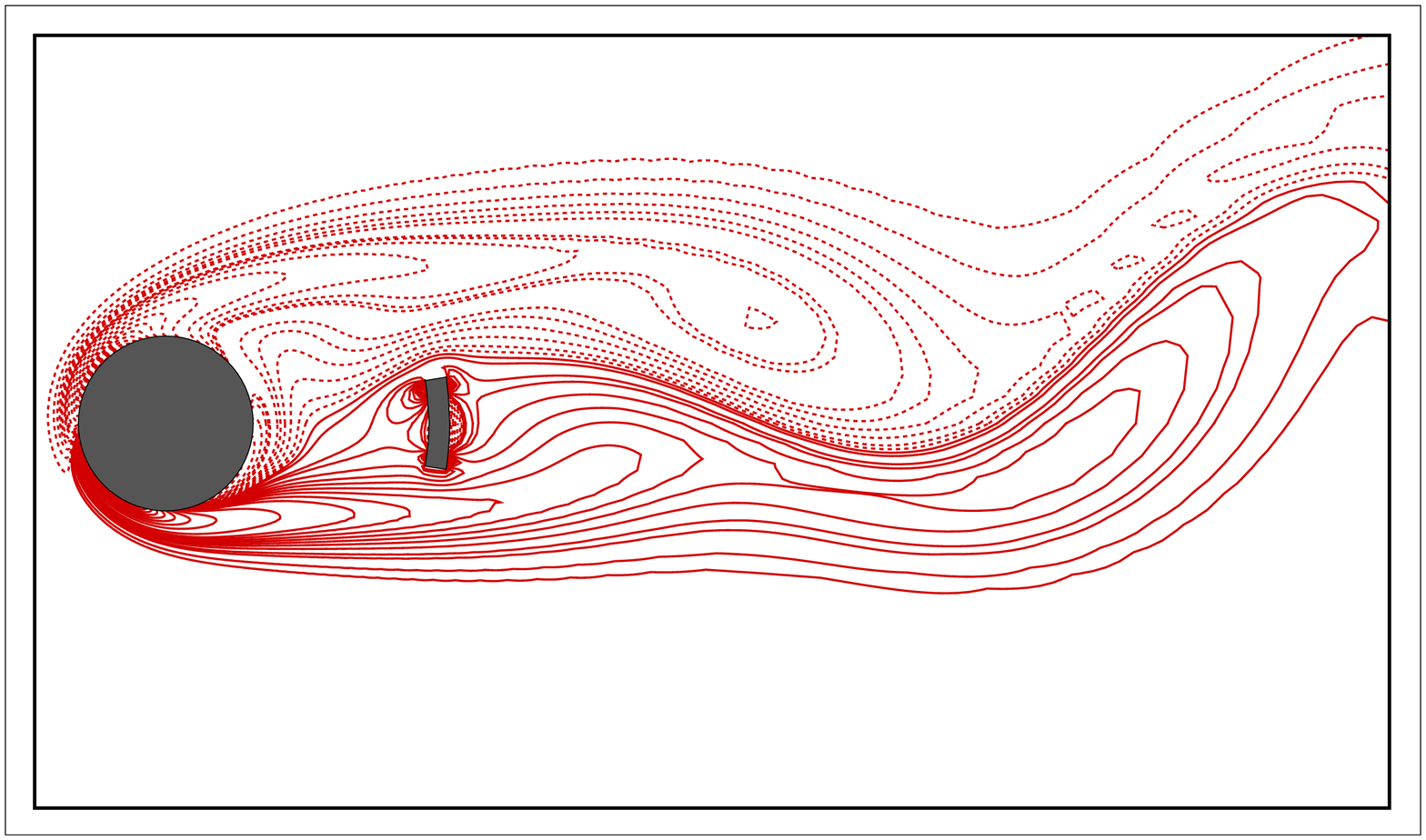}
\\
\hspace{0.5em}\scriptsize{$t=t_0+(1/4)T$}
\\
\includegraphics[width=0.29\textwidth,trim={0.5cm 0.3cm 0.5cm 0.3cm},clip]{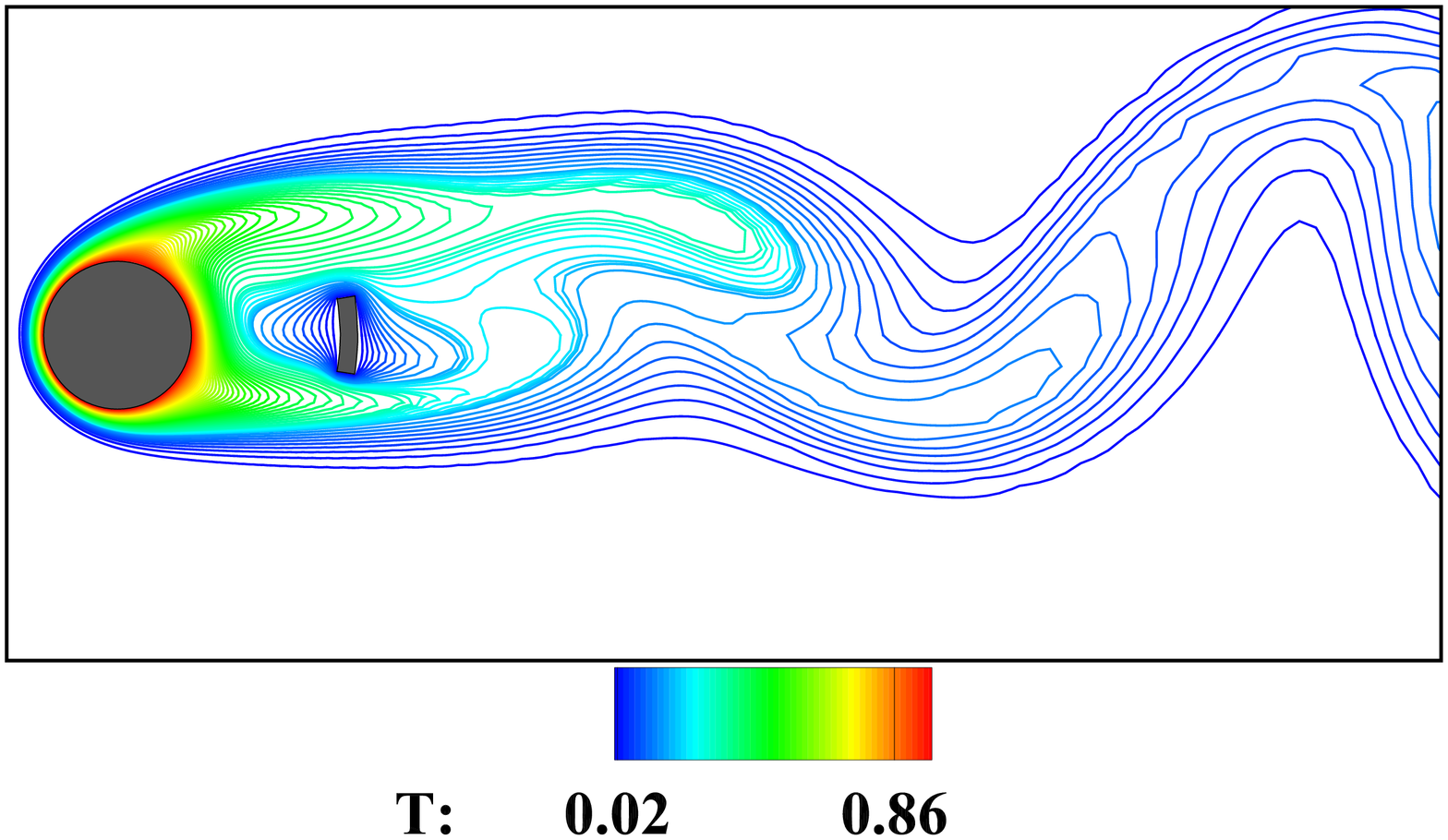}
\includegraphics[width=0.3\textwidth,trim={0.5cm 0.3cm 0.3cm 0.3cm},clip]{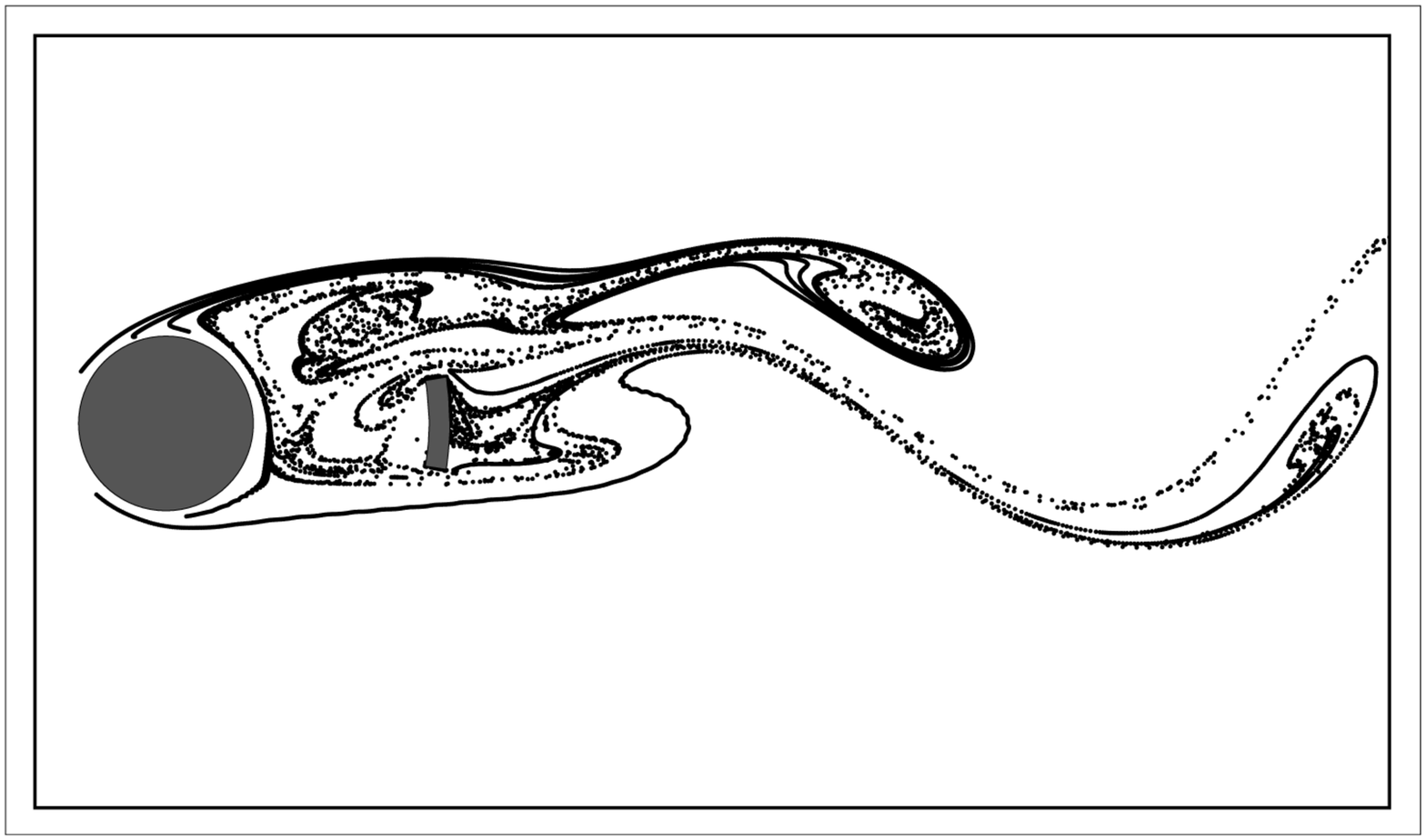}
\includegraphics[width=0.3\textwidth,trim={0.5cm 0.3cm 0.3cm 0.3cm},clip]{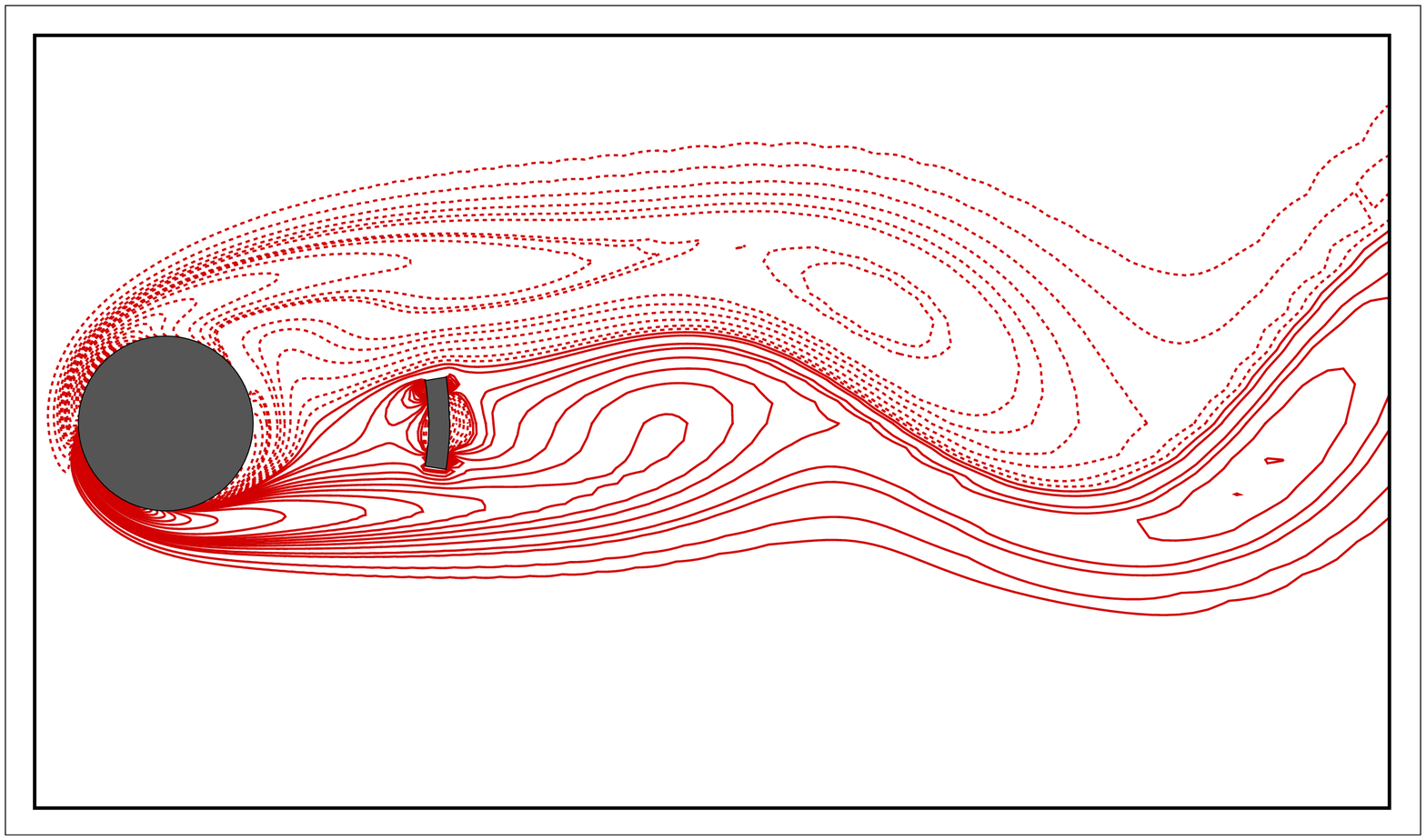}
\\
\hspace{0.5em}\scriptsize{$t=t_0+(1/2)T$}
\\
\includegraphics[width=0.29\textwidth,trim={0.5cm 0.3cm 0.5cm 0.3cm},clip]{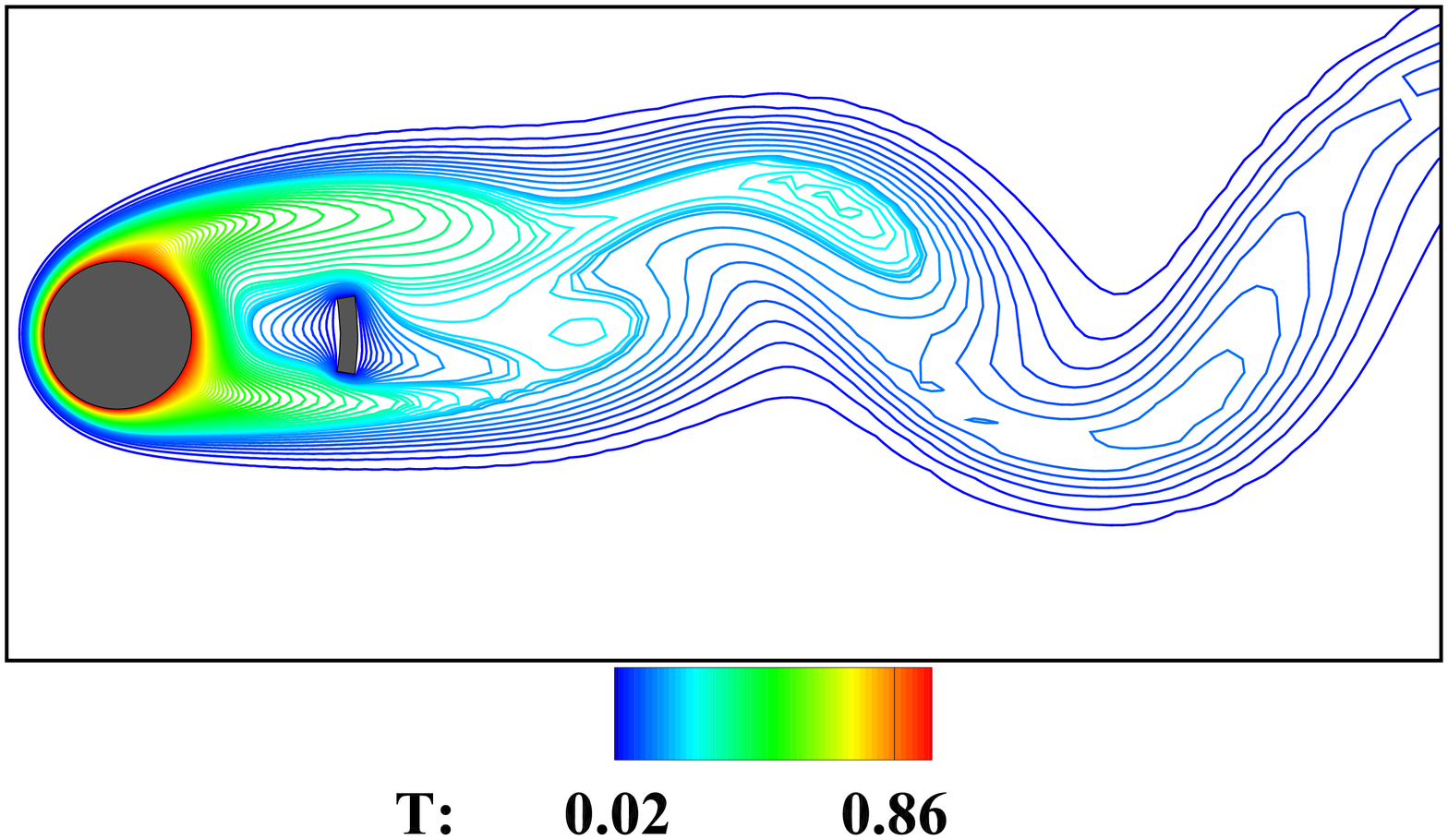}
\includegraphics[width=0.3\textwidth,trim={0.5cm 0.3cm 0.3cm 0.3cm},clip]{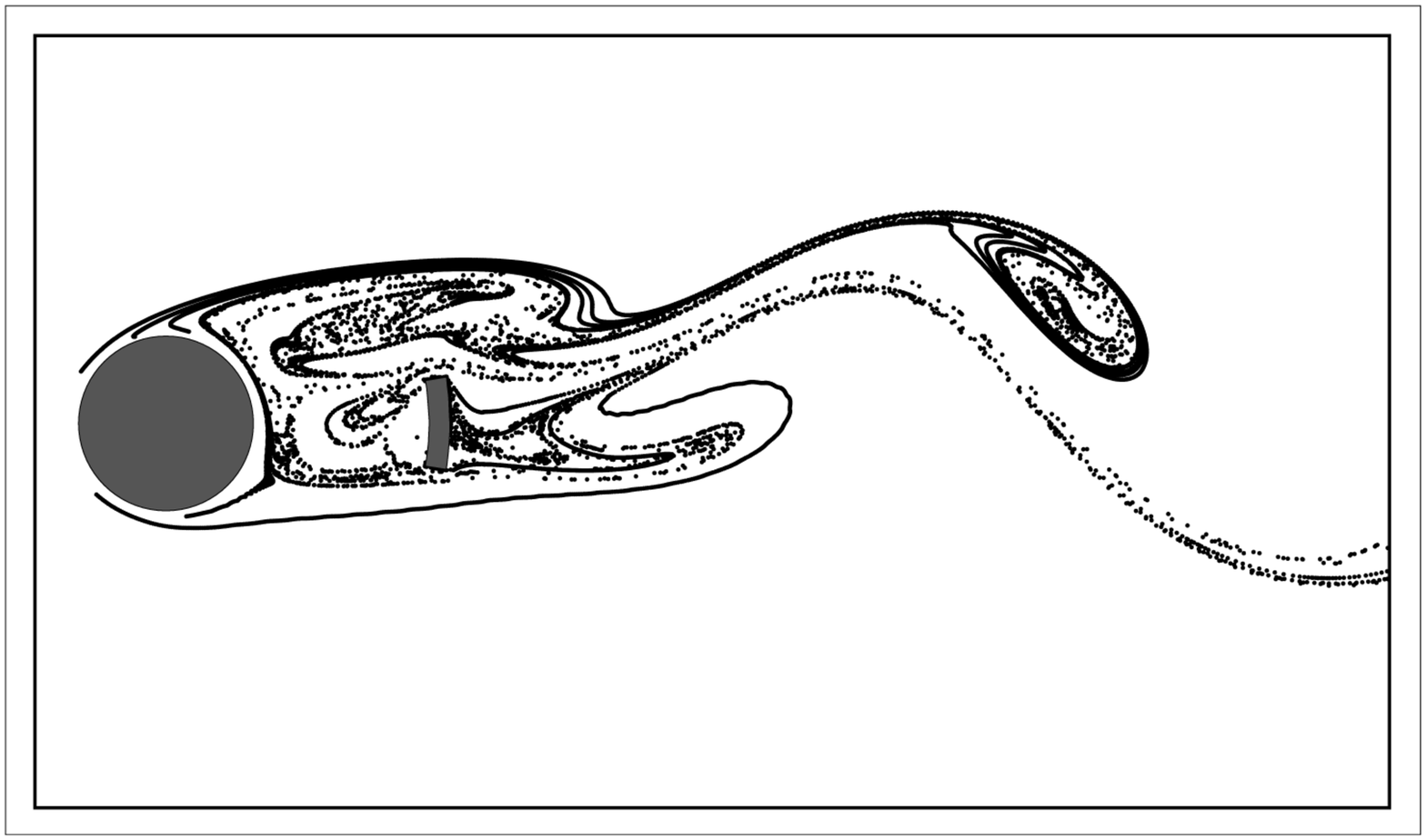}
\includegraphics[width=0.3\textwidth,trim={0.5cm 0.3cm 0.3cm 0.3cm},clip]{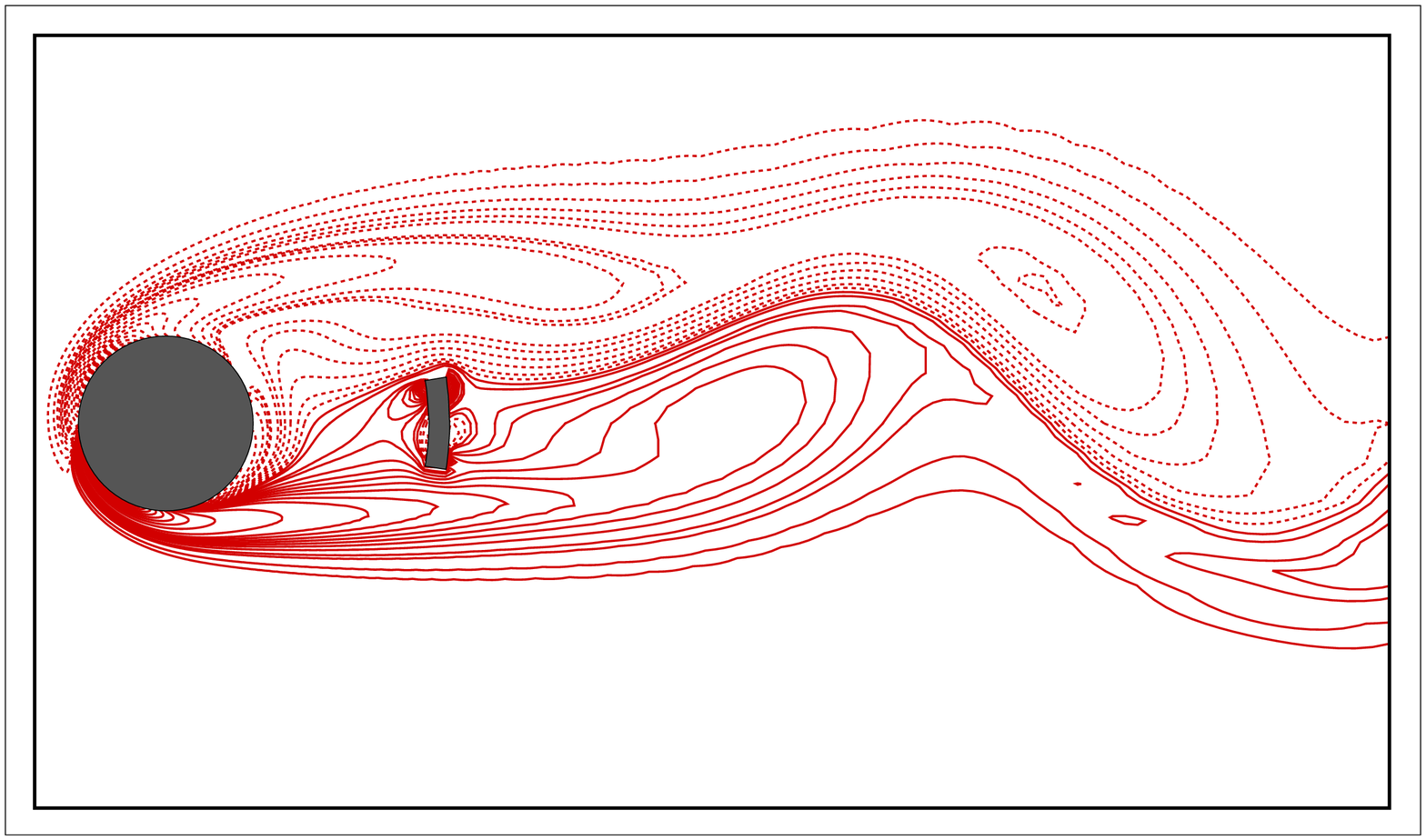}
\\
\hspace{0.5em}\scriptsize{$t=t_0+(3/4)T$}
\\
\includegraphics[width=0.29\textwidth,trim={0.5cm 0.3cm 0.5cm 0.3cm},clip]{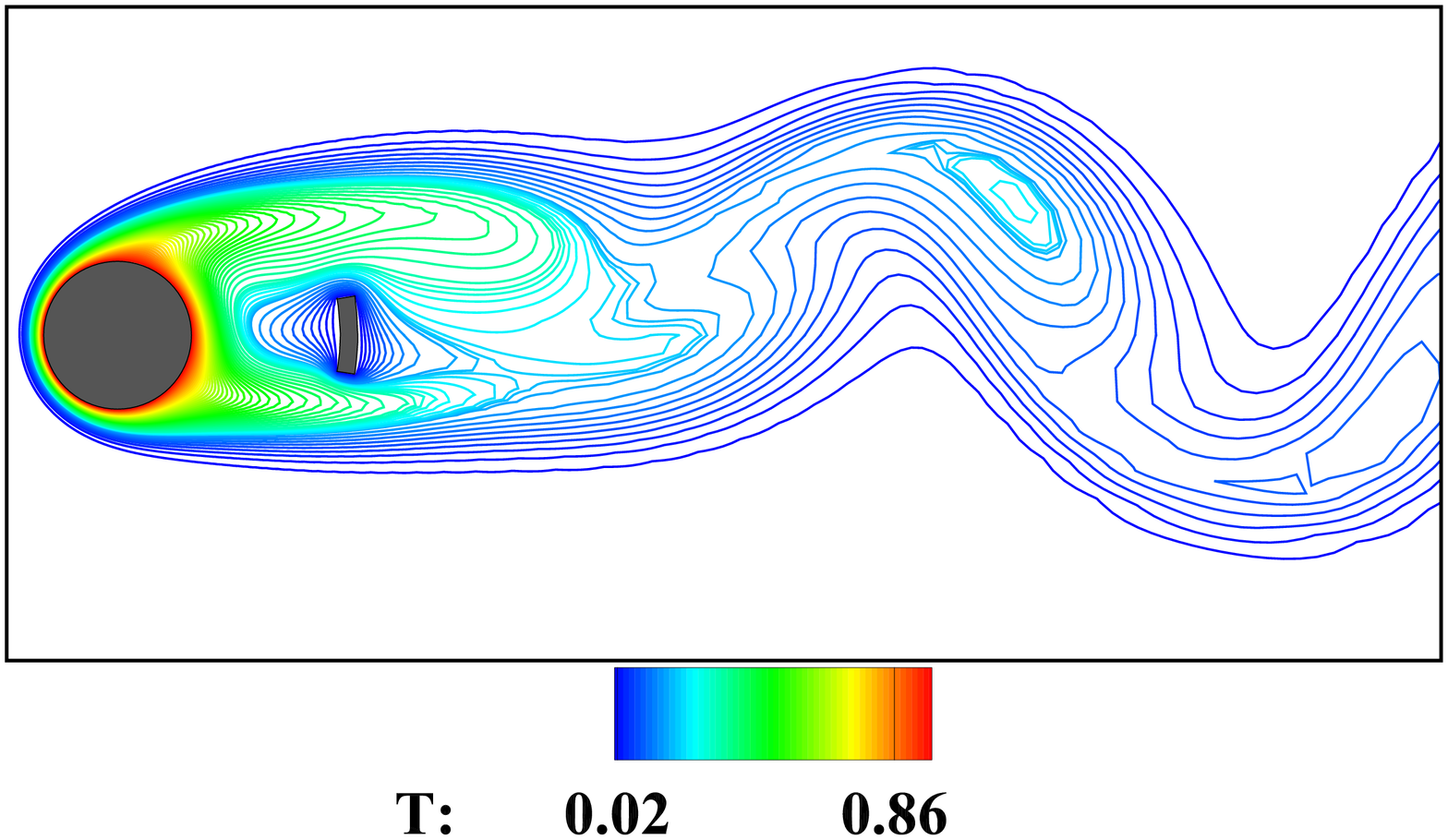}
\includegraphics[width=0.3\textwidth,trim={0.5cm 0.3cm 0.3cm 0.3cm},clip]{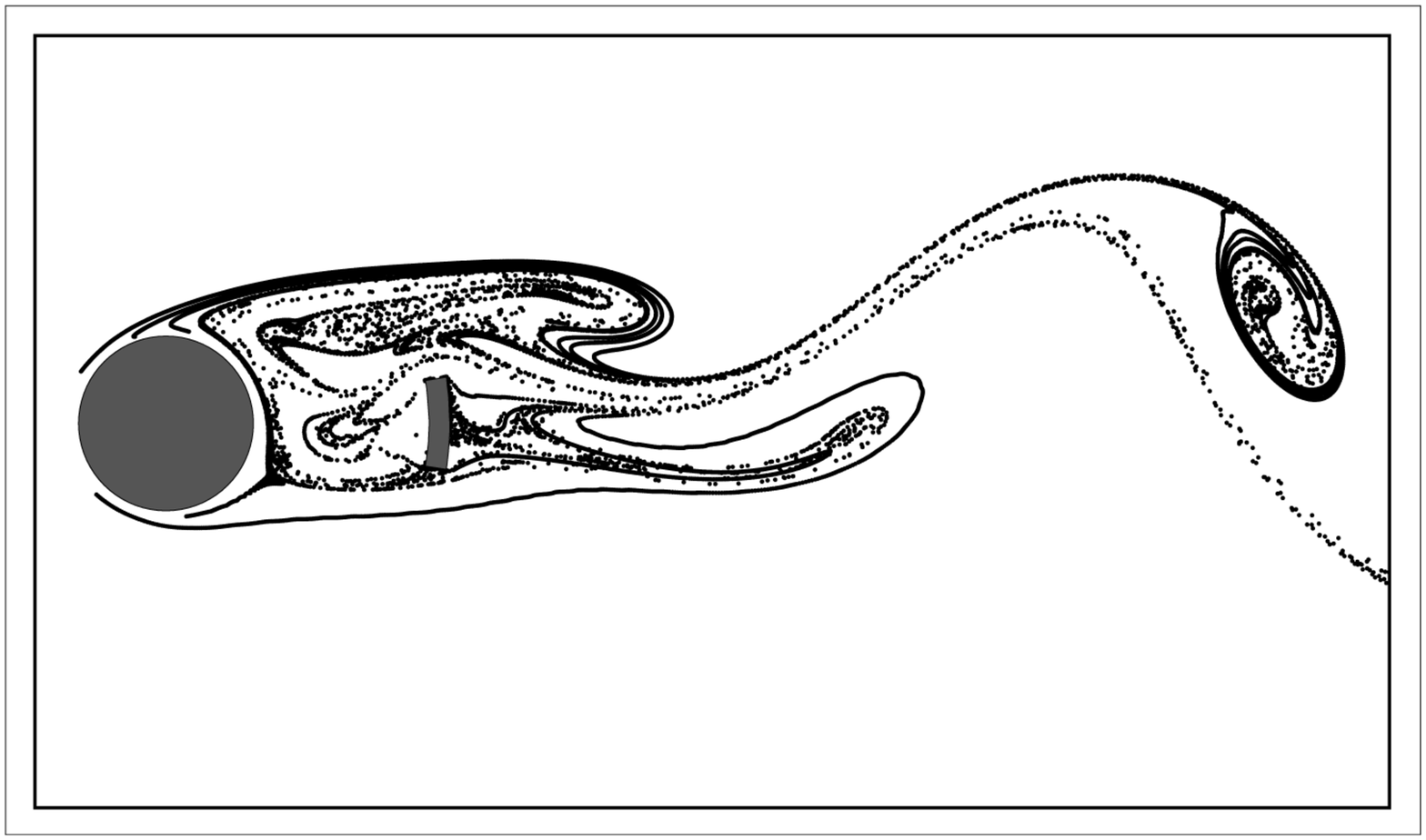}
\includegraphics[width=0.3\textwidth,trim={0.5cm 0.3cm 0.3cm 0.3cm},clip]{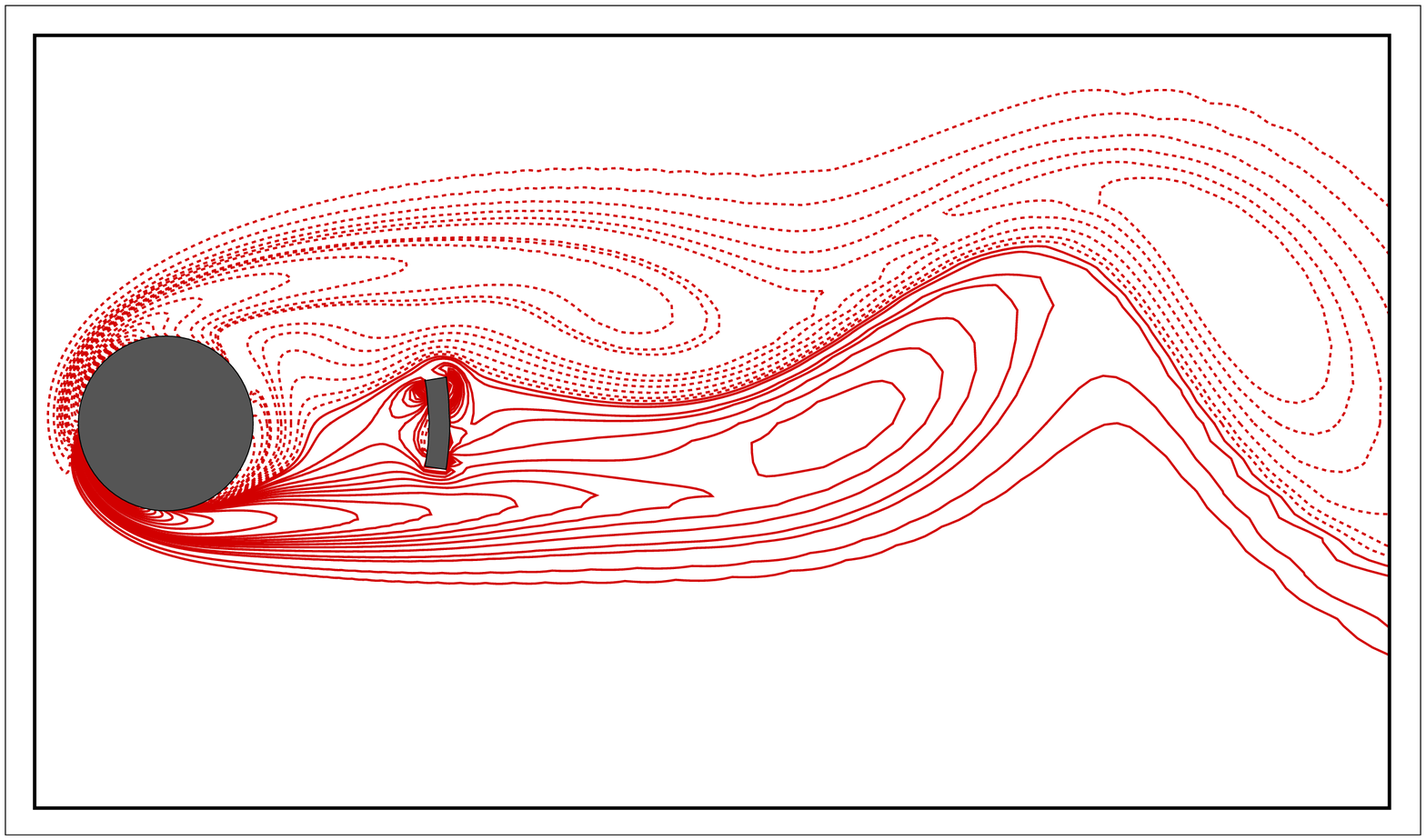}
\\
\hspace{0.5em}\scriptsize{$t=t_0+(1)T$}
\\
\includegraphics[width=0.29\textwidth,trim={0.5cm 0.3cm 0.5cm 0.3cm},clip]{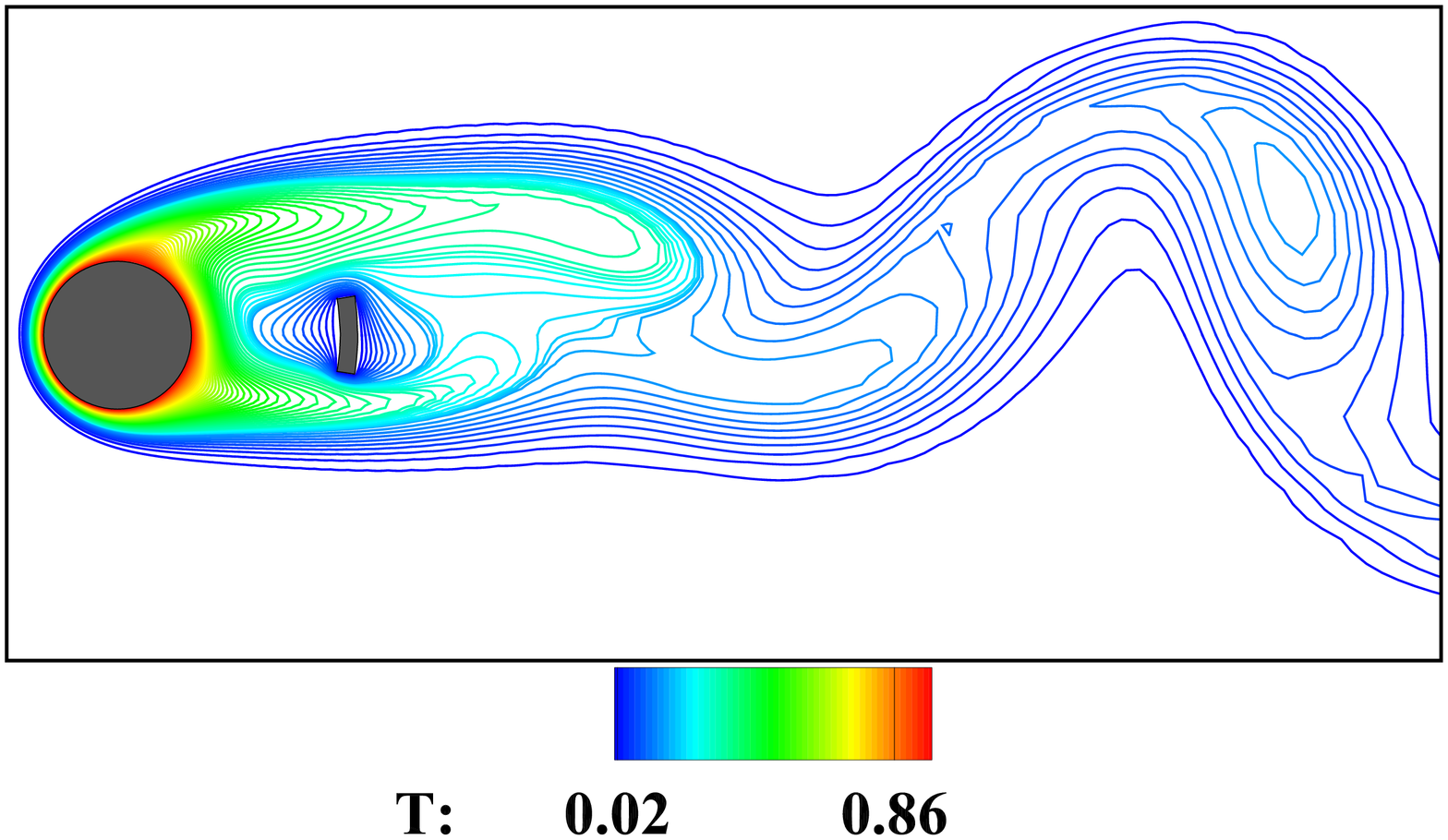}
\includegraphics[width=0.3\textwidth,trim={0.5cm 0.3cm 0.3cm 0.3cm},clip]{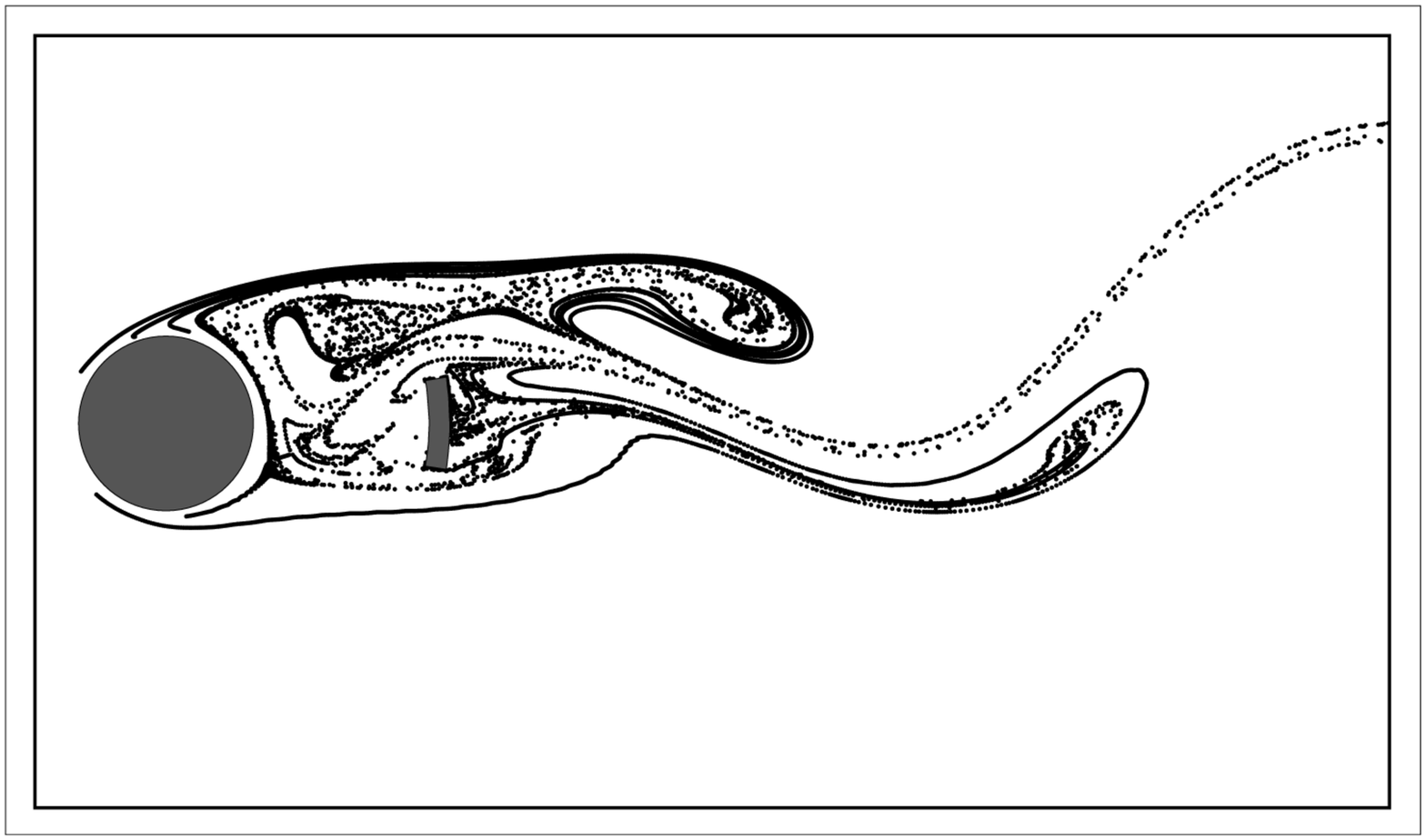}
\includegraphics[width=0.3\textwidth,trim={0.5cm 0.3cm 0.3cm 0.3cm},clip]{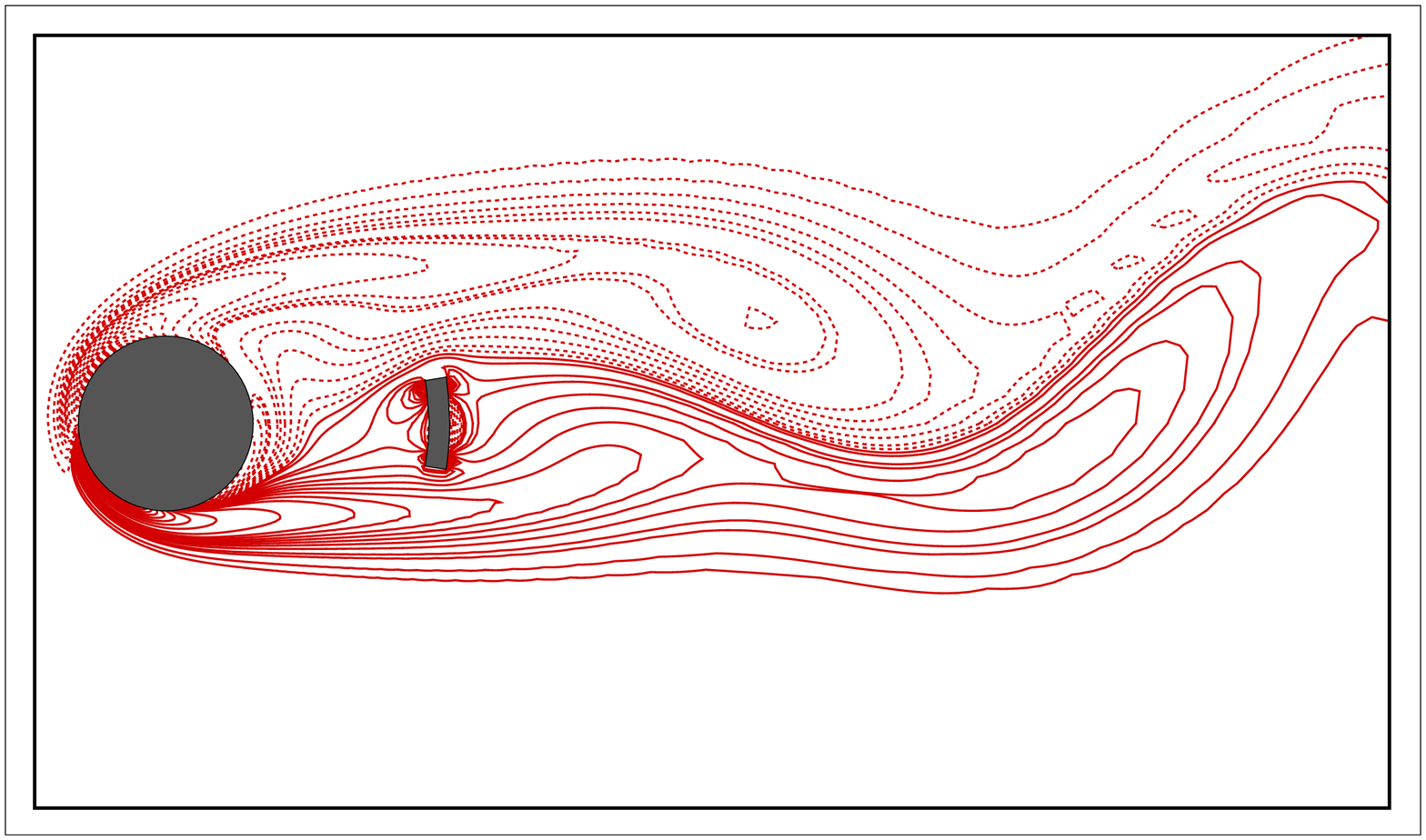}
\\
\hspace{2cm}(a) \hspace{4cm}(b) \hspace{4cm}(c)\hspace{2cm}
 \caption{(a) Isotherm, (b) streakline and (c) vorticity contour for $Pr=0.7$, $Re=150$, $\alpha=0.5$ and $d/R_0=2$ at different phases.}
 \label{fig:d_2_a_0-5}
\end{figure*}

\begin{figure*}[!t]
\centering
\scriptsize{$t=t_0+(0)T$}
\\
\includegraphics[width=0.3\textwidth,trim={0.5cm 0.3cm 0.5cm 0.3cm},clip]{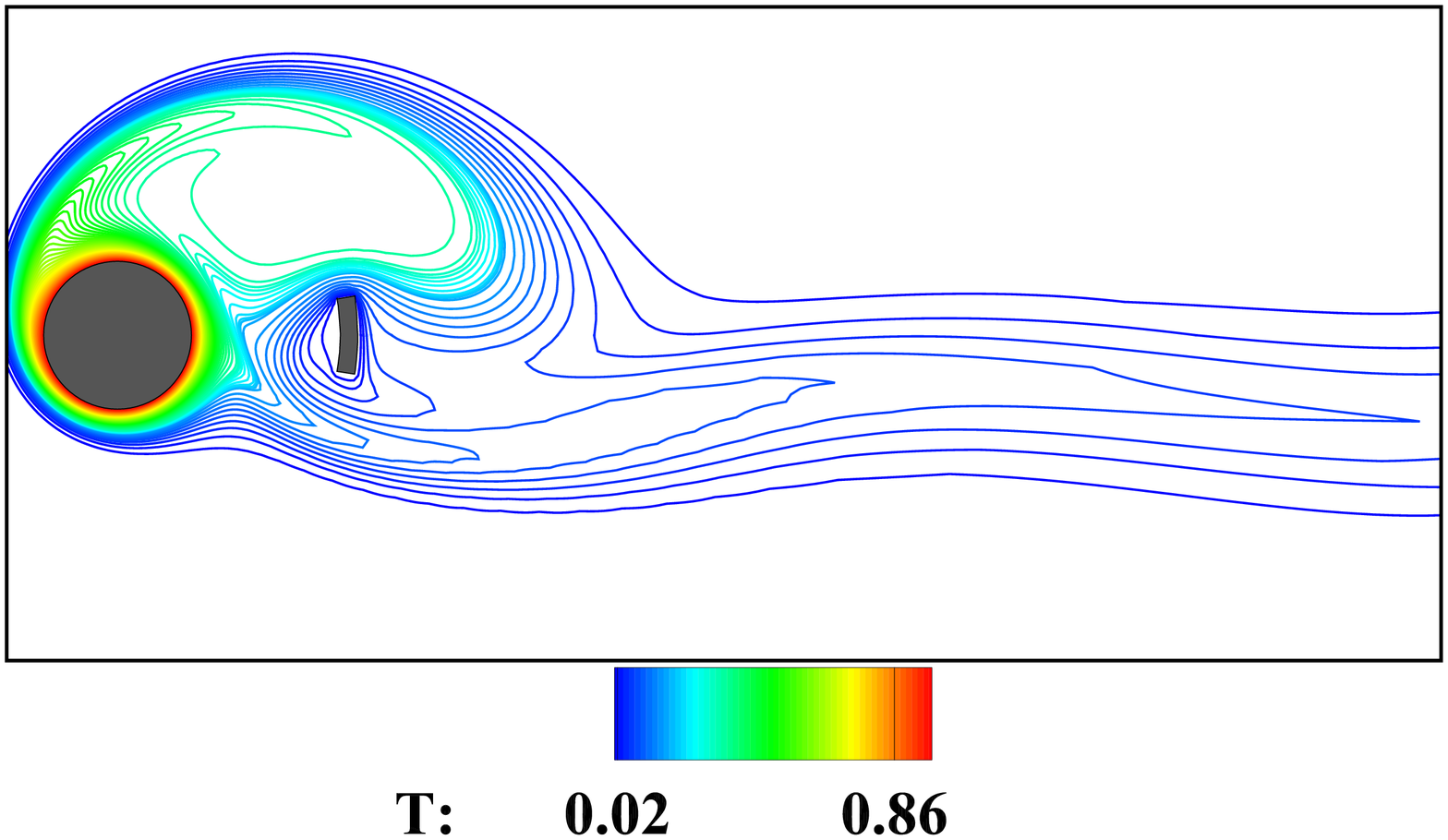}
\includegraphics[width=0.3\textwidth,trim={0.5cm 0.3cm 0.3cm 0.3cm},clip]{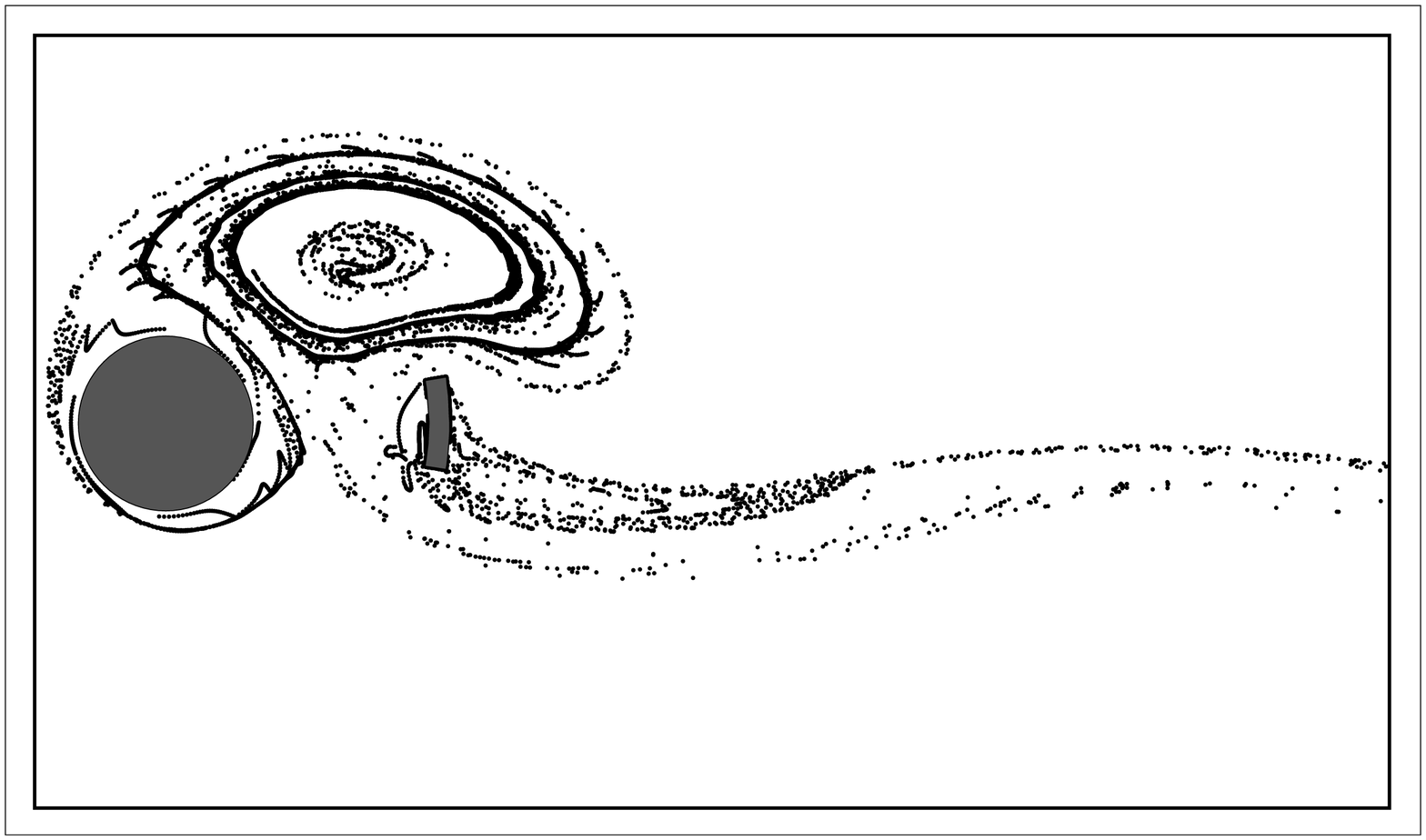}
\includegraphics[width=0.3\textwidth,trim={0.5cm 0.3cm 0.3cm 0.3cm},clip]{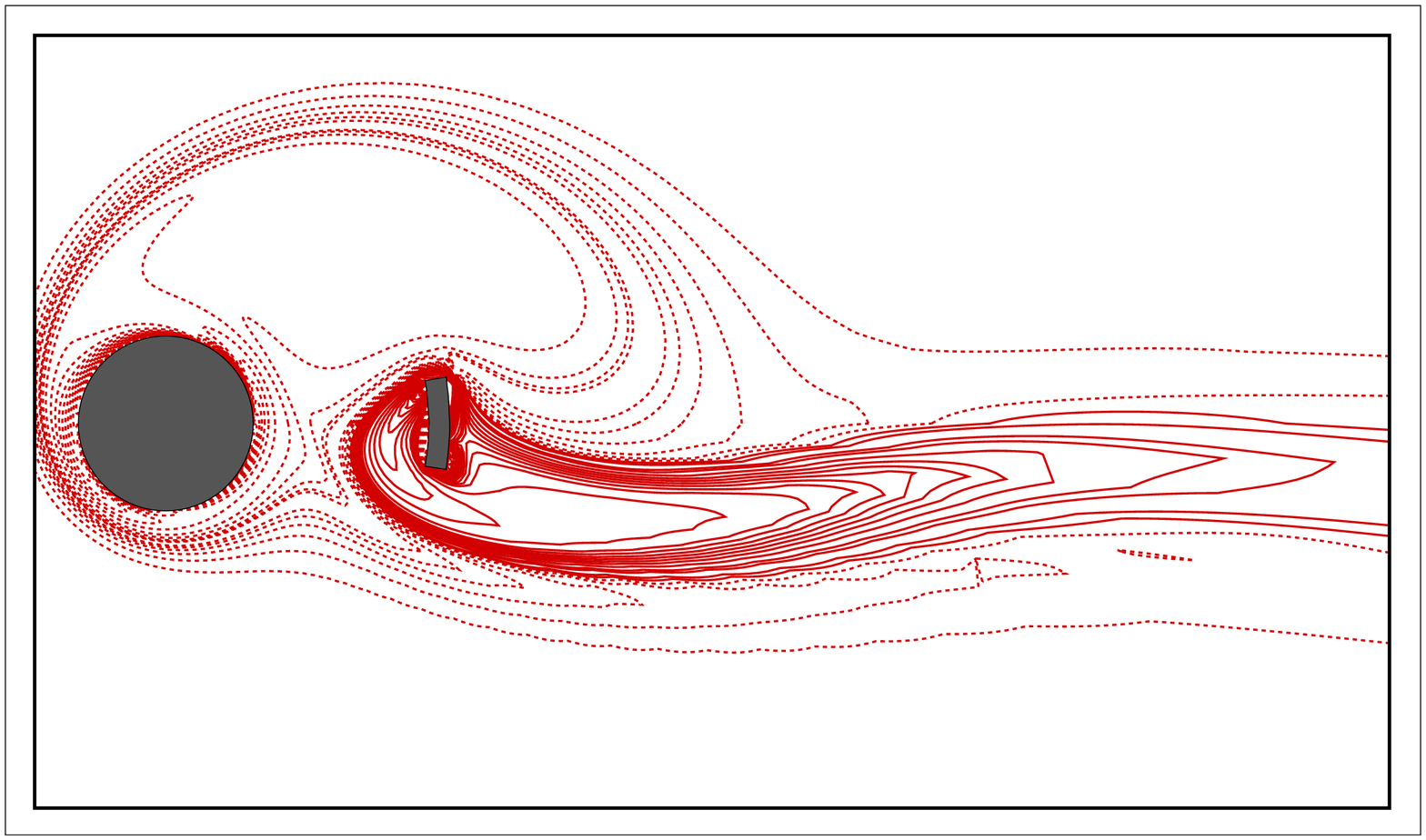}
\\
\hspace{0.5em}\scriptsize{$t=t_0+(1/4)T$}
\\
\includegraphics[width=0.29\textwidth,trim={0.5cm 0.3cm 0.5cm 0.3cm},clip]{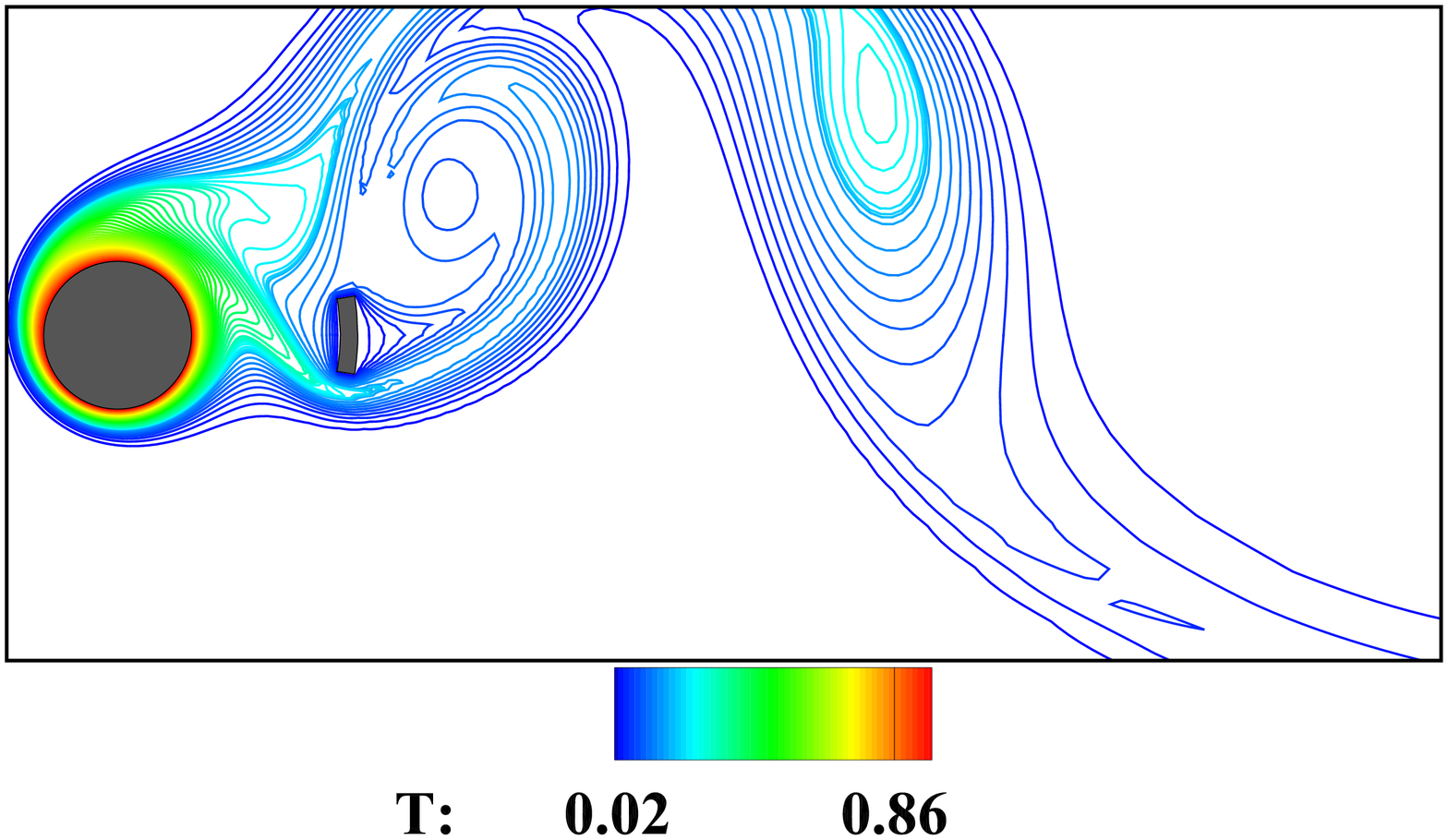}
\includegraphics[width=0.3\textwidth,trim={0.5cm 0.3cm 0.3cm 0.3cm},clip]{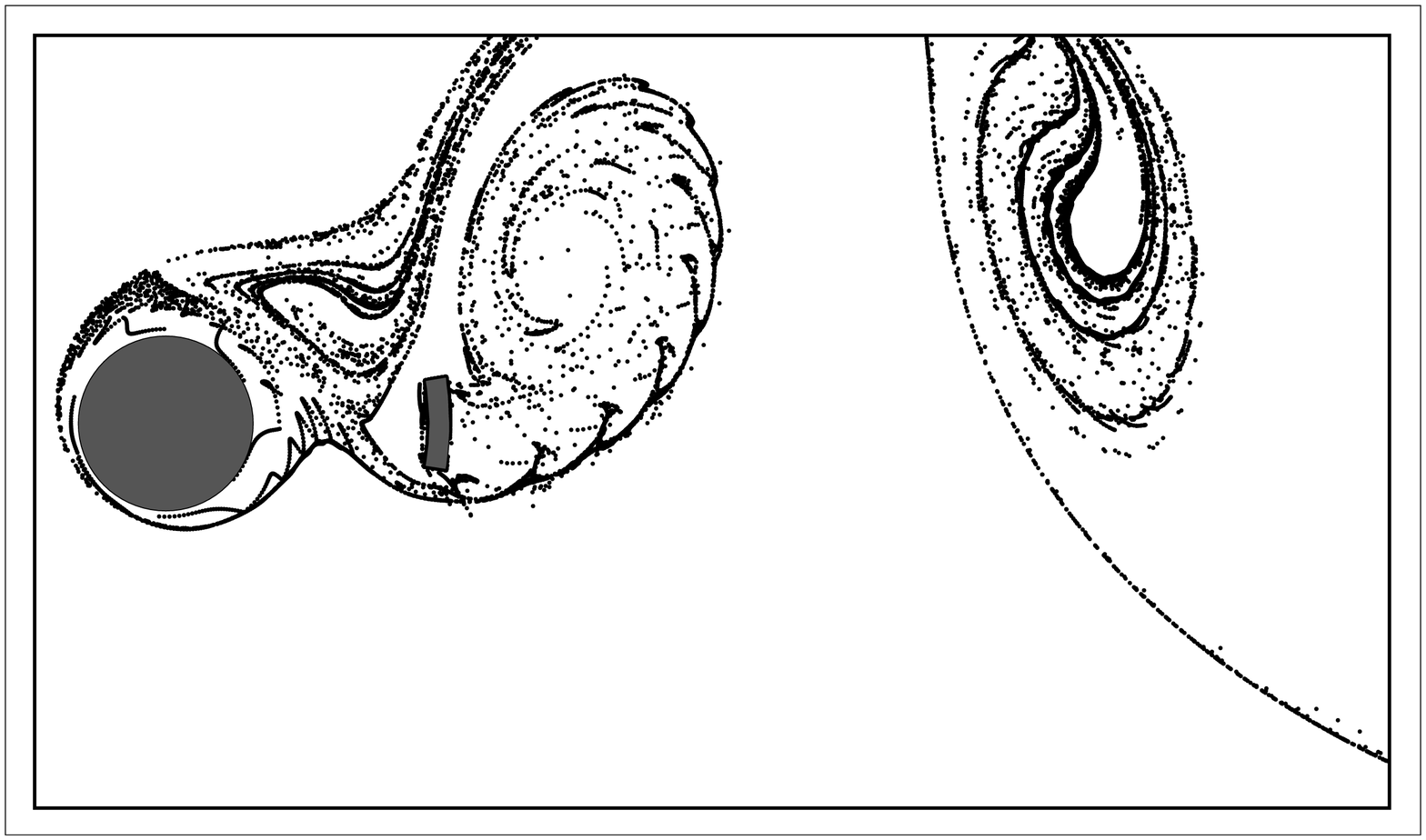}
\includegraphics[width=0.3\textwidth,trim={0.5cm 0.3cm 0.3cm 0.3cm},clip]{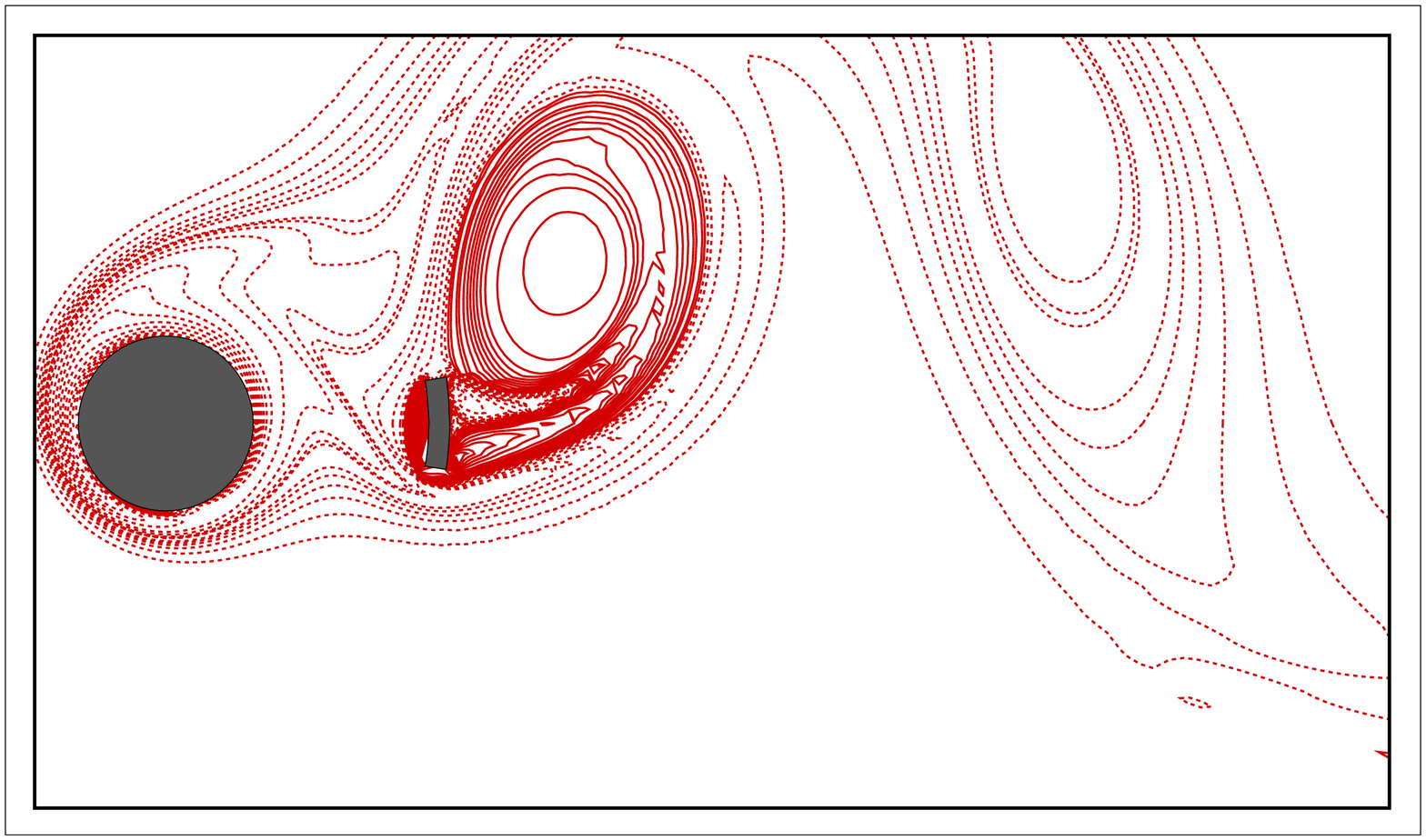}
\\
\hspace{0.5em}\scriptsize{$t=t_0+(1/2)T$}
\\
\includegraphics[width=0.29\textwidth,trim={0.5cm 0.3cm 0.5cm 0.3cm},clip]{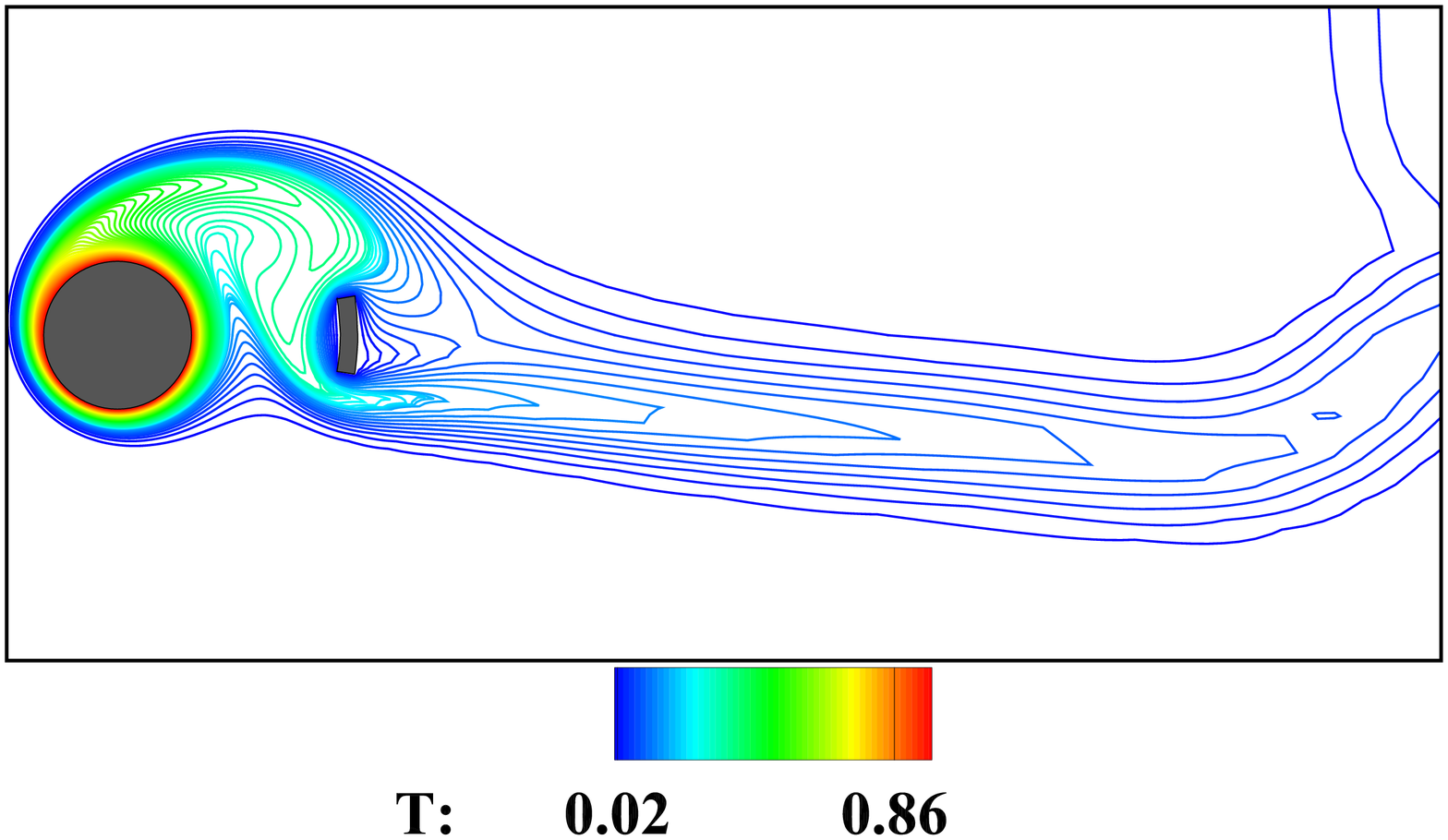}
\includegraphics[width=0.3\textwidth,trim={0.5cm 0.3cm 0.3cm 0.3cm},clip]{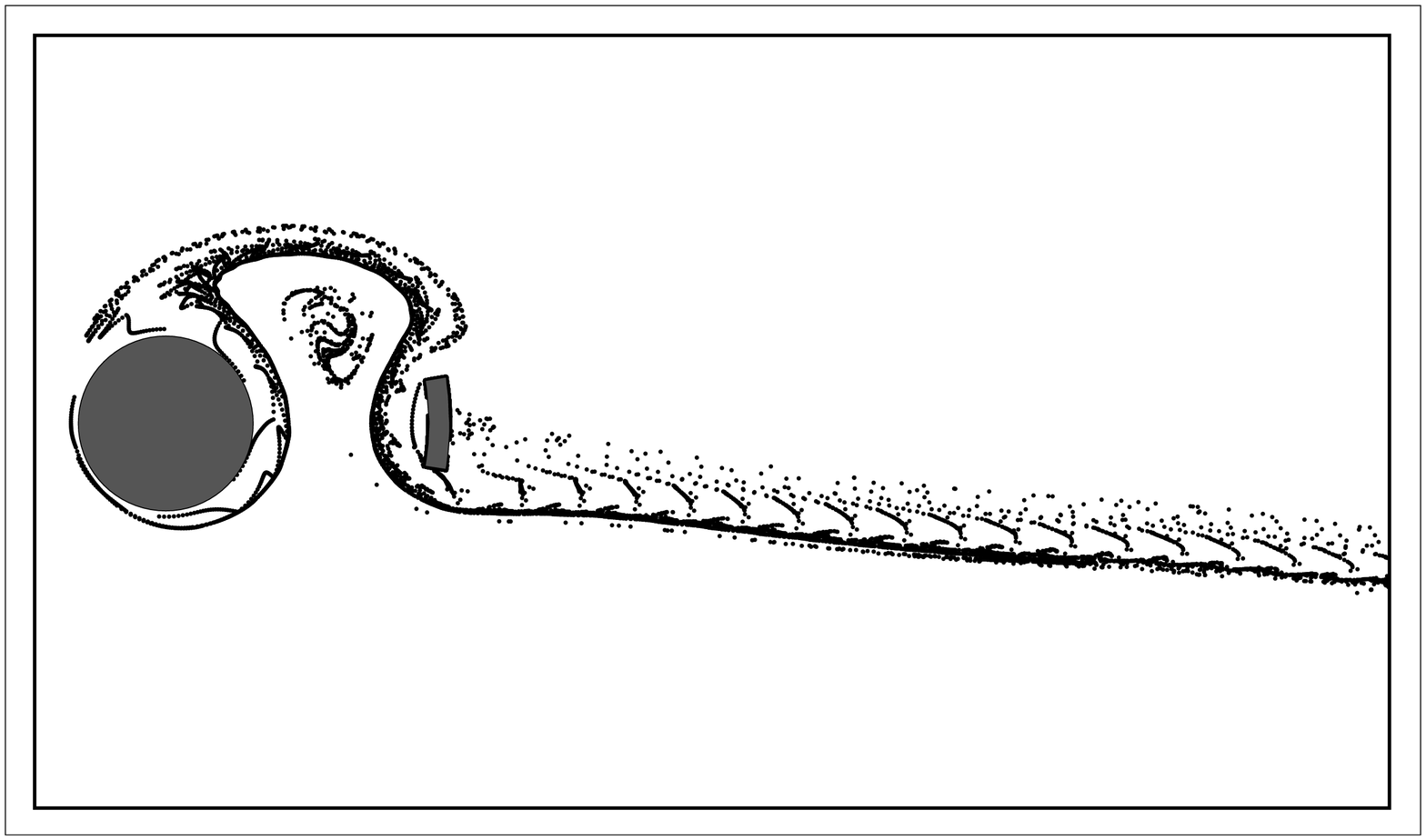}
\includegraphics[width=0.3\textwidth,trim={0.5cm 0.3cm 0.3cm 0.3cm},clip]{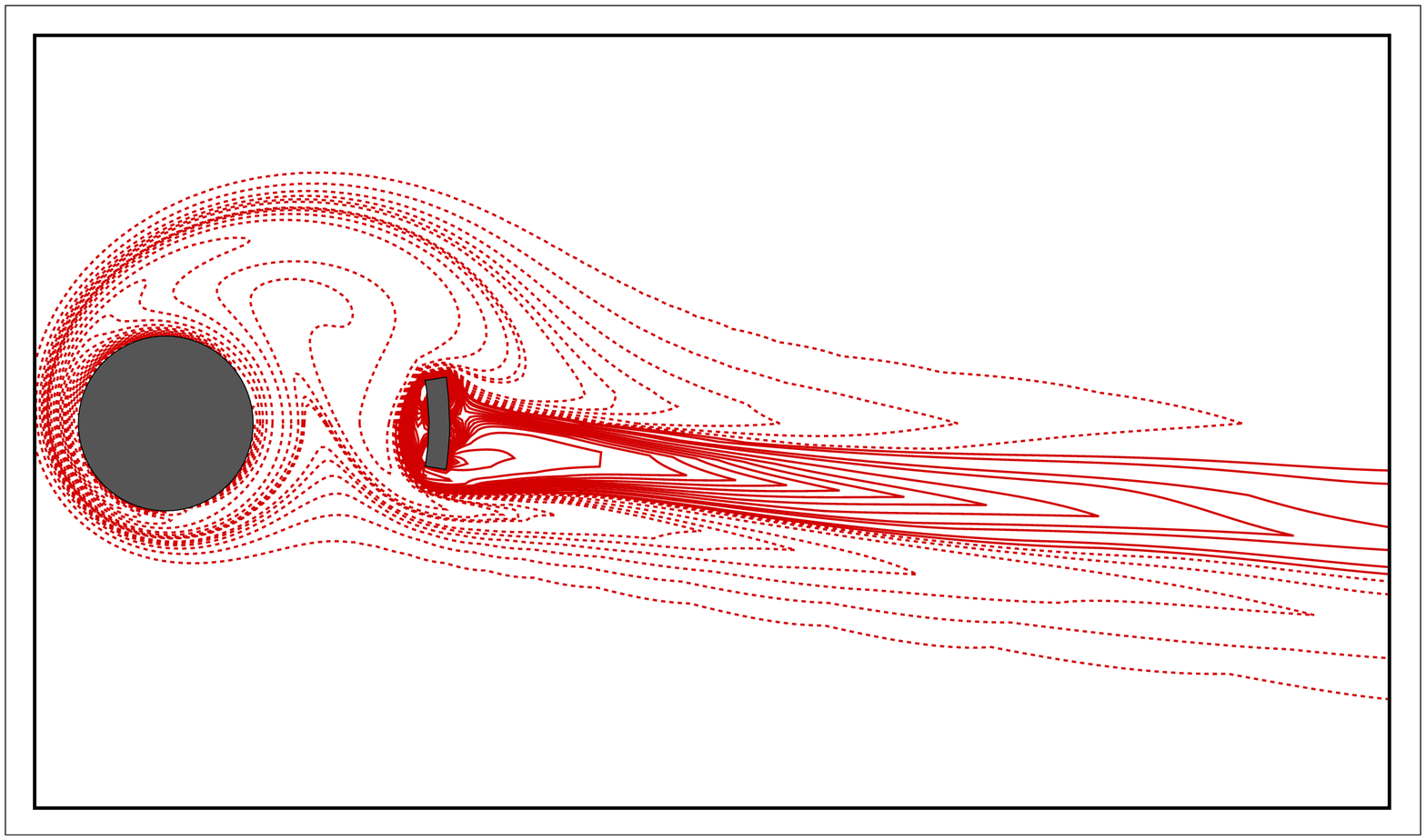}
\\
\hspace{0.5em}\scriptsize{$t=t_0+(3/4)T$}
\\
\includegraphics[width=0.29\textwidth,trim={0.5cm 0.3cm 0.5cm 0.3cm},clip]{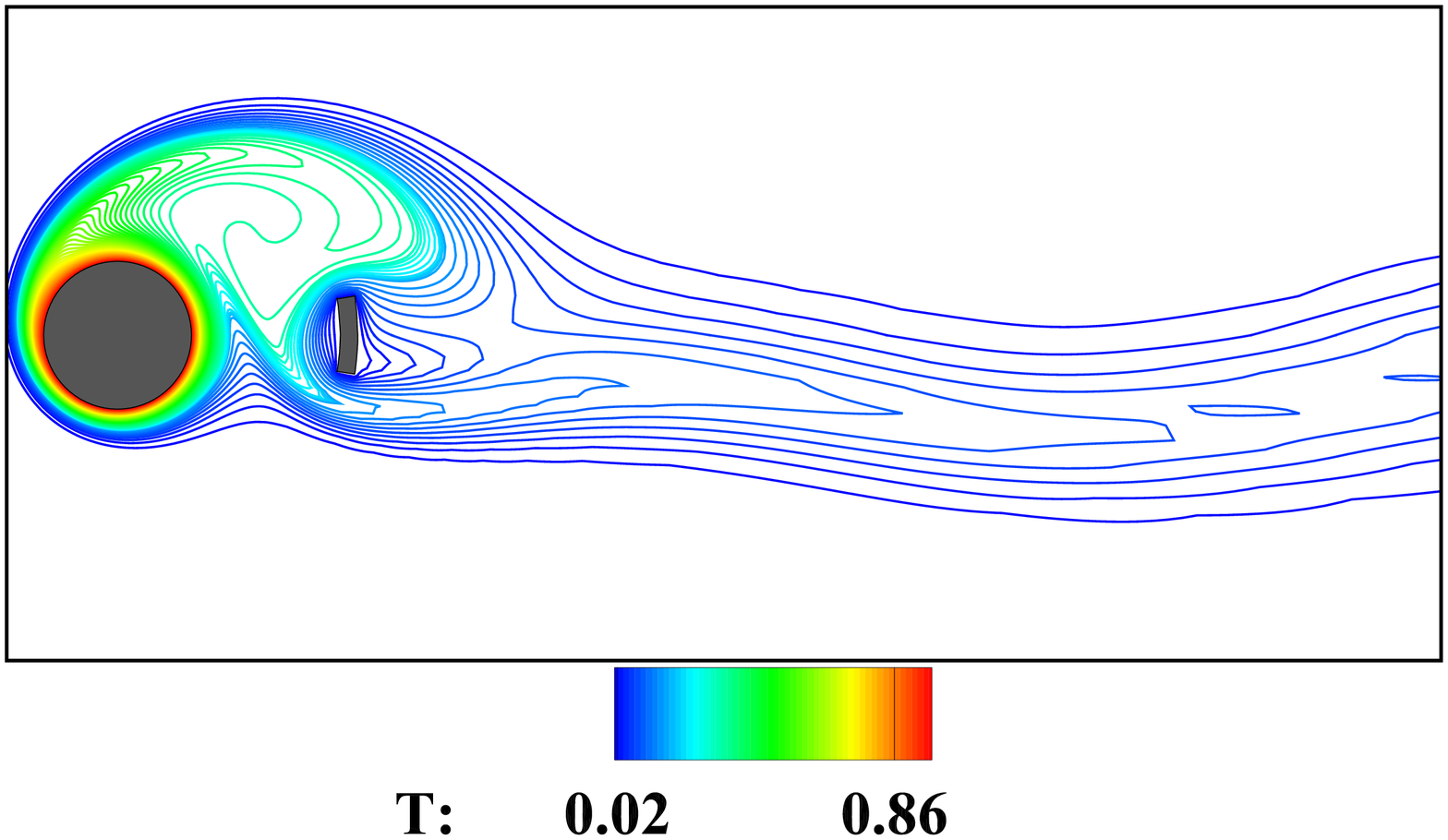}
\includegraphics[width=0.3\textwidth,trim={0.5cm 0.3cm 0.3cm 0.3cm},clip]{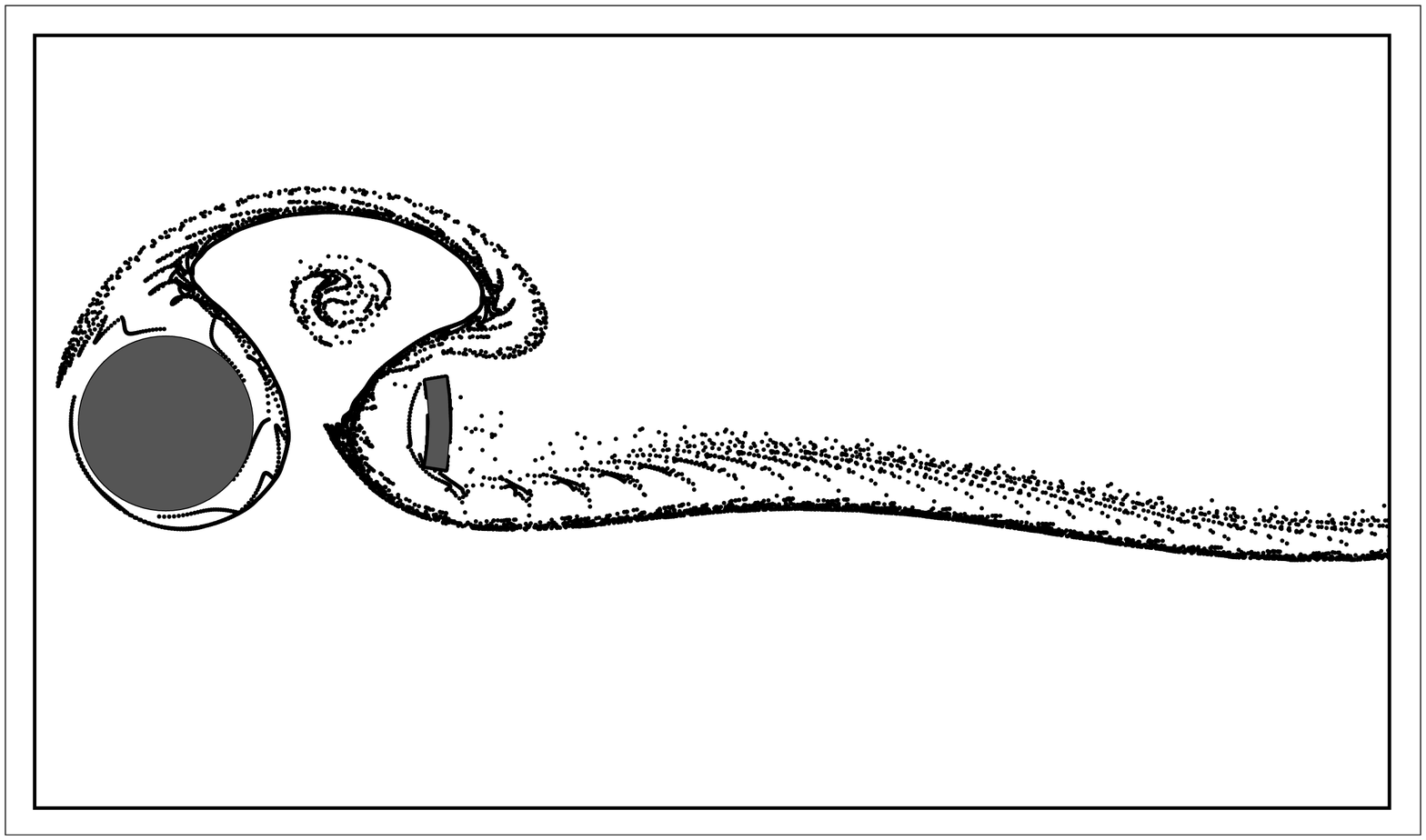}
\includegraphics[width=0.3\textwidth,trim={0.5cm 0.3cm 0.3cm 0.3cm},clip]{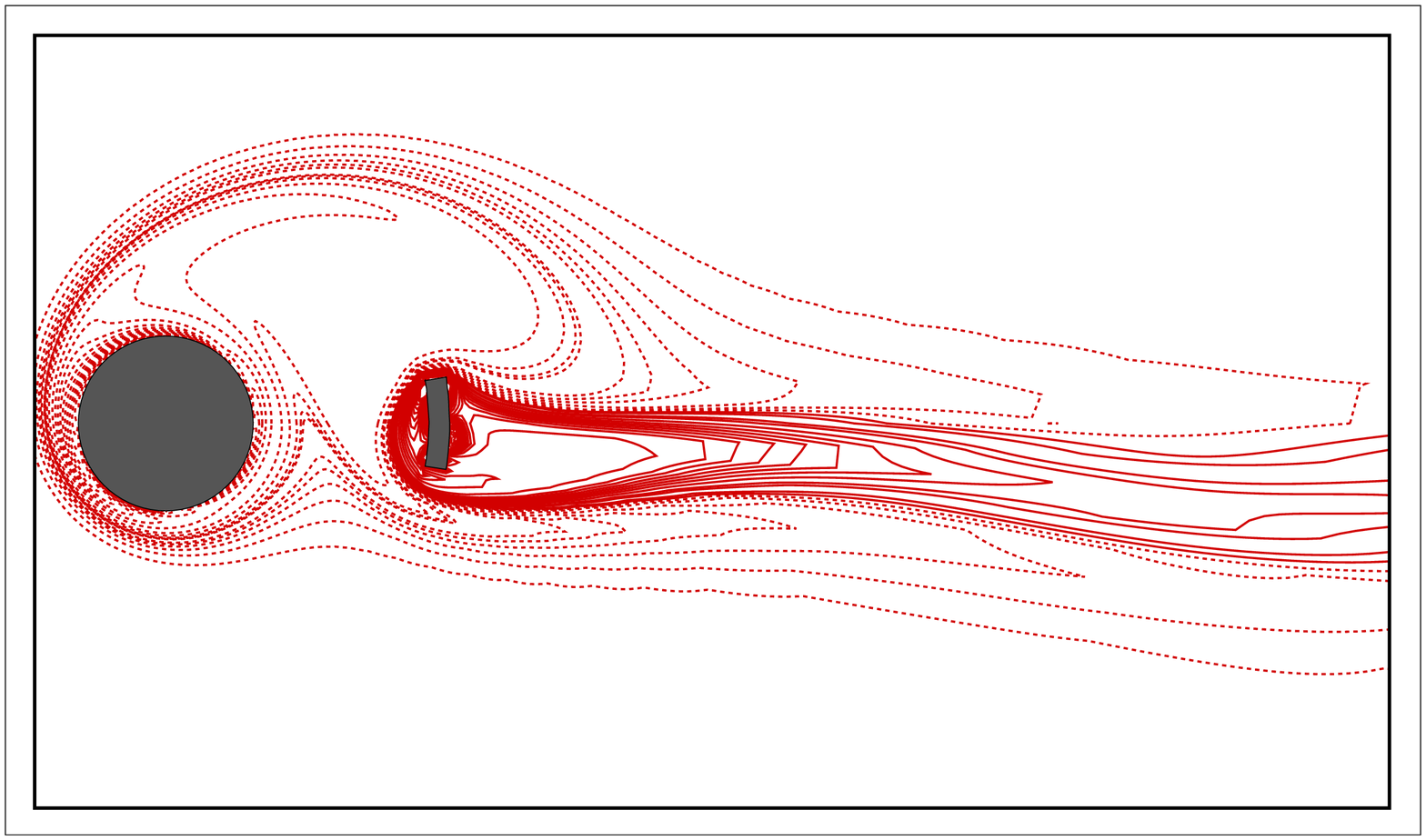}
\\
\hspace{0.5em}\scriptsize{$t=t_0+(1)T$}
\\
\includegraphics[width=0.29\textwidth,trim={0.5cm 0.3cm 0.5cm 0.3cm},clip]{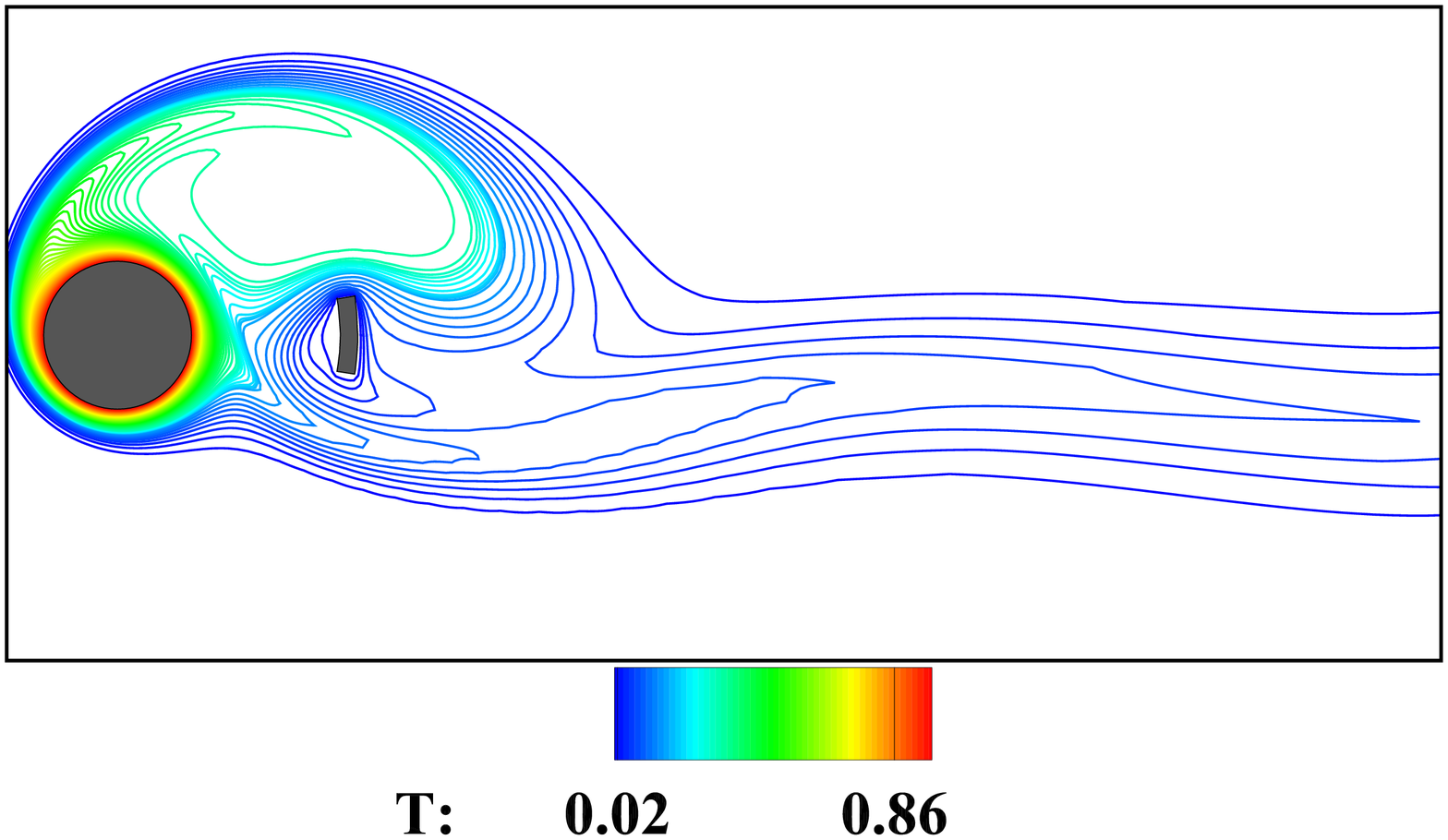}
\includegraphics[width=0.3\textwidth,trim={0.5cm 0.3cm 0.3cm 0.3cm},clip]{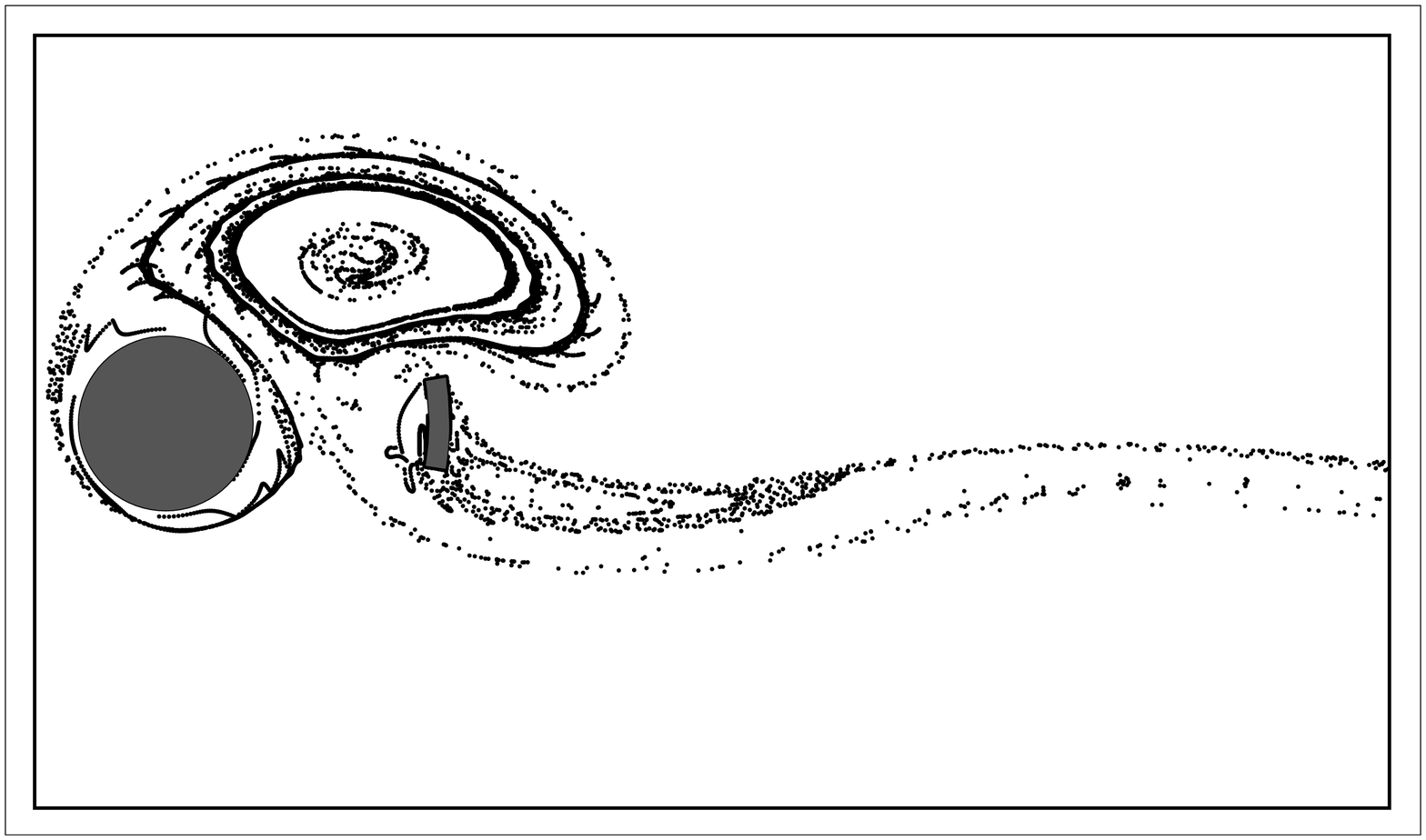}
\includegraphics[width=0.3\textwidth,trim={0.5cm 0.3cm 0.3cm 0.3cm},clip]{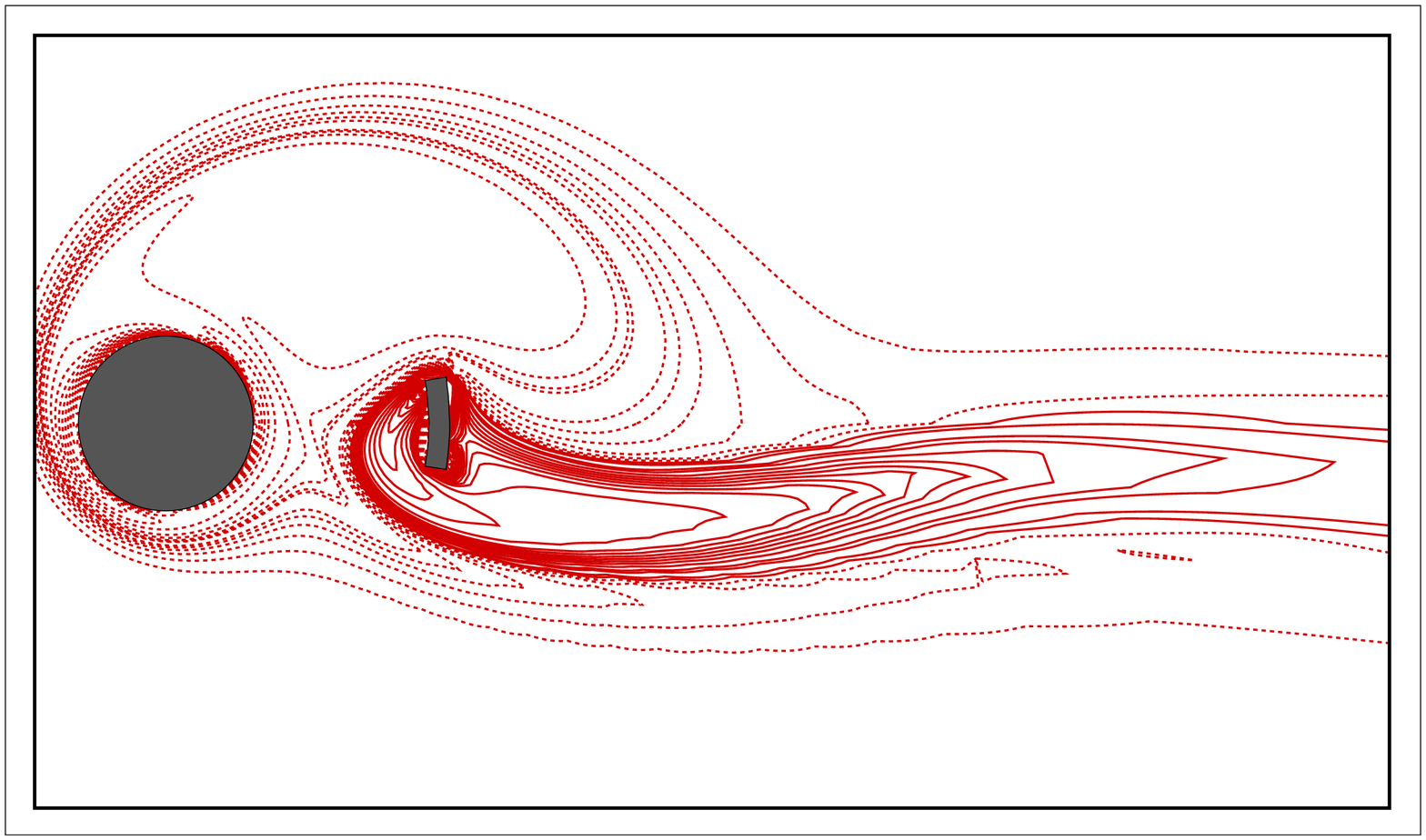}
\\
\hspace{2cm}(a) \hspace{4cm}(b) \hspace{4cm}(c)\hspace{2cm}
 \caption{(a) Isotherm, (b) streakline and (c) vorticity contour for $Pr=0.7$, $Re=150$, $\alpha=3.25$ and $d/R_0=2$ at different phases.}
 \label{fig:d_2_a_3-25}
\end{figure*}

\cref{fig:d_2_a_0-5} shows the isotherm, streakline and vorticity for $\alpha=0.5$ and $d/R_0=2$. Two vortices are shed periodically from the upper and lower sides of the cylinder, according to the streakline and vorticity. One recirculation zone is formed by the interaction of the shear layers between the cylinder and the plate, near the top of the plate. Positive equi-vorticity lines originated from the cylinder partially covers the control plate. The isotherm contours show that two warm blobs convect away with the shedding vortices. The vortex shedding plane is slightly higher than the centerline by approximately $\theta=15\degree$. This angle of the vortex shedding plane is slightly lower than that of \cref{fig:d_1_a_0-5} due to the increase in $d/R_0$. This happens due to the interaction of shear layers around the control plate. \cref{fig:d_2_a_3-25} exhibits the isotherm, streakline and vorticity for $\alpha=3.25$ and $d/R_0=2$. Here, two vortices are periodically shed. One is shed from the top of the cylinder, and the other one is shed from behind the plate. One temporary recirculation zone is formed between the cylinder and the plate, which gradually merges with the upper vortex. Due to the high rotational rate, the bottom vortex is pulled upwards and pushes the upper vortex. As a result, the upper vortex is shed much earlier. The negative equi-vorticity lines cover the positive equi-vorticity lines that originated from the control plate. After the negative vortex is shed, the shear layer is split into two by the positive vorticity contour. The shear layers from the top and bottom of the control plate are squeezed together by the negative vorticity contour to form the positive vortex. The vortex shedding plane is shifted by approximately $\theta=40\degree$ from the centerline, and this angle is also slightly lower than that of \cref{fig:d_1_a_3-25}. The widths of vortices are much larger than those at \cref{fig:d_1_a_3-25}. The interaction between the shear layer and the boundary layer of the cylinder thickens the thermal boundary layer near the front stagnation point and increases the density of the isotherm contour near the rear stagnation point and at the bottom of the cylinder. It leads to the reduction of heat transfer near the front stagnation point and an increase in heat transfer rate near the rear stagnation point and bottom of the cylinder. \cref{fig:d_2_a_0-5,fig:d_2_a_3-25} show that as $\alpha$ increases from $0.5$ to $3.25$ for $d/R_0=2$, the vortices increase in size.\\

\begin{figure*}[!t]
\centering
\scriptsize{$t=t_0+(0)T$}
\\
\includegraphics[width=0.3\textwidth,trim={0.5cm 0.3cm 0.5cm 0.3cm},clip]{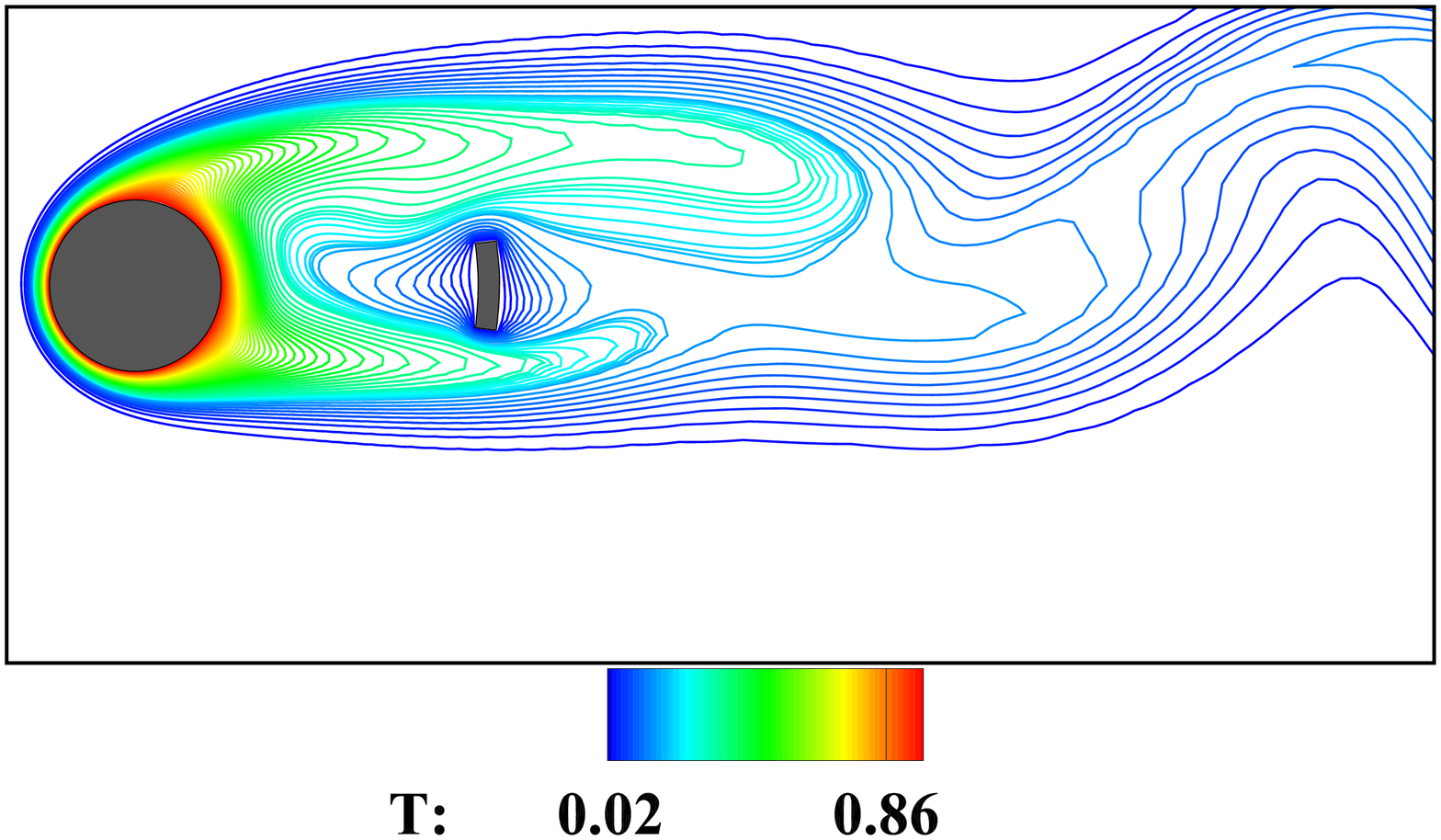}
\includegraphics[width=0.3\textwidth,trim={0.5cm 0.3cm 0.3cm 0.3cm},clip]{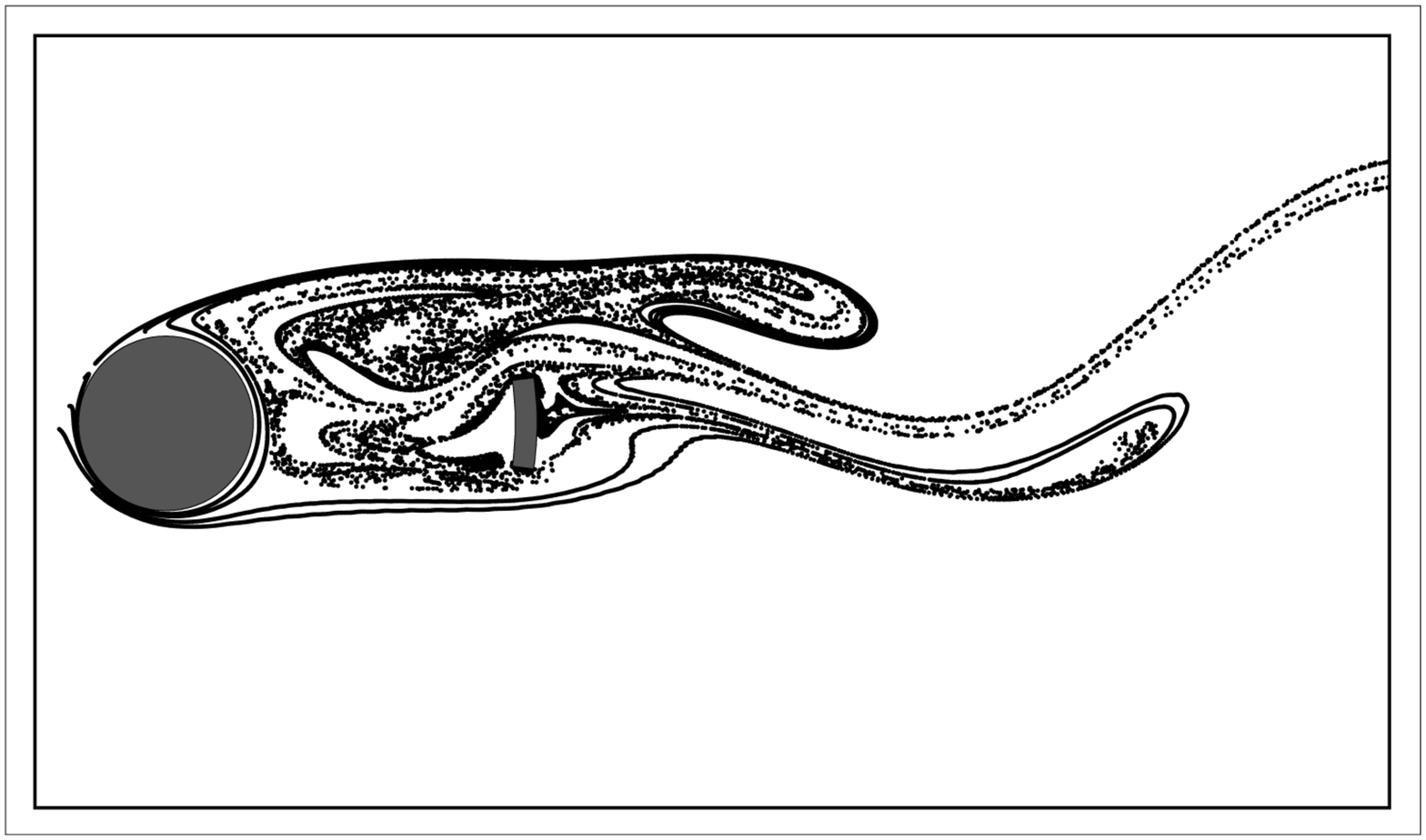}
\includegraphics[width=0.3\textwidth,trim={0.5cm 0.3cm 0.3cm 0.3cm},clip]{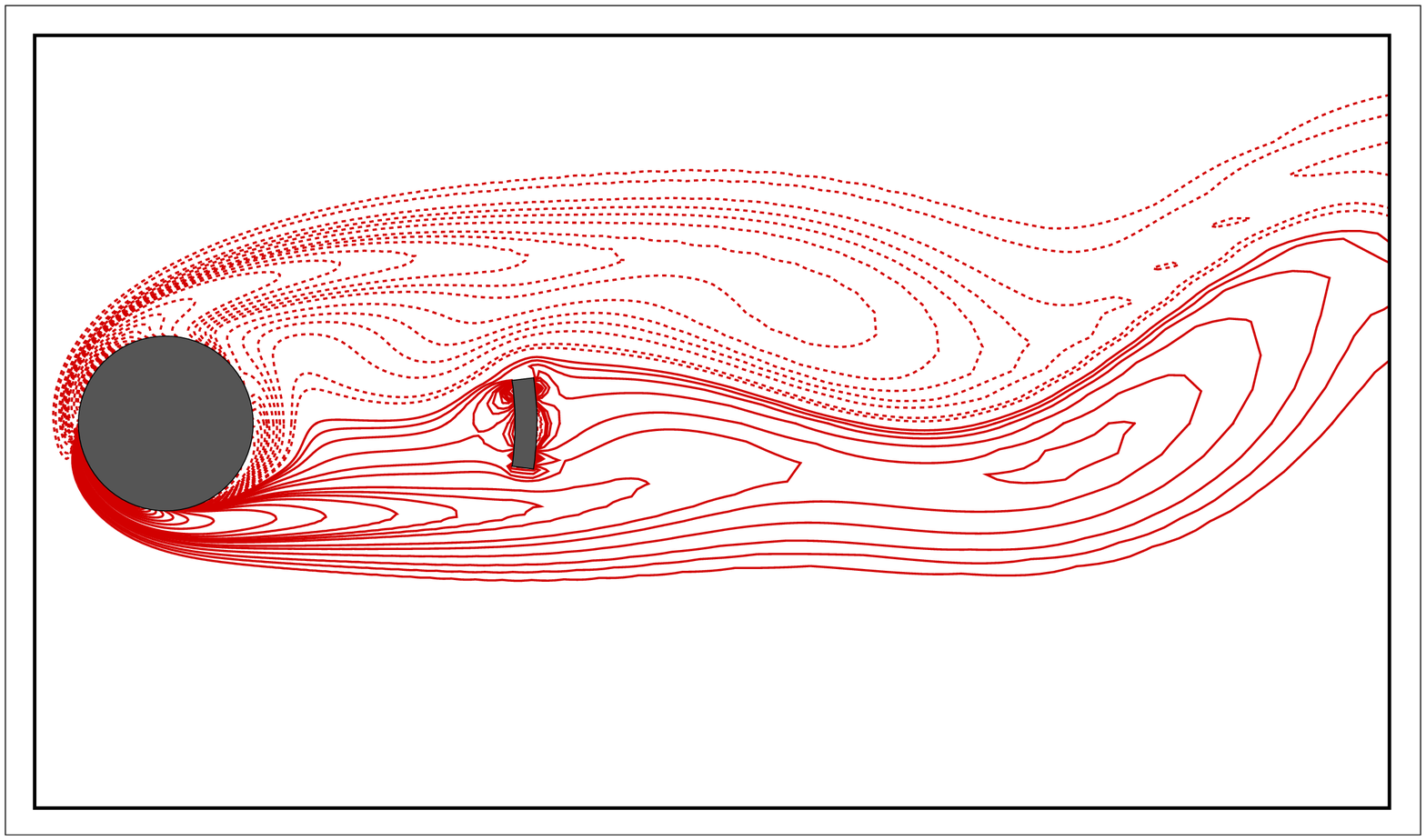}
\\
\hspace{0.5em}\scriptsize{$t=t_0+(1/4)T$}
\\
\includegraphics[width=0.29\textwidth,trim={0.5cm 0.3cm 0.5cm 0.3cm},clip]{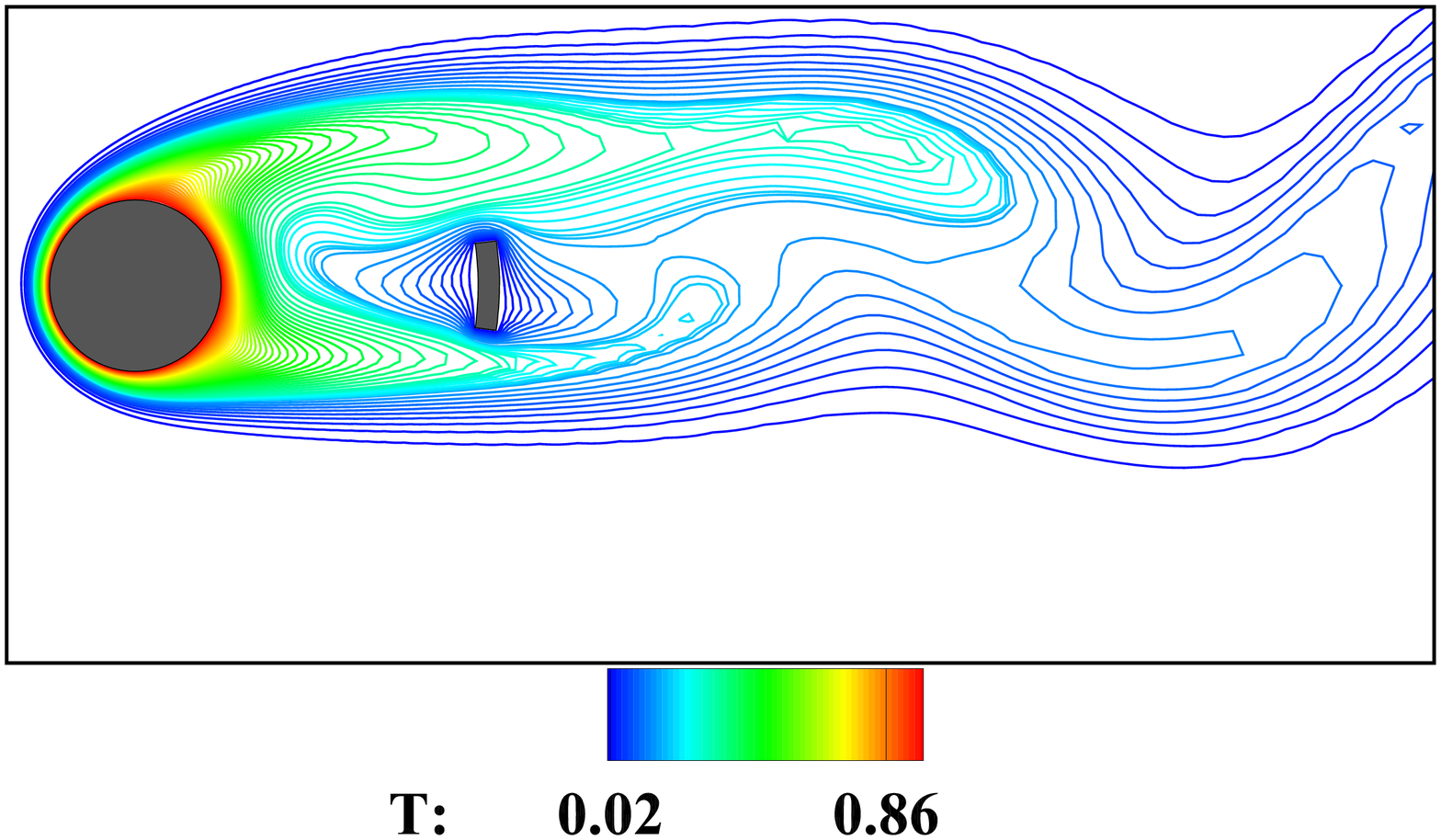}
\includegraphics[width=0.3\textwidth,trim={0.5cm 0.3cm 0.3cm 0.3cm},clip]{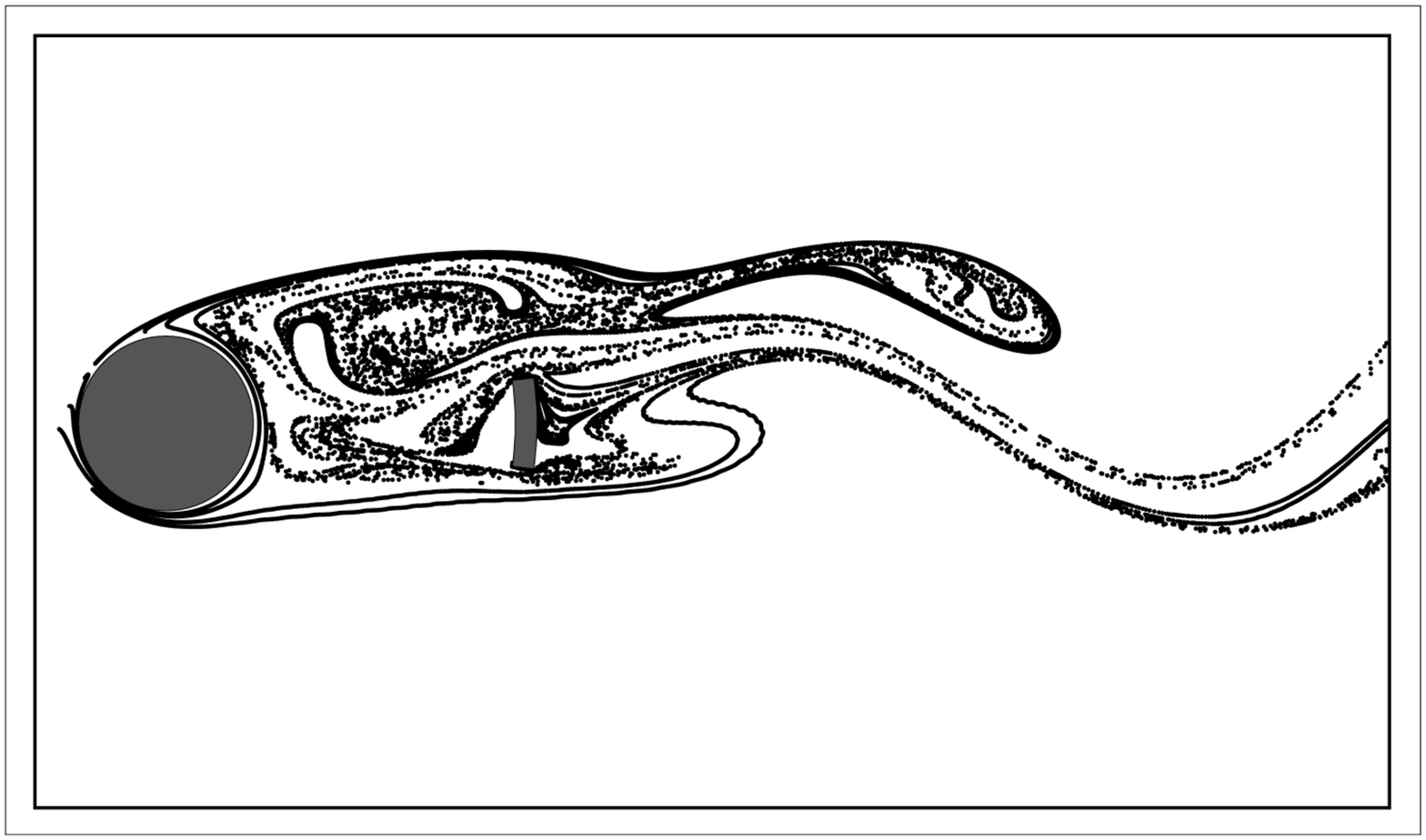}
\includegraphics[width=0.3\textwidth,trim={0.5cm 0.3cm 0.3cm 0.3cm},clip]{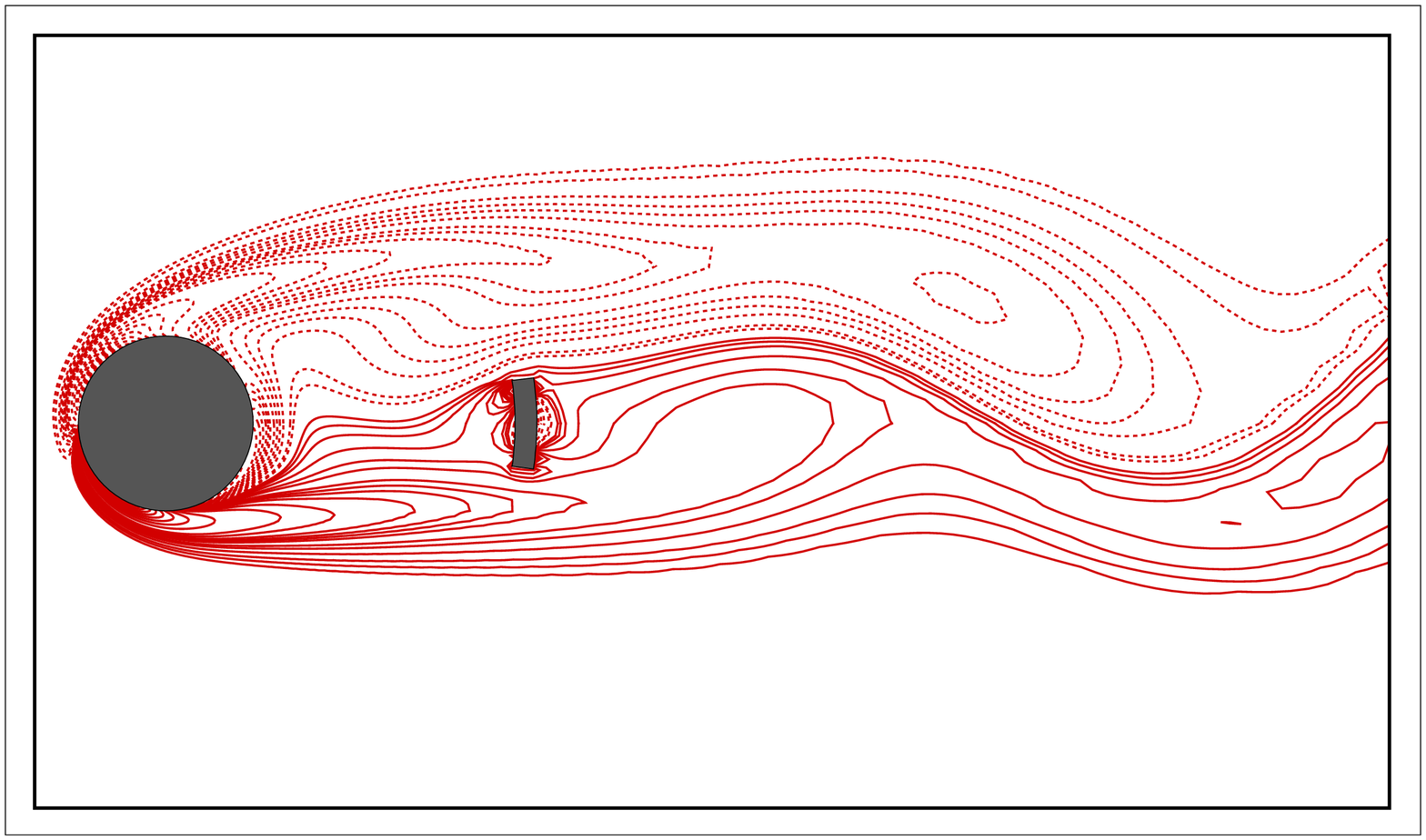}
\\
\hspace{0.5em}\scriptsize{$t=t_0+(1/2)T$}
\\
\includegraphics[width=0.29\textwidth,trim={0.5cm 0.3cm 0.5cm 0.3cm},clip]{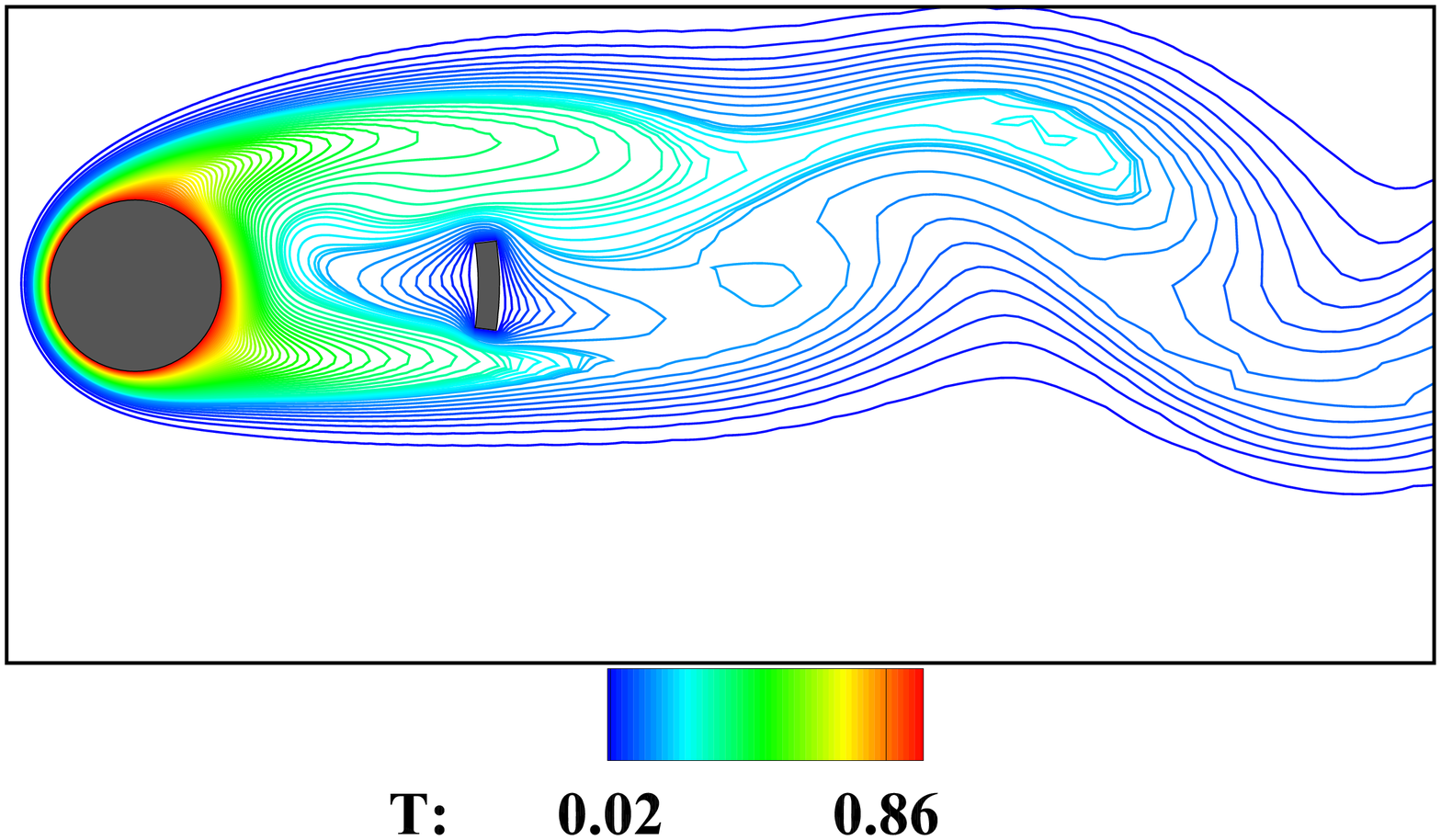}
\includegraphics[width=0.3\textwidth,trim={0.5cm 0.3cm 0.3cm 0.3cm},clip]{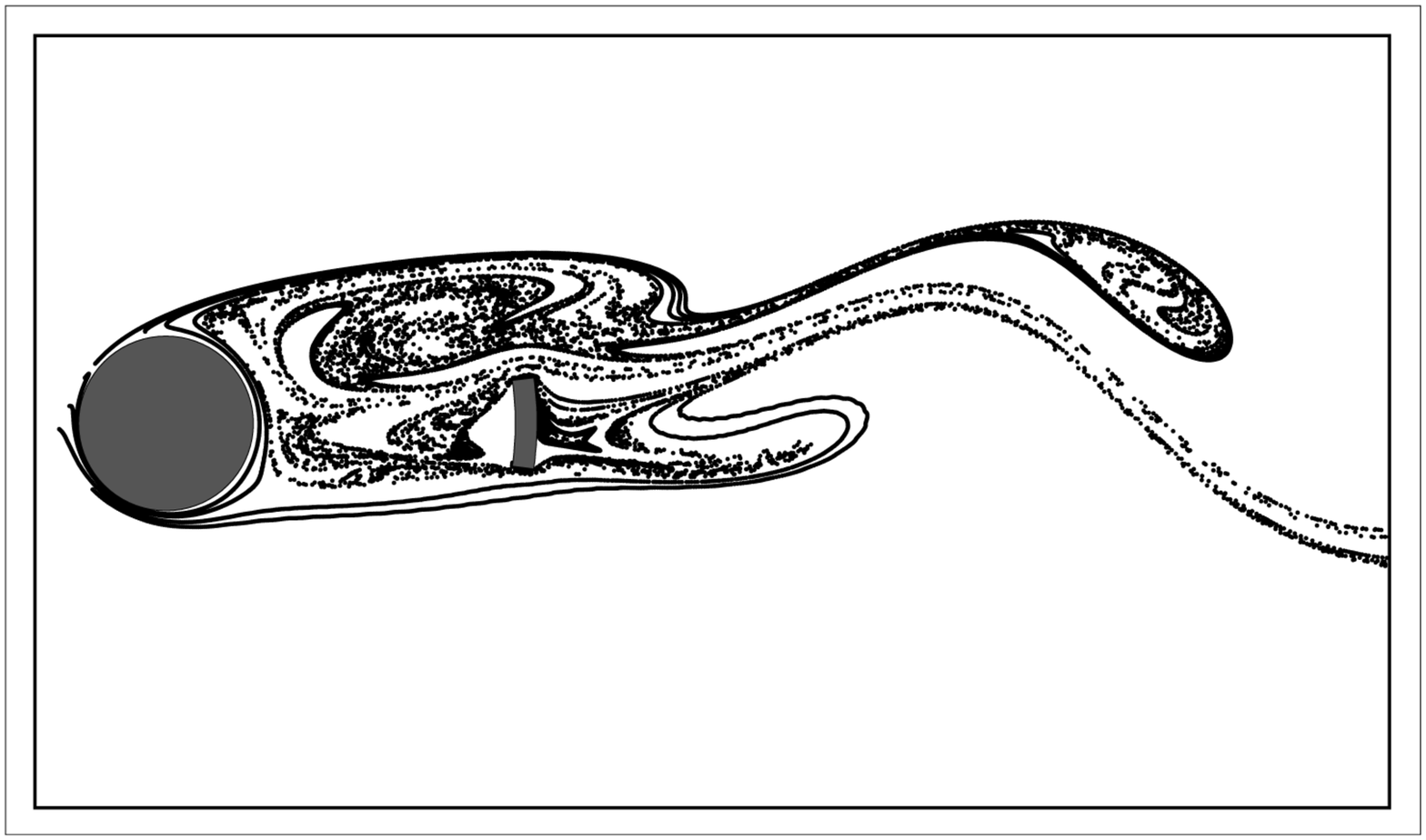}
\includegraphics[width=0.3\textwidth,trim={0.5cm 0.3cm 0.3cm 0.3cm},clip]{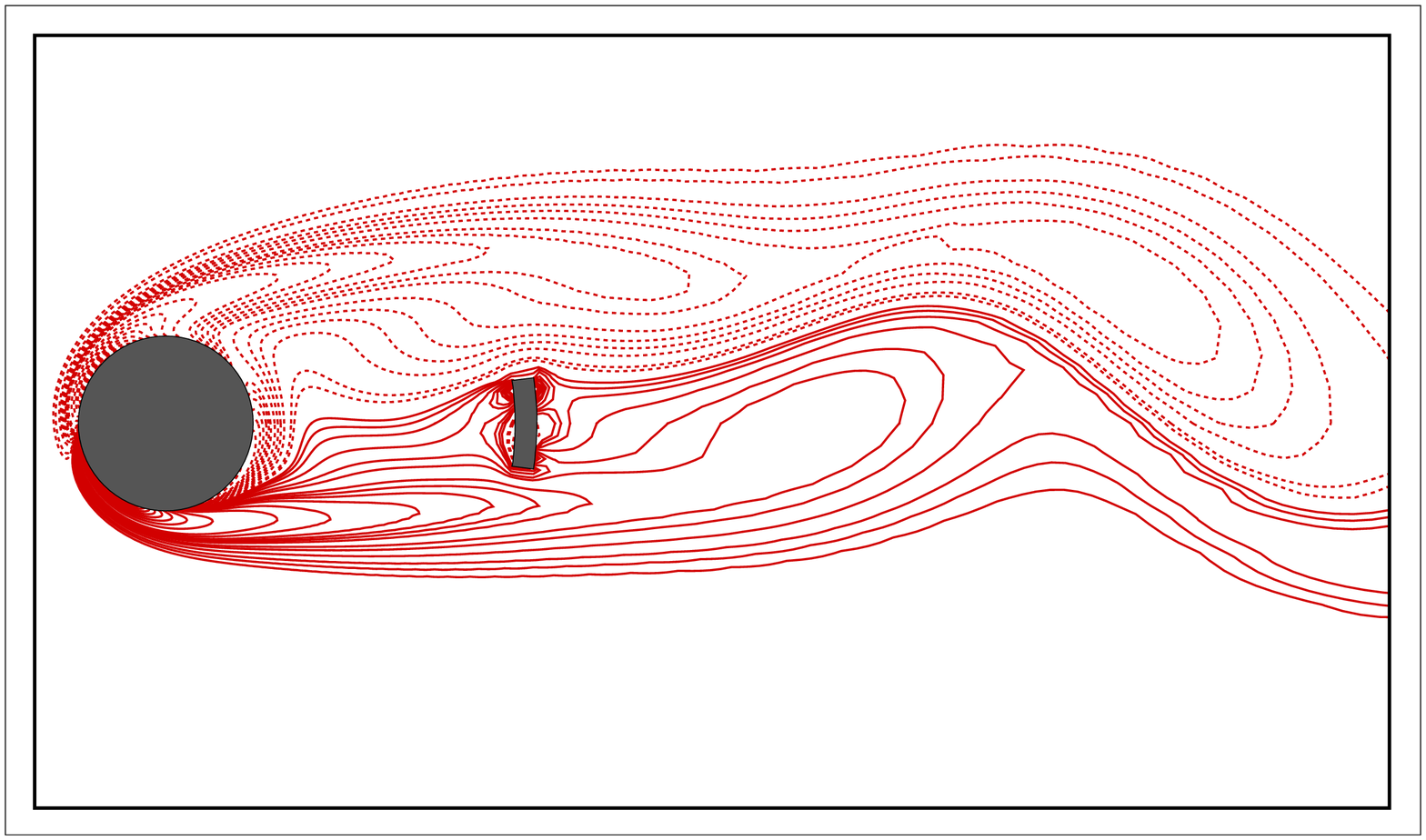}
\\
\hspace{0.5em}\scriptsize{$t=t_0+(3/4)T$}
\\
\includegraphics[width=0.29\textwidth,trim={0.5cm 0.3cm 0.5cm 0.3cm},clip]{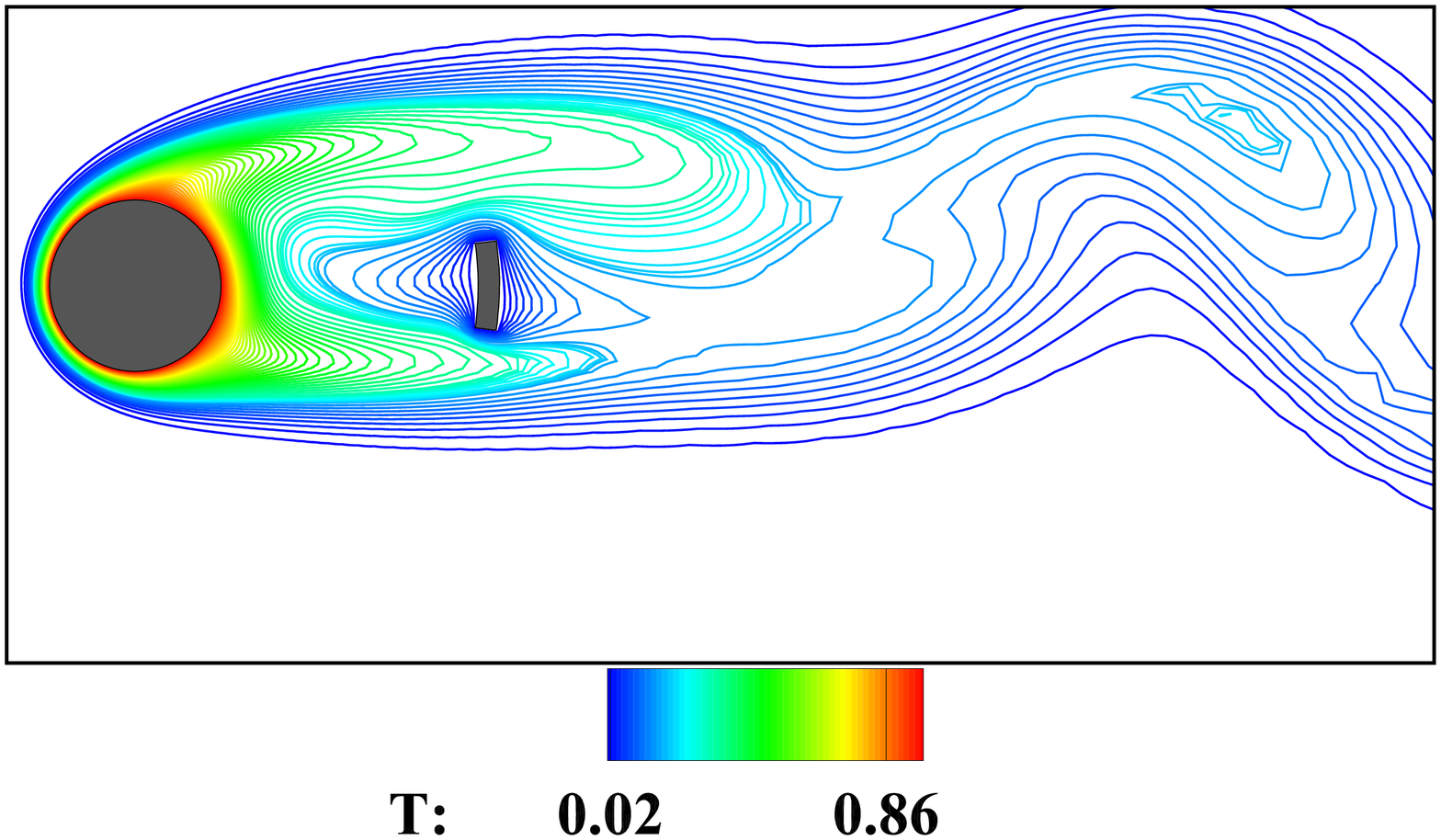}
\includegraphics[width=0.3\textwidth,trim={0.5cm 0.3cm 0.3cm 0.3cm},clip]{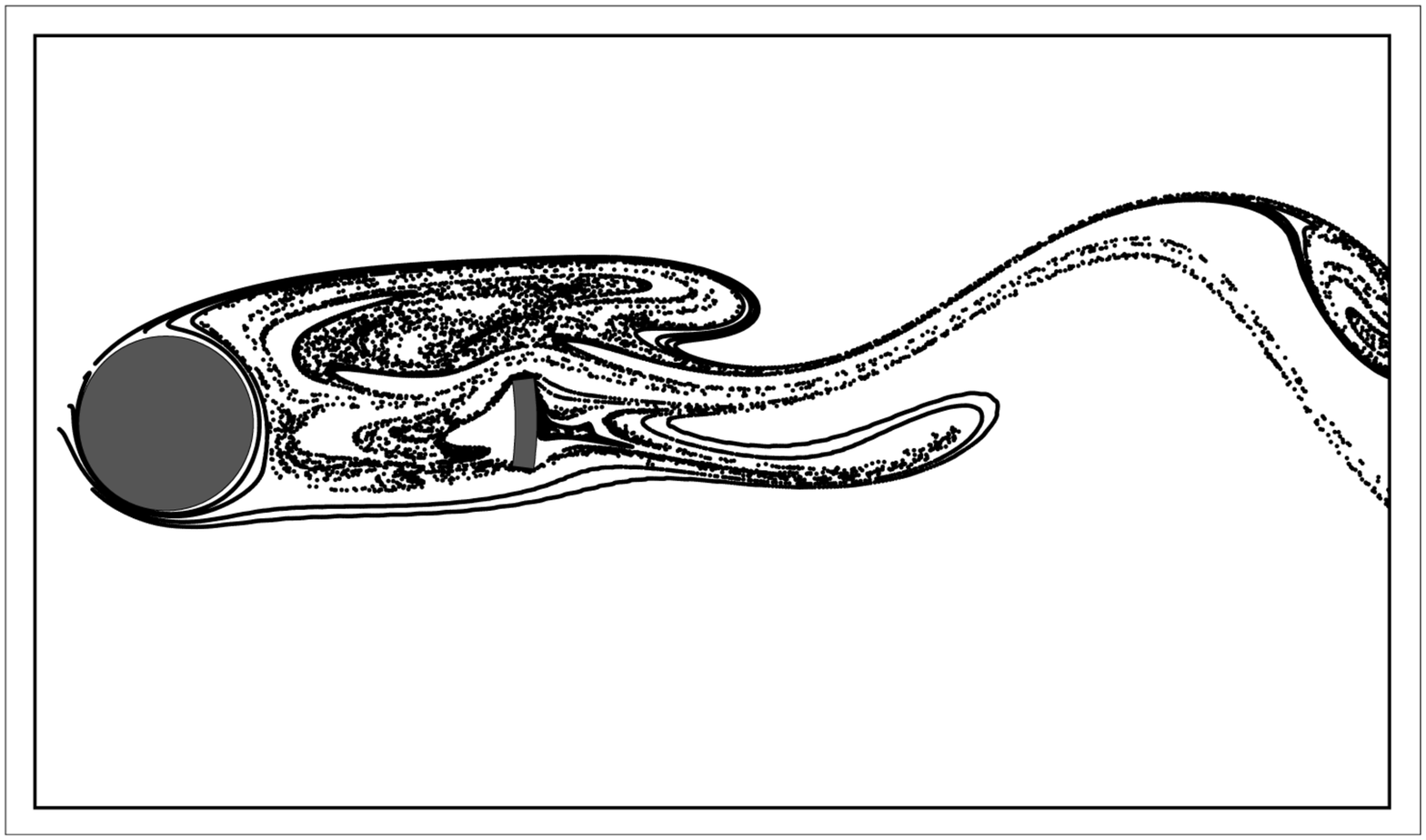}
\includegraphics[width=0.3\textwidth,trim={0.5cm 0.3cm 0.3cm 0.3cm},clip]{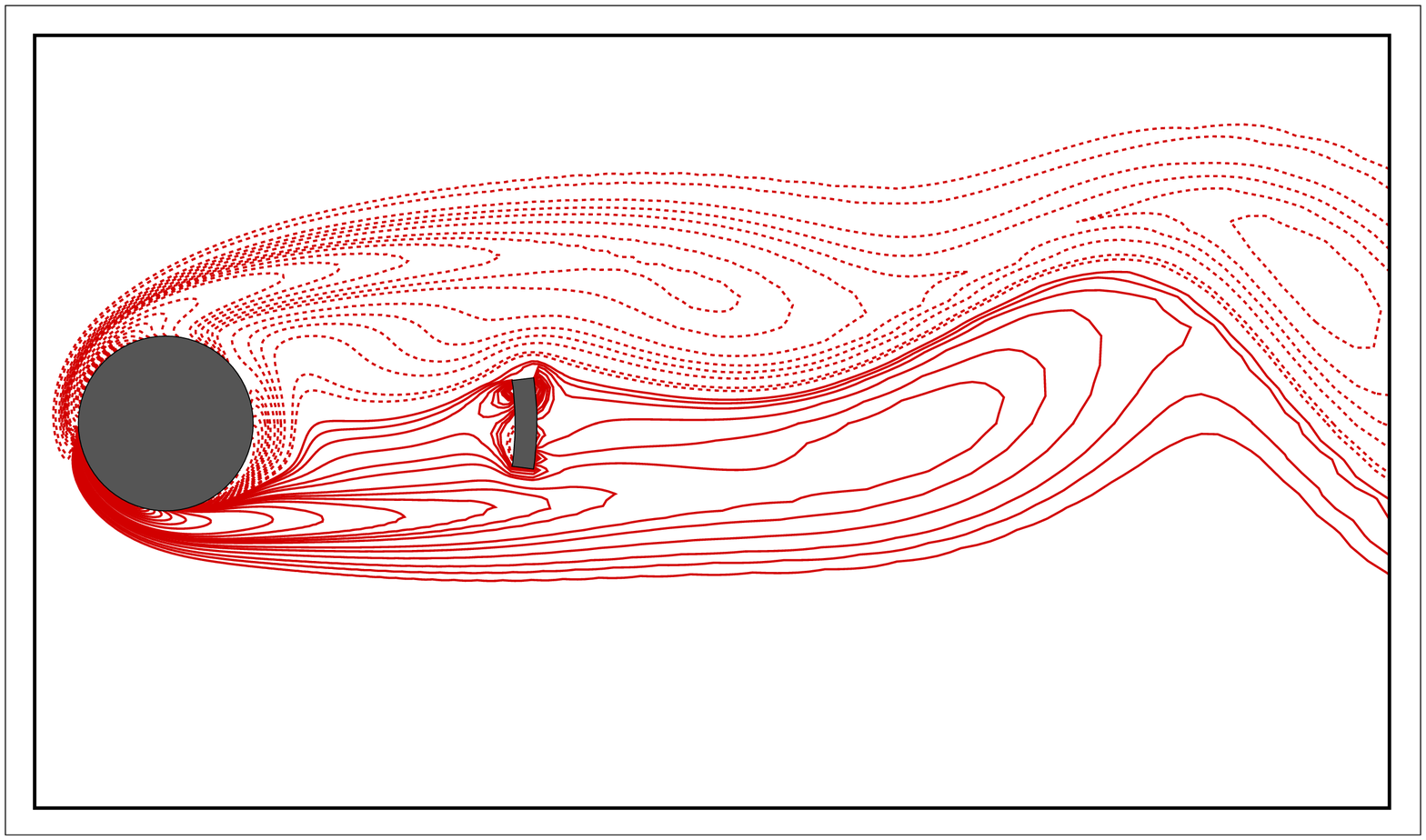}
\\
\hspace{0.5em}\scriptsize{$t=t_0+(1)T$}
\\
\includegraphics[width=0.29\textwidth,trim={0.5cm 0.3cm 0.5cm 0.3cm},clip]{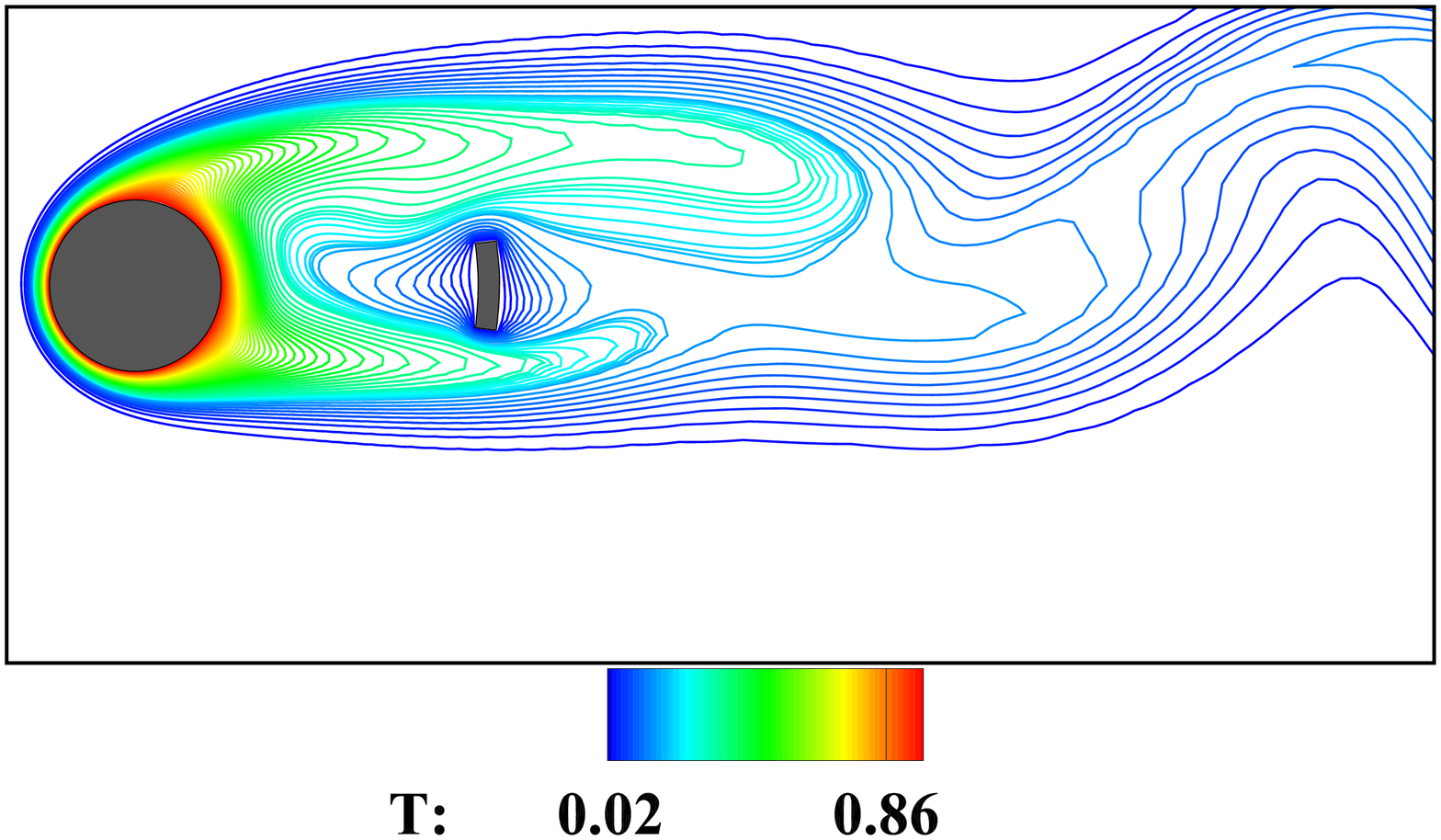}
\includegraphics[width=0.3\textwidth,trim={0.5cm 0.3cm 0.3cm 0.3cm},clip]{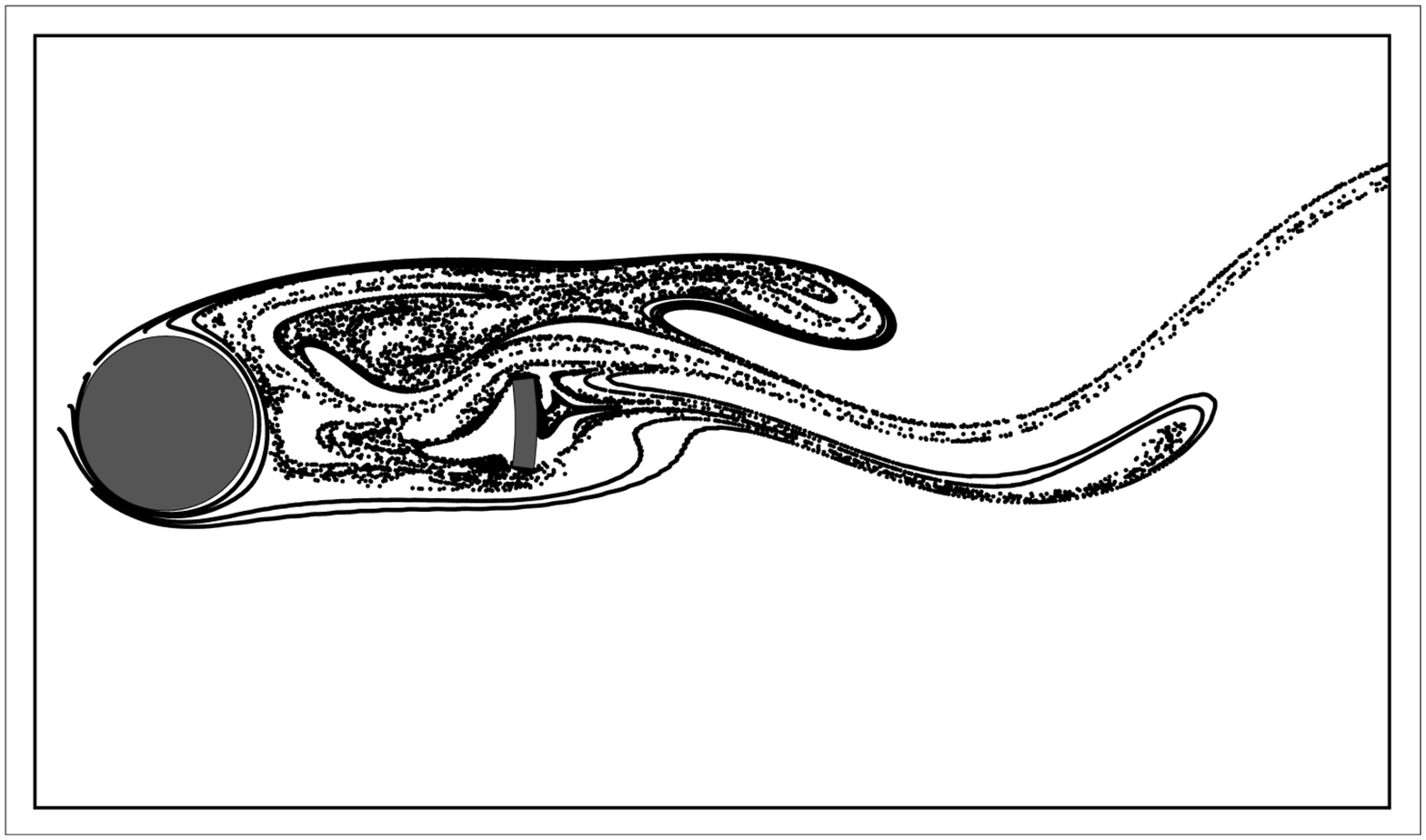}
\includegraphics[width=0.3\textwidth,trim={0.5cm 0.3cm 0.3cm 0.3cm},clip]{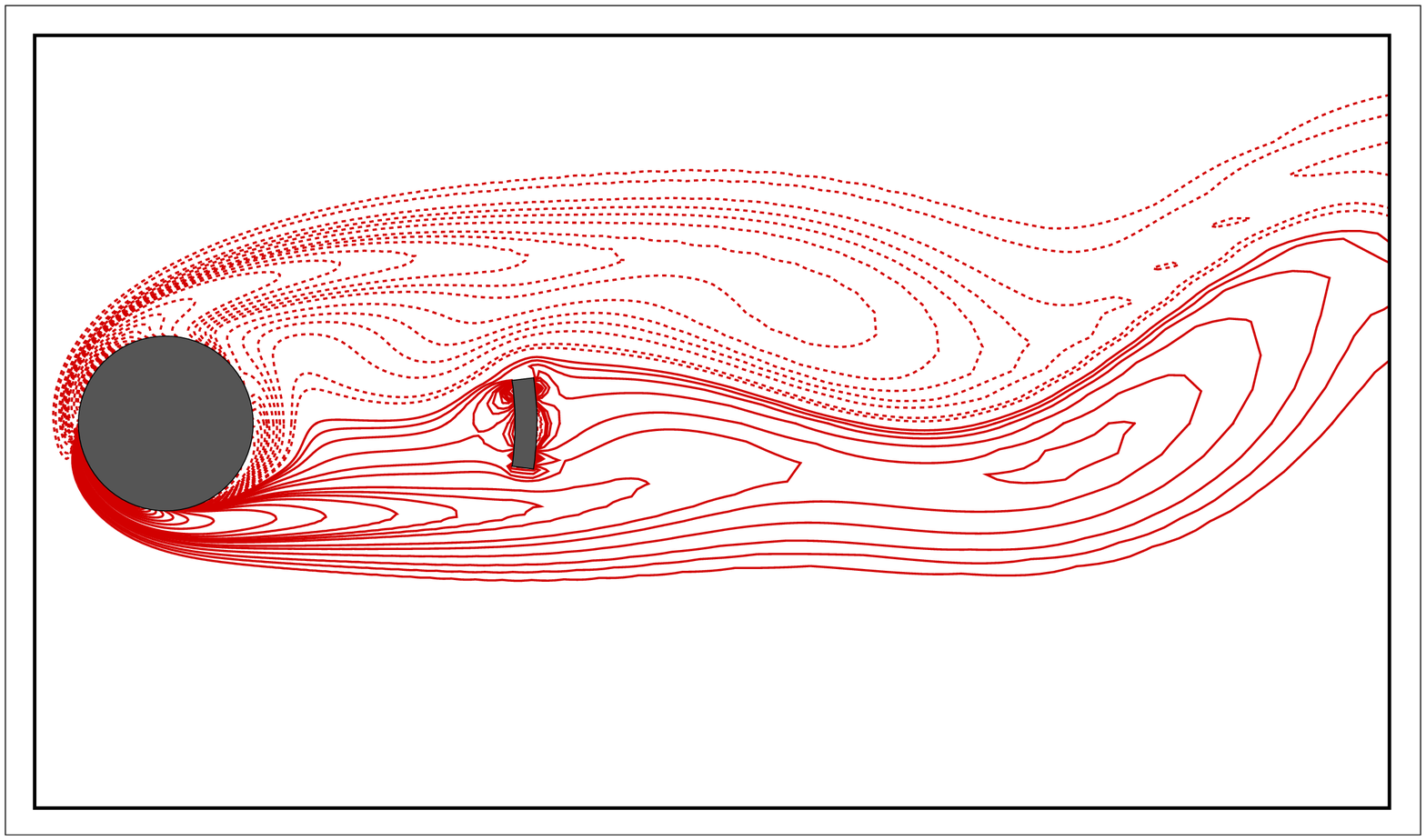}
\\
\hspace{2cm}(a) \hspace{4cm}(b) \hspace{4cm}(c)\hspace{2cm}
 \caption{(a) Isotherm, (b) streakline and (c) vorticity contour for $Pr=0.7$, $Re=150$, $\alpha=0.5$ and $d/R_0=3$ at different phases.}
 \label{fig:d_3_a_0-5}
\end{figure*}

\begin{figure*}[!t]
\centering
\scriptsize{$t=t_0+(0)T$}
\\
\includegraphics[width=0.3\textwidth,trim={0.5cm 0.3cm 0.5cm 0.3cm},clip]{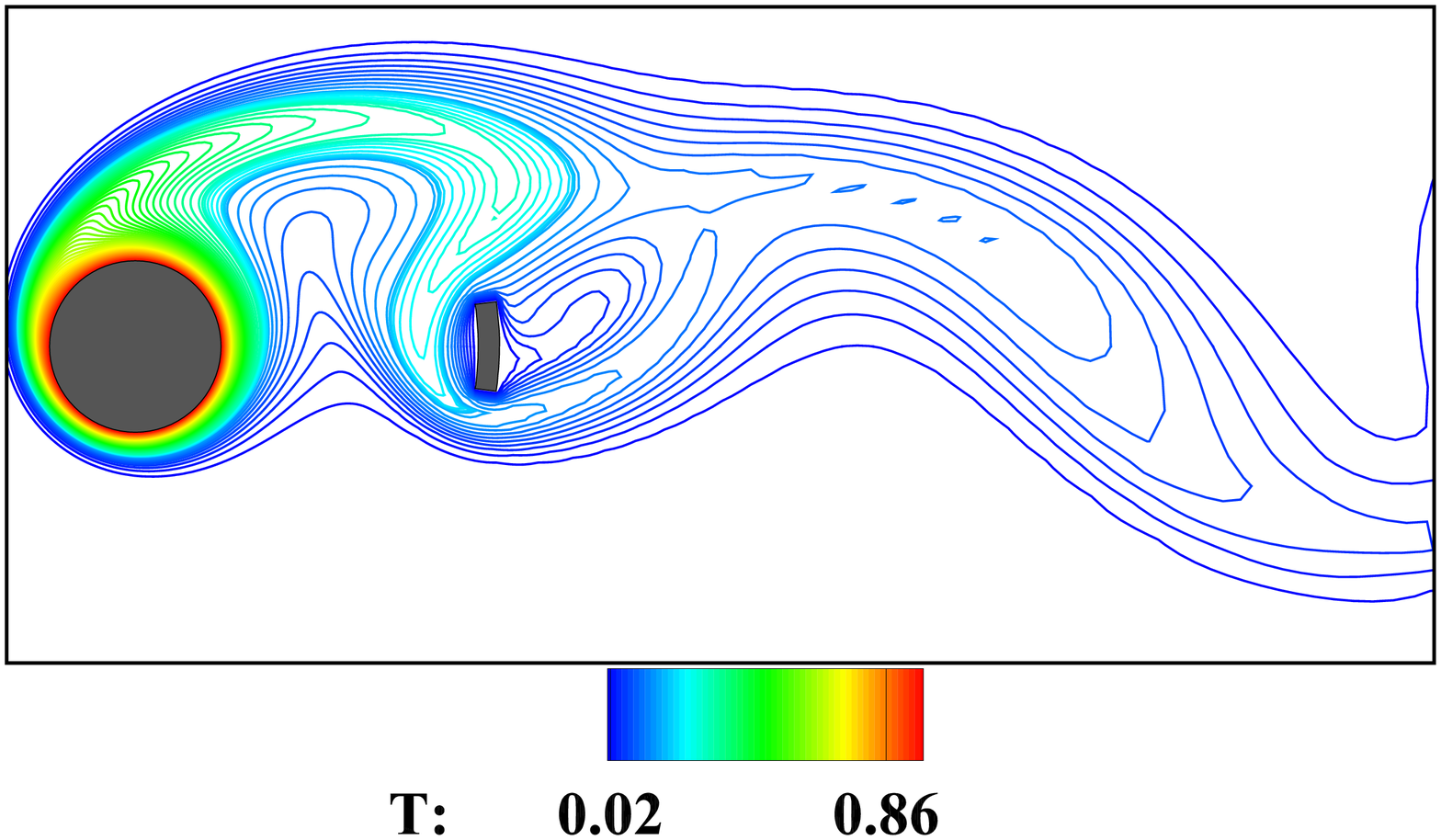}
\includegraphics[width=0.3\textwidth,trim={0.5cm 0.3cm 0.3cm 0.3cm},clip]{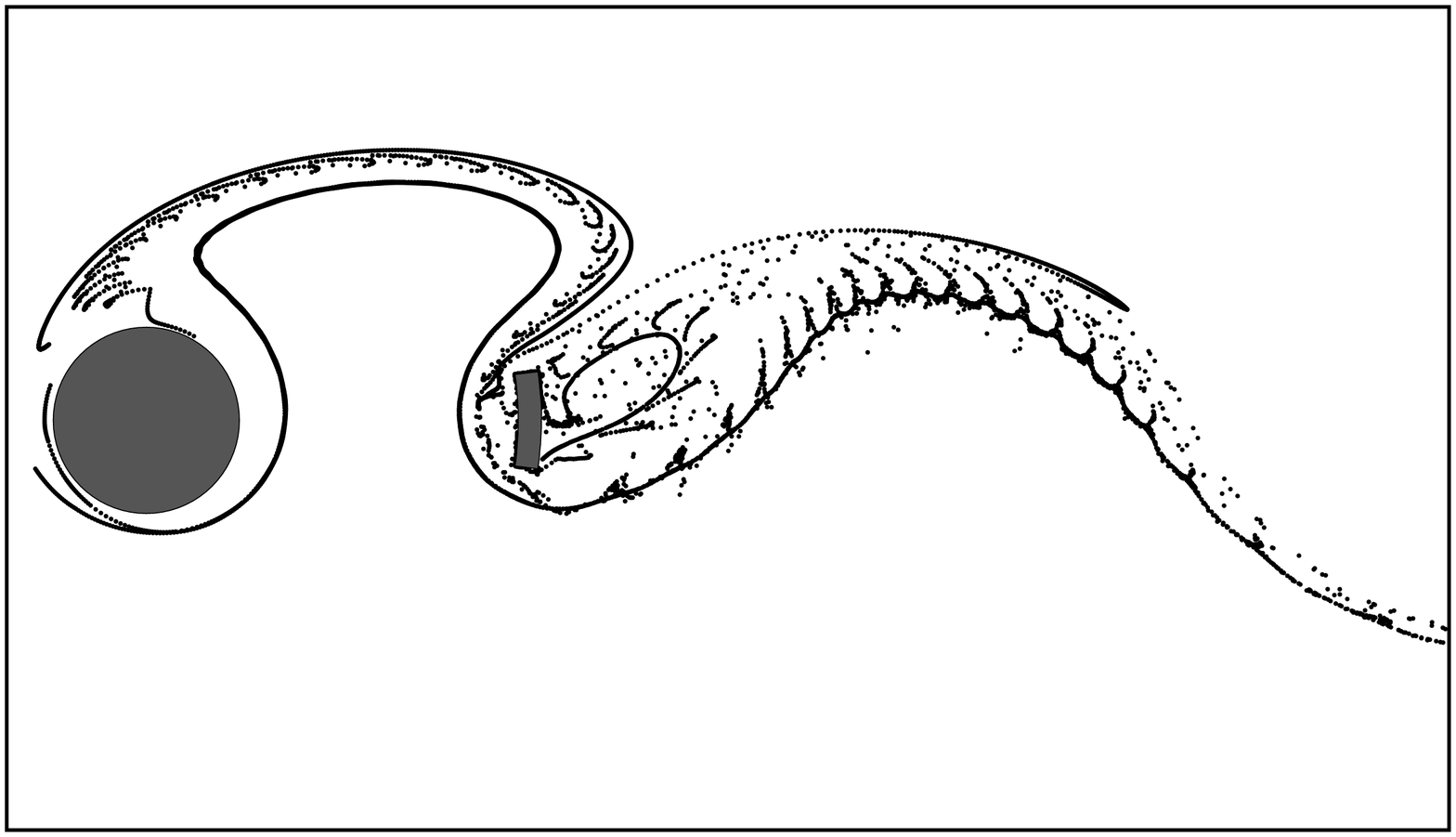}
\includegraphics[width=0.3\textwidth,trim={0.5cm 0.3cm 0.3cm 0.3cm},clip]{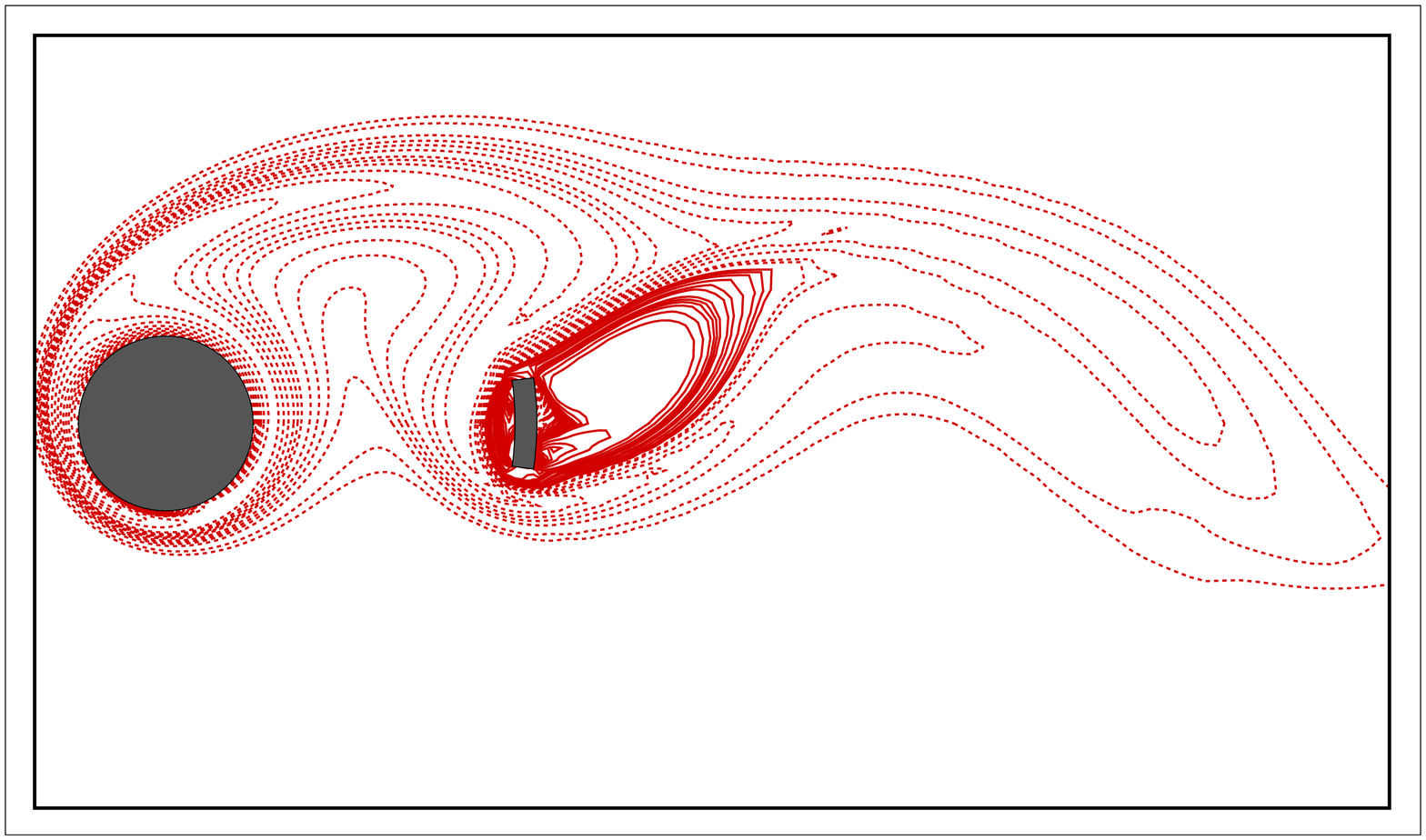}
\\
\hspace{0.5em}\scriptsize{$t=t_0+(1/4)T$}
\\
\includegraphics[width=0.29\textwidth,trim={0.5cm 0.3cm 0.5cm 0.3cm},clip]{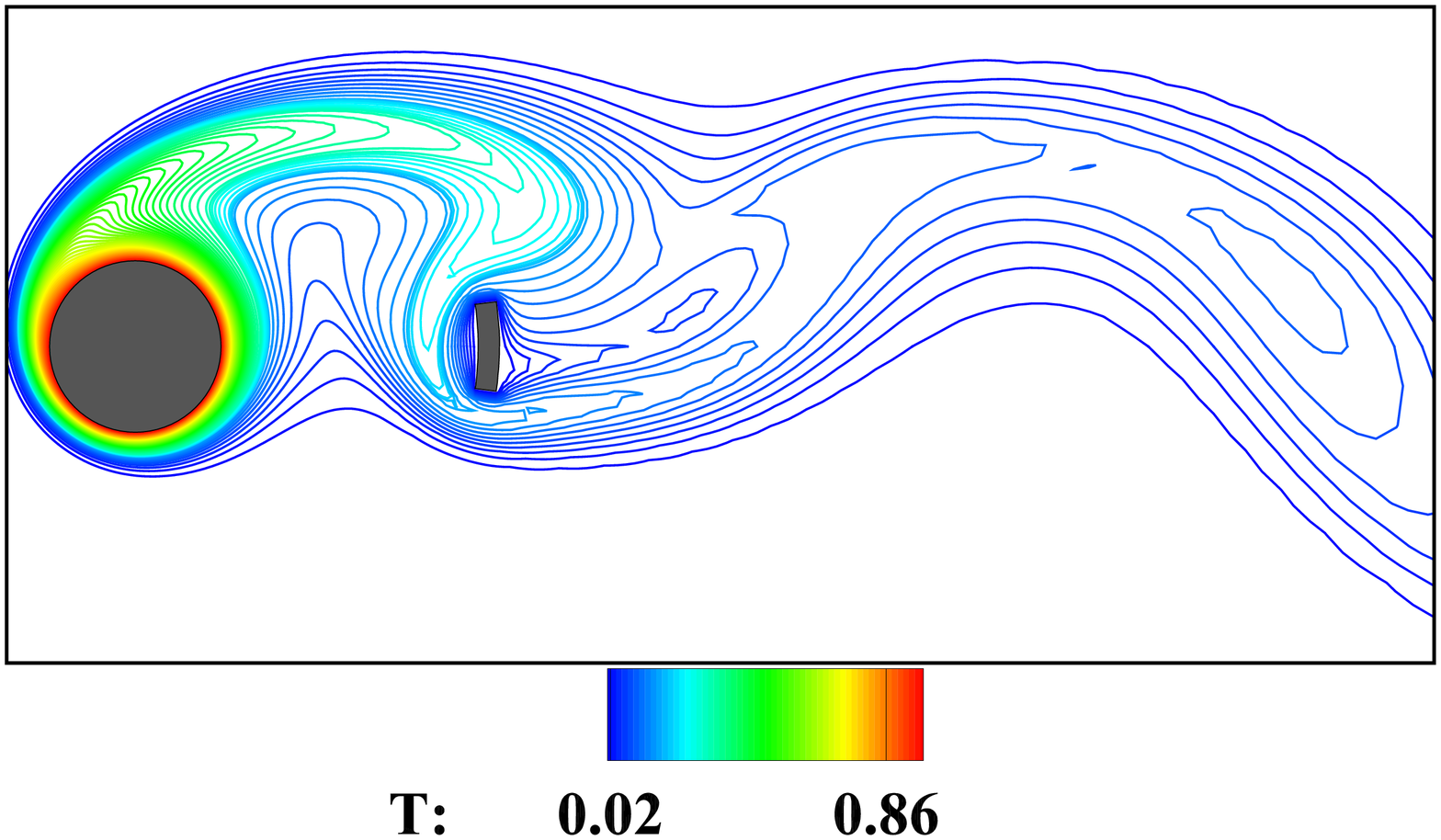}
\includegraphics[width=0.3\textwidth,trim={0.5cm 0.3cm 0.3cm 0.3cm},clip]{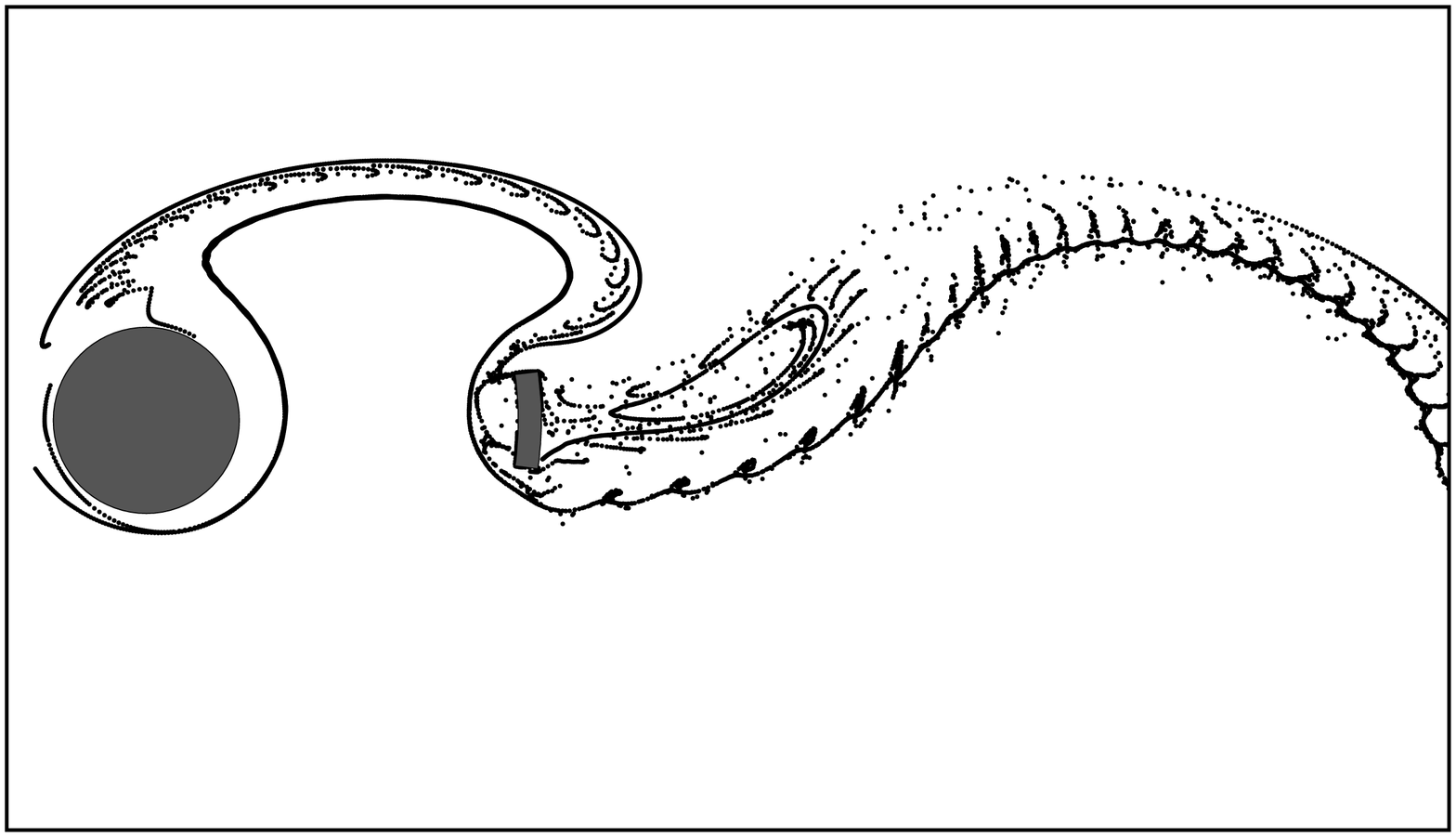}
\includegraphics[width=0.3\textwidth,trim={0.5cm 0.3cm 0.3cm 0.3cm},clip]{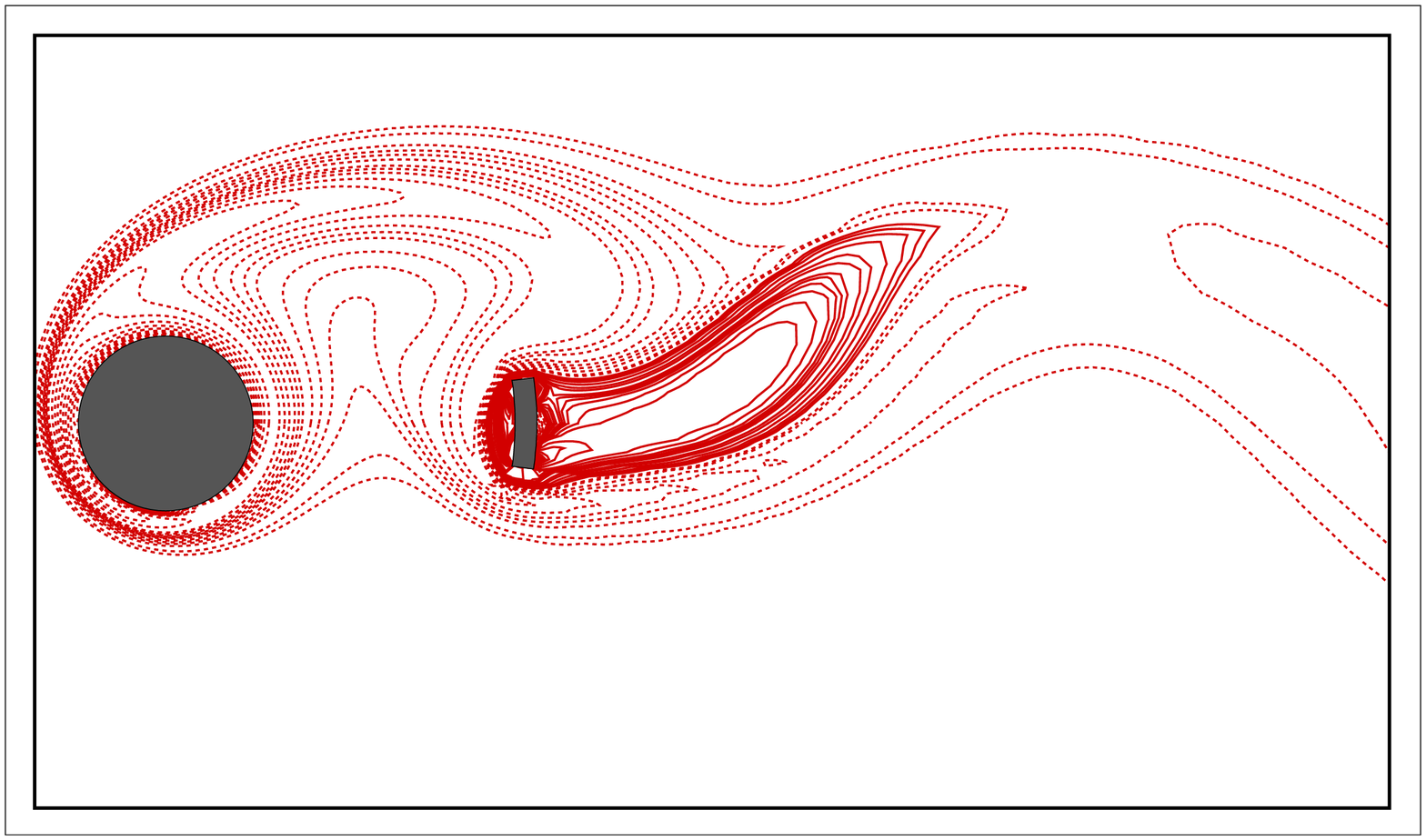}
\\
\hspace{0.5em}\scriptsize{$t=t_0+(1/2)T$}
\\
\includegraphics[width=0.29\textwidth,trim={0.5cm 0.3cm 0.5cm 0.3cm},clip]{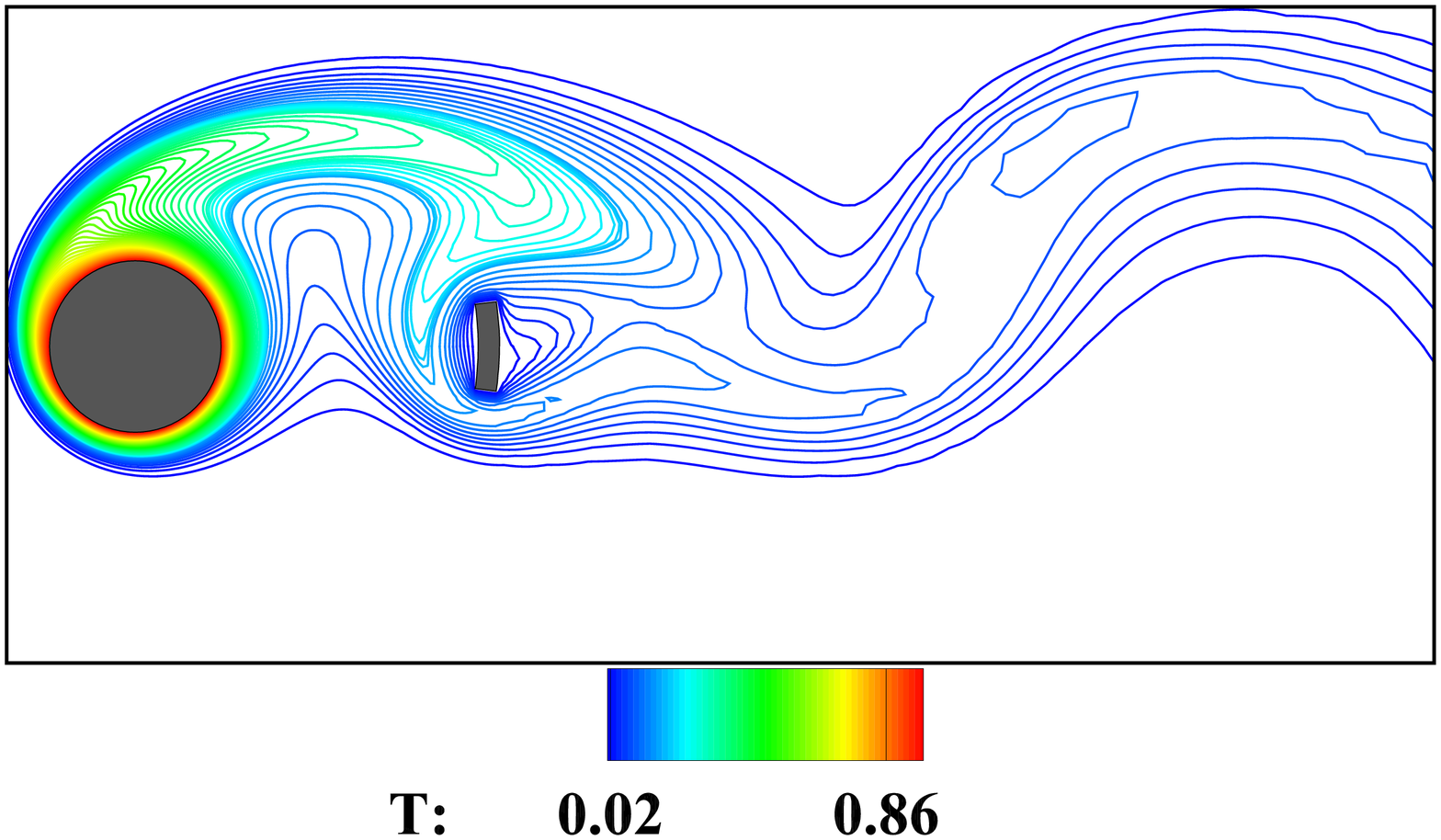}
\includegraphics[width=0.3\textwidth,trim={0.5cm 0.3cm 0.3cm 0.3cm},clip]{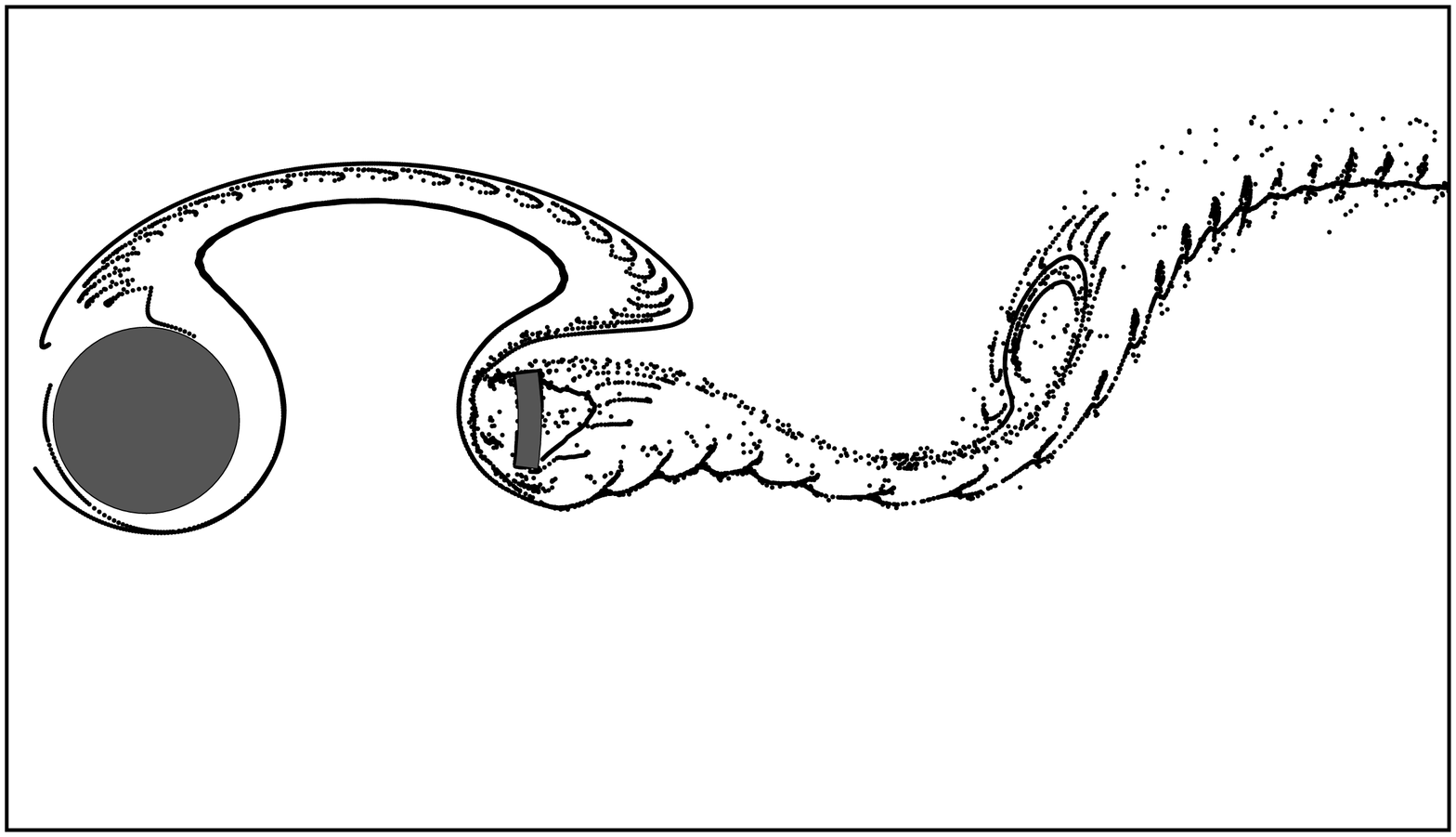}
\includegraphics[width=0.3\textwidth,trim={0.5cm 0.3cm 0.3cm 0.3cm},clip]{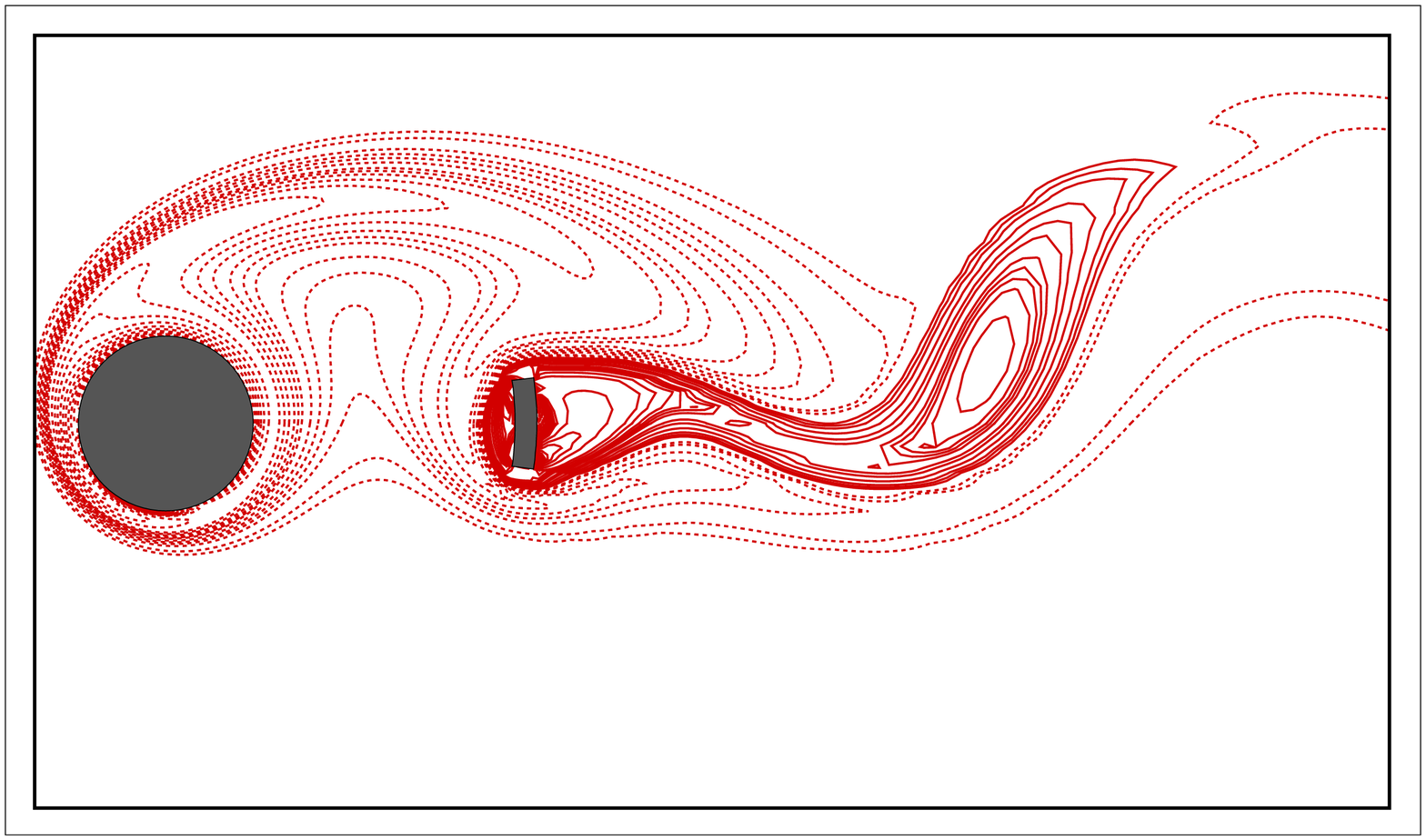}
\\
\hspace{0.5em}\scriptsize{$t=t_0+(3/4)T$}
\\
\includegraphics[width=0.29\textwidth,trim={0.5cm 0.3cm 0.5cm 0.3cm},clip]{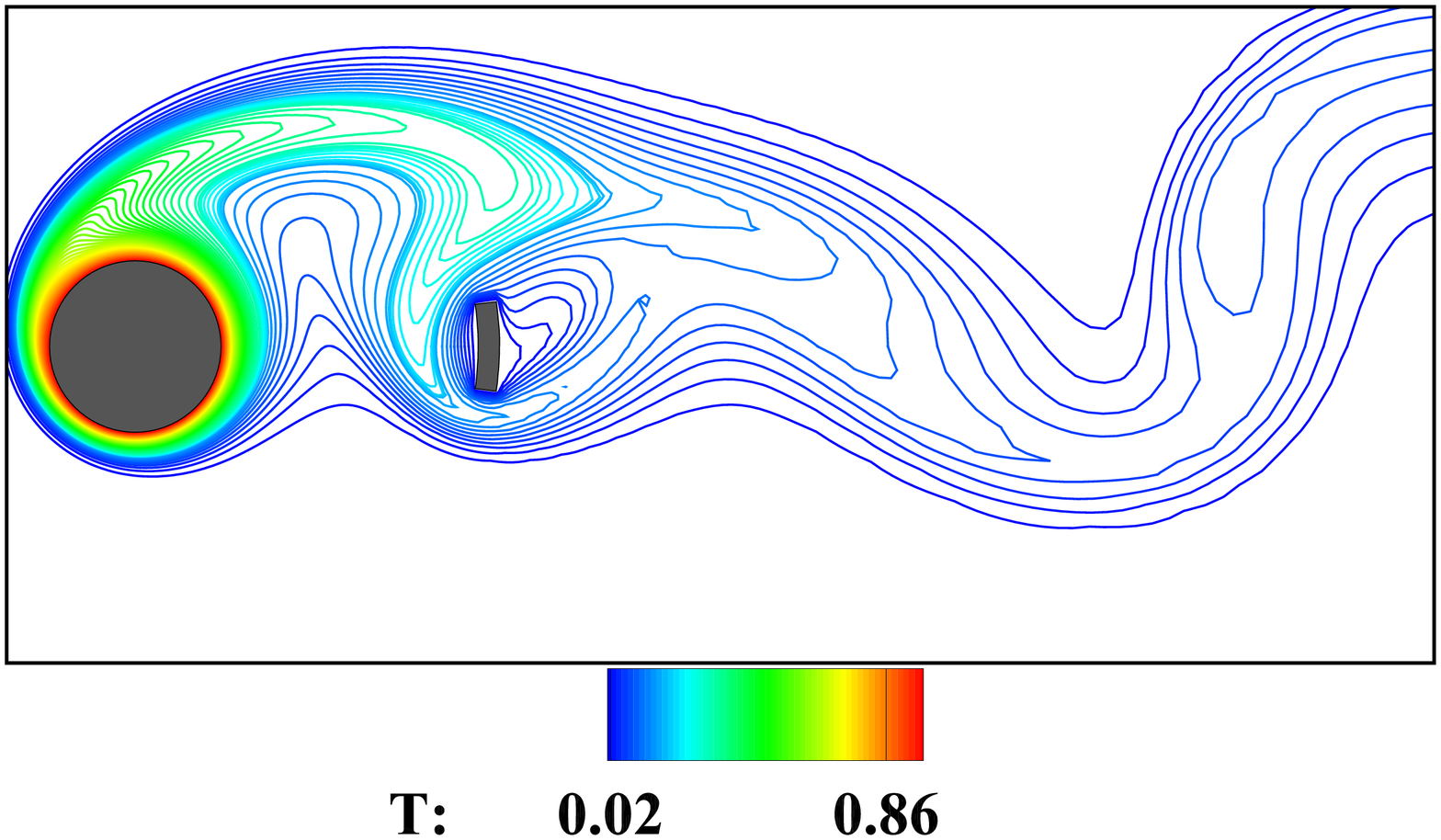}
\includegraphics[width=0.3\textwidth,trim={0.5cm 0.3cm 0.3cm 0.3cm},clip]{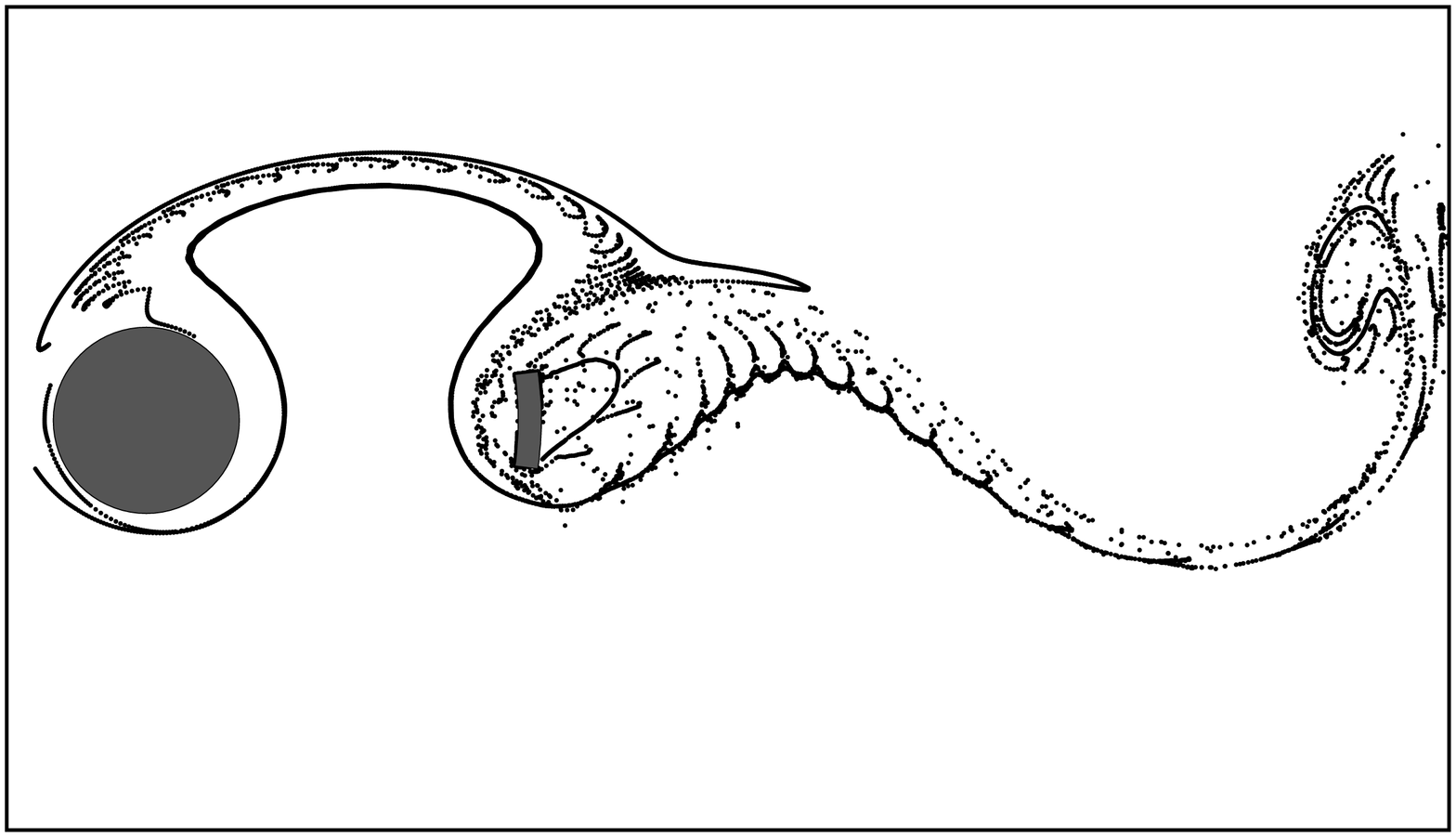}
\includegraphics[width=0.3\textwidth,trim={0.5cm 0.3cm 0.3cm 0.3cm},clip]{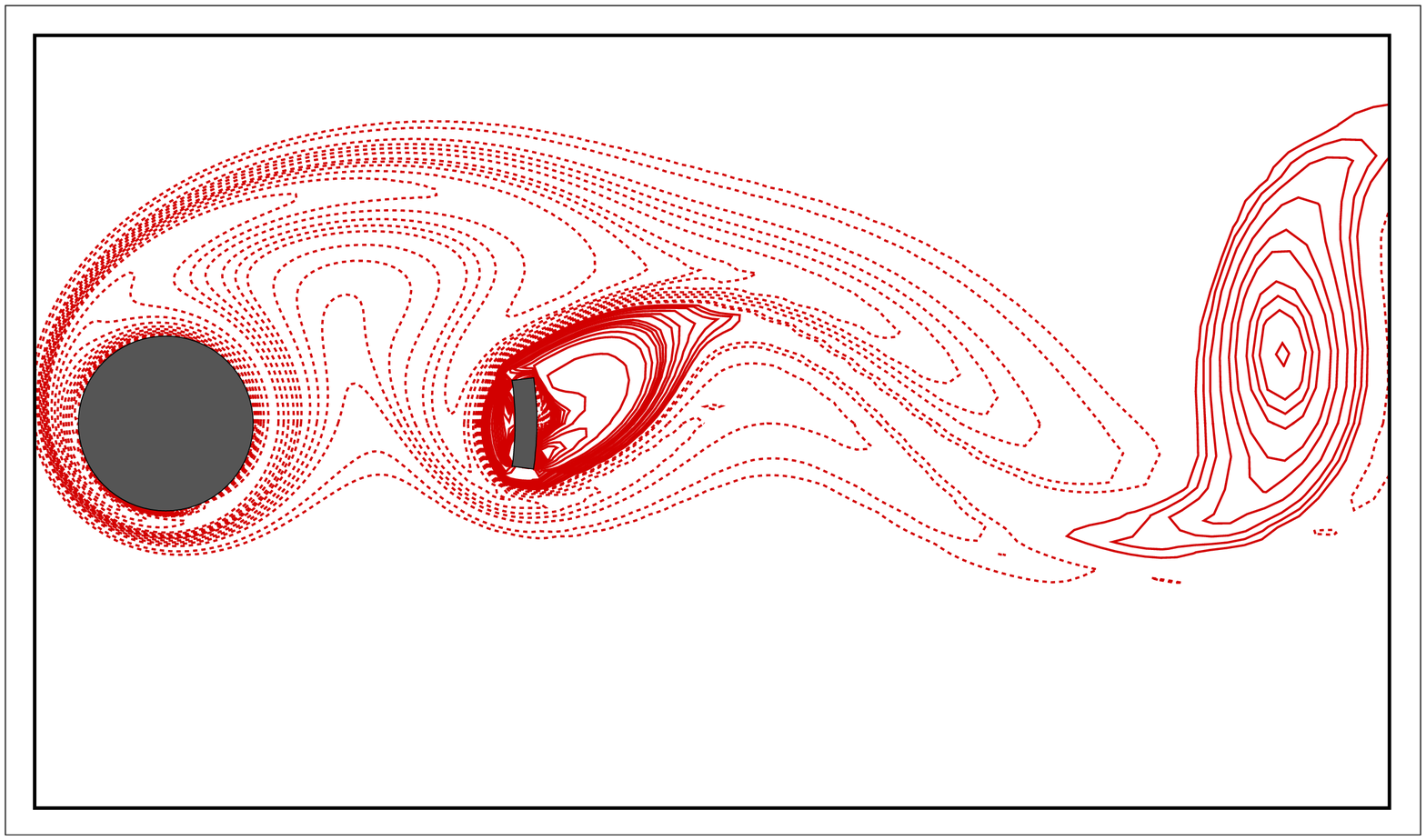}
\\
\hspace{0.5em}\scriptsize{$t=t_0+(1)T$}
\\
\includegraphics[width=0.29\textwidth,trim={0.5cm 0.3cm 0.5cm 0.3cm},clip]{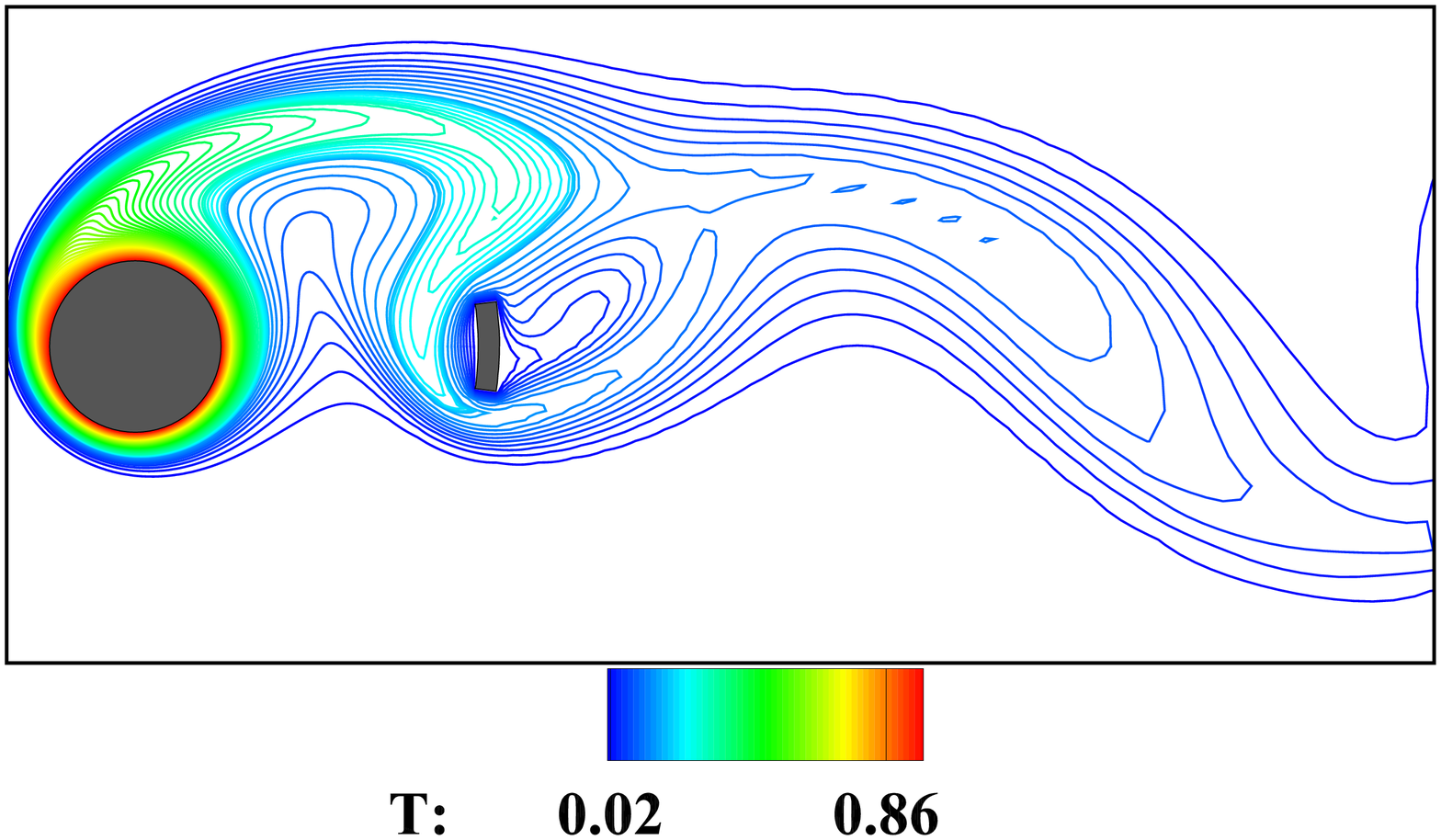}
\includegraphics[width=0.3\textwidth,trim={0.5cm 0.3cm 0.3cm 0.3cm},clip]{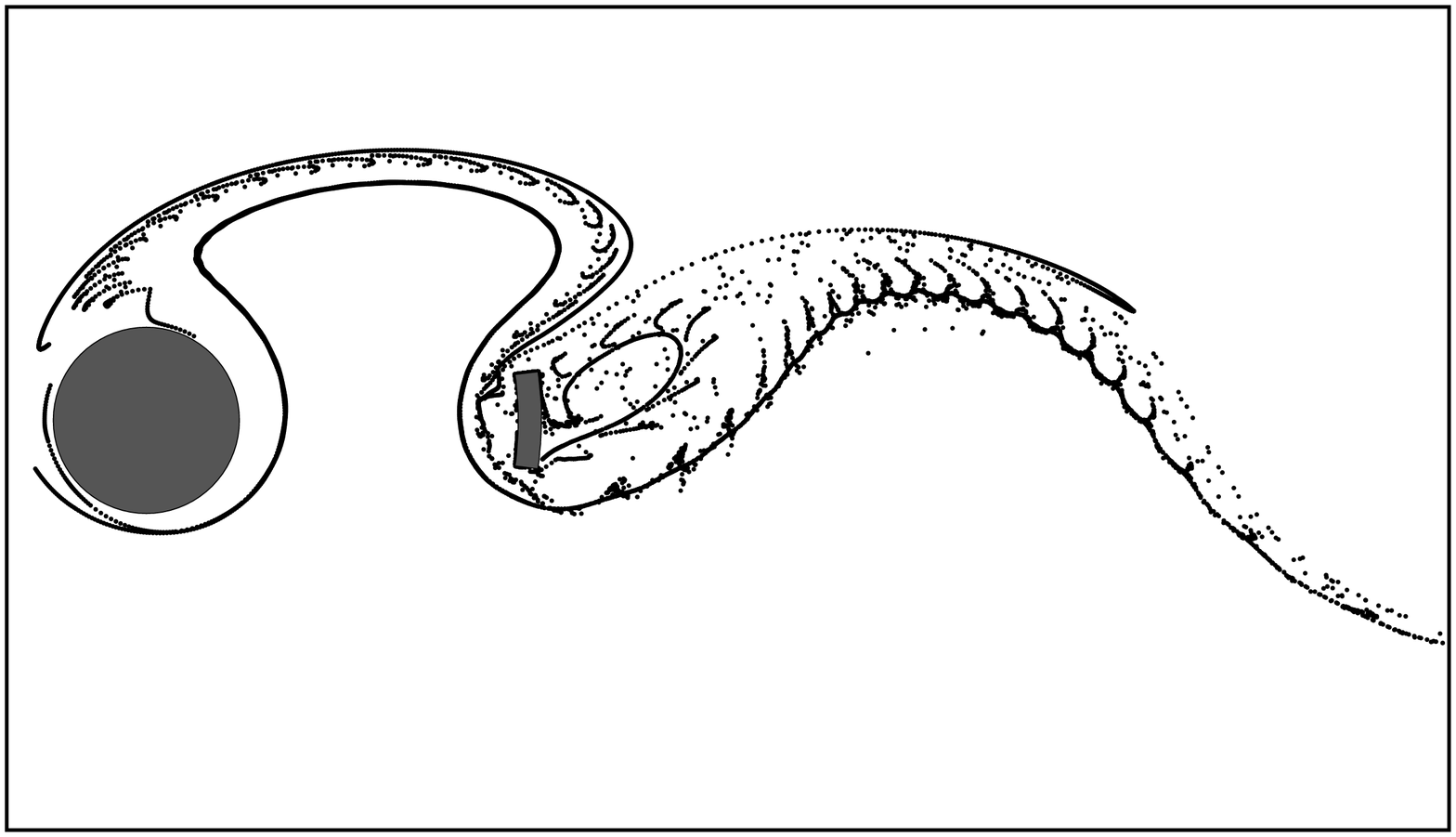}
\includegraphics[width=0.3\textwidth,trim={0.5cm 0.3cm 0.3cm 0.3cm},clip]{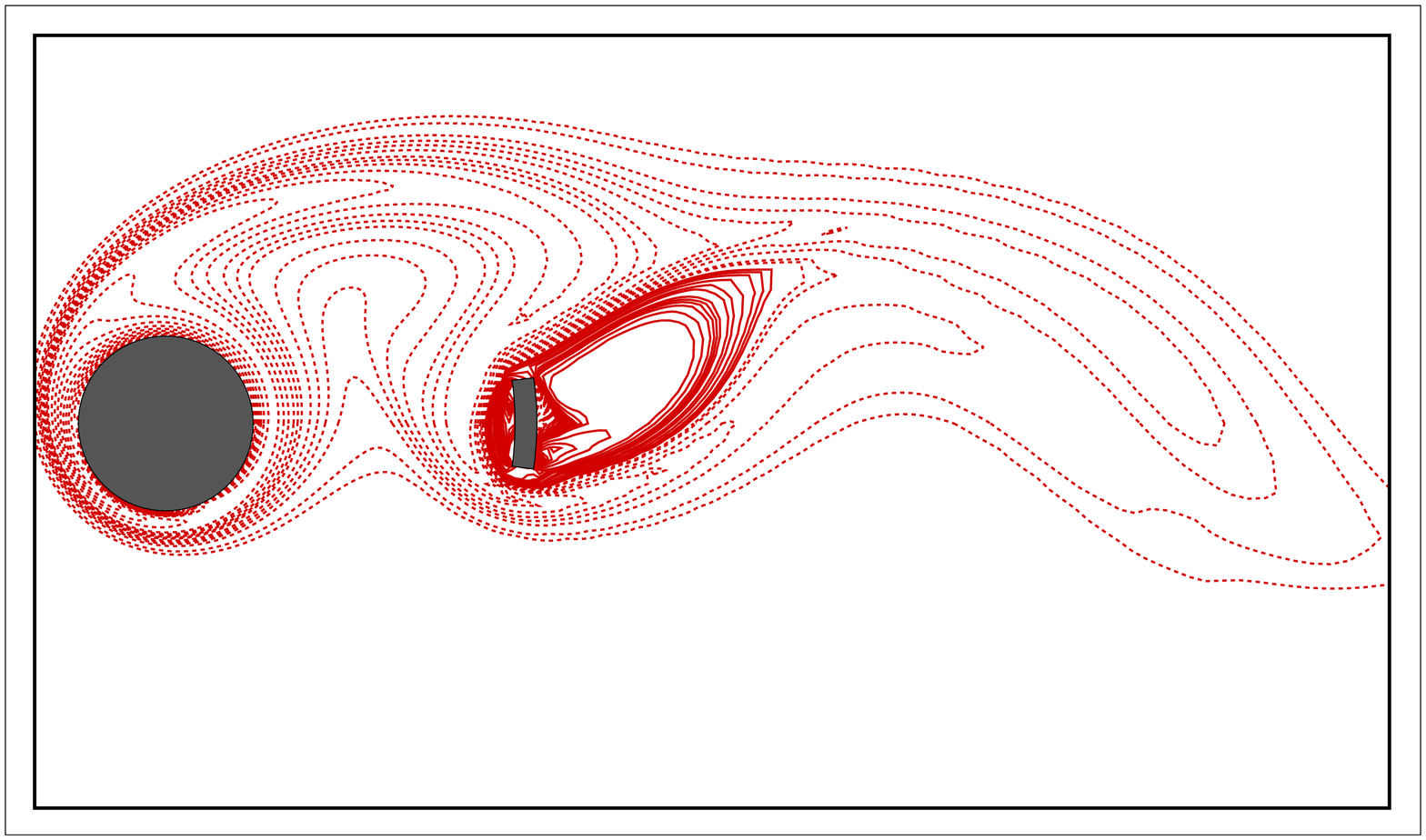}
\\
\hspace{2cm}(a) \hspace{4cm}(b) \hspace{4cm}(c)\hspace{2cm}
 \caption{(a) Isotherm, (b) streakline and (c) vorticity contour for $Pr=0.7$, $Re=150$, $\alpha=3.25$ and $d/R_0=3$ at different phases.}
 \label{fig:d_3_a_3-25}
\end{figure*}

Isotherm, streakline and vorticity are displayed in \cref{fig:d_3_a_0-5} for $\alpha=0.5$ and $d/R_0=3$. Two vortices shed periodically from the upper and lower sides of the cylinder. The bottom vortex is slightly sleeker than the upper one. The positive equi-vorticity lines coming from the cylinder, partially cover the control plate, and the interaction between the shear layers sheds the positive vortex. One recirculation zone is formed between the cylinder and the plate, which gradually merges with the upper vortex. The density of the isotherm contour is higher near the front stagnation point, which means the rate of heat transfer is much higher in this region. Also, two warm blobs convect away periodically from the upper and lower sides of the cylinder. Also, the vortex shedding plane is at an angle of approximately $\theta=8.5\degree$ with the centerline, which is much lower than the previous placements of the control plate. It happens as the bottom shear layers are resisted by the control plate to freely move upwards. \cref{fig:d_3_a_3-25} exhibits the isotherm, streakline and vorticity are displayed for $\alpha=3.25$ and $d/R_0=3$. Two vortices shed periodically behind the control plate. The rotational motion of the fluid surrounding the cylinder causes the negative equi-vorticity lines to surround the cylinder as well as the positive equi-vorticity lines that originate from the control plate. The positive equi-vorticity  lines also cover the control plate. The shear layers that originate from the cylinder get split after interaction with the shear layer around the control plate, and they merge together during the shedding of the negative vortex. Most of the fluid particles that flow across the cylinder are sucked down and flow below the control plate. This complex flow dynamics is the combined effect of the high rotational rate and the placement of the control plate. It reduces the angle of the vortex shedding plane with the centerline to approximately $\theta=12\degree$, which is much less than the previous placements of the control plate with this high rotational rate. Also, the size of the negative vortex is drastically reduced and becomes extremely sleek due to the interaction of shear layers. The lower vortex grows from the bottom of the plate and moves upwards. The density of the isotherm contour around the cylinder is very low due to the high rotational rate. As a result, the boundary layer thickens around the cylinder and suppresses the rate of force convective heat transfer. The isotherm contours indicate that two warm blobs periodically convect away from the upper side of the cylinder and the lower end of the control plate. Therefore, the placement of the control plate, together with the rotational rate, considerably suppressed the vortex shedding process as well as the heat convection. \cref{fig:d_3_a_0-5,fig:d_3_a_3-25} illustrate that the size of the vortices grow as $\alpha$ increases from $0.5$ to $3.25$ for $d/R_0=3$. Also, \cref{fig:d_1_a_0-5,fig:d_2_a_0-5,fig:d_3_a_0-5} show that the wake length of vortices increases with increasing distance of the control plate from the cylinder surface at $\alpha=0.5$. It is also observed that the increasing distance of the control plate significantly decreases the angle of the vortex shedding plane with the centerline for respecting rotational rates.\\

\begin{figure*}[!t]
\centering
\scriptsize{$\alpha=0.5$}
\\
\includegraphics[width=0.3\textwidth,trim={0.4cm 0.25cm 0.5cm 0.5cm},clip]{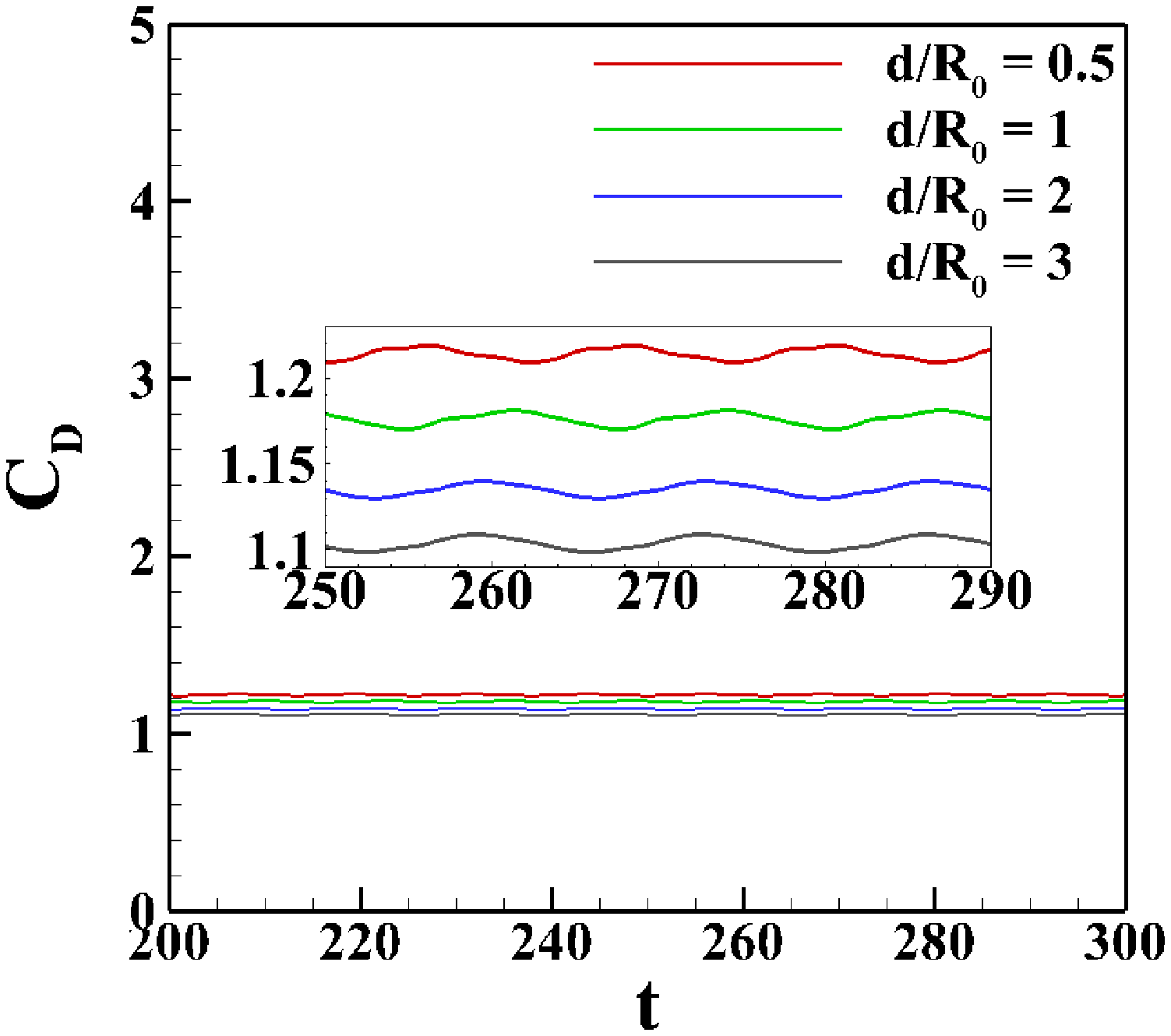}\hspace{1cm}%
\includegraphics[width=0.3\textwidth,trim={0.5cm 0.25cm 0.5cm 0.5cm},clip]{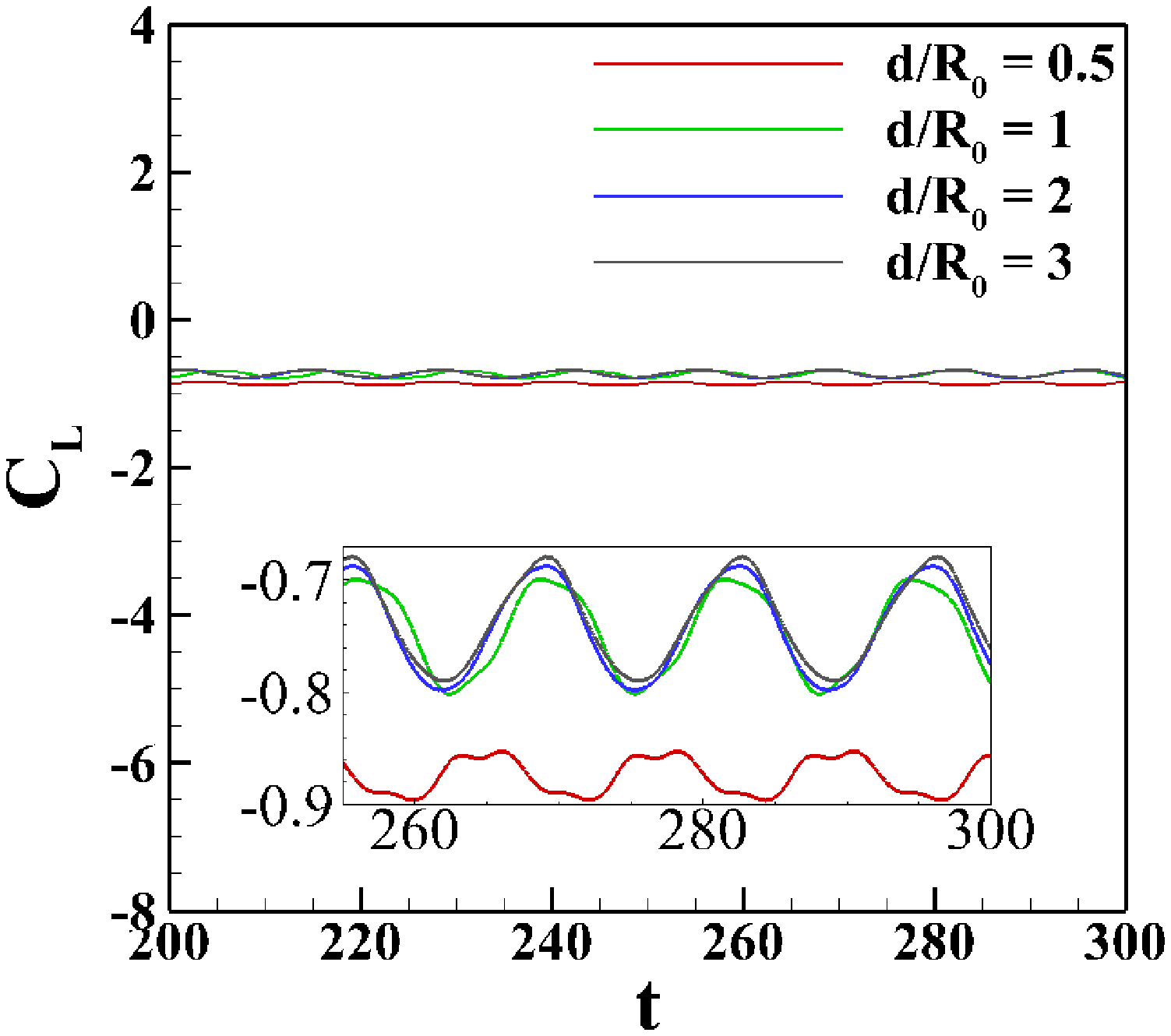}
\\
\hspace{0.5em}\scriptsize{$\alpha=1$}
\\
\includegraphics[width=0.3\textwidth,trim={0.5cm 0.25cm 0.5cm 0.5cm},clip]{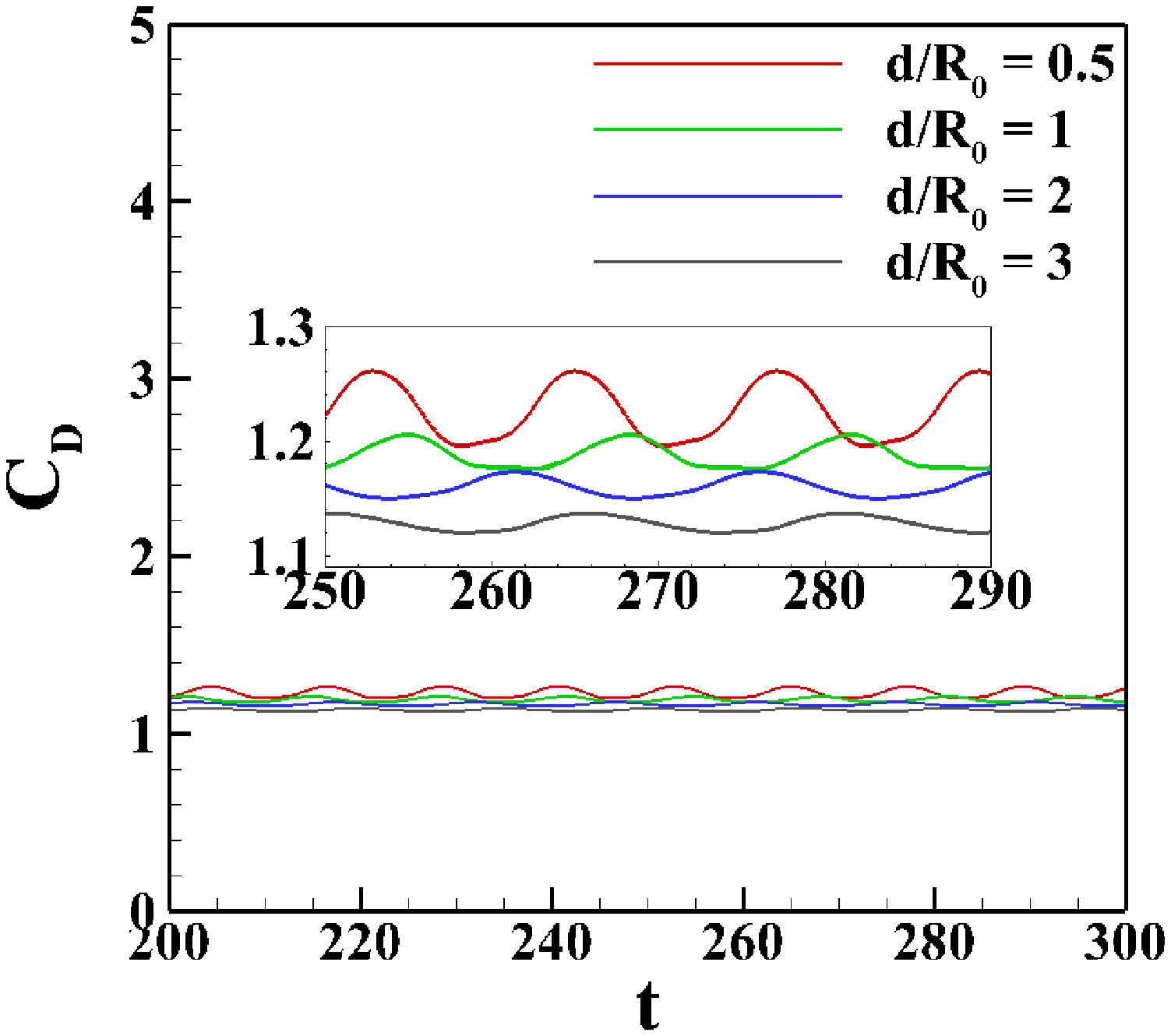}\hspace{1cm}%
\includegraphics[width=0.3\textwidth,trim={0.5cm 0.25cm 0.5cm 0.5cm},clip]{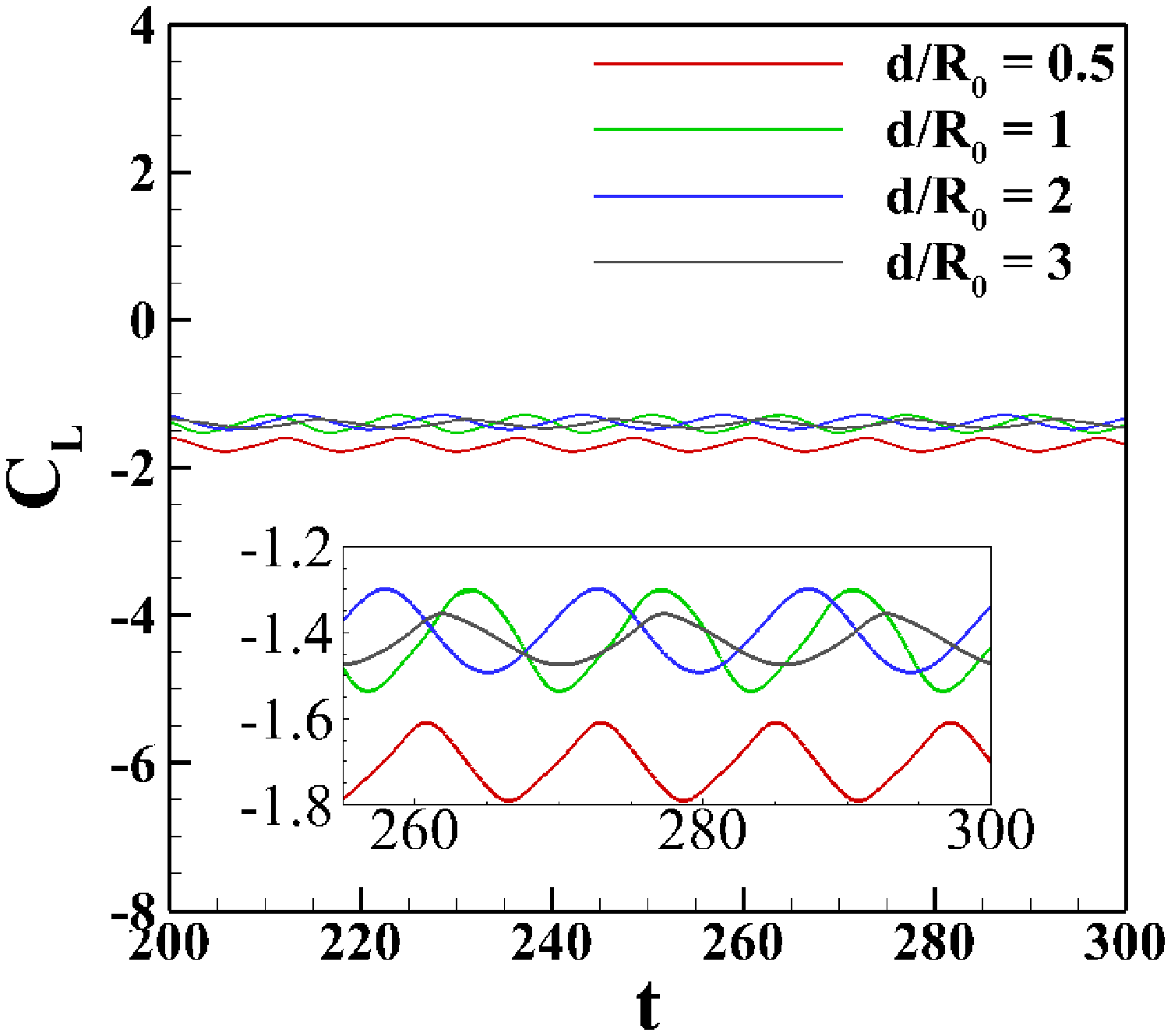}
\\
\hspace{0.5em}\scriptsize{$\alpha=2.07$}
\\
\includegraphics[width=0.3\textwidth,trim={0.5cm 0.25cm 0.5cm 0.5cm},clip]{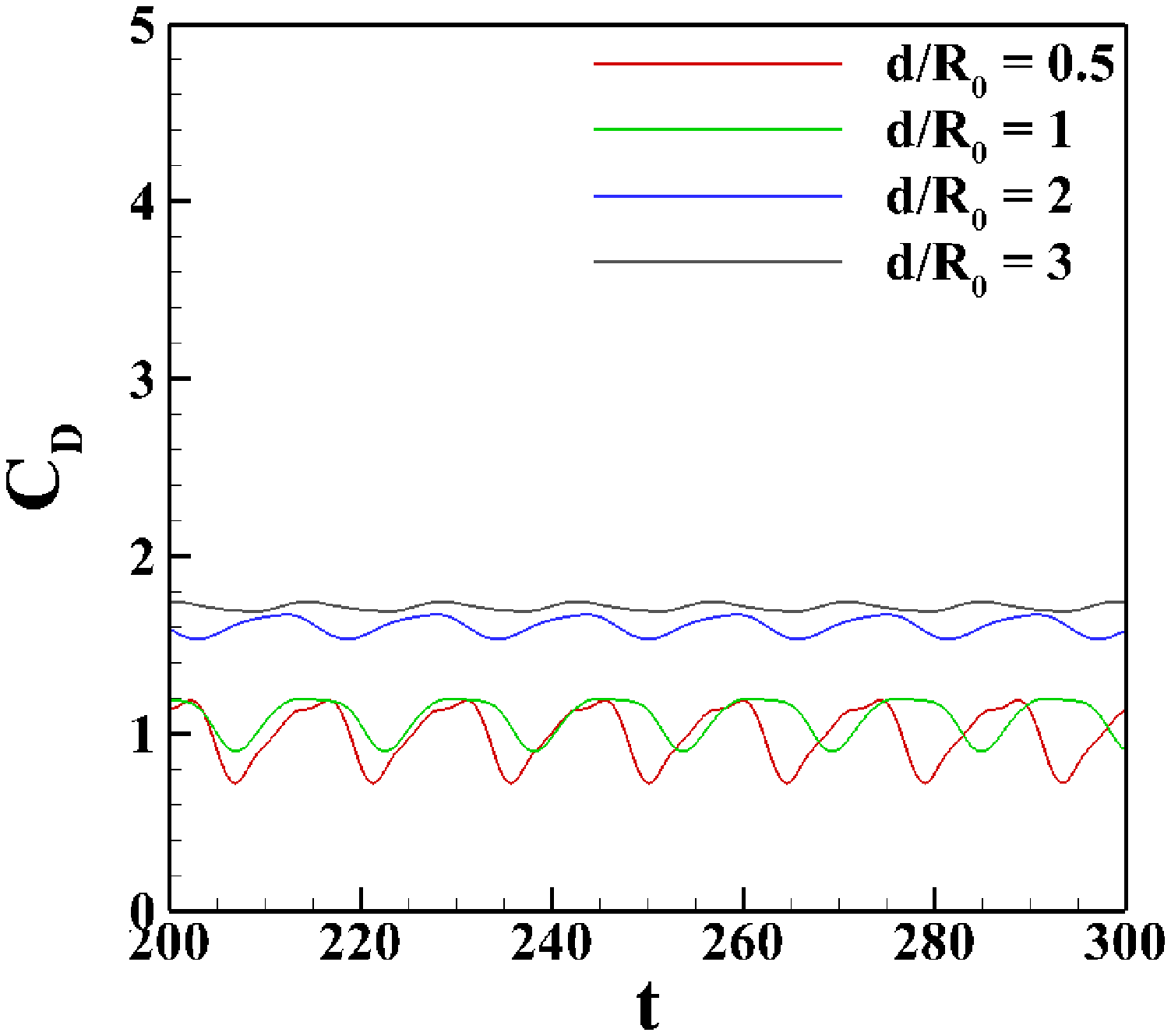}\hspace{1cm}%
\includegraphics[width=0.3\textwidth,trim={0.5cm 0.25cm 0.5cm 0.5cm},clip]{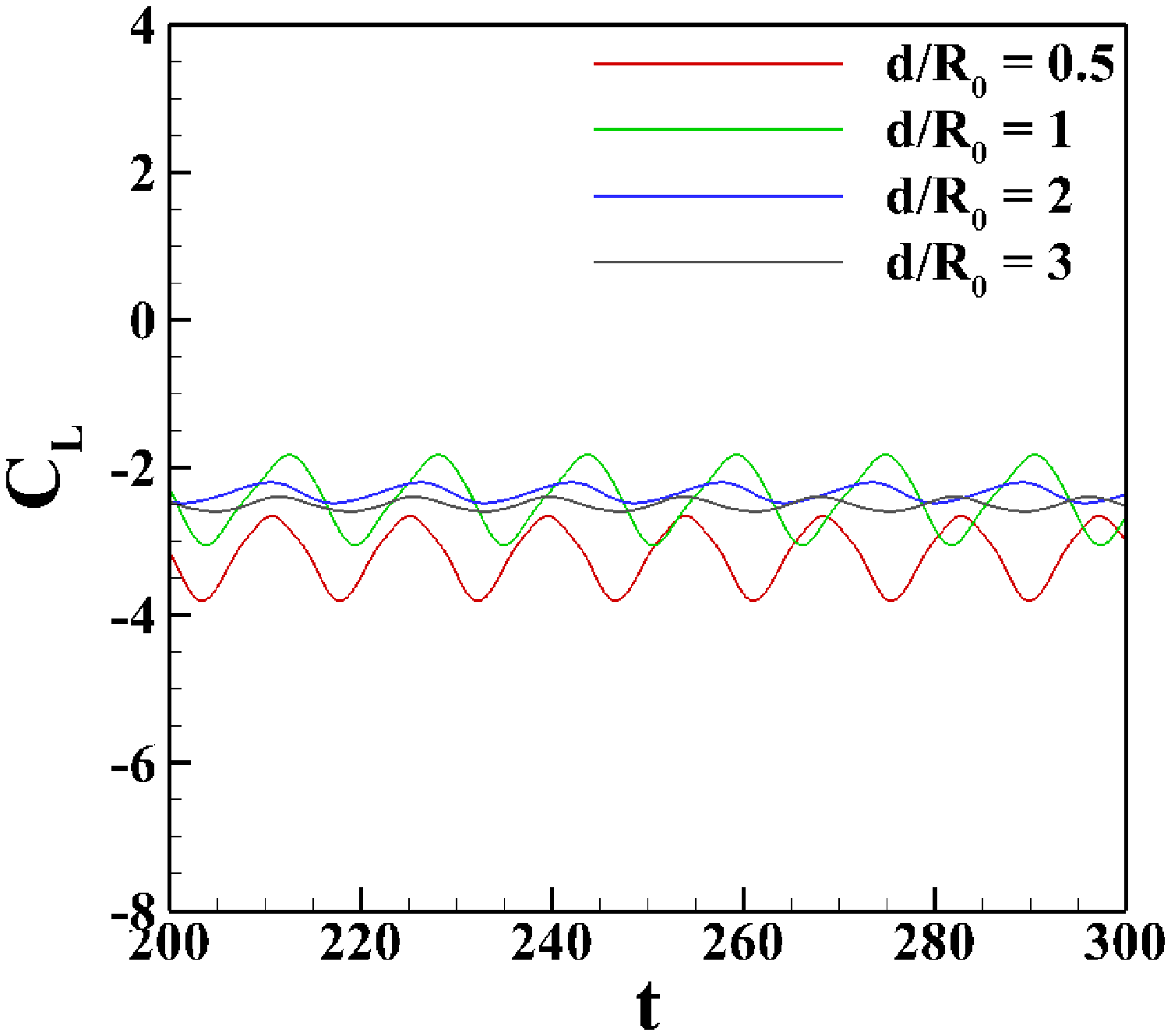}
\\
\hspace{0.5em}\scriptsize{$\alpha=3.25$}
\\
\includegraphics[width=0.3\textwidth,trim={0.5cm 0.25cm 0.5cm 0.5cm},clip]{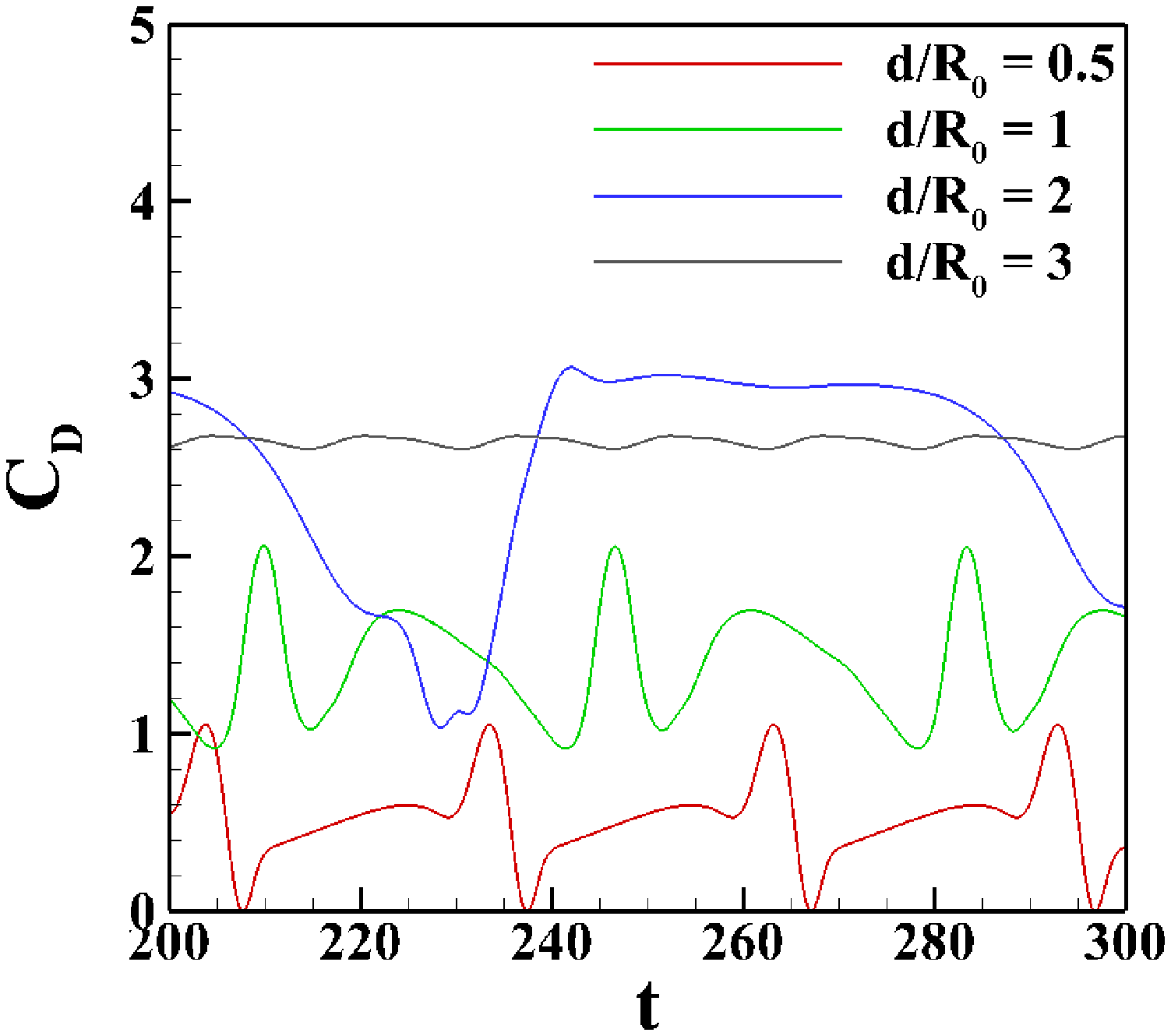}\hspace{1cm}%
\includegraphics[width=0.3\textwidth,trim={0.5cm 0.25cm 0.5cm 0.5cm},clip]{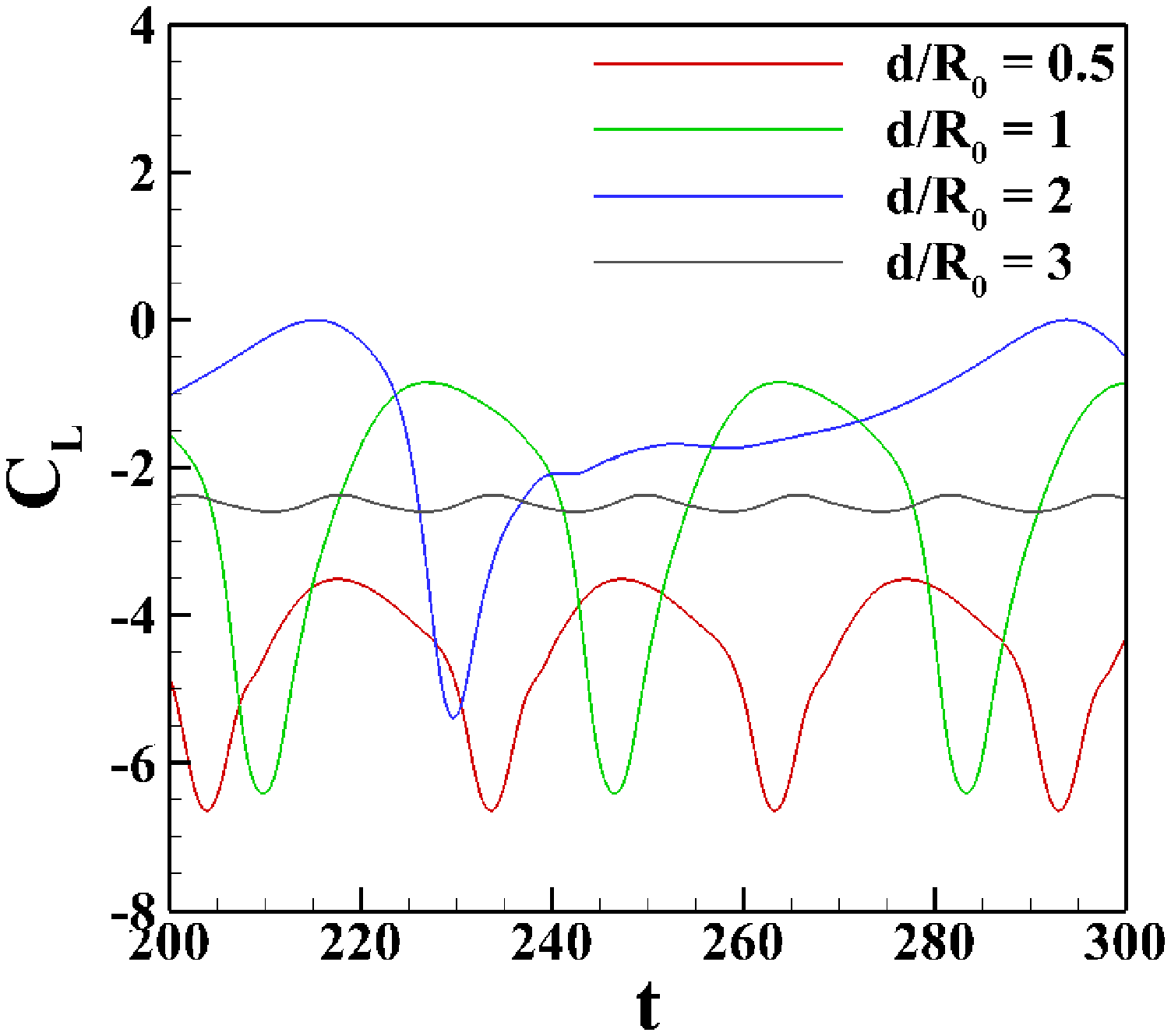}
\\
\hspace{2cm}(a) \hspace{4cm}(b)\hspace{2cm}
 \caption{(a) Drag coefficient $C_D$ and (b) lift coefficient $C_L$ with varying $d/R_0$.}
 \label{fig:lift-drag_a}
\end{figure*}

\begin{figure*}[!t]
\centering
\scriptsize{$d/R_0=0.5$}
\\
\includegraphics[width=0.3\textwidth,trim={0.5cm 0.25cm 0.5cm 0.5cm},clip]{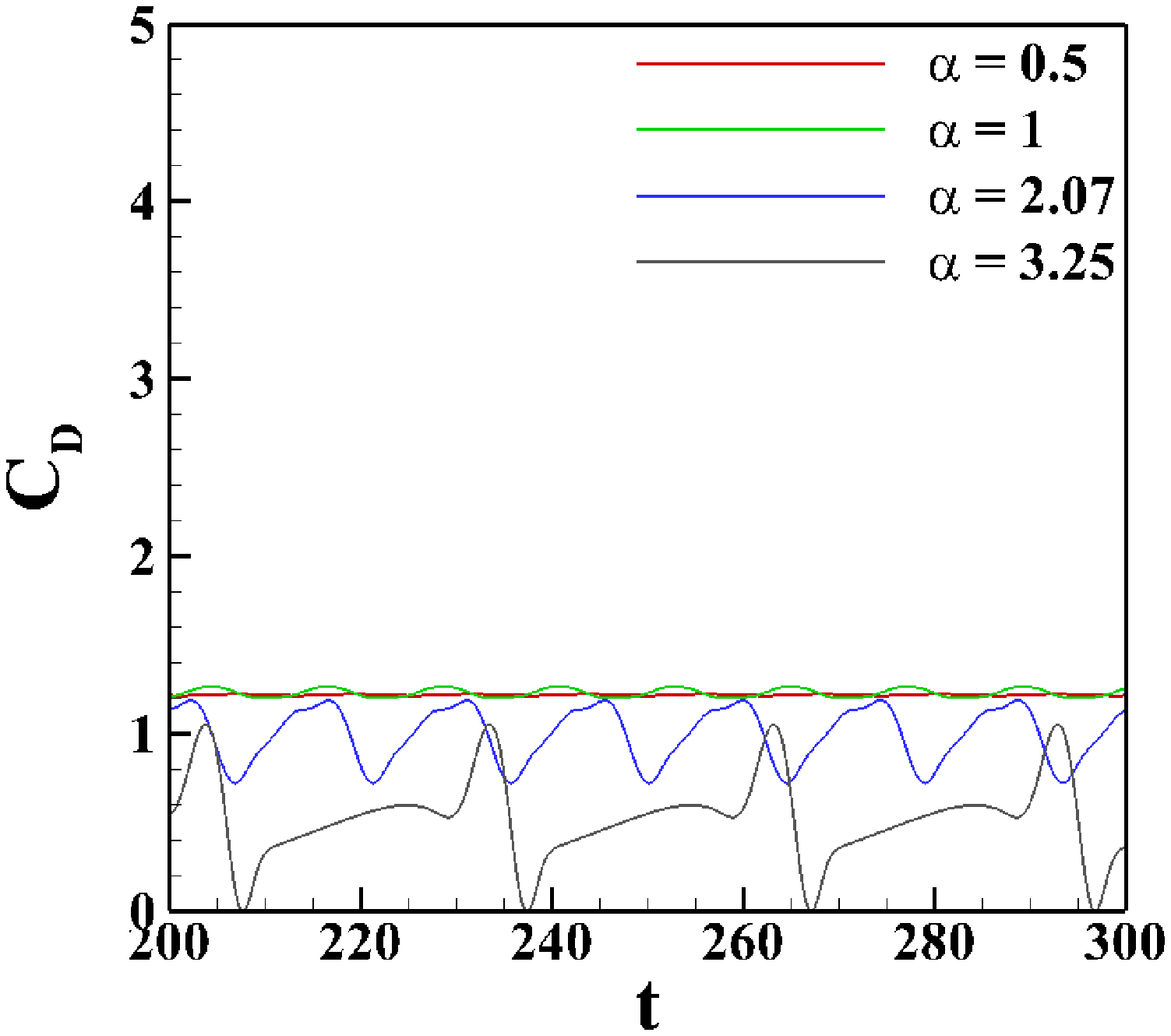}\hspace{1cm}%
\includegraphics[width=0.3\textwidth,trim={0.5cm 0.25cm 0.5cm 0.5cm},clip]{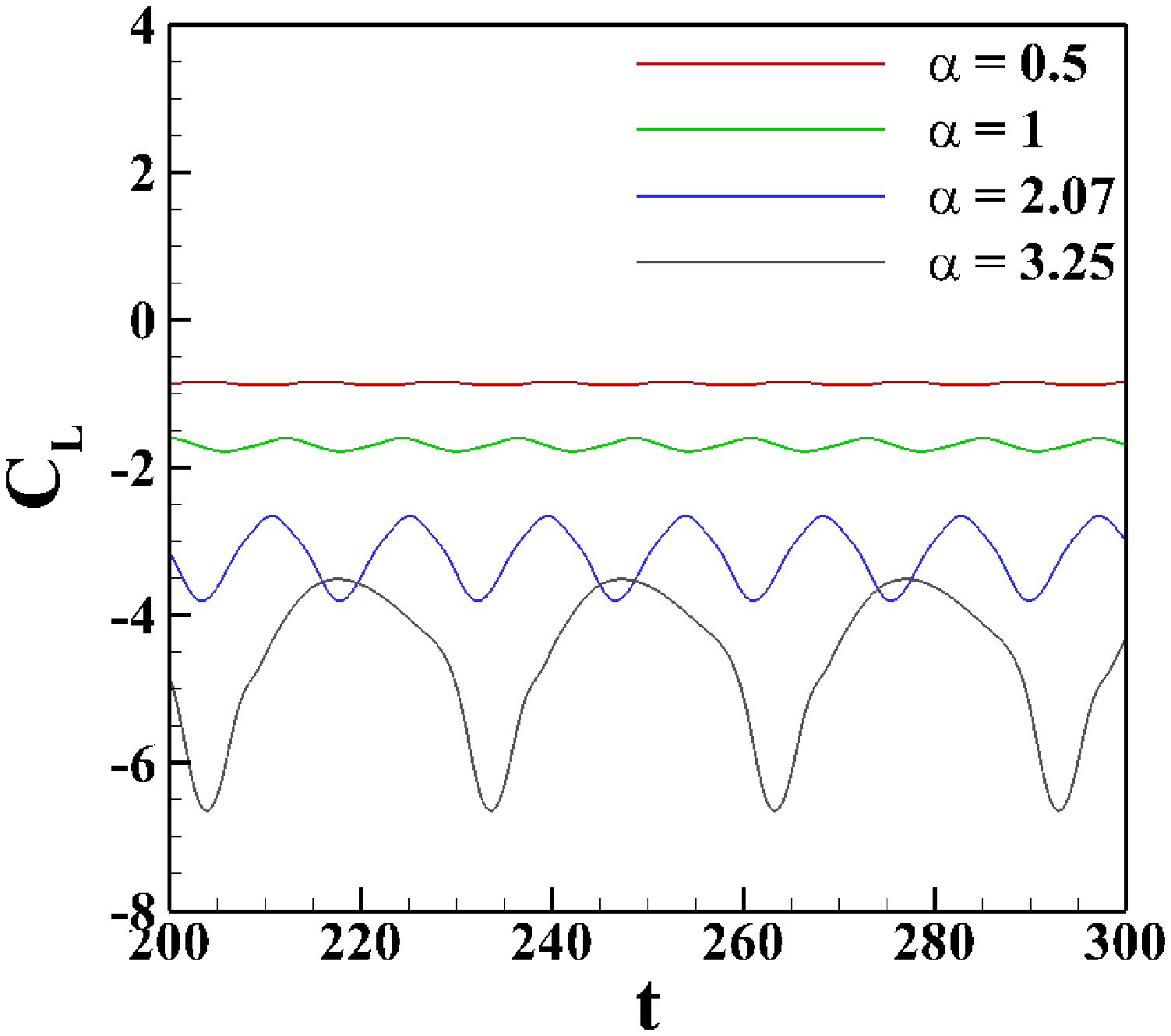}
\\
\hspace{0.5em}\scriptsize{$d/R_0=1$}
\\
\includegraphics[width=0.3\textwidth,trim={0.5cm 0.25cm 0.5cm 0.5cm},clip]{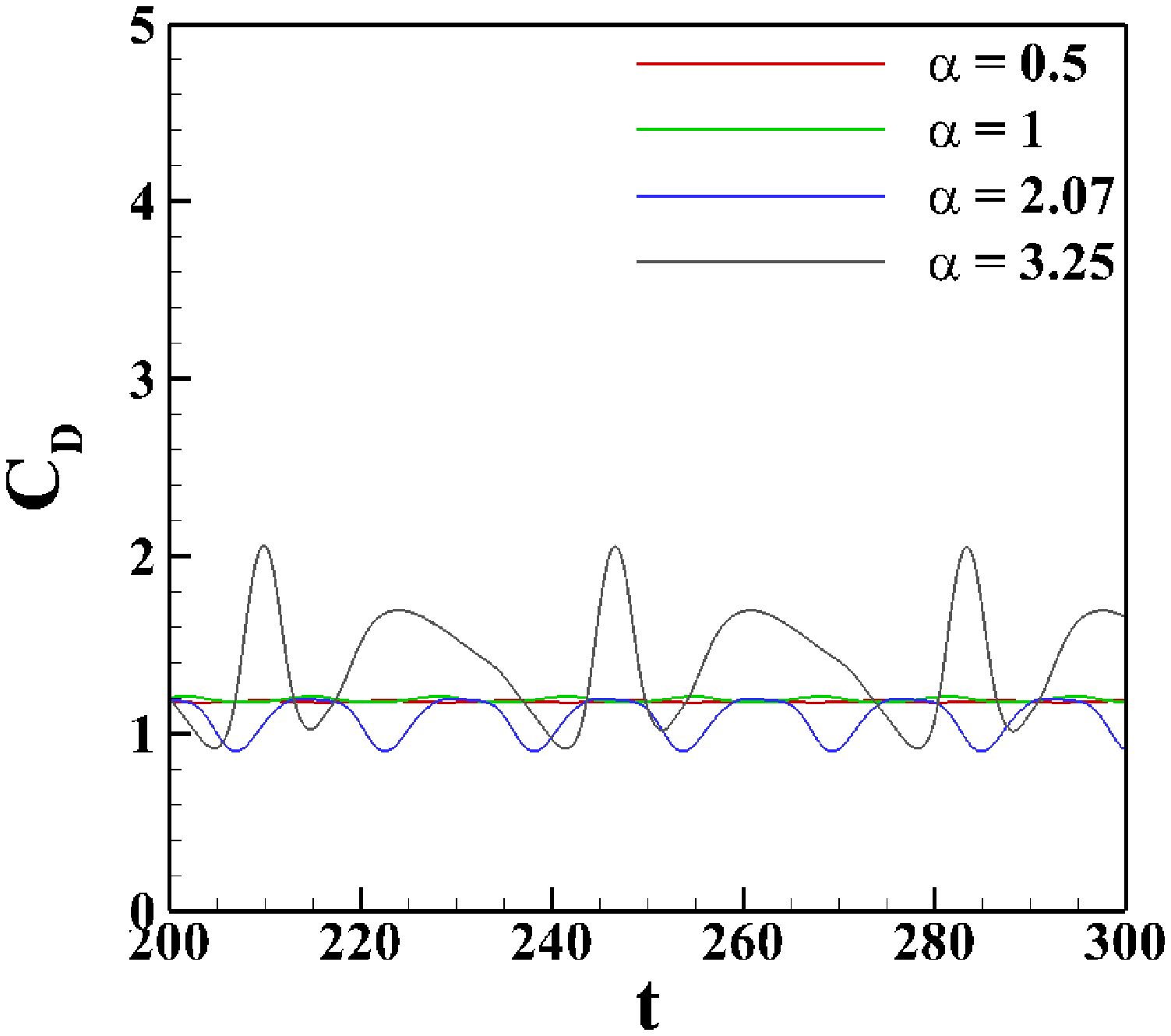}\hspace{1cm}%
\includegraphics[width=0.3\textwidth,trim={0.5cm 0.25cm 0.5cm 0.5cm},clip]{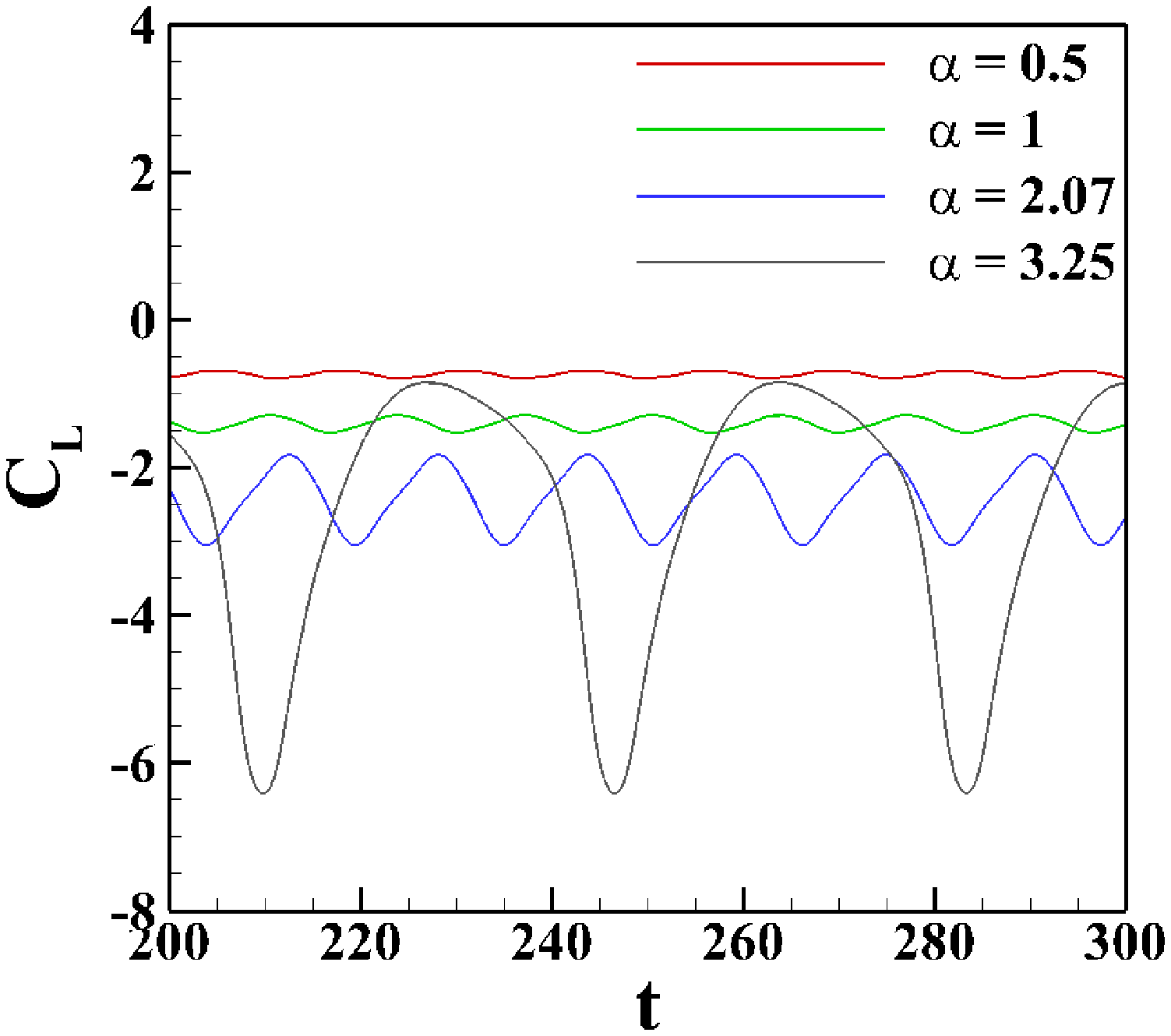}
\\
\hspace{0.5em}\scriptsize{$d/R_0=2$}
\\
\includegraphics[width=0.3\textwidth,trim={0.5cm 0.25cm 0.5cm 0.5cm},clip]{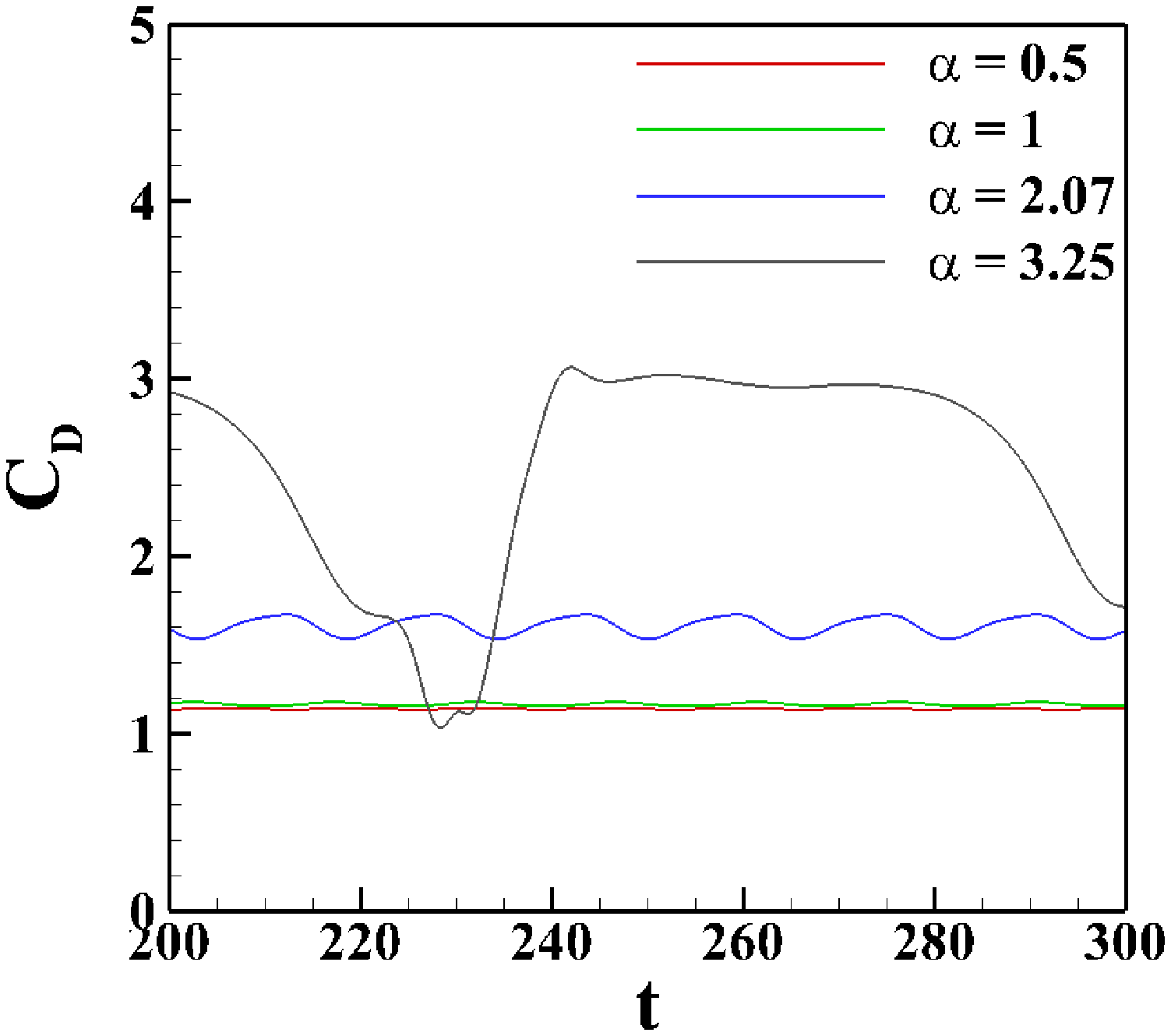}\hspace{1cm}%
\includegraphics[width=0.3\textwidth,trim={0.5cm 0.25cm 0.5cm 0.5cm},clip]{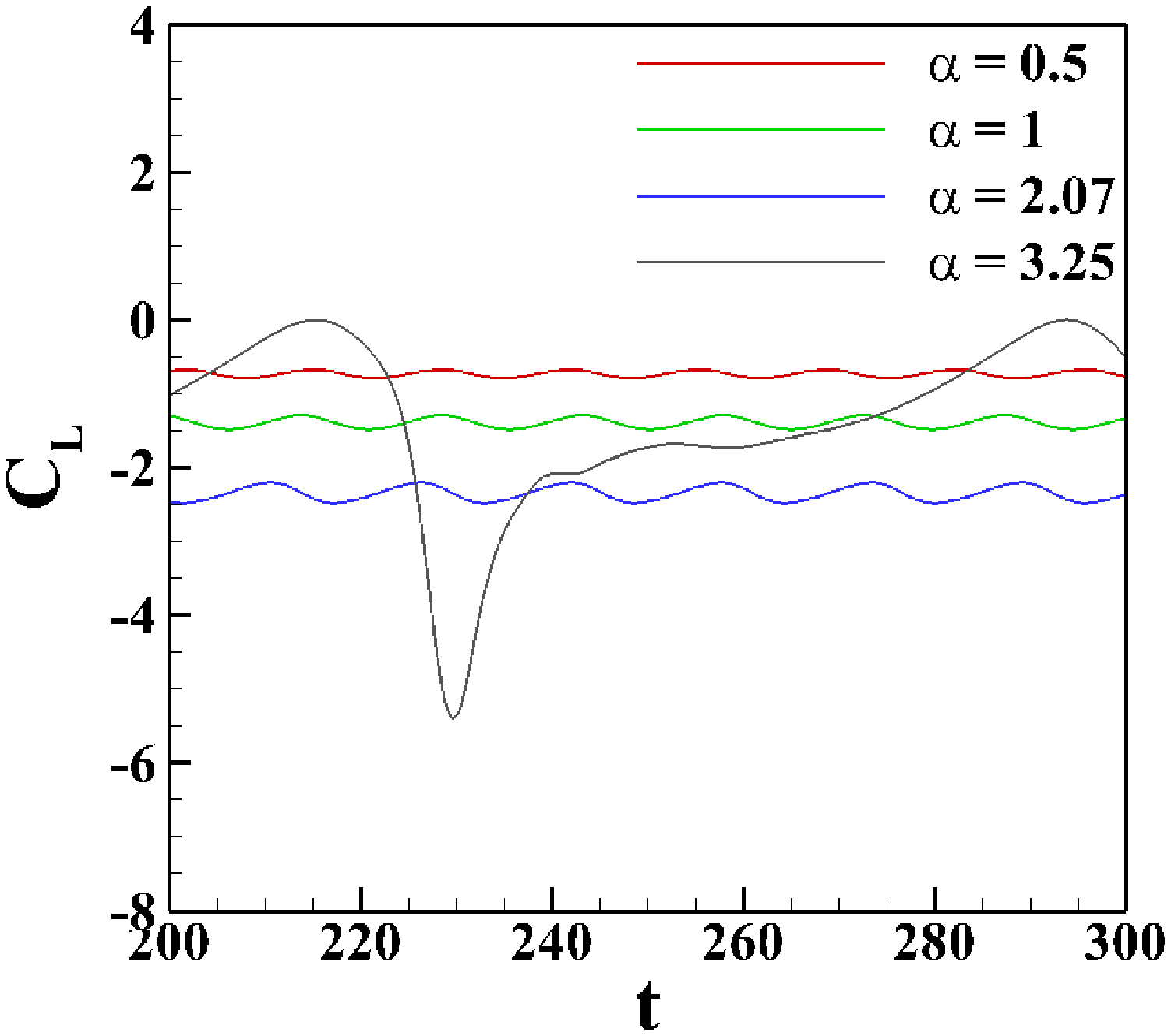}
\\
\hspace{0.5em}\scriptsize{$d/R_0=3$}
\\
\includegraphics[width=0.3\textwidth,trim={0.5cm 0.25cm 0.5cm 0.5cm},clip]{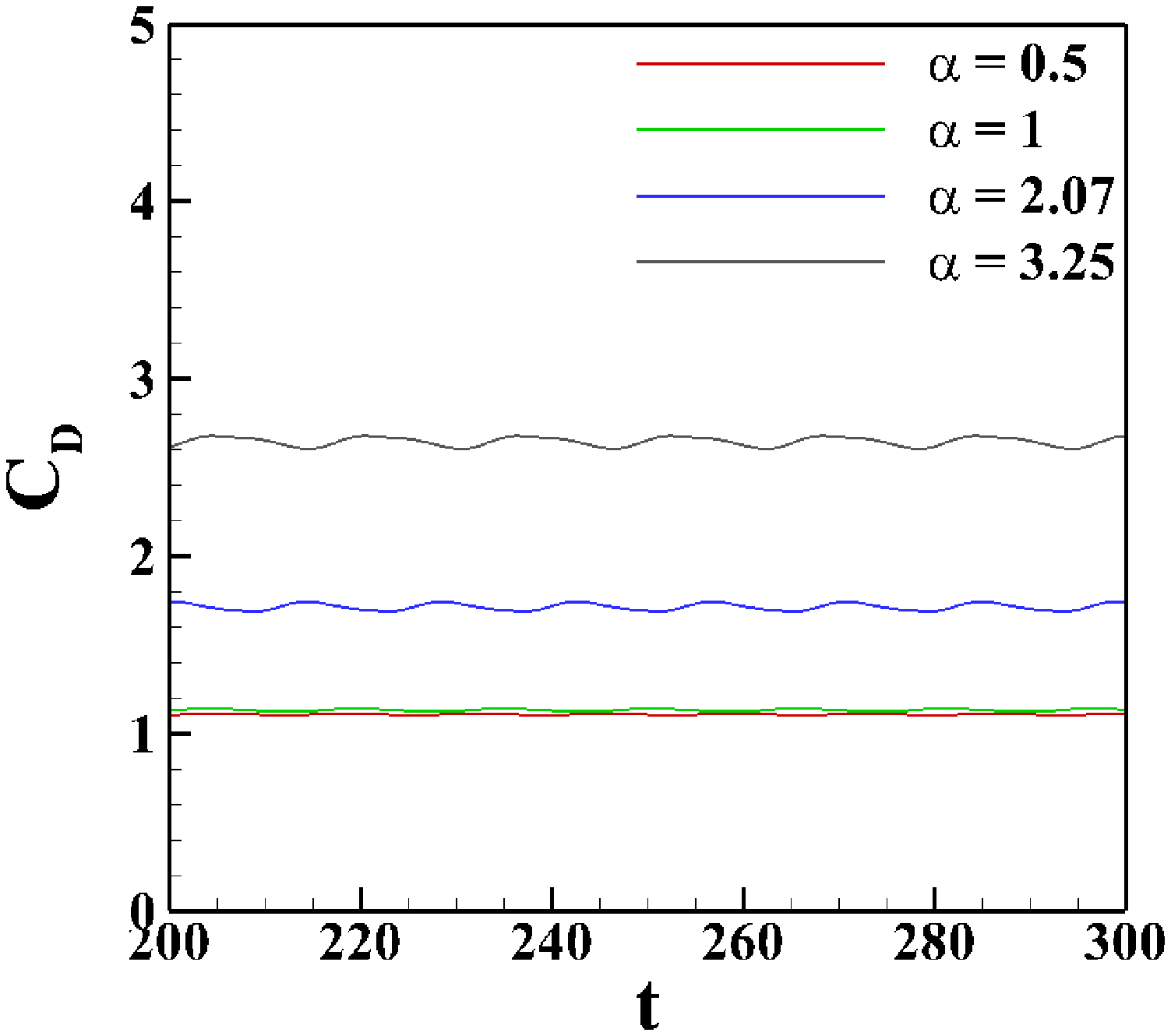}\hspace{1cm}%
\includegraphics[width=0.3\textwidth,trim={0.5cm 0.25cm 0.5cm 0.5cm},clip]{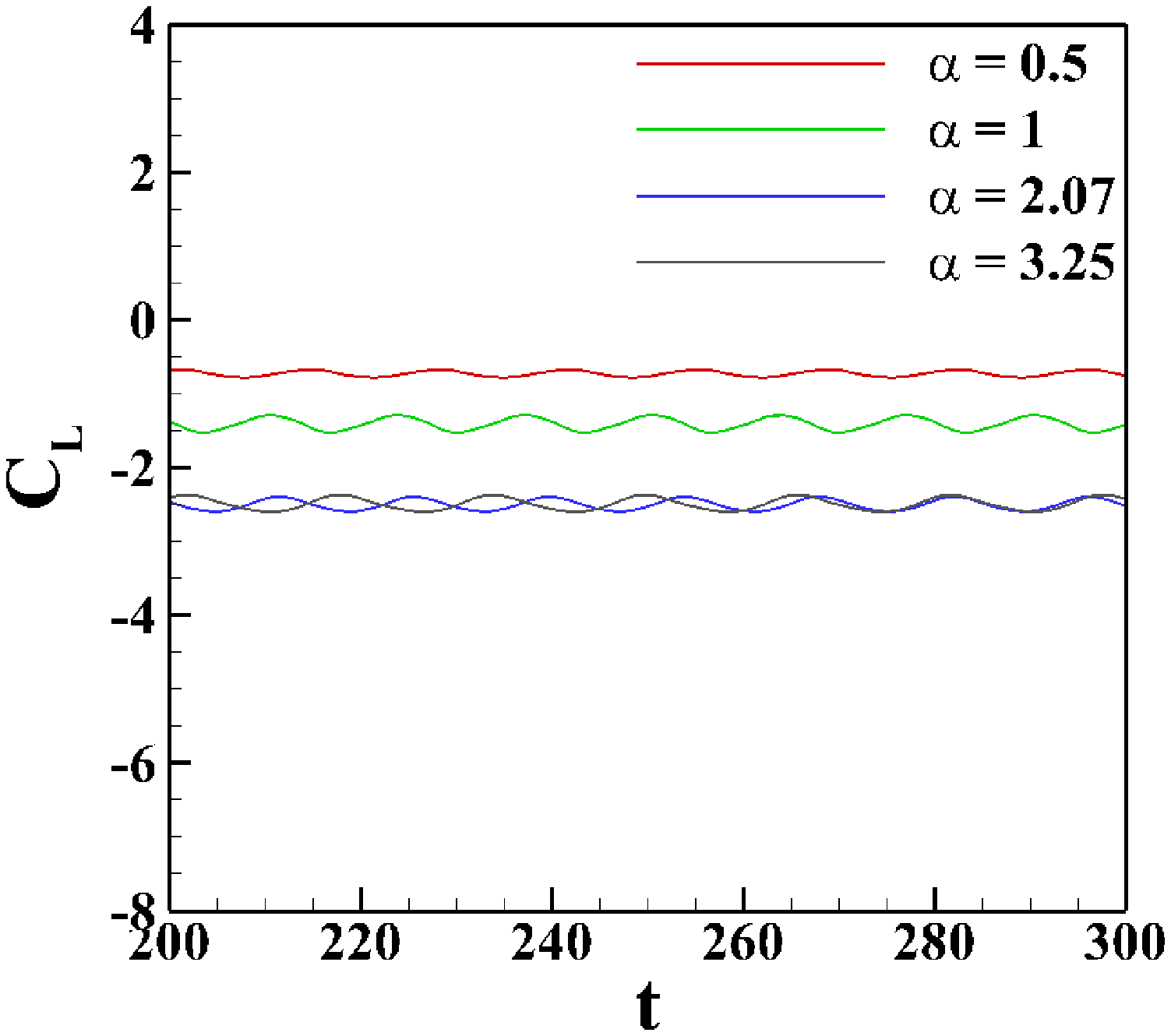}
\\
\hspace{2cm}(a) \hspace{4cm}(b)\hspace{2cm}
 \caption{(a) Drag coefficient $C_D$ and (b) lift coefficient $C_L$ with varying $\alpha$.}
 \label{fig:lift-drag_d}
\end{figure*}

The drag ($C_D$) and lift ($C_L$) coefficients at different $\alpha$ with varying $d/R_0$ are shown in \cref{fig:lift-drag_a}. The figures show that the drag and lift coefficients are periodic in nature. For $\alpha=0.5$, the drag coefficient gradually decreases with increasing $d/R_0$ and the lift coefficient is minimum at $d/R_0=0.5$. The differences in the drag coefficients for this $\alpha=0.5$ are very small. There is not much difference in lift coefficient for $d/R_0=1$, $2$, and $3$. When $\alpha=1$, gradual decrease in the drag coefficient is found with increasing $d/R_0$. The lift coefficient is found to be minimum for $d/R_0=0.5$. The maximum value of the lift coefficient is observed for $d/R_0=1$ and $2$. When $\alpha=2.07$, maximum value of the drag coefficient is found for $d/R_0=3$ and minimum value is found for $d/R_0=0.5$. Here, the maximum value of the lift coefficient is found for $d/R_0=1$ and the minimum value is found for $d/R_0=0.5$. When the rotation rate is at its maximum, i.e., $\alpha=3.25$, the amplitudes of the drag and lift coefficients increase drastically for all $d/R_0$. Here, the minimum values of lift and drag coefficients are found for $d/R_0=0.5$ and the minimum values are found $d/R_0=2$. For $d/R_0=3$, the amplitudes of the drag and lift coefficients are the smallest. So, the impact of various positionings of the arc-shaped control plate is significant at higher rotational rates. In \cref{fig:lift-drag_d}, the drag ($C_D$) and lift ($C_L$) coefficients at different $d/R_0$ with varying $\alpha$ are shown. At, $d/R_0=0.5$, the maximum value of the $C_D$ is found for $\alpha=1$ and the minimum value is found for $\alpha=3.25$. With increasing $\alpha$, the maximum value of $C_L$ gradually decreases while the amplitude of $C_L$ gradually increases. The highest amplitude of the drag and lift coefficients is observed for $\alpha=3.25$. When the plate distance is increased to $1$ and $2$, the maximum value of $C_D$ and the minimum value of $C_L$ are found for $\alpha=3.25$. Also, the amplitudes are maximum for the highest rotational rate. When $d/R_0=3$, $C_D$ gradually increases while $C_L$ gradually deceases as $\alpha$ increases. The lift coefficients suggest that the lock-on vortices are shed under all the considered rotational rates and distances of the control plate.\\

\begin{figure*}[!t]
\centering
\includegraphics[width=0.3\textwidth,trim={0.5cm 0.25cm 0.5cm 0.5cm},clip]{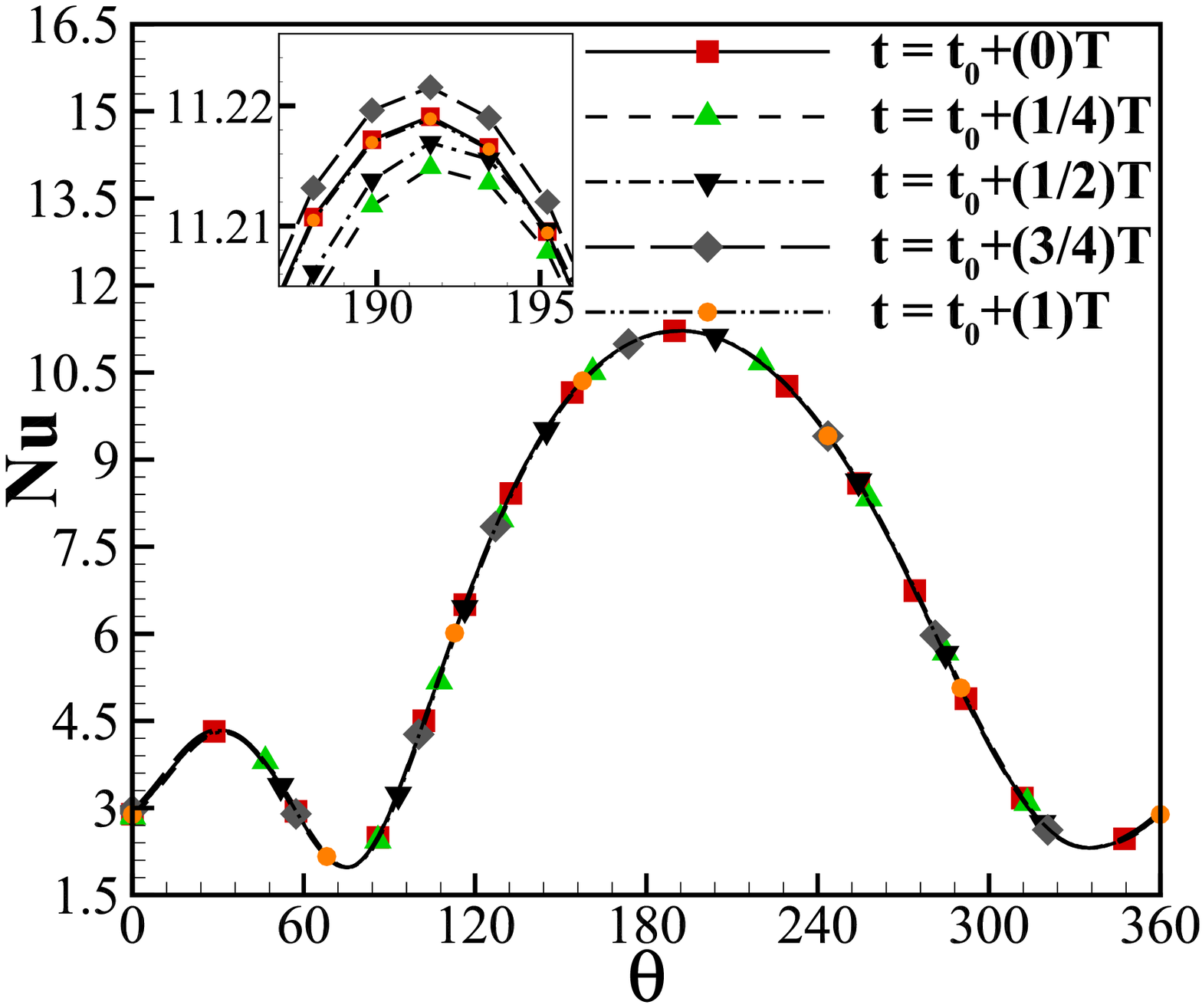}
\includegraphics[width=0.3\textwidth,trim={0.5cm 0.5cm 0.5cm 0.5cm},clip]{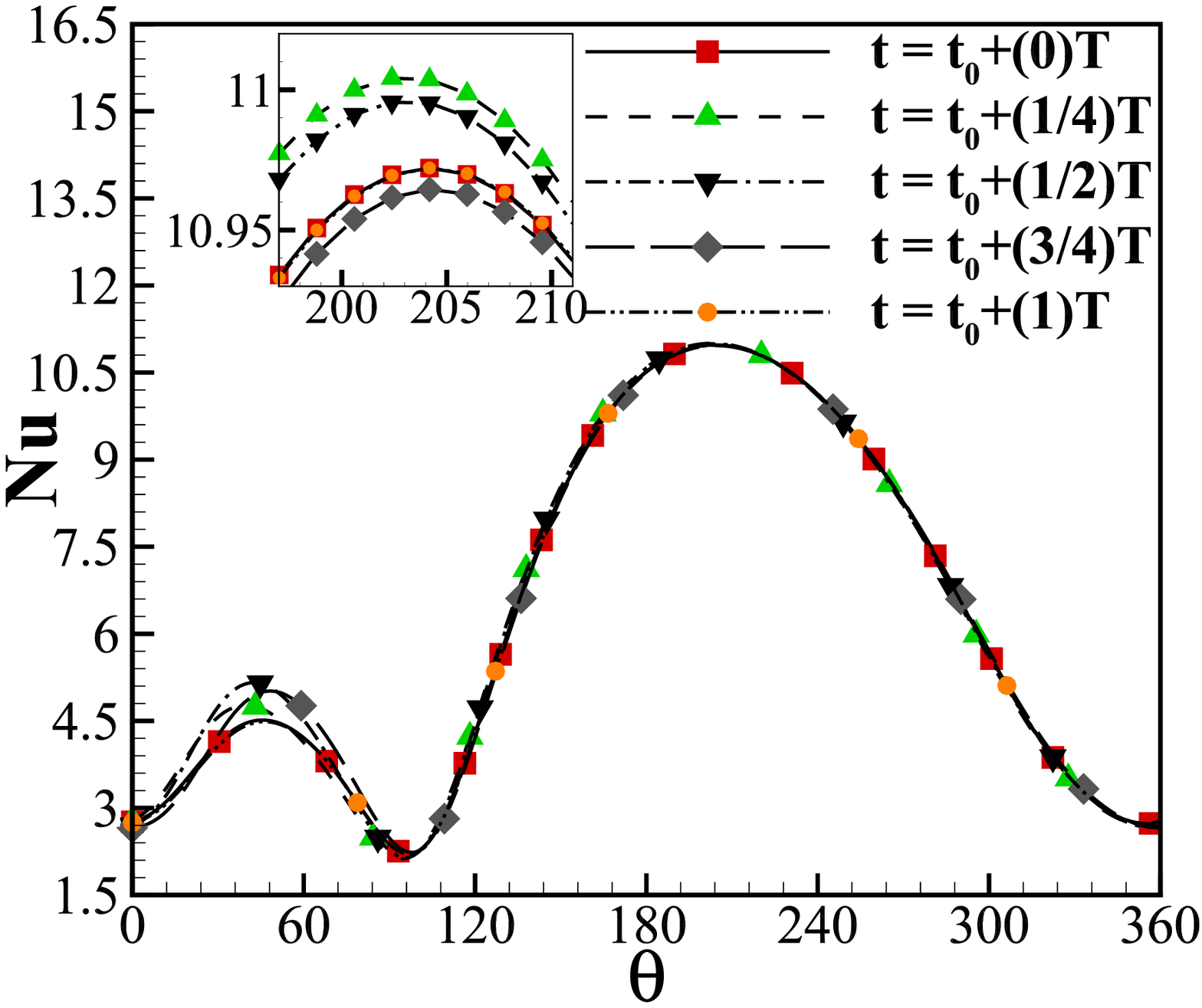}
\includegraphics[width=0.3\textwidth,trim={0.5cm 0.25cm 0.5cm 0.5cm},clip]{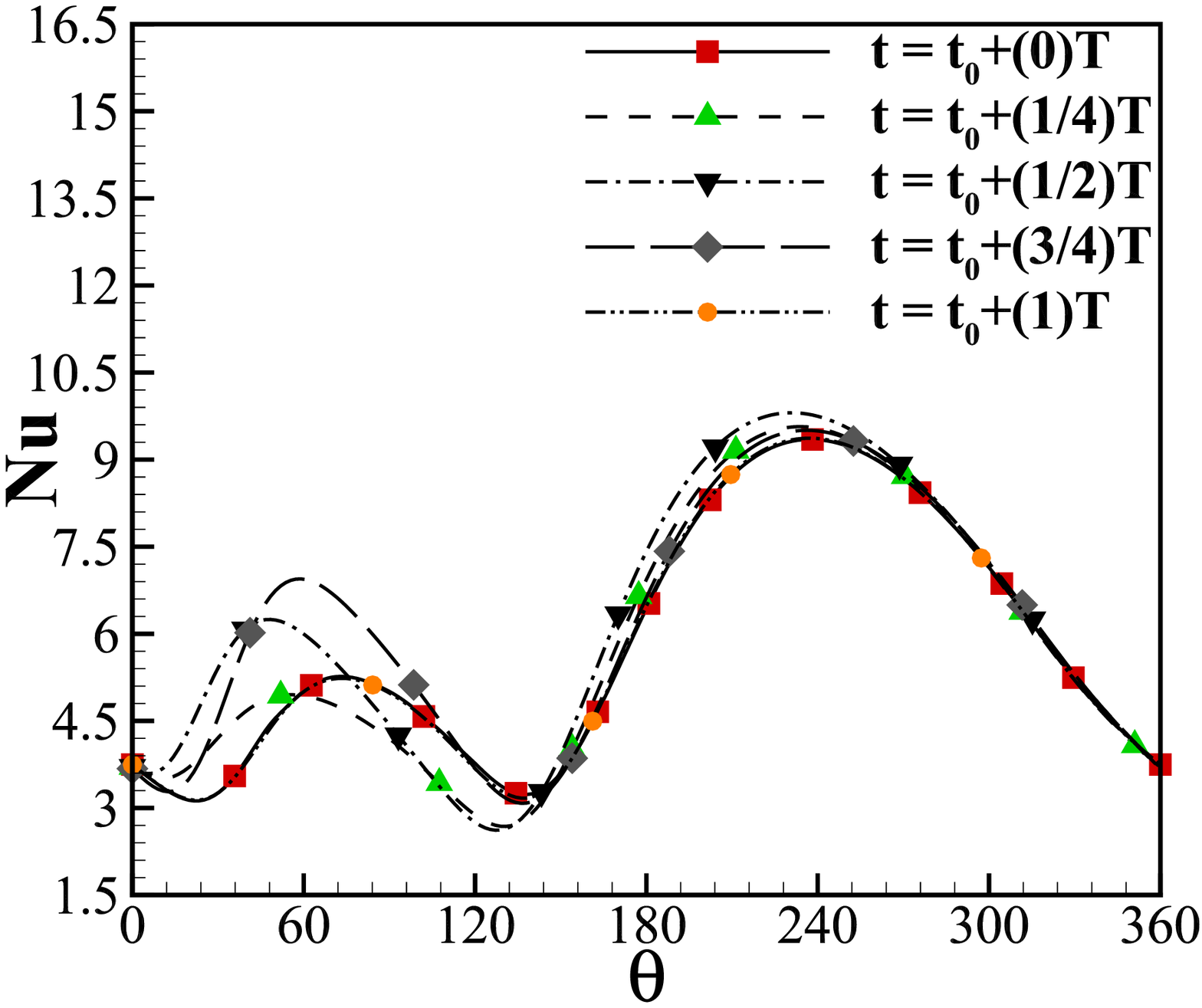}
\\
\hspace{1cm}(a)\hspace{4cm}(b)\hspace{4cm}(c)\hspace{4cm}
\\
\hspace{0.5em}
\\
\includegraphics[width=0.3\textwidth,trim={0.5cm 0.5cm 0.5cm 0.5cm},clip]{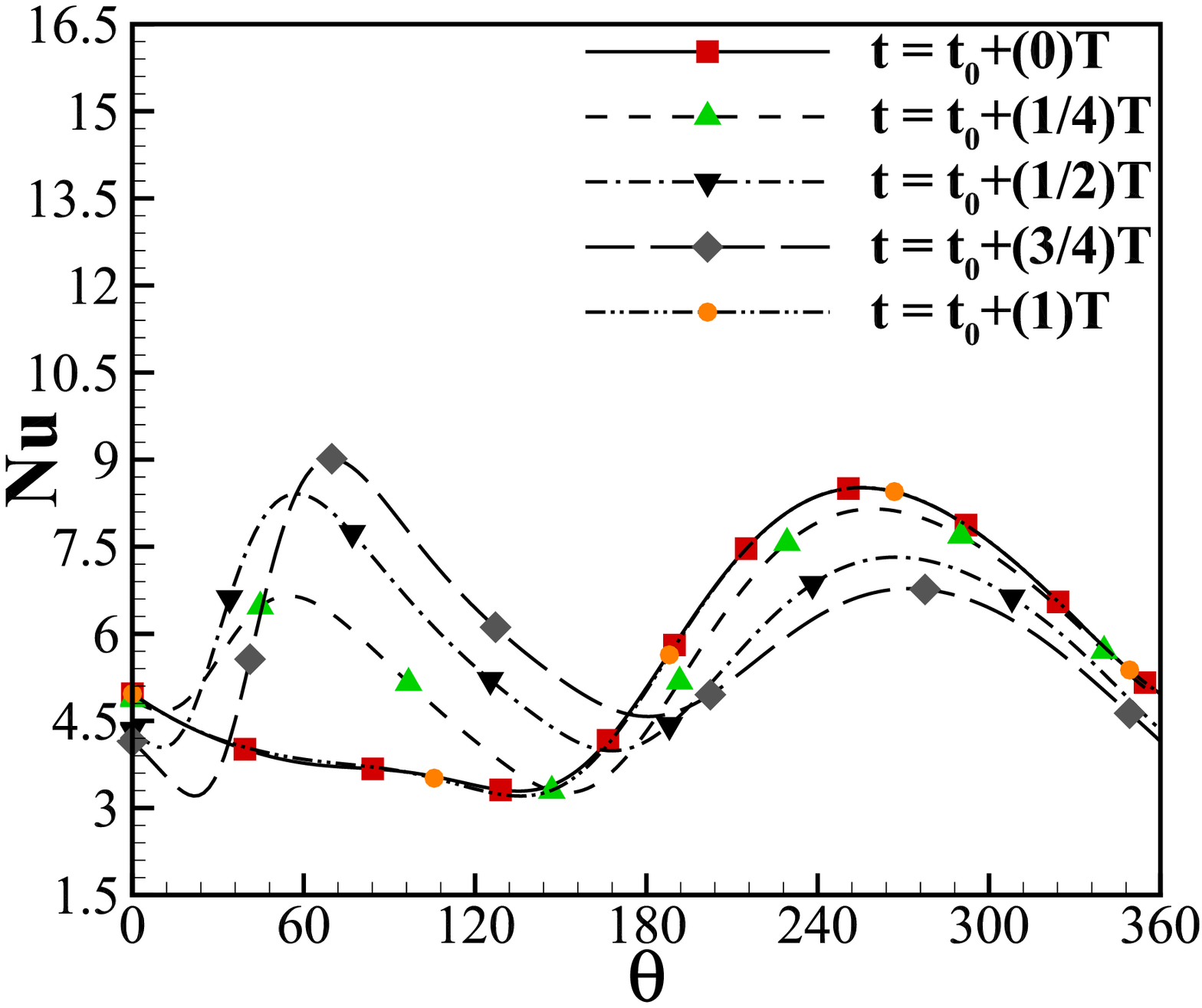}
\includegraphics[width=0.3\textwidth,trim={0.5cm 0.25cm 0.5cm 0.5cm},clip]{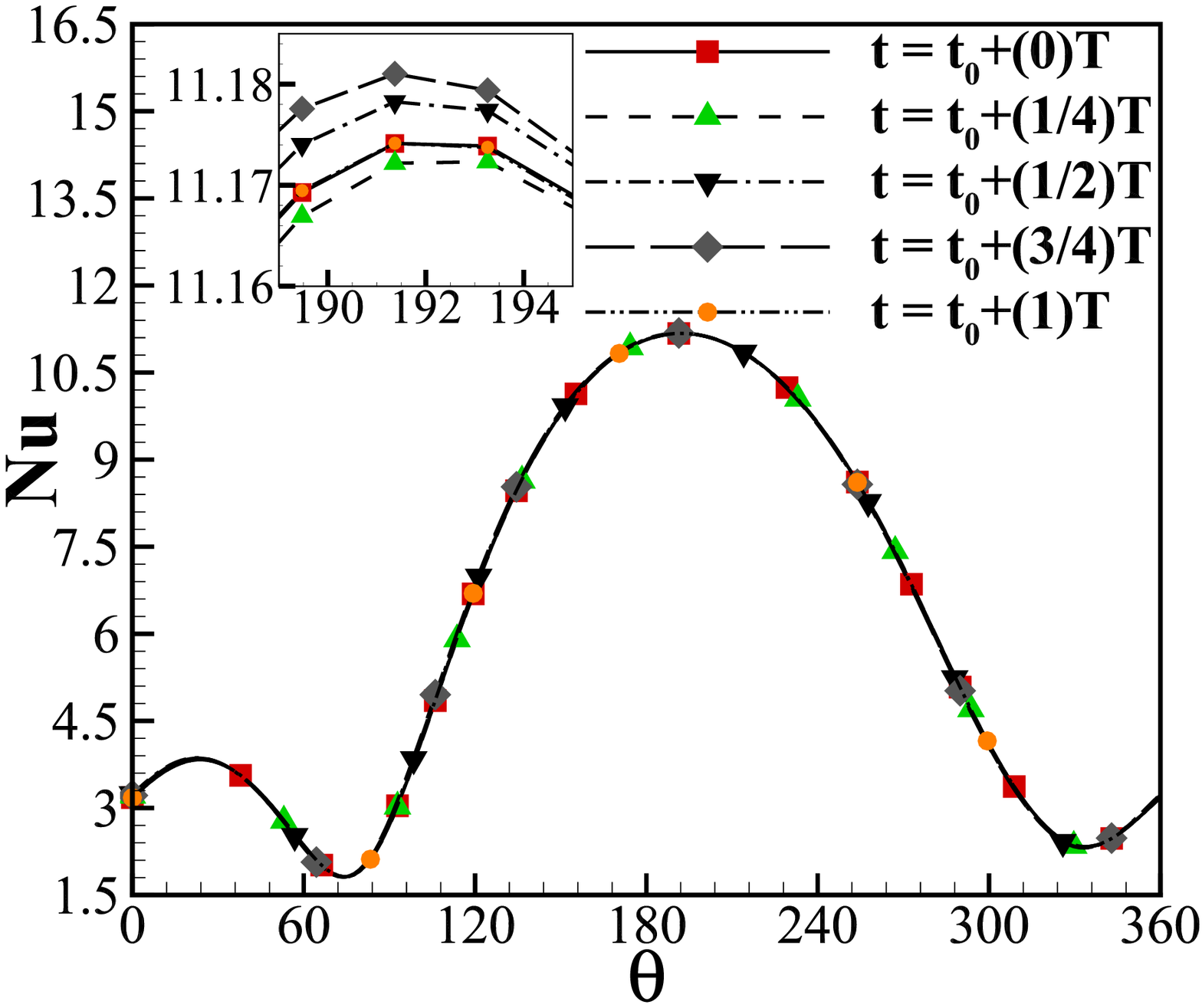}
\includegraphics[width=0.3\textwidth,trim={0.5cm 0.5cm 0.5cm 0.5cm},clip]{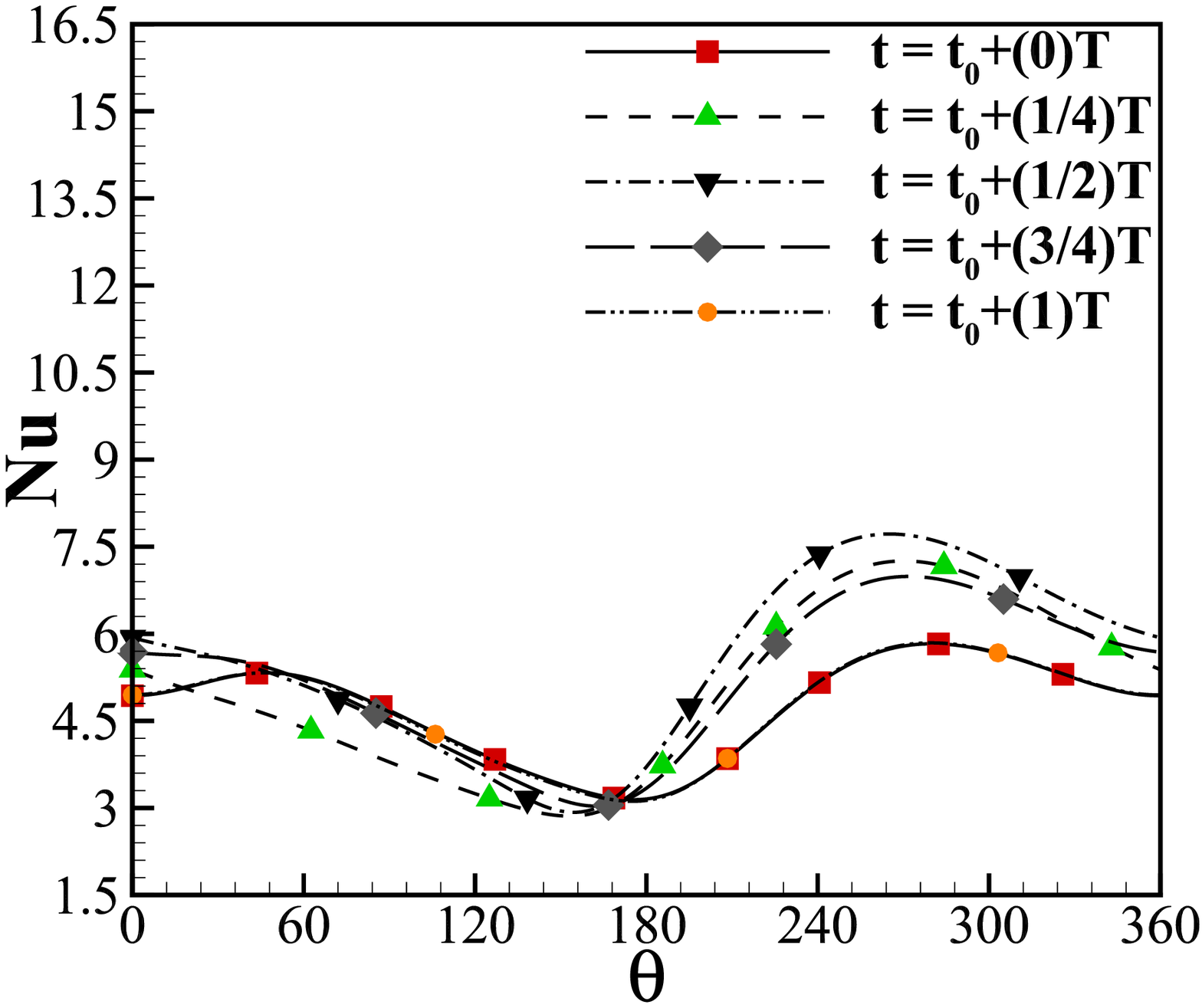}
\\
\hspace{1cm}(d)\hspace{4cm}(e)\hspace{4cm}(f)\hspace{4cm}
\\
\hspace{0.5em}
\\
\includegraphics[width=0.3\textwidth,trim={0.5cm 0.25cm 0.5cm 0.5cm},clip]{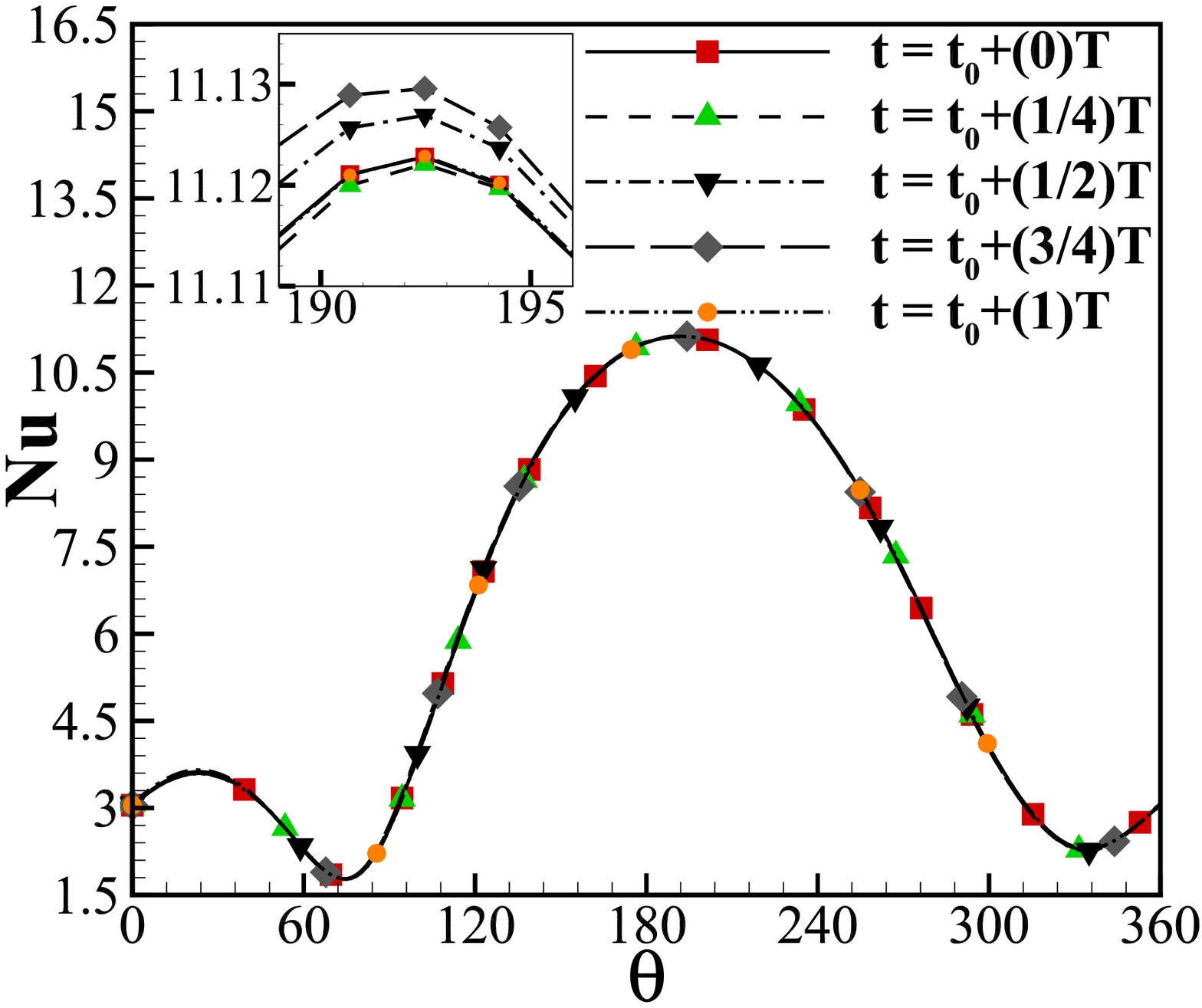}
\includegraphics[width=0.3\textwidth,trim={0.5cm 0.5cm 0.5cm 0.5cm},clip]{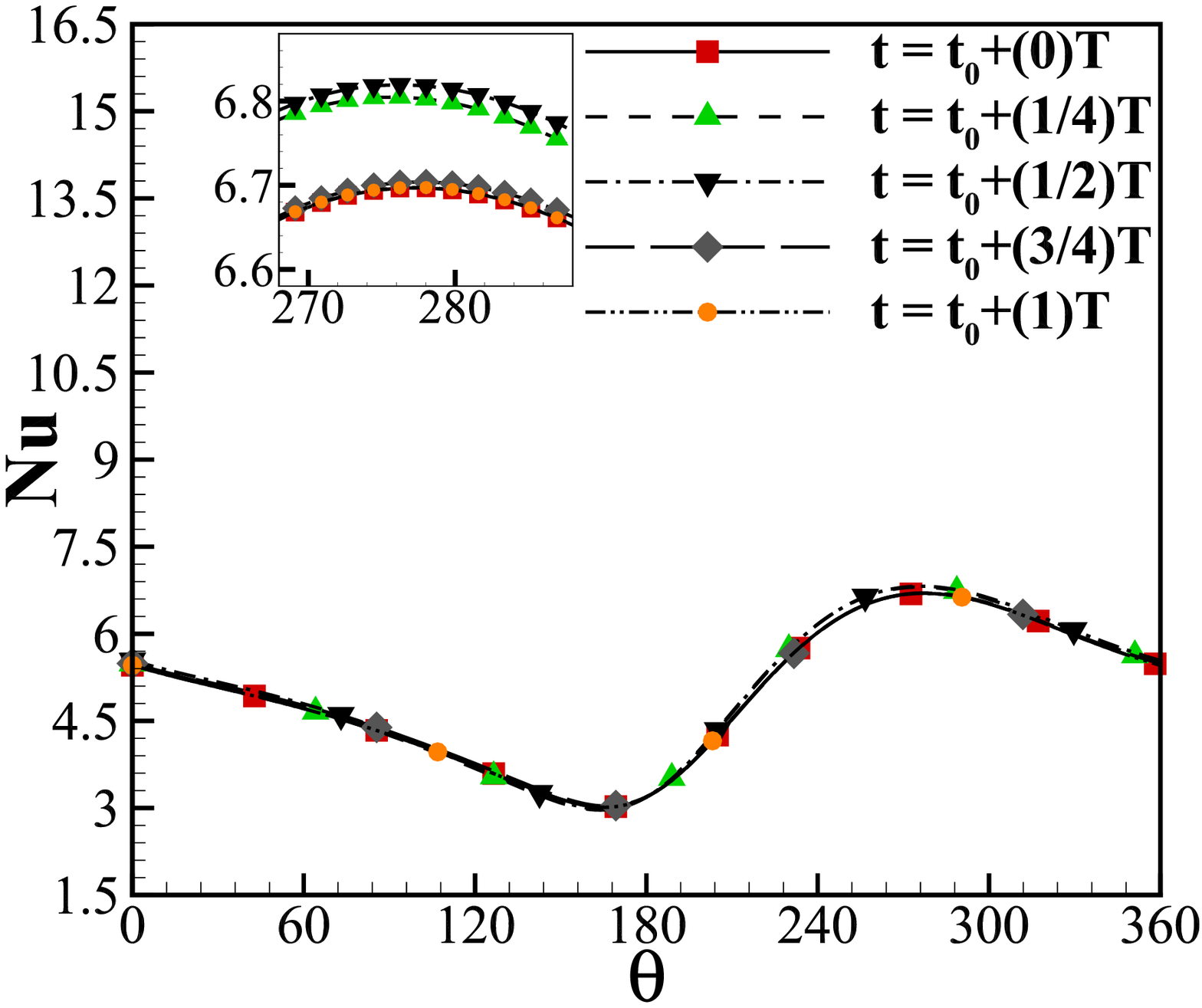}
\\
\hspace{2cm}(g) \hspace{4cm}(h)\hspace{2cm}
 \caption{Local Nusselt number variation at periodic phases for (a) $d/R_0=1$, $\alpha=0.5$; (b) $d/R_0=1$, $\alpha=1$; (c) $d/R_0=1$, $\alpha=2.07$; (d) $d/R_0=1$, $\alpha=3.25$; (e) $d/R_0=2$, $\alpha=0.5$; (f) $d/R_0=2$, $\alpha=3.25$; (g) $d/R_0=3$, $\alpha=0.5$; and (h) $d/R_0=3$, $\alpha=3.25$.}
 \label{fig:local_nusselt}
\end{figure*}

\cref{fig:local_nusselt} shows the variation of local Nusselt numbers at periodic phases for various rotational rates of the cylinder and different positioning of the plate. \cref{fig:local_nusselt}(a) shows the variation of $Nu$ for $d/R_0=1$ and $\alpha=0.5$. It can be seen that the maximum value of $Nu$ is slightly shifted downwards from the front stagnation point ($\theta=180\degree$) approximately to $\theta=192\degree$. It indicates the difference in heat transfer processes between the upper and lower half of the cylinder surface. The differences in values of $Nu$ between the periodic phases are very small. A local maximum peak of $Nu$ is found at $\theta\approx 30\degree$ which indicates the higher rate of heat convection in this area. This is supported by the concentrated isotherm contours in this area close to the cylinder surface shown in \cref{fig:d_1_a_0-5}. As the $\alpha$ increases to $2$ for $d/R_0=1$ in \cref{fig:local_nusselt}(b), the differences in values are increased at different phases. The highest point of $Nu$ is found around $\theta=204\degree$. It shows the difference in heat transfer mechanisms from the upper and lower surfaces. A local maximum peak of $Nu$ is found at $\theta\approx 42\degree$ indicating higher rate of heat convection in this area. It is also supported by the highly concentrated isotherm contours in \cref{fig:d_1_a_1-0}. When $\alpha=2.07$ for $d/R_0=1$, the maximum value of $Nu$ slightly decreases in \cref{fig:local_nusselt}(c) than that the previous cases and the maximum point of heat transfer is around $\theta=240\degree$. Local maximum peak is found to be changing position between $\theta\approx 42\degree$ and $\theta\approx 78\degree$ at different periodic phases due to the complex vortex shedding phenomenon. These areas convect a large amount of heat into the fluid. The asymmetric $Nu$-distribution around the front stagnation point shows that the heat transfer process from the upper part of the cylinder surface is far different from the heat transfer process from the lower part of the cylinder surface. \cref{fig:local_nusselt}(d) shows the variation of $Nu$ with maximum rotation rate, $\alpha=3.25$ for $d/R_0=1$ and the maximum value of $Nu$ drastically decreases and occurs at $\theta\approx 72\degree$ i.e. near the rear stagnation point. As many researchers previously mentioned, here too, large rotational rates significantly reduce the maximum heat transfer rate from the cylinder \cite{badr1985laminar,paramane2009numerical,sufyan2015free}. A local maximum peak is found at $\theta\approx 252\degree$. The reduction of the maximum peak value of $Nu$ at front stagnation point with increasing $\alpha$ hints to the fact that more heat is transferred under conduction in this area. This happens due to the thickening of the boundary layer around the cylinder surface at the high rotational rate. \cref{fig:local_nusselt}(e) shows the variation of $Nu$ with $\alpha=0.5$ and $d/R_0=2$. It shows that the maximum point of heat transfer is around $\theta=191\degree$. Also, the peak value is slightly lower than that of $d/R_0=1$ due to the vortex shedding process. A local maximum peak is found at $\theta\approx 24\degree$ i.e. the heat transfer is higher in this area. It is also supported by the respective dense isotherm contours. In \cref{fig:local_nusselt}(f), $\alpha$ is increased to $3.25$ for $d/R_0=2$ and it is found that the highest value of $Nu$ significantly reduced than the previous case. The highest value of $Nu$ is observed around $\theta=264\degree$ i.e. maximum heat transfer under convection occurs in this area. This is supported by the respective dense isotherm contour in this region close to the cylinder surface. A local maximum value of $Nu$ is found at $\theta\approx 60\degree$ at periodic phases $t=t_0+(0)T,\ t_0+(1)T$, i.e. the heat transfer is enhanced in this area under convection by the complex vortex shedding process. The $Nu$-distribution at $180\degree\leq\theta\leq 0\degree$ is significantly different than the $Nu$-distribution at $360\degree\leq\theta\leq 180\degree$. This demonstrates that the lower half of the cylinder surface convects more heat than the upper half. \cref{fig:local_nusselt}(g) shows the variation of $Nu$ for $\alpha=0.5$ and $d/R_0=3$. The highest value of $Nu$ is observed around $\theta=192\degree$. The maximum value is slightly lower than that of $d/R_0=1,\ 2$. A local maximum value of $Nu$ distribution is found at $\theta\approx24\degree$. The maximum heat transfer under convection occurs in these areas. The highest value of $Nu$-distribution curve is significantly reduced in \cref{fig:local_nusselt}(h) where $\alpha=3.25$ and $d/R_0=3$ as compared to \cref{fig:local_nusselt}(d) for $d/R_0=1$ and \cref{fig:local_nusselt}(f) for $d/R_0=2$. This indicates that the increasing distance of the control plate significantly reduced the heat transfer under convection for the fixed $\alpha$. The highest value of $Nu$ is shifted to $\theta\approx276\degree$ due to the complex vortex shedding. The lowest value of $Nu$-distribution curve at the front stagnation point indicates that a large amount of heat is transferred by conduction at this place. Also, the distribution curve at $180\degree\leq\theta\leq 0\degree$ is significantly different than the curve at $360\degree\leq\theta\leq 180\degree$, which shows that the lower half of the cylinder surface convects more heat than the upper half.\\

\begin{figure}[!t]
\centering
\includegraphics[width=0.45\textwidth,trim={0.2cm 0.25cm 0.5cm 0.5cm},clip]{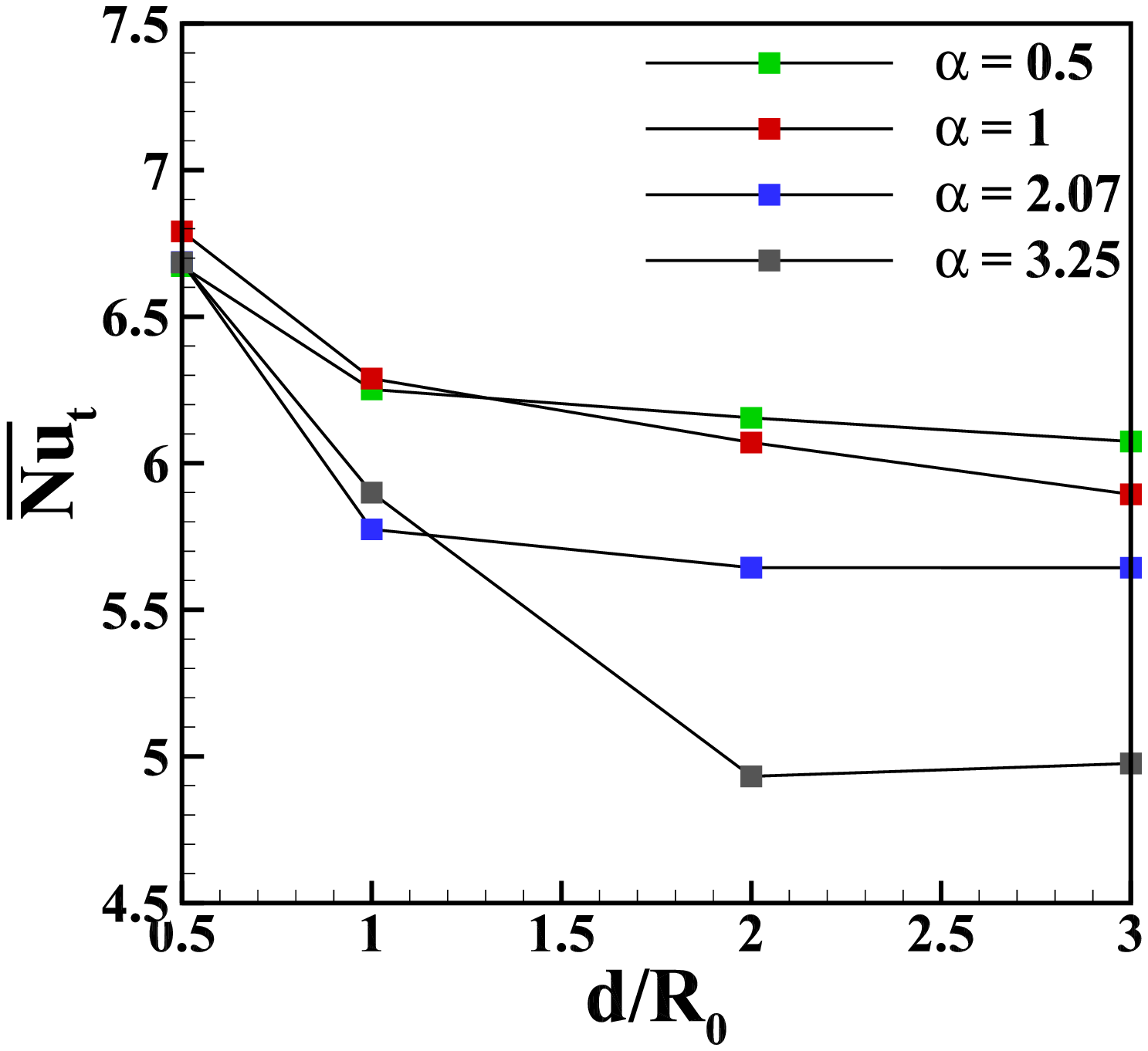}
\hspace{0.5em}%
\includegraphics[width=0.45\textwidth,trim={0.2cm 0.5cm 0.5cm 0.5cm},clip]{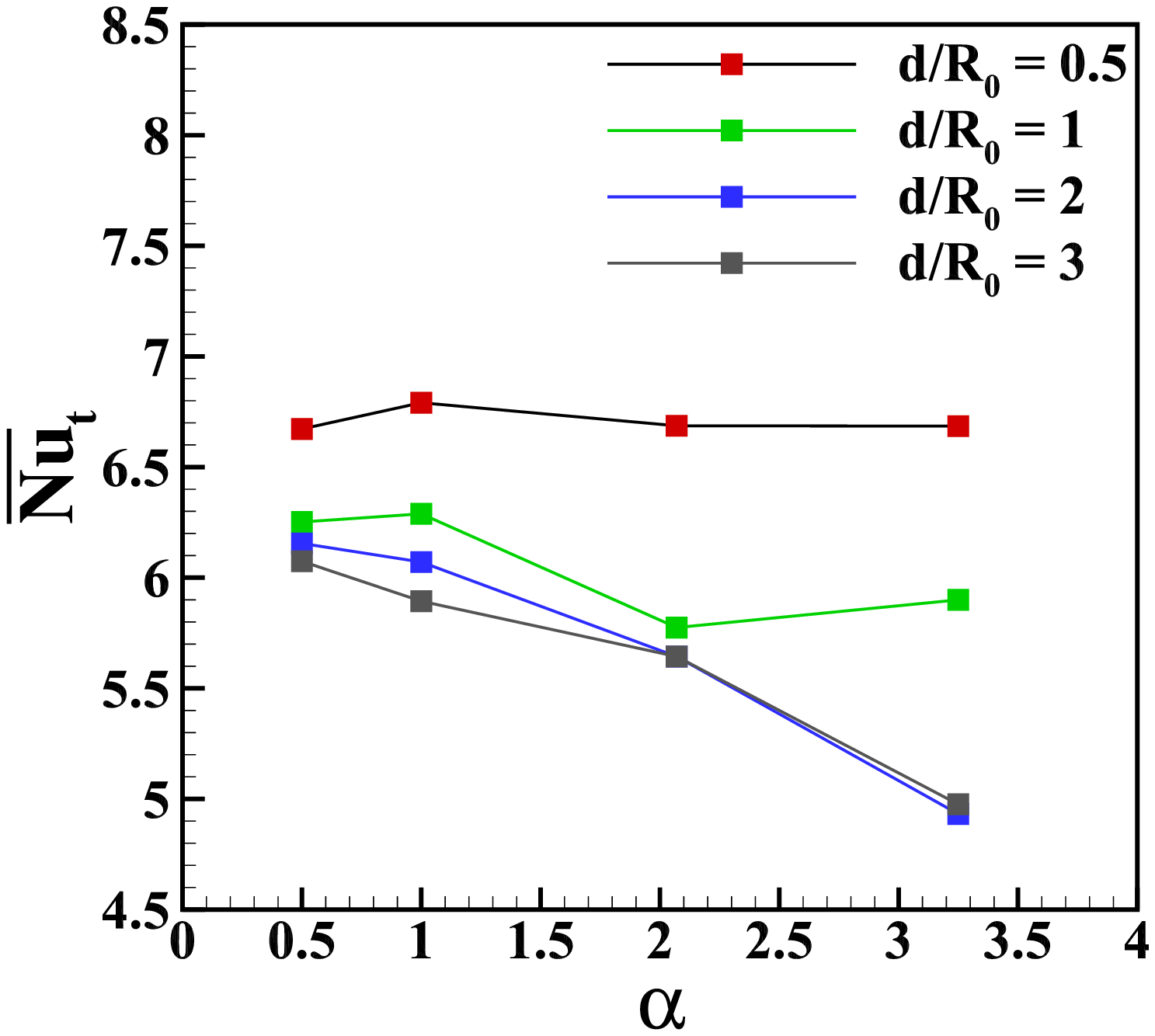}
\\
(a) \hspace{7cm}(b)
\caption{(a) $\overline{Nu}_t$ for varying $\alpha$, and (b) $\overline{Nu}_t$ for varying $d/R_0$.}
\label{fig:total_nu}
\end{figure}

\cref{fig:total_nu} exhibits the variation of time-averaged total Nusselt number ($\overline{Nu}_t$) with \cref{fig:total_nu}(a) varying $\alpha$ and \cref{fig:total_nu}(b) varying $d/R_0$. The values of $\overline{Nu}_t$ for $\alpha=0.5$ are $6.672265$, $6.251865$, $6.154835$ and $6.074185$ with $d/R_0=0.5$, $1$, $2$ and $3$ respectively. It means that the increasing distance of control plate reduces the heat transfer rate at $\alpha=0.5$. The values of $\overline{Nu}_t$ for $\alpha=1$ are $6.790804$, $6.28877$, $6.07076$ and $5.89388$ with $d/R_0=0.5$, $1$, $2$ and $3$ respectively. It means that the increasing distance of control plate also reduces the heat transfer rate at $\alpha=1$. The values of $\overline{Nu}_t$ for $\alpha=2.07$ are $6.686615$, $5.774501$, $5.6436$ and $5.643545$ with $d/R_0=0.5$, $1$, $2$ and $3$ respectively. Here also, the increasing distance of control plate reduces the heat transfer rate. The values of $\overline{Nu}_t$ for $\alpha=3.25$ are $6.68507$, $5.899766$, $4.931795$ and $4.9757$ with $d/R_0=0.5$, $1$, $2$ and $3$ respectively. Again the increasing distance of the control plate reduces the heat transfer rate except for $d/R_0=3$. This occurs due to the interaction of high rotation and the large distance of the control plate. \cref{fig:total_nu}(a) shows that $\overline{Nu}_t$ gradually deceases with increasing $\alpha$ at $d/R_0=2,\ 3$ and the maximum value of $\overline{Nu}_t$ is found for $d/R_0=0.5$, $\alpha=0.5$. \cref{fig:total_nu}(b) shows that increasing $d/R_0$ significantly reduces $\overline{Nu}_t$ within the range of $0.5\leq d/R_0 \leq 2$ for all rotational rates. However, if we place the plate further at a distance $d/R_0=3$, not much change occurs. It is found from the comparison of maximum and minimum values of $\overline{Nu}_t$ that certain positioning of the control plate and rotational rate can enhance the heat transfer rate by $37.69\%$.


\section{Conclusion\protect}\label{Conclusion}
We numerically examined the control of a uniform, viscous fluid flow past circular cylinder by an arc-shaped plate positioned in the normal direction behind an isothermally heated circular cylinder rotating in the cross stream. The governing equations are discretized using a HOC finite difference technique, and the system of algebraic equations obtained by the HOC discretization is solved using the Bi-conjugate gradient stabilised iterative method. According to the research, the distance between the control plate and the cylinder surface has a considerable impact on fluid flow along with the rotation of the cylinder. The structure of the wake changes depending on the position of the plate. When $\alpha$ is less than $1$ with $d/R_0=1$, two vortices as lumps of hot fluid are shed periodically from either side of the cylinder; when $\alpha$ is greater than $2.07$ with $d/R_0=1$, a large negative vortex of heated fluid is shed from the upper side of the cylinder and another positive vortex of hot fluid is shed behind the control plate on a periodic basis. The increasing rotational rates increase the size of vortices and decrease the wake length for all positions of the control plate. The vortex shedding plane is shifted from the centerline by the cylinder's rotational motion. For all rotational rates, the increased distance of the control plate decreases the angle of the vortex shedding plane with the centerline, but the angle is increased with increasing rotational rates for all positions of the control plate. At higher rotational rates, the positive vortex is pulled upwards due to the interaction of fluid, and it pushes the negative vortex, causing an early shedding of it. Placing the control plate at $d/R_0=3$ along with a high rotational rate is found to significantly reduce the size of vortices. It is also found that the impact of various positionings of the arc-shaped control plate is significant at higher rotational rates. An additional recirculation zone is found for ($d/R_0=2,\ \alpha=0.5,\ 3.25$) and ($d/R_0=3,\ \alpha=3.25$). Drag and lift coefficients for all $0.5\leq d/R_0\leq3$ and $0.5\leq\alpha\leq3.25$ have a periodic nature . The values of drag and lift coefficients can be reduced or increased by utilising the rotation of the cylinder and the placement of the plate. The maximum value of drag coefficient is achieved for $d/R_0=2$ and $\alpha=3.25$ which is about $3$. All vortices shed are locked-on under the scope of considered parameters. It is found that the rotational rates relocate the highest point of heat transfer further from the front stagnation point, i.e., increasing the heat transfer by conduction in this region. The increasing distance of the control plate significantly reduced the heat transfer under convection for the fixed $\alpha$. The combined effect of rotation and the positioning of the control plate causes a different heat transfer mechanism at the upper half of the cylinder surface than at the lower half. For fixed $d/R_0=2$ and $\alpha=3.25$, the maximum point of heat transfer is shifted towards the rear stagnation point from the front stagnation point due to the complex vortex shedding.

\section*{Author Declarations}
The authors have no conflicts to disclose.

\section*{Data Availability Statement}
The data that support the findings of this study are available from the corresponding author upon reasonable request.

\bibliographystyle{IEEEtran}
 \bibliography{biblio_modified}





\end{document}